\documentclass[a4paper, 11pt, twoside, openright]{memoir}

\usepackage{template_preamble}

\usepackage{etoolbox}
\makeatletter
\newcommand{\extraPartText}[1]{\def\@extraPartText{#1}}
\pretocmd{\@endpart}{\vspace{8ex}\begingroup\@extraPartText\par\endgroup\let\@extraPartText\relax}{}{}
\makeatother

\usepackage[top = 4.2cm, bottom = 4.5cm, inner = 4.5cm, outer = 3.7cm]{geometry}

\colorlet{mdtRed}{red!50!black}

\usepackage{bm}
\usepackage{bbold}

\begin{document}

\thispagestyle{titlepage}
\newgeometry{margin=0.5cm}

\begin{center}
  \includegraphics[width=0.25\linewidth]{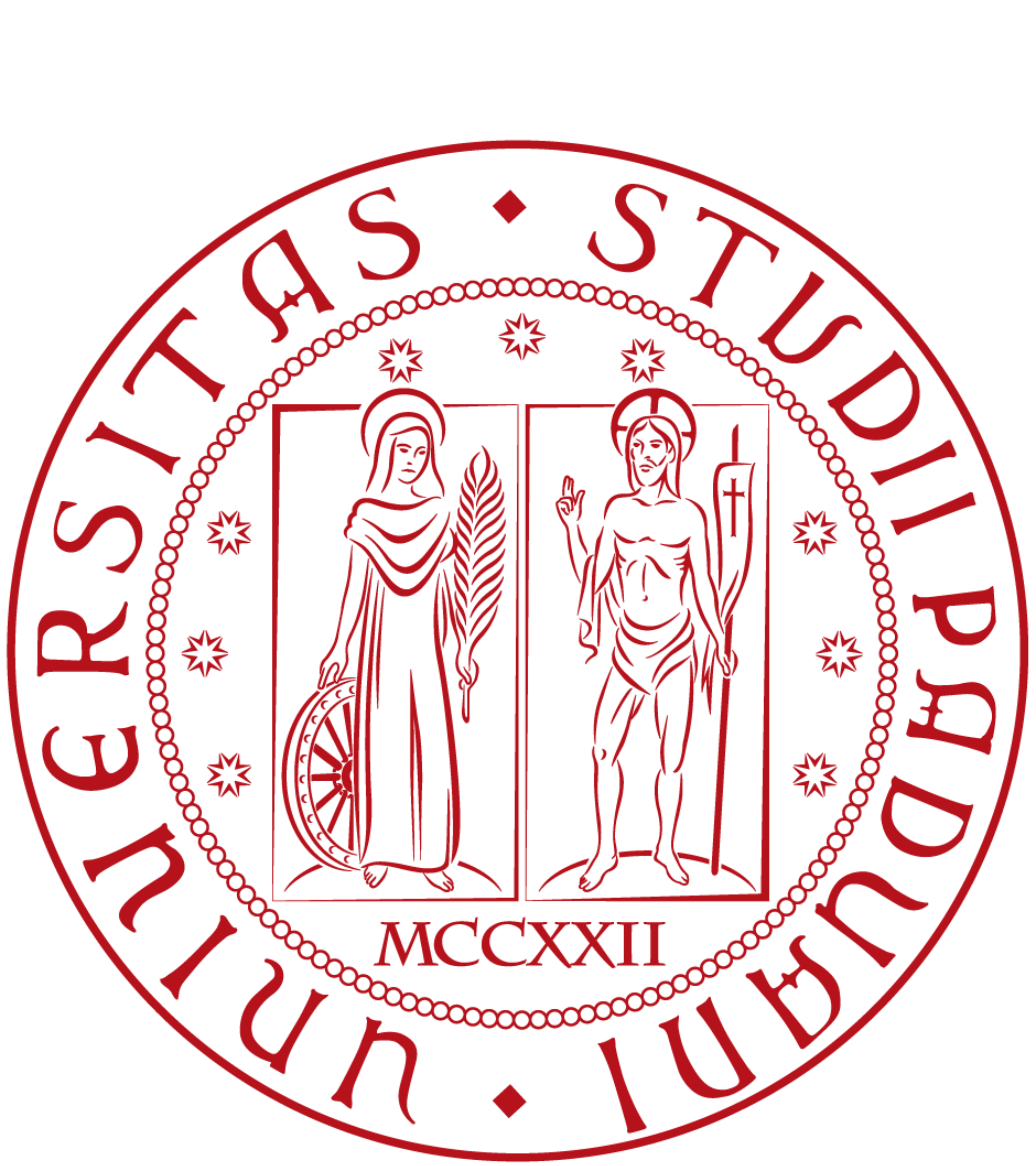}\\
  \vspace{1cm}
  \fontsize{20pt}{18pt} \textsc{University of Padova}\\
  \vspace{1.5cm}
  \fontsize{16pt}{14pt} \normalfont{Department of Physics and Astronomy ``Galileo Galilei''}\\
  \vspace{1.5cm}
  \fontsize{16pt}{14pt} \normalfont{Ph.D. course in} \textsc{Physics}\\
  \vspace{0.5cm}
  \fontsize{16pt}{14pt} \textsc{XXXV cycle}\\
  \vspace{1cm}
  \fontsize{16pt}{14pt} \textsc{Doctoral Thesis}\\
  \vspace{1cm}
  {\fontsize{30pt}{30pt}\selectfont\titleColor\textsc{Information and Criticality}}\\
  \vspace{0.3cm}
  {\fontsize{30pt}{30pt}\selectfont\titleColor\textsc{in Complex Stochastic Systems}}\\

  \vspace{2cm}  \begin{minipage}[t]{0.10\textwidth}
    \hfill
  \end{minipage}~
  \begin{minipage}[t]{0.40\textwidth}
    \textbf{\textsc{Ph.D. candidate:}}\\

    \textsc{\titleColor Giorgio Nicoletti}\\

    \vspace{0.5cm}
    \textbf{\textsc{Coordinator:}}\\

    prof. Giulio Monaco
  \end{minipage}~
  \begin{minipage}[t]{0.40\textwidth}
    \raggedleft
    \textbf{\textsc{Supervisors:}}\\
    
    \vspace*{0.5cm}
    prof. Amos Maritan\\
    \vspace*{0.5cm}
    prof. Samir Suweis
  \end{minipage}~
  \begin{minipage}[t]{0.10\textwidth}
    \hfill
  \end{minipage}\\
\vspace{2cm}
\fontsize{16pt}{14pt} \normalfont{Academic year 2021/2022}
\end{center}
\restoregeometry

\newpage
\thispagestyle{empty}
\mbox{}

\newpage
\thispagestyle{empty}
\begin{flushright}
\vspace*{\fill}
\emph{More is different.}\\
\textsc{- Philip W. Anderson}
\end{flushright}
\vspace{3cm}
\begin{flushright}
\emph{The full appreciation of natural phenomena must go beyond physics in the usual \\ sense [...] because the separation of fields is merely a human convenience.\\ Nature is not interested in our separations, and many of the interesting phenomena bridge the gaps between fields.}\\
\textsc{- Richard Feynman}
\end{flushright}
\vspace{3cm}
\begin{flushright}
\emph{La creatività è soprattutto la capacità di porsi continuamente delle domande.}\\
\textsc{- Piero Angela}
\vspace*{\fill}
\end{flushright}

\newpage
\thispagestyle{empty}


\frontmatter

\pagenumbering{gobble}
\thispagestyle{empty}
\chapter*{Abstract}
\lettrine{I}{n this Thesis}, we will explore how tools from Statistical Physics and Information Theory can help us describe and understand complex systems. We will often deal with minimal and paradigmatic models characterized by stochastic processes, employing both analytical techniques and numerical simulations.

In the first part of this Thesis, we will focus on the interplay between internal interactions and environmental changes and how they shape the properties of the many degrees of freedom of a complex system. We model the environment as an independent stochastic process that affects the parameters of the system, a fruitful paradigm in several fields - such as changing carrying capacities in ecosystems, switching environments in chemical systems, mutating strategies in microbial communities, different regimes of neural activity, and diffusion in disordered or inhomogeneous media. At any time, the environmental state is not known, so we are interested in the joint distribution of the internal degrees of freedom marginalized over the environmental ones. By computing its mutual information in different scenarios, we explicitly describe the internal and effective dependencies arising in the system. We study the properties of environmental information in different scenarios and how mutual information encodes internal and environmental processes, as well as how it can help us disentangle them. Yet, often we are not able to observe and describe all the internal dependencies of a complex system, and typically we resort to effective representations. By building information-preserving projections and focusing on the paradigmatic case of underdamped systems, we show these optimal effective representations may be unexpectedly singular and undergo abrupt changes. These results help us understand the fundamental limits of approximating complex models with simpler effective ones while attempting to preserve their dependencies, that may arise from unknown environmental changes.

In the second part of this Thesis, we leverage this approach and ideas from criticality and apply them to neural systems at different scales. We build on the hypothesis that brain dynamics may be poised near a critical point, finding seemingly scale-free correlations and power-law distributed neuronal avalanches in data from the somatosensory barrel cortex of rats. However, we show that an environmental-like latent variable that models unobserved neural activity may lead to power-law neuronal avalanches in the absence of criticality. Remarkably, the properties of mutual information suggest that, whereas avalanches may emerge from an external stochastic modulation, interactions between neural populations are the fundamental biological mechanism that gives rise to seemingly scale-free correlations. We further explore the role of the structure of interactions at a different but relevant scale - whole-brain dynamics. We develop a stochastic continuous-time formulation of a well-known cellular automaton, showing how the mean-field limit predicts a bistability region. Yet, a continuous transition appears with an interaction network, with localized oscillations that provide a dynamical mechanism for the emergence of clusters of activity known to generate functional networks. Our results shed light on the role of the interaction network, its topological features, and unobserved modulation in the emergence of collective patterns of brain activity.

Finally, these ideas will lead us to a phenomenological coarse-graining procedure for neural timeseries. We test it in equilibrium and non-equilibrium models, as well as models of conditionally independent variables in a stochastic environment - which, as we have shown throughout this Thesis, often display non-trivial and unexpected features. In doing so, we further test the fascinating hypothesis that the Statistical Physics of phase transition may serve as a powerful and universal framework for understanding biological and living systems.

\thispagestyle{empty}
\chapter*{Acknowledgments}
\lettrine{T}{here are many people} without whom I would have not gotten this far. First and foremost, I wish to thank my family for making this possible in the first place. To my mum and dad: your love, your support, and what you taught me throughout my life are the reasons I was able to get where I am today and pursue a scientific career. I could not have asked for anything better. To my brother, my grandmother, my aunt, and my uncle: thank you for the never-ending encouragement, the never-ending love, and the life you make me live every day. To Ilaria: I cannot wait to see what lies ahead, and to share it with you. It is perhaps too easy to take all of this for granted, but words cannot express how grateful I am for the family I have. 

Of course, I owe my deepest gratitude to Amos and Samir for guiding me through this journey. Thank you for your constant support and encouragement, and for what you taught me during these years. Working with you has been an invaluable experience, both from a scientific and a human perspective. I will never forget what I have learned from you. A special thank goes to Daniel, for working and thinking together, for trying to see ``information'' everywhere, and for teaching me so many things. But also, in no particular order: to Prajwal, Clelia, Emanuele, Benedetta, Giacomo, Elisa, Jacopo, Fabio, Davide, to Stefano and Leonardo, to Sandro, and to all the members of the LIPh group, past and present - thank you! This would not have been such an amazing and memorable journey without you. We had a lot of fun, and I will cherish all the time we spent together.

Finally, I could not forget all my friends who, one way or another, made these past three years - and many, many years before them - as wonderful as they have been. Even if our lives may lead us to distant places, the memories we made together and those we will make will always stick around. I have been lucky to meet you.

\vspace*{0.2cm}
\noindent PS: to Gea, I am sorry that you have not been able to diligently watch me at my desk while I was writing this thesis. I promise that I will carefully explain its contents to you very soon, and perhaps to Giotto too.

\cleardoublepage

\newgeometry{top = 4.2cm, bottom = 4.5cm, inner = 4.5cm, outer = 3.7cm}
\pagenumbering{roman}
\begin{KeepFromToc}
 \tableofcontents
\end{KeepFromToc}
\restoregeometry

\chapter*{Preface}
\addcontentsline{toc}{chapter}{Preface}\markboth{Preface}{}
\lettrine{I}{n the first part of this Thesis}, we will focus on the interplay between internal interactions and environmental changes and how they shape the properties of complex systems. In Chapter \ref{ch:PRL_PRE_1} and Chapter \ref{ch:PRL_PRE_2}, we consider models where an independent stochastic process - the environment - changes the parameters of the system in time. This approach has several applications and can describe changing carrying capacities in ecosystems, switching environments in chemical systems, mutating strategies in microbial communities, different regimes of neural activity, and diffusion in disordered or inhomogeneous media. We assume that, at any time, the environmental state is not known, and that we are interested in the properties of the joint distribution of the internal degrees of freedom marginalized over the environmental ones. Even from a data-driven perspective, this is a relevant scenario - as a growing wealth of data is populating the realm of biological, chemical, and neural systems, direct extrapolation of some properties belonging to the underlying dynamics has become prominent, and one might ask whether these reconstructed features may arise from nothing but our ignorance about the unobserved environment in which the system lives. In Chapter \ref{ch:PRL_PRE_1}, we compute the mutual information between the internal degrees of freedom in the absence of interactions. By doing so, we explicitly describe the effective dependencies arising from the shared environment in different scenarios, from discrete environments to continuous ones, in equilibrium and non-equilibrium settings. We find that a stochastic environment in suitable limits may give rise to non-trivial dependencies in the system, associated with a corresponding environmental information. 

The description of these emerging dependencies is useful in Chapter \ref{ch:PRL_PRE_2} to understand the complex interplay between internal interactions and changing environments. In the case of linearized interactions, i.e., of systems close to a potential minimum, we find that the mutual information of the whole system encodes the internal and environmental processes as distinct contributions. Furthermore, such contributions can always be fully disentangled in suitable limits. Non-linear interactions, on the other hand, lead to a much richer structure, giving rise to a new term that models the interference between the dependencies induced by internal interactions and those stemming from the environment. This interference can be either constructive - with the dependency between the particles amplified by the presence of both interactions and environment at once - or destructive, depending on the shape of the internal interactions. Remarkably, although we focused on paradigmatic but rather comprehensive physical models, these ideas have a much larger scope. In particular, in machine learning and artificial neural network, disentangled representations of the data and generative models with latent variables are widely used.

These are archetypal examples of the fact that we may not be able to observe the many degrees of freedom of a complex system. Indeed, we may not have experimental access to all of the degrees of freedom or it may be more useful to study the behavior of coarse-grained variables. Thus, we rather seek an effective representation of the observed evolution. To this end, in Chapter \ref{ch:PRE}, we build information-preserving projections of a possibly unknown complex dynamics. By focusing on the paradigmatic case of underdamped systems, where we are interested in the position subspace rather than the whole phase space, we show that optimal effective models may have a singular optimal parameter space. Crucially, this discontinuity in the optimal parameter space is triggered by the minimization of the information loss which induces abrupt changes in the effective model. Our results pose fundamental challenges to the ambition of inferring underlying parameters from effective low-dimensional models, as the appearance of this transition in paradigmatic systems translates into an alarming warning signal for more general cases. Furthermore, they help us understand the fundamental limits of approximating complex models with simpler effective ones while attempting to preserve their dependencies.

In the second part of this Thesis,  we leverage these results and ideas from phase transitions and criticality applied to biological and in particular neural systems. In Chapter \ref{ch:scirep}, we analyze local field potentials from the somatosensory barrel cortex of rats. We find seemingly scale-free correlations and power-law distributed neuronal avalanches, building on the hypothesis that brain dynamics may be poised near a critical point. To describe these features, we consider a general class of stochastic processes describing an archetypal evolution of neural activity driven by another, but unobserved, external process. We show how the properties of this environmental-like latent variable are crucial in producing power-law neuronal avalanches. Further, we exploit the previously derived properties of the mutual information to study the interplay between internal and extrinsic activity, allowing us to understand how the underlying dependencies shape the observed properties of neural activity. In particular, our work suggests that, whereas avalanches may emerge from an external stochastic modulation that affects all degrees of freedom in the same way, interactions between neural populations are the fundamental biological mechanism that gives rise to seemingly scale-free correlations.

Hence, we further explore the role of the structure of interactions in Chapter \ref{ch:jphys}. We turn our attention to a different but relevant scale, and in particular to models for whole-brain dynamics. We develop a stochastic continuous-time formulation of a well-known cellular automaton via a master equation approach. We show analytically how the mean-field limit predicts a discontinuous transition with a bistable region in the control parameter space. Yet, when we go beyond mean-field by adding interaction networks connecting different brain regions, the picture is drastically different. The transition becomes continuous and localized oscillations emerge, providing a dynamical mechanism able to sustain clusters of activity that are known to generate functional networks. Our results shed light on the role of the underlying network structure in the emergence of collective patterns of brain activity, highlighting that critical-like transitions may be also driven by topological features of functional connections at the whole-brain level.

Finally, in Chapter \ref{ch:PRR} ideas from phase transitions and criticality will lead us to a model-free and phenomenological coarse-graining procedure recently introduced to deal with neural timeseries. We test concepts from this phenomenological Renormalization Group in well-known equilibrium and non-equilibrium models, studying different interaction topologies and whether they affect the coarse-graining outcomes. We also consider models of conditionally independent variables in a stochastic environment - which, as we have shown throughout this Thesis, often display non-trivial and unexpected features. Overall, we highlight the efficacy and limitations of these approaches.

Building from these results, future perspectives are manifold - from studying how biological and living systems harvest and process information, to how phase transitions and criticality may help them achieve optimal solutions and strategies in this direction. And all these ideas have a common framework behind them - Statistical Physics.

\chapter*{Publications}
\addcontentsline{toc}{chapter}{Publications}\markboth{Publications}{}

\lettrine{T}{he} following scientific articles that I have co-authored have been published by a peer-reviewed journal. The list is in chronological order.\\

\begin{itemize}
  \item[{\large $\blacktriangleright$}] \textbf{Giorgio Nicoletti}, Samir Suweis, Amos Maritan. \textit{Scaling and criticality in a phenomenological renormalization group}. \textcolor{mdtRed}{\textsc{Physical Review Research}} 2, 023144 (2020).\\

  \item[{\large $\blacktriangleright$}] Benedetta Mariani, \textbf{Giorgio Nicoletti}, Marta Bisio, Marta Maschietto, Oboe Roberto, Alessandro Leparulo, Samir Suweis, Stefano Vassanelli. \textit{Neuronal avalanches across the rat somatosensory barrel cortex and the effect of single whisker stimulation}. \textcolor{mdtRed}{\textsc{Frontiers in Systems Neuroscience}} 15:709677 (2021).\\
  
  \item[{\large $\blacktriangleright$}] \textbf{Giorgio Nicoletti} and Daniel M. Busiello. \textit{Mutual information disentangles interactions from changing environments}. \textcolor{mdtRed}{\textsc{Physical Review Letters}} 127, 228301 (2021)\footnote[2]{This work was selected as a Physical Review Letters Editors' Suggestion and as a highlight in PRL's weekly tip sheet for reporters. The magazine ``Physics'' of the American Physical Society dedicated the viewpoint ``\href{https://physics.aps.org/articles/v14/162}{Distinguishing Noise Sources with Information Theory}'' to our paper, with the aim of illustrating our results to the non-technical public (\emph{Physics 14, 162}).}.\\
  
  \item[{\large $\blacktriangleright$}] Benedetta Mariani, \textbf{Giorgio Nicoletti}, Marta Bisio, Marta Maschietto, Stefano Vassanelli, Samir Suweis. \textit{Disentangling the critical signatures of neural activity}. \textcolor{mdtRed}{\textsc{Scientific Reports}} 12, 10770 (2022).\\
  
  \item[{\large $\blacktriangleright$}] Giacomo Barzon\footnote[1]{These authors contributed equally.}, \textbf{Giorgio Nicoletti}$^*$, Benedetta Mariani, Marco Formentin, Samir Suweis. \textit{Criticality and network structure drive emergent oscillations in a stochastic whole-brain model}. \textcolor{mdtRed}{\textsc{Journal of Physics: Complexity}} 3 025010 (2022).\\
  
  \item[{\large $\blacktriangleright$}] \textbf{Giorgio Nicoletti}, Amos Maritan, Daniel M. Busiello. \textit{Information-driven transitions in projections of underdamped dynamics}. \textcolor{mdtRed}{\textsc{Physical Review E}} 106, 014118 (2022).\\

  \item[{\large $\blacktriangleright$}] \textbf{Giorgio Nicoletti} and Daniel M. Busiello. \textit{Mutual information in changing environments: Nonlinear interactions, out-of-equilibrium systems, and continuously-varying diffusivities}. \textcolor{mdtRed}{\textsc{Physical Review E}} 106, 014153 (2022).\\
  
  \item[{\large $\blacktriangleright$}] \textbf{Giorgio Nicoletti}, Leonardo Saravia, Ferdinando Momo, Amos Maritan, Samir Suweis. \textit{The emergence of scale-free fires in Australia}. \textcolor{mdtRed}{\textsc{iScience}} 26, 106181 (2023).\\
\end{itemize}

\noindent The following works are currently available as preprints. \vspace*{0.3cm}
\begin{itemize}
  \item[{\large $\blacktriangleright$}] \textbf{Giorgio Nicoletti} and Daniel M. Busiello. \textit{Information dynamics emerging from memory and adaptation in non-equilibrium sensing of living systems}. \textcolor{mdtRed}{\textsc{arXiv}} 2301.12812 (2023).\\
\end{itemize}


\mainmatter

\chapter{Introduction}
Statistical Physics was born to describe large systems with many degrees of freedom. For these systems, it is often the case in which we are not necessarily interested in their microscopic details - but rather in their macroscopic, collective behavior that emerges from the interactions of their individual parts.

One of the most striking features of Statistical Physics is the study of phase transitions, sudden changes in the behavior of a system as a result of small changes in external conditions. In the last century, powerful theoretical tools have been developed to understand these phenomena, from the scaling hypothesis to the Renormalization Group. At the edge of a second-order phase transition, macroscopic properties that are not predictable from the microscopic details emerge - fluctuations become long-range correlated both in space and time even with short-range interactions, the susceptibility to external perturbations peaks, and a newfound scale invariance reflects into the appearance of complex spatiotemporal patterns. These emergent features are often connected to trade-offs between order and disorder, robustness and flexibility, everlasting silence and never-ending activity. As such, a fascinating yet debated hypothesis was formulated in the last decades - that being poised close to a critical point might be an optimal strategy for biological and living systems \cite{mora2011biological, munoz2018colloquium}. Properties of critical systems may be desirable for information storage and processing \cite{nykter2008critical, dearcangelis2010learning, palo2017criticallike, hidalgo2014information}, collective motion, behavior, and response \cite{cavagna2010scalefree, cavagna2017dynamic, roli2018dynamical, khajehabdollahi2022when, kinouchi2006optimal}, as well as computational capabilities and information transmission \cite{marinazzo2014information, legenstein2007edge, rossert2015edge, stoop2016auditory, vanni2011criticality, boedecker2012information}. And the key idea is that these properties need not be present at the microscopical level, but rather emerge spontaneously as the system approaches the edge of a second-order phase transition.

These concepts are tightly related to the idea that ``more is different'', as Philip W. Anderson wrote fifty years ago \cite{anderson1972more}. Complex systems are often characterized by emergent properties - properties that are observed only at the level of the system as a whole, but that are not present at the level of their individual components. Arguably, phase transitions are one of the most emblematic examples of emergence, where the very same microscopic system can lead to qualitatively different behaviors at different values of the control parameter, e.g., the temperature. It is even more striking that phase transitions formally exist only in the thermodynamic limit, where complexity arises from simple interactions only when an infinite number of degrees of freedom is at play - with the words of Leo Kadanoff, for Statistical Physics and our paradigmatic models ``infinitely more is different'' \cite{kadanoff2009more}. Clearly, real-world systems are not infinite, and no thermodynamic limit exists in nature. But it is not hard to imagine that this story is telling us something quite profound about the relation between our models and the phenomena we want to model. Complexity is ubiquitous in the natural world at all scales, from microbial communities to entire ecosystems, from neural and brain dynamics to social interactions. Notwithstanding, our models are most often quite simple instead - with the goal of capturing minimal but essential features of what we are observing. Then, criticality and phase transitions may be one of the mechanisms with which we are able to generate complexity out of otherwise simple individual rules, and ultimately understand what the fundamental pieces of our model are.

Indeed, as Anderson puts it, it is undeniable that new and fundamental questions arise at different scales. A flawless description of the dynamical evolution of a single degree of freedom does not explain the behavior of many, interconnected ones, which is instead profoundly dependent on the structure of interactions among them. This is where our quest for minimal models becomes paramount - quantum mechanics is not useful to describe ecosystems, but a minimal description of species interactions is fundamental if we want to capture their behaviors. Ultimately, this is because the interactions between the components give rise to emergent phenomena, which are fundamentally different from the behavior of the components themselves.

In the course Thesis, we will explore these ideas in different contexts. In the first part, we will try to address an apparently orthogonal question - what happens in paradigmatic systems that are affected by an unobserved but changing environment? Yet, as we will see, this question is relevant in the context of complex systems, where understanding where the interdependencies between the individual parts emerge from is of prime importance. In particular, we will show how tools from Information Theory can help us disentangle effective dependencies arising due to the shared environment at the level of information encoded in the probability distributions describing the system as a whole. Then, we will further explore what Information Theory can tell us about the relation between simple effective models and the complex dynamics they are trying to approximate. In the second part, instead, we will explicitly use ideas coming from phase transitions and criticality to study the properties of neural activity, both in data and in simple models. We will see how interactions and external, environmental-like modulation shape neural activity, and how the underlying structure of interactions can qualitatively change the properties of our model. To this end, in this introductory Chapter, we will review how criticality comes about in archetypal models, what are its main features, as well as how we can describe them. Phase transitions in dynamical models, in particular, will lead us to stochastic processes and probability, where we will have the opportunity to introduce the main quantities from Information Theory that we will use.


\section{Equilibrium and dynamical phase transitions}
In this Section, we will briefly recall the main ideas behind equilibrium and dynamical second-order phase transitions. To do so, we will consider two archetypal models - the Ising model and the contact process - which will allow us to sketch the fundamental features of such transitions. Remarkably, the concept of scale invariance and the sudden lack of a characteristic scale that appears near critical points have been argued to be a crucial aspect found in natural systems and complex systems in general \cite{mora2011biological, munoz2018colloquium}. In the second part of this Thesis, we will be interested in how such signatures are found in neural activity in particular, and how they may be connected to dynamical phase transitions and in particular absorbing ones.

\begin{figure*}
    \centering
    \begin{minipage}{0.3\textwidth}
        \centering
        \includegraphics[height=4cm]{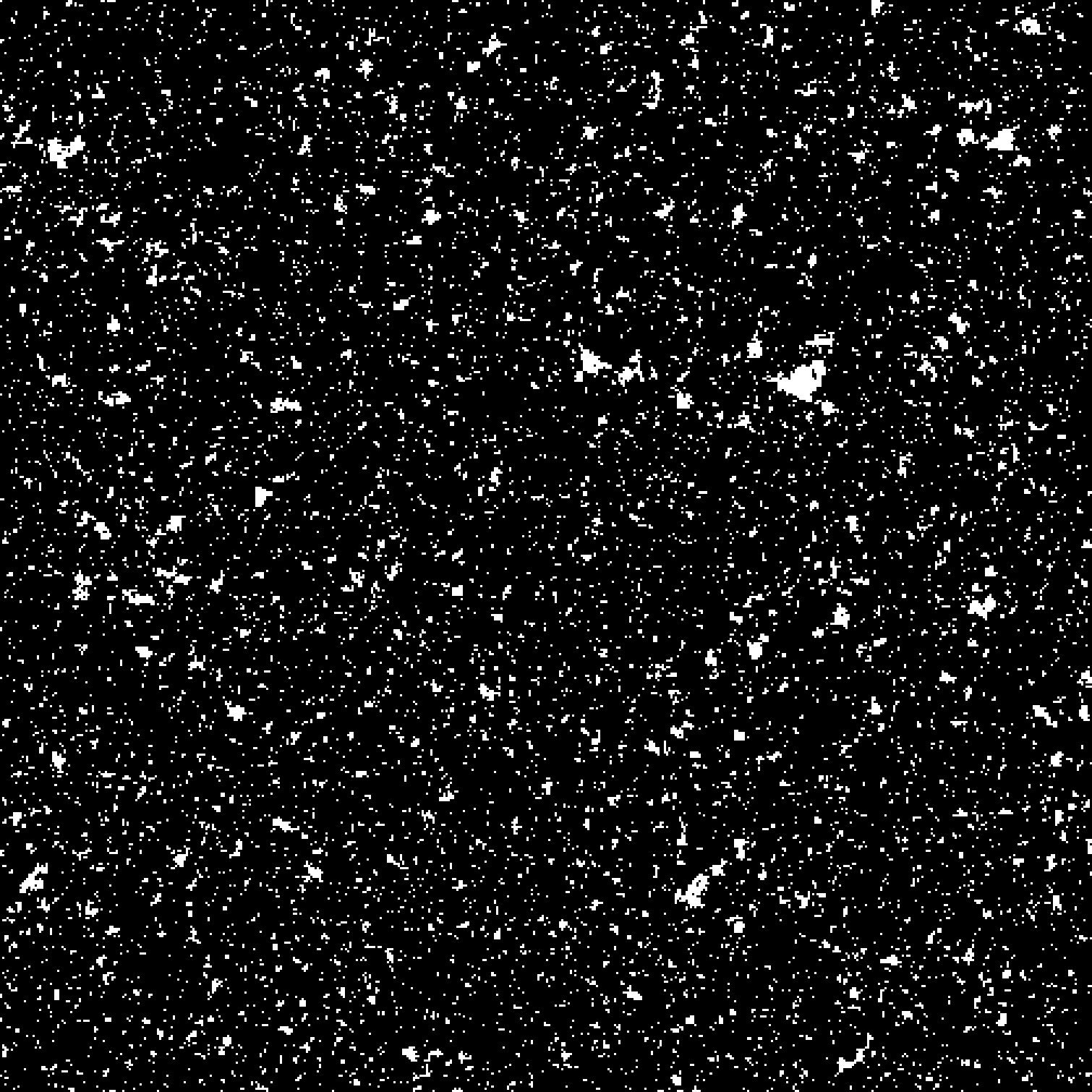}
        \subcaption{$T = 0.95 T_c$}
    \end{minipage}
    \hfill
    \begin{minipage}{0.3\textwidth}
        \centering
        \includegraphics[height=4cm]{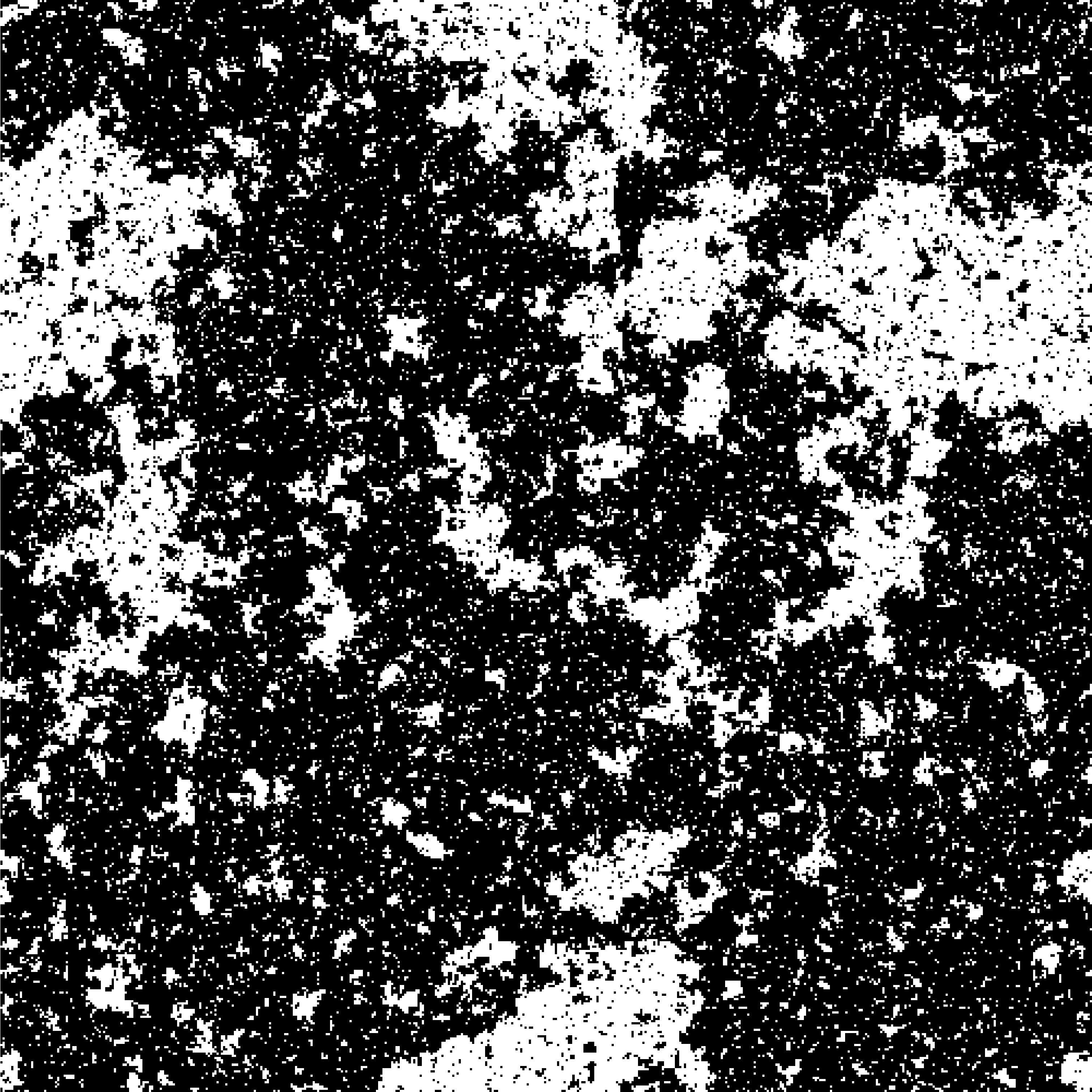}
        \subcaption{$T = T_c$}
    \end{minipage}
    \hfill
    \begin{minipage}{0.3\textwidth}
        \centering
        \includegraphics[height=4cm]{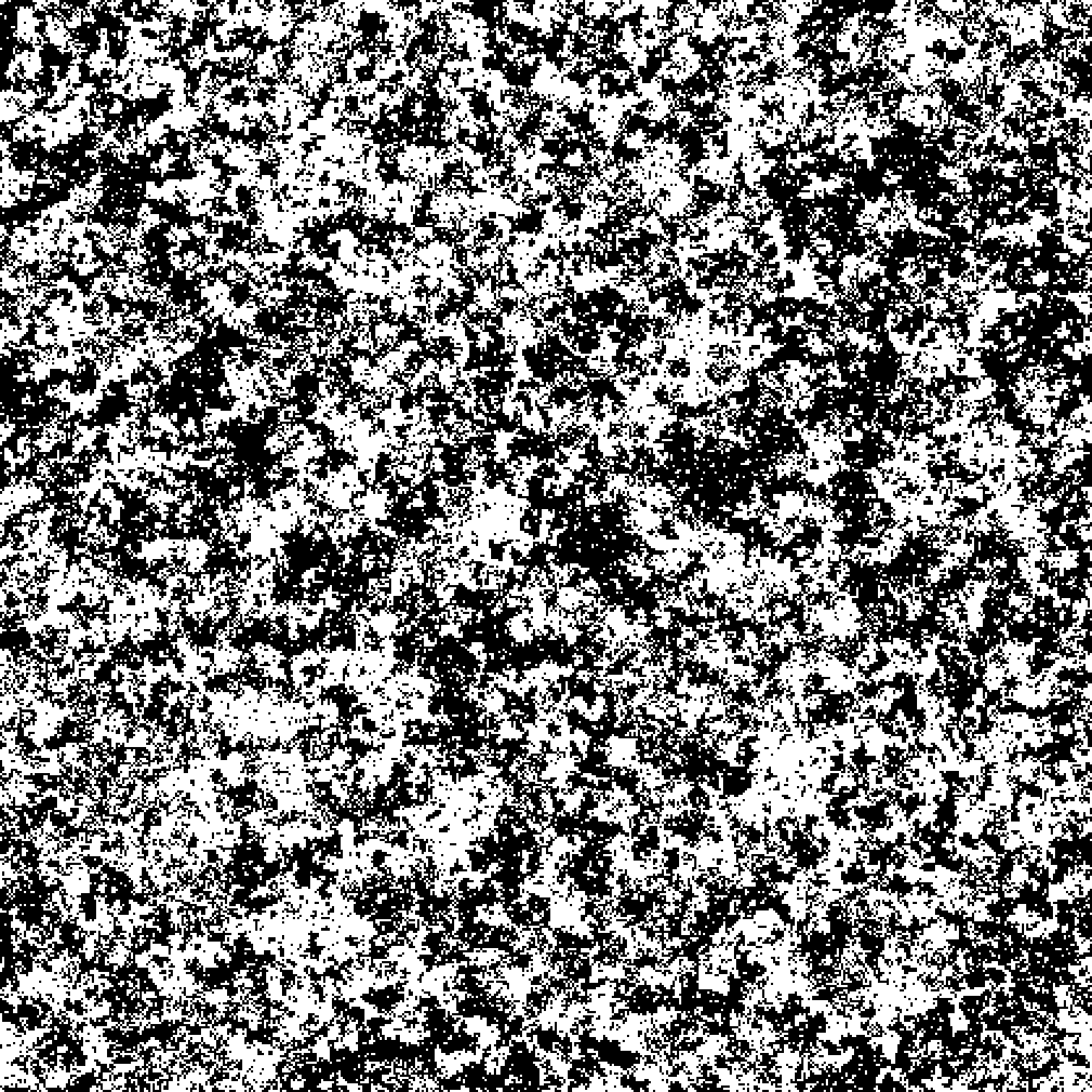}
        \subcaption{$T = 1.05 T_c$}
    \end{minipage}
    \caption{Configurations of the two-dimensional Ising model in a $500\times 500$ lattice below, at, and above the critical temperature. These and all the other configurations in this Thesis are obtained from a Monte Carlo simulation performed using the Wolff algorithm}
    \label{fig:T_ising}
\end{figure*}

Notably, a prominent framework to understand how natural and biological systems may tune themselves to a critical point is that of self-organized criticality (SOC) \cite{bak2013how}. Although we will not enter into the details here, the core idea is the presence of a dynamical feedback on the control parameter, i.e., on the parameter that determines the phase of the system, by either dissipation or driving. With the proper assumptions - for instance, timescale separation between the system's dynamics and the feedback's one - the system self-organizes to the vicinity of the critical point, without any need for fine-tuning. Besides SOC, it has been argued that biological systems may have evolved to achieve criticality - or, at least, to be described by models close to the edge of a phase transition - due to its inherent functional advantages, for instance via adaptation and plasticity \cite{sole1999criticality, torres-sosa2012criticality, levina2007dynamical, levina2009phase, buendia2020feedback, tu2018adaptation, khajehabdollahi2020evolution}.

\subsection{The Ising model}
Scale invariance is at the heart of critical phenomena. A system is said to be scale invariant if its physical properties remain unchanged at different scales - for instance, the structure of self-similar geometrical objects like fractals does not change at different spatial scales. They are effectively scale-free. Formally, a scale-invariant scalar function needs to obey
\begin{align}
\label{eqn:intro:scale_invariant}
\frac{f(y)}{f(x)} = \phi \left(\frac{y}{x}\right) \quad \forall x, y,
\end{align}
for some function $\phi$. Eq.~\eqref{eqn:intro:scale_invariant} tells us that the function $f$ varies in the same way at all scales. The class of differentiable functions that satisfy this condition can be easily found, and they are power laws of the form
\begin{align}
\label{eqn:intro:power_laws}
f(x) = c x^{a}
\end{align}
where $a = \phi'(1)$ is a constant. In fact, a change of scale of the form $x \to c x$ gives $f(cx) = c^a f(x)$. A function $g(x)$ with a given characteristic scale $\xi$, say $g(x) = \exp(-x/\xi)$, would transform differently at different scales. This is the reason why power-laws are regarded as signatures of scale invariance.

In physical systems, power-laws prominently arise in the presence of second-order phase transitions. Here, we first focus on a paradigmatic equilibrium case - that of a transition between an ordered state and a disordered one. Consider a lattice of $N$ spins, i.e., of binary variables $S_i = \pm 1$ with $i = 1, \dots, N$, described by the Ising Hamiltonian
\begin{align}
\label{eqn:intro:ising_hamiltonian}
H_\text{Ising} = -J\sum_{\langle ij \rangle}S_iS_j
\end{align}
where the sum runs over the lattice's nearest neighbors, and $J$ is the coupling strength. It is well known that in dimension $d > 1$ the Ising model described by Eq.~\eqref{eqn:intro:ising_hamiltonian} displays qualitatively different behaviors depending on the temperature $T$ (see Figure \ref{fig:T_ising}). The interaction described by Eq.~\eqref{eqn:intro:ising_hamiltonian} is a short-range one, and indeed at high temperatures the model displays short-range correlations and disorder, with spins randomly oriented. On the other hand, although correlations are still short-range at low $T$, long-range order appears because all the spins tend to align to minimize the energy. More precisely, the magnetization
\begin{align*}
m = \frac{1}{N}\ev{\sum_{i=1}^{N} S_i}
\end{align*}
vanishes if in the disordered phase at $T>T_c$, whereas it is different from zero in the ordered one at $T<T_c$. $T_c$ defines a critical temperature, at which the system undergoes a second-order phase transition as it begins to order \cite{onsager1944crystal}, and $m$ is a local order parameter that allows us to discriminate between the two different phases. It is around the critical temperature that the physical properties of the model are the most interesting.

\begin{figure*}
    \centering
    \begin{minipage}{0.3\textwidth}
        \centering
        \includegraphics[height=4cm]{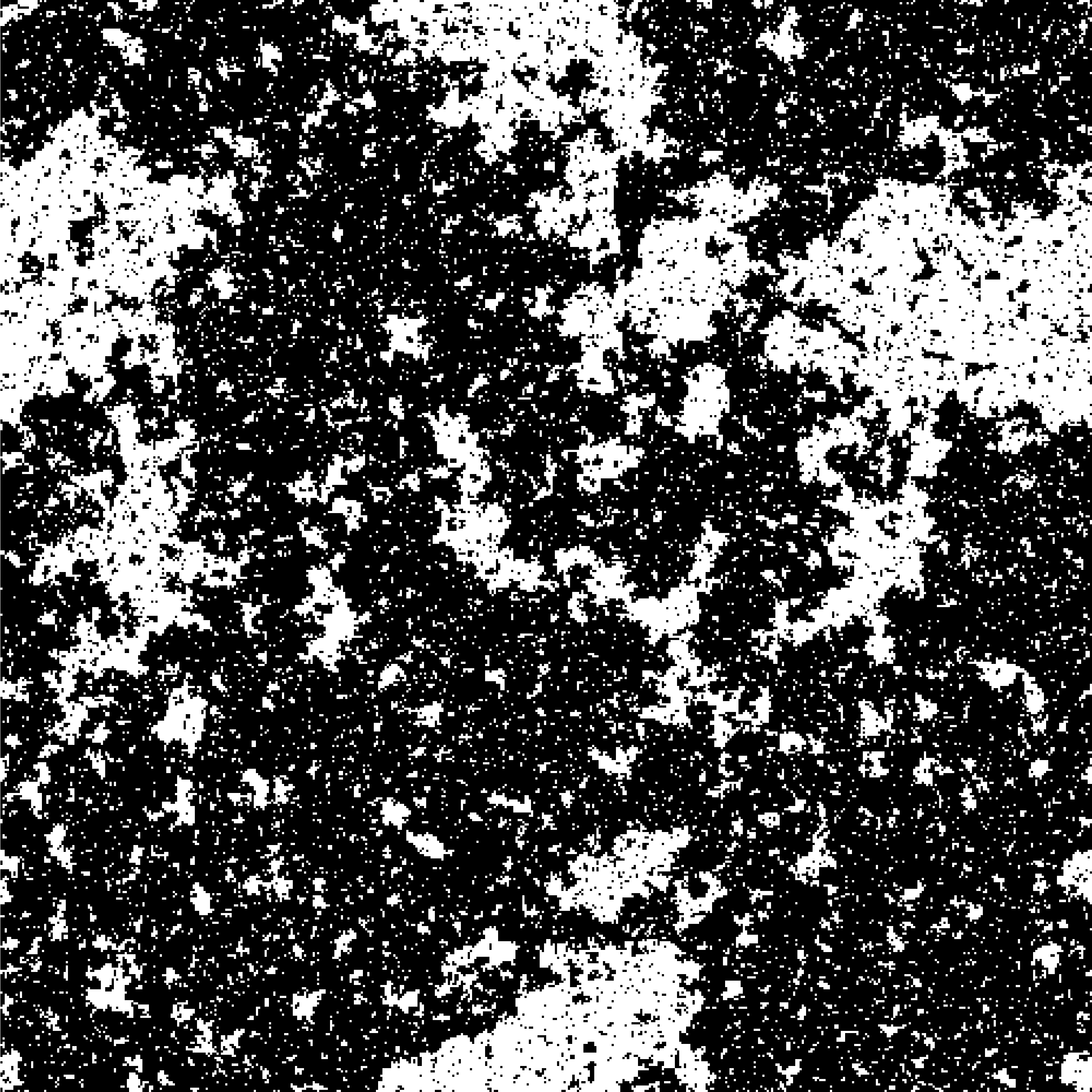}
        \subcaption{$N_\text{spins}=250000$}
    \end{minipage}
    \hfill
    \begin{minipage}{0.3\textwidth}
        \centering
        \includegraphics[height=4cm]{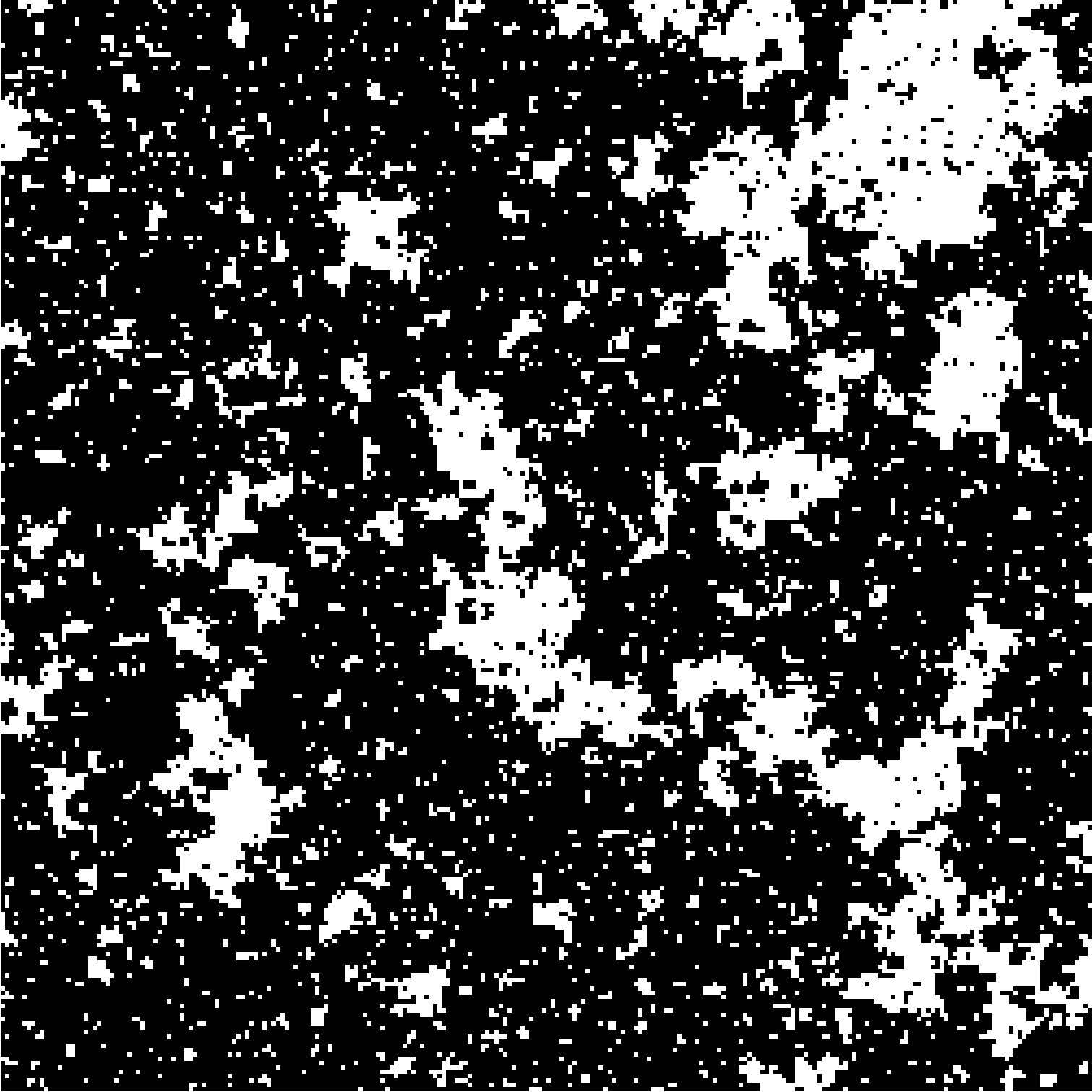}
        \subcaption{$N_\text{spins}=62500$}
    \end{minipage}
    \hfill
    \begin{minipage}{0.3\textwidth}
        \centering
        \includegraphics[height=4cm]{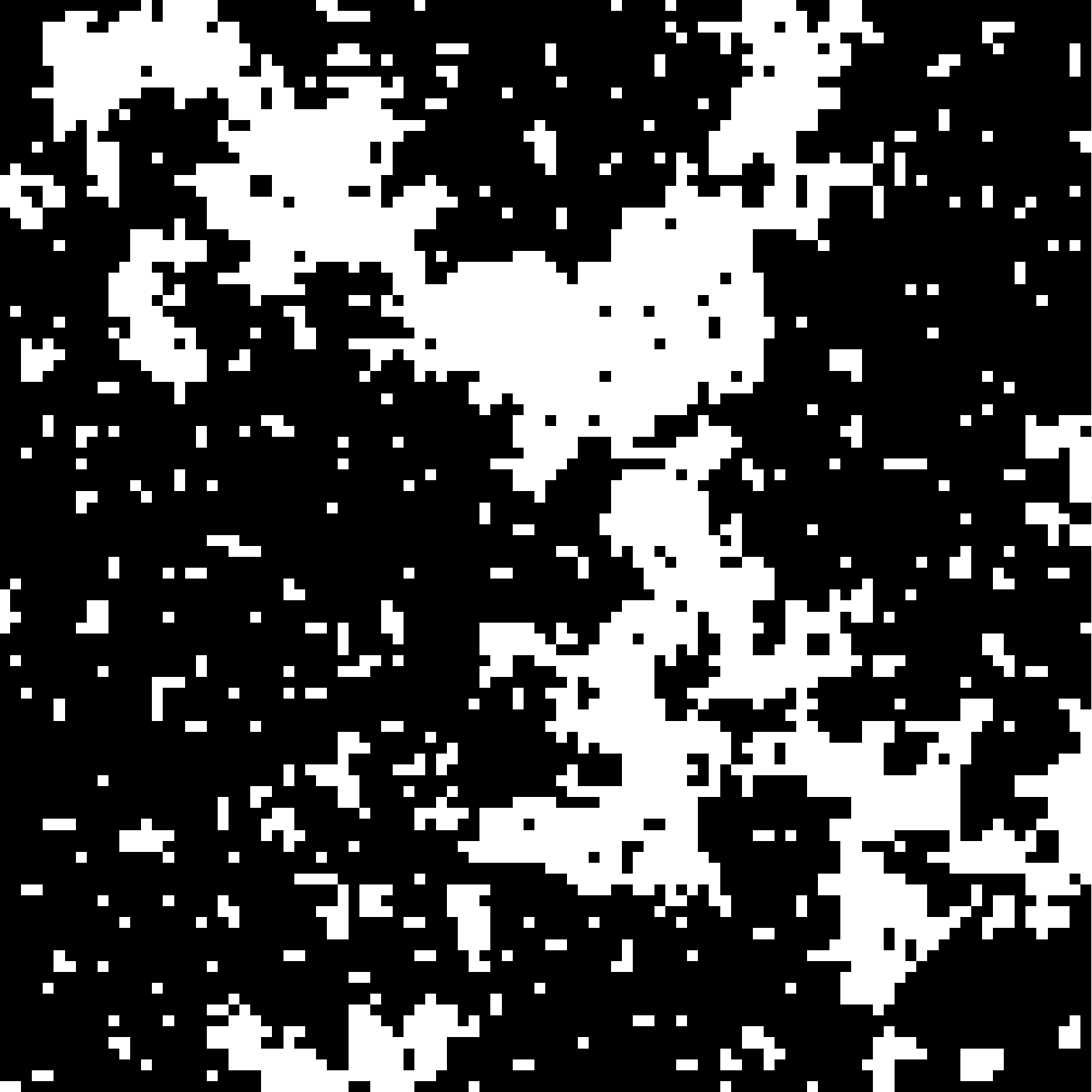}
        \subcaption{$N_\text{spins}=10000$}
    \end{minipage}
    \caption{A configuration of the 2D Ising model at criticality. (a) A $500\times500$ lattice. (b) A $250\times250$ sublattice. (c) A $100\times100$ sublattice. Notice how islands of spins exist at all scales}
    \label{fig:critical_ising}
\end{figure*}

As we approach $T_c$ something remarkable happens - the correlations between the spins become more and more long-range, leading to the formation of islands of aligned spins. Exactly at criticality, these islands exist at all scales and the configuration of the Ising model becomes fractal-like. Figure \ref{fig:critical_ising} shows how sublattices at different scales resemble the full lattice. This is perhaps one of the most striking features of criticality: at the critical point, the correlations between the microscopic degrees of freedom become relevant at all scales, regardless of the nature of the starting interaction. Since no relevant scale can be defined, the physical properties of the model turn out to be scale-free and described by power-laws. For instance, the order parameter scales as
\begin{align}
\label{eqn:intro:beta_exp}
m \sim |t|^{\beta} \quad (t<0)
\end{align}
where $t = (T-T_c)/T_c$ and $\beta$ is called a critical exponent. A number of these exponents can be found in the thermodynamic limit $N\to\infty$, where ergodicity is broken. For instance, the zero-field susceptibility is divergent
\begin{align}
\label{eqn:intro:gamma_exp}
\chi = \pdv{m}{h}\biggl|_{h=0} \sim |t|^{-\gamma}
\end{align}
where $h$ is an external pinning field - with a corresponding Hamiltonian $H_h = - h\sum_i S_i$ - that selects one of the two ergodic regions of the phase space defined by the symmetry group $\mathbb{Z}^2$ of the Ising model, and that become disjointed in the thermodynamic limit \cite{goldenfeld1992lectures}. Similarly, away from criticality, the connected correlation function $G(r) = \ev{S_r S_0}-\ev{S_r}\ev{S_0} \sim e^{-r/\xi}$ defines a correlation length $\xi$ due to the short-range nature of the microscopic interaction. However, close to the critical point, the correlation length diverges as
\begin{align}
\label{eqn:intro:nu_exp}
\xi \sim |t|^{-\nu}
\end{align}
so that, exactly at $T = T_c$, the decay of the correlations becomes algebraic
\begin{align}
\label{eqn:intro:corr_algeb}
G(r) \sim r^{-(d-2+\eta)}.
\end{align}
This is precisely what we mean when we say that at criticality the system is scale-free - Eq.~\eqref{eqn:intro:nu_exp} together with Eq.~\eqref{eqn:intro:corr_algeb} implies that $\xi \to \infty$ as we approach the critical point, so all scales are equally relevant.

\begin{figure*}
    \centering
    \begin{minipage}{0.45\textwidth}
        \centering
        \includegraphics[height=4cm]{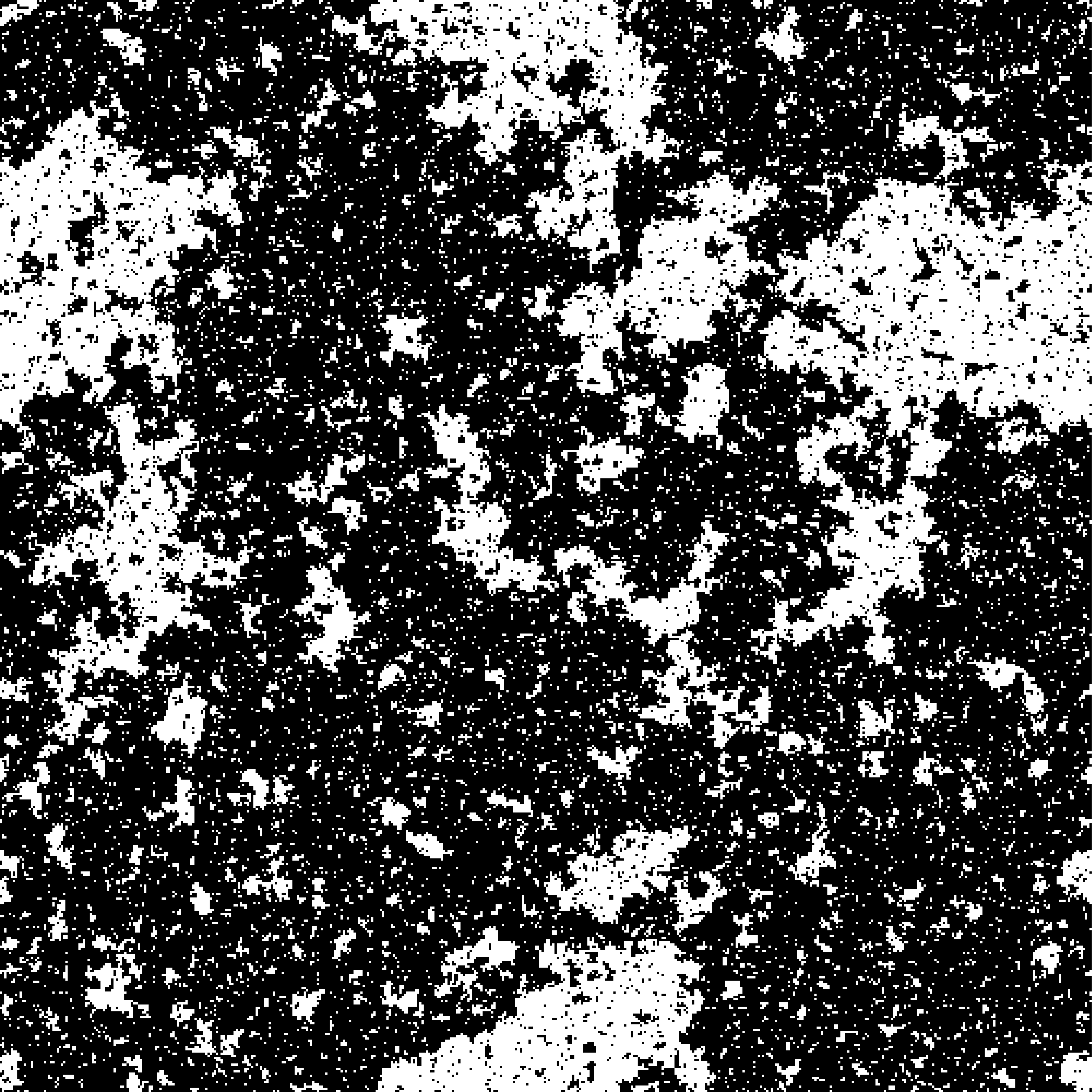}
        \subcaption{Original lattice}
    \end{minipage}
    \begin{minipage}{0.45\textwidth}
        \centering
        \includegraphics[height=4cm]{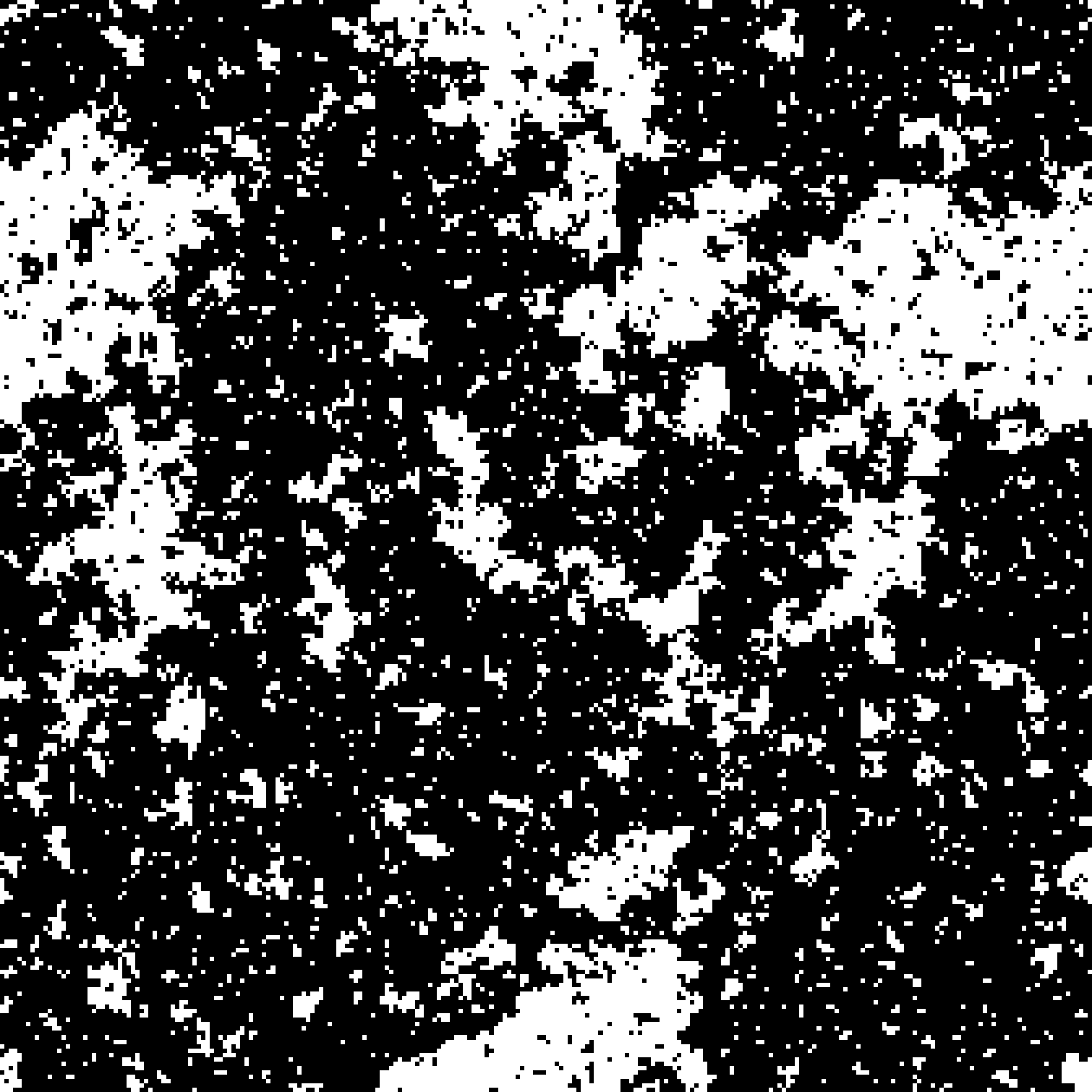}
        \subcaption{First generation of spins}
    \end{minipage}
    
    \vspace{0.5cm}
    
    \begin{minipage}{0.45\textwidth}
        \centering
        \includegraphics[height=4cm]{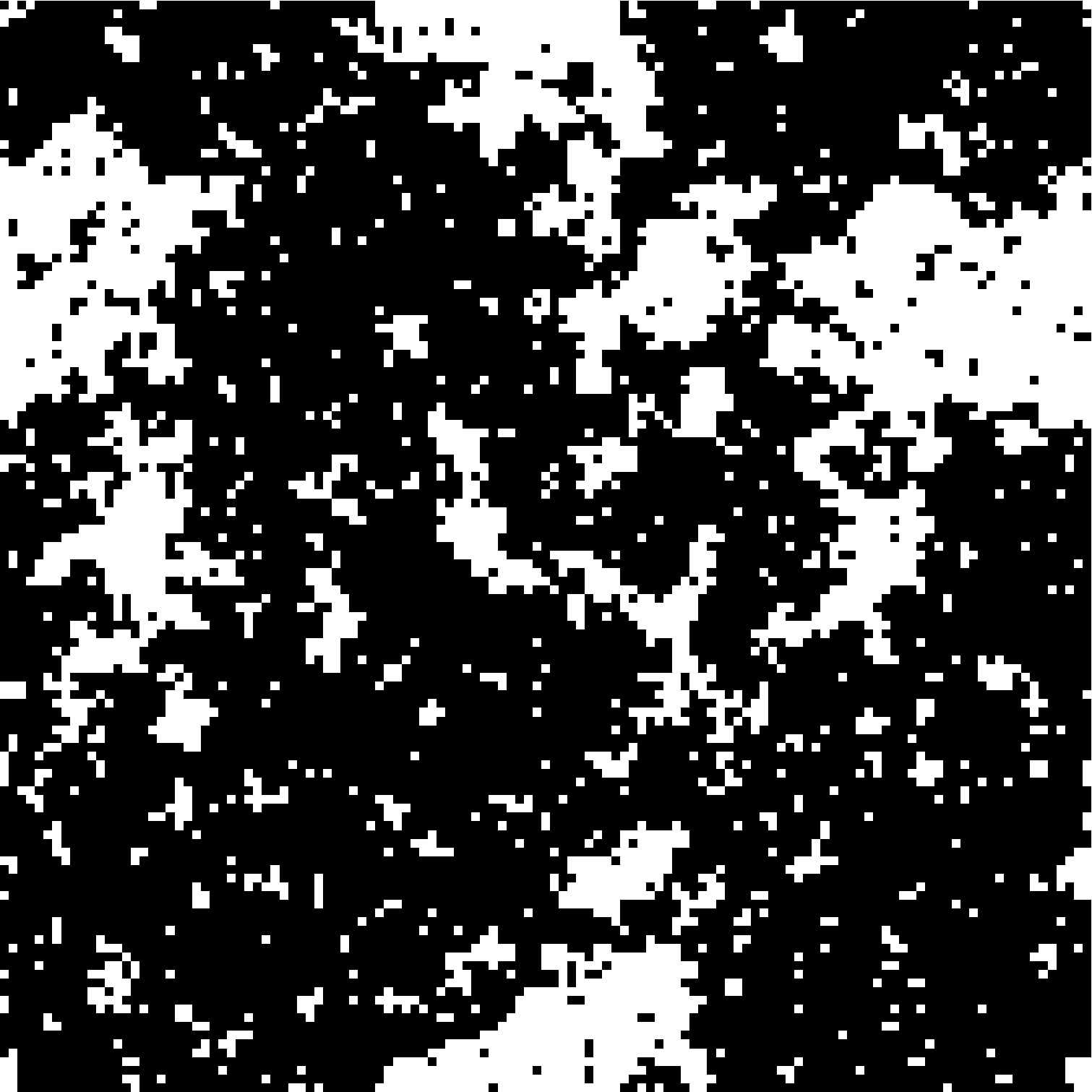}
        \subcaption{Second generation of spins}
    \end{minipage}
    \begin{minipage}{0.45\textwidth}
        \centering
        \includegraphics[height=4cm]{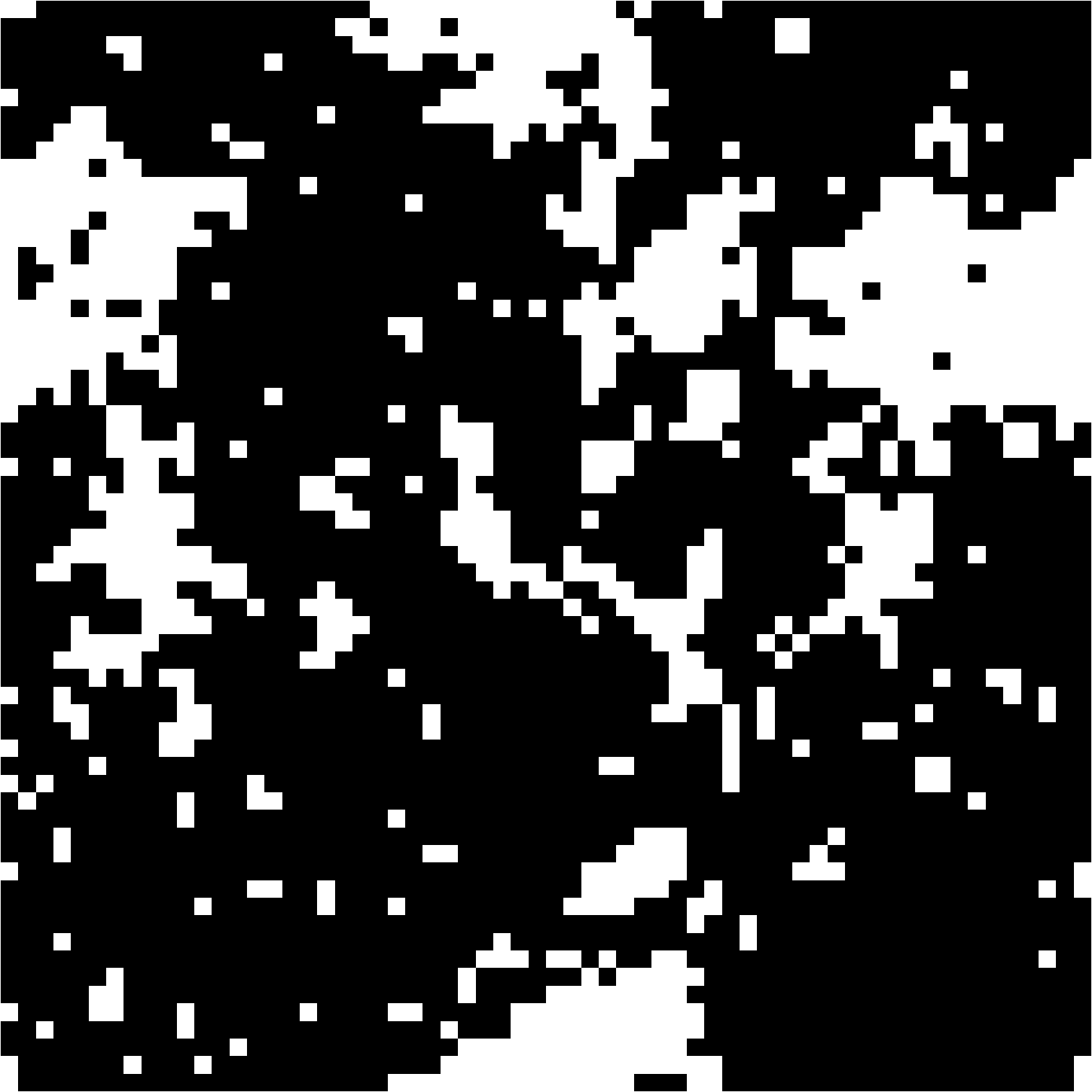}
        \subcaption{Third generation of spins}
    \end{minipage}
    \caption{A critical configuration of the 2D Ising model is left invariant under a coarse-graining transformation due to its underlying scale invariance. The coarse-graining procedure we used here is a $2x2$ block-spin transformation}
    \label{fig:coarse_grain_ising}
\end{figure*}

The fact that the collective behavior emerging at criticality becomes independent of the microscopic interaction leads us to another feature of critical systems: universality. Indeed, the critical behavior depends only on a few key properties, like the symmetry of the Hamiltonian, the dimensionality of the system, and the range of interactions. These properties define a universality class, leading to different systems sharing the same set of exponents. From a theoretical point of view, this is well understood in the context of the Renormalization Group, but we can get an intuition already at the level of coarse-graining. Loosely speaking, a coarse-graining transformation amounts to changing the scale at which we look at the system - an example of a spatial coarse-graining in a lattice would be to group together plaquettes of nearby spins via a majority rule, known as a block-spin transformation. In the case of the Ising model, for $T<T_c$ after enough coarse-graining steps the system ends up in one of its two ground states with all the spins aligned. On the other hand, for $T>T_c$ the system ends up in a fully disordered configuration. At criticality, fluctuations exist at all scales and the system is described by power-laws. Due to such intrinsic scale invariance, the coarse-graining yields no change at the critical point, as we can see in Figure \ref{fig:coarse_grain_ising}. Qualitatively, this is where universality comes to be - as the coarse-graining procedure unravels the long-range properties of the system by smoothing out its short-range fluctuations.

The Ising model serves as an archetypal example of which kind of properties we expect to find a criticality, from long-range correlations without the need for long-range interactions, to a diverging susceptibility that allows for a large sensitivity to external stimuli. A number of these properties have been argued to be optimal for biological systems \cite{cavagna2010scalefree, cavagna2017dynamic, roli2018dynamical, khajehabdollahi2022when, kinouchi2006optimal, mora2011biological, munoz2018colloquium, marinazzo2014information, vanni2011criticality, palo2017criticallike, cocchi2017criticality, strogatz2022fifty, niebur2014criticality, massobrio2015criticality, dearcangelis2010learning, nykter2008critical, zeraati2021self, boedecker2012information, legenstein2007edge, chialvo2010emergent}, and in particular for the brain, as we will see in detail in Chapter \ref{ch:scirep} and Chapter \ref{ch:jphys}. At the same time, critical and abrupt transitions may be detrimental to ecological systems, for instance. Although not included in this thesis, in \cite{nicoletti2021emergence} we studied how isotropic percolation - another kind of equilibrium transition - can help us understand the emergence of scale-free fires, that are invariant under spatial coarse-graining, during the devastating 2019-2020 bushfire season in Australia. Yet, complex systems are often described by dynamical processes. In the next Sections, we will briefly show how phase transitions arise in non-equilibrium settings, described by stochastic processes.

\subsection{The contact process}
The contact process is one of the simplest non-equilibrium models, describing the spreading of activity, and was first introduced as a toy model for epidemic spreading without immunization \cite{harris1974contact}. Much like the Ising model for equilibrium systems, the contact process is an archetypal example of non-equilibrium phase transitions. Consider, in general, a collection of $i = 1, \dots, N$ nodes of a given network. Each site can be either active or inactive, and we identify its state by means of a binary variable $\sigma_i (t) = 1, \, 0$, respectively. The activity spreads via a nearest-neighbors interaction, and it depends on the number of active neighbors,
\begin{align*}
n_i(t) = \sum_{j\in\ev{i}} \sigma_j(t),
\end{align*}
whereas each active site is emptied at a given rate $\mu$. Thus the rates $w(\sigma_i(t)\to\sigma_i(t+dt), n_i(t))$ are given by
\begin{equation}
\label{eqn:intro:cp_rates}
\begin{gathered}
w(0\to1, n_i) = \frac{\lambda n_i}{k_i} \\
w(1\to0, n_i) = \mu
\end{gathered}
\end{equation}
where $\lambda$ is the spreading rate, and $k_i$ is the number of neighbors of the $i$-th site, e.g., the coordination number in a lattice. As we will see, it exists a critical value of the control parameter $\lambda = \lambda_c$ that separates two qualitatively different global behaviors. If $\lambda>\lambda_c$, the spreading rate is high and a site does not recover fast enough, so at large times an active fluctuating phase permeates the system. On the other hand, if $\lambda<\lambda_c$, the activity is not able to spread fast enough and the system eventually finds itself in a state where all the sites are empty. From its definition, the local dynamical rules in Eq.~\eqref{eqn:intro:cp_rates} do not allow the system to escape from such a configuration. These two phases appear only in the long time limit - that is, they are stationary solutions of the model - and the transition to the absorbing state is dynamically irreversible.

In this Section, we will briefly describe in which sense this transition to an absorbing state is in fact a phase transition, and how we can recover a notion of criticality similar to that of equilibrium systems. If we are to identify a phase transition, we need to introduce an order parameter first. One possible choice is the density of active site $\rho(t)$ defined as
\begin{align}
\label{eqn:intro:density_site}
\rho(t) = \lim_{N\to\infty}\frac{1}{N}\ev{\sum_{i=1}^N \sigma_i(t)}
\end{align}
where the average $\ev\cdot$ is performed over many realizations of the stochastic process. This seems to be a good candidate - it is clearly zero in the absorbing state, and is different from zero if the activity is proliferating. Then, the equation of motion for the local order parameter $\rho_i(t)$, defined as
\begin{align*}
\rho_i(t) = \mathbb{P}[\sigma_i(t)=1],
\end{align*}
is the master equation
\begin{align*}
\mathbb{P}[\sigma_i(t+\Delta t) = 1] = \mathbb{P}[\sigma_i(t) = 1][1-\mu\Delta t] + \frac{\lambda \Delta t}{k_i}\sum_{j\in\partial i} \mathbb{P}[\sigma_i(t)=0,\sigma_j(t)=1]
\end{align*}
and the continuous time limit yields
\begin{align}
\label{eqn:intro:cp_mastereq_cont}
\dv{t}\rho_i(t) = -\mu \rho_i(t) + \frac{\lambda}{k_i}\sum_{j\in\partial i} \mathbb{P}[\sigma_i(t)=0,\sigma_j(t)=1]
\end{align}
where $\partial i$ is the set of neighbors of the $i$-th site. This is still an exact relation, but the last term in the r.h.s. makes it impossible to solve it exactly. Thus, to understand the nature of the phase diagram of the contact process we consider a mean-field approach, which amounts to assuming that the joint probability in the last term of Eq.~\eqref{eqn:intro:cp_mastereq_cont} is factorizable, i.e., $\mathbb{P}[\sigma_i(t)=0,\sigma_j(t)=1] = \mathbb{P}[\sigma_i(t)=0]\mathbb{P}[\sigma_j(t)=1]$. This is equivalent to assuming that the underlying network is fully-connected, which further implies a homogeneity assumption $\rho_i = \rho$ and $k_i = k$. We end up with
\begin{align}
\label{eqn:intro:cp_mf}
\dv{\rho}{t} & = -\mu\rho + \frac{\lambda}{Nk} \left[Nk(1-\rho)\rho\right] \nonumber \\
& = \rho(\lambda-1)-\lambda\rho^2.
\end{align}
where we rescaled the control parameter to restore extensivity.

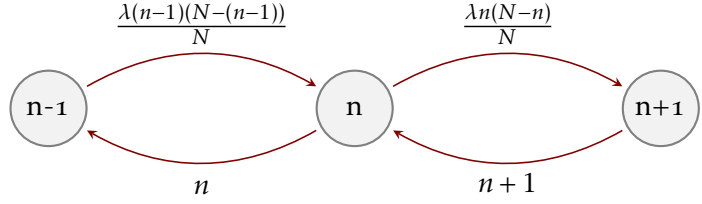
\begin{figure*}
\centering
\usetikzlibrary{arrows, automata, positioning}
\tikzstyle{me}=[circle,draw=black!50,fill=black!5,thick,
inner sep=3pt,minimum size=10mm]
\tikzstyle{arr}=[->,shorten >=2pt,shorten <=2pt,>=stealth,semithick]
\begin{tikzpicture}
		\node[me] (n) {n};
		\node[me] (n-1) [left = 3cm of n] {n-1};
		\node[me] (n+1) [right = 3cm of n] {n+1};
		\draw[arr, mdtRed] (n+1) to [bend left = 30] node[midway, below] {\color{black}$n+1$} (n) ;
		\draw[arr, mdtRed] (n) to [bend left = 30] node[midway, above] {\color{black}$\frac{\lambda n (N-n)}{N}$} (n+1);
		\draw[arr, mdtRed] (n) to [bend left = 30] node[midway, below, yshift=-0.1cm] {\color{black}$\,n\,$} (n-1);
		\draw[arr, mdtRed] (n-1) to [bend left = 30] node[midway, above] {\color{black}$\frac{\lambda (n-1)(N-(n-1))}{N}$} (n);
\end{tikzpicture}
\caption{The transition scheme for the mean-field contact process, where $n$ is the number of active sites, $N-n$ is the number of empty sites, and the spreading rate is rescaled $\lambda \to \lambda/N$. The spontaneous emptying rate is unitary}
\label{fig:cp_master_equation}
\end{figure*}

Notice that we can derive the same equation of motion directly from the master equation for the mean-field contact process. The rates are outlined in Figure \ref{fig:cp_master_equation}, where $n$ is the number of active sites and $N-n$ the number of empty sites. Hence the master equation is
\begin{align}
\label{eqn:intro:cp_master_equation}
\dot{p}_n = -p_n\left[n+\frac{\lambda n}{N}(N-n)\right] + p_{n+1}(n+1) + p_{n-1}\frac{\lambda(n-1)}{N}\left[N-(n-1)\right]
\end{align}
where $p_n(t)$ is the probability of having $n$ active sites at time $t$, and we set $\mu = 1$ for conciseness. We are interested in the average value of active sites $\ev{n}$, whose evolution obeys
\begin{align*}
\ev{\dot{n}(t)} & = \ev{-n^2+\frac{\lambda n^2(N-n)}{N} + n(n-1) + \frac{n(n+1)(N-n)}{N}} \\
& = \ev{n}(\lambda-1) - \frac{\lambda \ev{n^2}}{N}.
\end{align*}
Since the density of active sites is given by
\begin{align}
\rho(t) = \lim_{N\to\infty}\frac{\ev{n(t)}}{N},
\end{align}
and in the large-$N$ limit we expect fluctuations to be negligible, so that $\ev{n}^2$ coincides with $\ev{n^2}$, we have 
\begin{align*}
\dot{\rho}(t) = \rho(t)(\lambda-1)-\lambda\rho^2(t)
\end{align*}
which is exactly Eq.~\eqref{eqn:intro:cp_mf}. However, in this framework, one could in principle look at next-to-leading order solutions performing a system-size expansion \cite{gardiner2004handbook}.

We immediately see that Eq.~\eqref{eqn:intro:cp_mf} has two possible stationary solutions, namely the vacuum solution and the active solution
\begin{align*}
\rho_\text{st}^v = 0, \qquad \rho_\text{st}^a = \frac{\lambda-\mu}{\lambda}.
\end{align*}
For $\rho_\text{st}^a$ to be positive, we need $\lambda>\mu$. A linear stability analysis with $\rho(t) = \rho_\text{st}+\epsilon(t)$ for $\epsilon \ll 1$ yields
\begin{align*}
\dv{\epsilon}{t} = (\lambda-\mu)\epsilon -2\lambda\rho_\text{st}\epsilon + \rho_\text{st}(\lambda-\mu) - \lambda \rho^2_\text{st}+O(\epsilon^2)
\end{align*}
so that $\dot\epsilon = \pm (\lambda-\mu)\epsilon$ in the absorbing and active state, respectively. Thus, we find a bifurcation point at
\begin{align}
\label{eqn:intro:cp_critical}
\lambda_c = \mu
\end{align}
which is the mean-field critical point of the contact process. If $\lambda>\lambda_c$ the stable solution is the active one, whereas for $\lambda<\lambda_c$ the vacuum is stable. Notice that there is no discontinuity in the order parameter, similar to the case of second-order phase transitions in equilibrium Statistical Mechanics.

\subsection{An absorbing phase transition}
The contact process at its critical point is described by scale-free properties. Its spatiotemporal structure can be probed in terms of both an equal time correlation function,
\begin{align}
\label{eqn:intro:et_cf}
c_\perp(r, t) = \ev{\sigma_i(t)\sigma_{i+r}(t)} - \ev{\sigma_i(t)}\ev{\sigma_{i+r}(t)} ,
\end{align}
and a time autocorrelation function,
\begin{align}
\label{eqn:intro:t_acf}
c_\parallel(t) = \ev{\sigma_i(t)\sigma_i(0)} - \ev{\sigma_i(t)}\ev{\sigma_i(0)},
\end{align}
where averages are over sites. Hence, we can independently introduce a correlation length in space $\xi_\perp$ and a correlation length in time $\xi_\parallel$. Away from criticality, we expect that for large values of their arguments both these correlation functions decay exponentially, namely
\begin{align*}
c_\perp(r, t) \sim e^{-r/\xi_\perp}, \quad c_\parallel(t) \sim e^{-t/\xi_\parallel}.
\end{align*}
In order to understand how these correlations behave, we let go of spatial homogeneity while still considering independent sites, that is, we write
\begin{align*}
\pdv{t}\rho(\vb{x},t) & = -\rho(\vb{x},t) + \frac{\lambda}{q}\sum_{\vb y} \mathbb{P}[\sigma_{\vb{x}}(t)=0]\mathbb{P}[\sigma_{\vb{y}}(t)=1] \\
& = -\rho(\vb{x},t) + \frac{\lambda}{q}\sum_{\vb y} \rho(\vb y, t)[1-\rho(\vb x, t)]
\end{align*}
where $\vb x \in \mathbb{R}^d$ denotes the position in a $d$-dimensional continuous space. Then, we expand $\rho(\vb{y},t)$ about the point $\vb{x}$,
\begin{align*}
\rho(\vb{y}, t) \approx \rho(\vb{x}, t) + (y_i-x_i)\partial_i \rho(\vb{x}, t) + \frac{1}{2} (y_i-x_i)(y_j-x_j)\partial_i\partial_j\rho(\vb{x}, t) + \dots
\end{align*}
where we use the Einstein convention. In a finite volume $V$ we have $\sum_{\vb y} \approx q/V \int_V d^dy$, so that
\begin{align}
\label{eqn:intro:cp_expansion}
\pdv{t}\rho(\vb{x},t) = -\rho(\vb{x},t) + \frac{\lambda}{V} \int d^dy \biggl[ &  \rho(\vb{x},t) + \frac{1}{2} |\vb{y}-\vb{x}|^2 \nabla^2\rho(\vb x, t) \\
& - \rho(\vb x, t)^2 - \frac{1}{2} |\vb{y}-\vb{x}|^2 \rho(\vb x,t) \nabla^2\rho(\vb x, t) + \dots \biggl] \nonumber
\end{align}
which depends in principle on all possible couplings between the order parameter and its derivatives. If we stop at the lowest order and consider only diffusive coupling, we end up with
\begin{align}
\label{eqn:intro:cp_coarse_grain}
\pdv{t}\rho(\vb{x},t) = (\lambda - \mu)\rho(\vb{x},t)-\lambda\rho^2(\vb{x},t) + D \nabla^2\rho(\vb{x},t)
\end{align}
where $D$ is the constant factor appearing in Eq.~\eqref{eqn:intro:cp_expansion}.

If we consider the sub-critical case, $\lambda < \lambda_c$, we expect $\rho(\vb x, t)$ to be small at large times - thus, we can neglect the non-linear term. This is clearly not true in the super-critical regime. Hence, we consider deviations from the homogeneous steady state by introducing the field
\begin{align*}
\psi(\vb x,t) = \begin{cases}
\rho(\vb x, t) & \lambda < \lambda_c \\
\rho(\vb x, t)-\frac{\lambda - \lambda_c}{\lambda} & \lambda > \lambda_c
\end{cases}
\end{align*}
and its equation of motion,
\begin{align*}
\pdv{\psi}{t} = -|\lambda - \lambda_c|\psi-\lambda\psi^2+D\nabla^2\psi \approx -|\lambda - \lambda_c|\psi+D\nabla^2\psi.
\end{align*}
Here, we assumed that at large times deviations from the steady state are small. If we consider, without loss of generality, a one-dimensional system, a simple Fourier transform $\psi(x,t) = \frac{1}{2\pi} \int dk e^{ikx} f(k,t)$
gives
\begin{align*}
f(k,t) = A(k) \exp[-(|\lambda - \lambda_c|+Dk^2)t].
\end{align*}
The solution is exactly a Green function $G_0(x,t)$ if we impose $\psi(x,0) = \delta(x)$, which implies $A(k) = 1$. Hence
\begin{align}
\label{eqn:intro:cp_green_function}
G_0(x,t) = \frac{1}{\sqrt{4\pi Dt}} e^{-|\lambda - \lambda_c|t-\frac{x^2}{4Dt}}.
\end{align}
Since $G_0(x,t)$ describes how a perturbation at the origin and at time zero propagates through the system, heuristically we immediately find that the lifetime of such fluctuations is $\propto |\lambda - \lambda_c|^{-1}$, and propagates at a characteristic distance of $\approx \sqrt{Dt} \sim |\lambda - \lambda_c|^{-1/2}$. Both these quantities diverge at criticality, showing how fluctuations become correlated at arbitrarily large distances and arbitrarily large times. Hence, at the critical point, we have
\begin{equation}
\label{eqn:intro:cp_nu}
\begin{gathered}
\xi_\perp \sim |\lambda-\lambda_c|^{-\nu_\perp} \\
\xi_\parallel \sim |\lambda-\lambda_c|^{-\nu_\parallel}
\end{gathered}
\end{equation}
giving rise to scale-free correlations both in space and in time. Similarly, other physical observables display the same behavior. For instance, we can define a susceptibility by adding an external field $h$, i.e., a spontaneous activation rate $w_h[0\to1] = h$. The mean field equation of motion becomes
\begin{align*}
\dv{\rho}{t} = \rho(\lambda-\mu)-\lambda\rho^2+h(1-\rho)
\end{align*}
and the susceptibility diverges as the power-law
\begin{align*}
\chi = \pdv{\rho_\text{st}}{h}\biggl|_{h=0} = \frac{1}{\lambda(\lambda-1)} \sim |\lambda - \lambda_c| ^{-1},
\end{align*}
signaling that the system becomes infinitely sensible to external perturbations. Furthermore, in absorbing phase transitions we also find critical slowing-down. If we solve Eq.~\eqref{eqn:intro:cp_mf} for $\rho(0) = \rho_0$, we find at large times
\begin{align*}
\rho(t) \simeq
\begin{cases}
\Delta\left[\frac{\Delta}{\rho_0}-\lambda\right]^{-1}e^{\Delta t} & \Delta < 0 \\
\frac{\Delta}{\lambda}+\frac{\Delta}{\lambda^2}\left[\lambda-\frac{\Delta}{\rho_0}\right]e^{-\Delta t} & \Delta >0
\end{cases}
\end{align*}
where $\Delta = \lambda - \lambda_c$, and for $\Delta \ne 0$. This implies that, away from criticality, the system approaches the stationary solution exponentially with a relaxation time $\tau = |\Delta|^{-1}$. As we approach the critical point, the relaxation time $\tau$ diverges. That is, at criticality it takes an infinite time to reach the steady state, defining a new critical exponent.

\begin{figure*}[t]
    \centering
    \includegraphics[width=.9\textwidth]{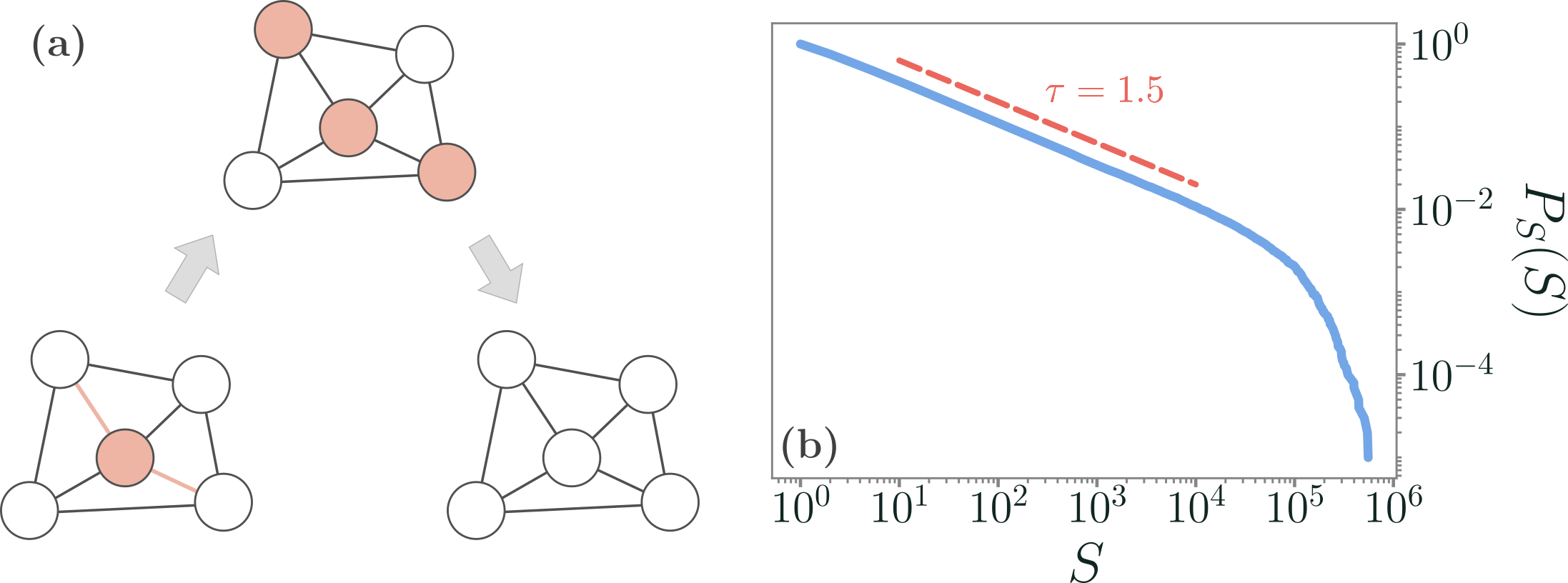}
    \caption{Avalanches in the mean-field contact process. (a) In the presence of an absorbing phase transition, an avalanche is defined as the cascade of activity that follows a seed perturbation on top of the absorbing configuration and ends when all sites have fallen back to the silent state once more. The size of an avalanche corresponds to the number of activations. (b) In the mean-field directed percolation universality class, at criticality the size of an avalanche $S$ decays as a power-law $\sim S^{-\tau}$ with an exponent $\tau = 3/2$. The plotted distribution is estimated from simulations performed with the Gillespie algorithm (see Appendix \ref{app:computational})}
    \label{fig:intro:cp_avalanches}
\end{figure*}

Finally, systems with absorbing phase transitions are quite interesting because they produce scale-free avalanches (Figure \ref{fig:intro:cp_avalanches}). An avalanche is the spread of activity produced by perturbing the absorbing state with a single seed of activity. In the subcritical regime, this cascade of activity will be short-lived and spatially localized close to the perturbation. On the other hand, in the super-critical phase activity will spread indefinitely, since the fixed point is the active one. At criticality, instead, both the duration and the number of activations lack a characteristic scale - we will observe large and long events, as well as short and small ones. If we call $S$ the number of activated sites during an avalanche, and $T$ the avalanche duration, we have
\begin{equation}
    \label{eqn:intro:avalanche_distributions}
    \begin{gathered}
        p_S(S)\sim S^{-\tau} \\
        p_T(T)\sim T^{-\tau_T}.
    \end{gathered}
\end{equation}
A formal connection can be drawn with the power-law scaling of the survival probability of the contact process at criticality \cite{munoz1999avalanche}. Remarkably, scale-free avalanches are observed in neural activity, most prominently at the mesoscopic scale \cite{beggs2003avalanches, petermann2009spontaneous, yu2011higherorder, hahn2010avalanches, gireesh2008avalanches, mazzoni2007dynamics, pasquale2008self, fontenele2019criticality, dehghani2012avalanche, gireesh2008neuronal, hahn2010neuronal}, and are often considered signs that the underlying dynamics might be tuned to a critical point, albeit from a heuristic perspective. In particular, in finite-size systems, we expect to see a cutoff that corresponds to large events that are limited by the system size itself. Notably, although we expect scale-free avalanches to be a prominent feature of models with absorbing states, a number of other mechanisms to produce them have been proposed, as we will see in Chapter \ref{ch:scirep}.

\section{From stochastic processes to information theory}
Many dynamical models are characterized by master equations, like the contact process in Eq. \eqref{eqn:intro:cp_master_equation}, which describe the stochastic evolution of a population of individuals that follows a given set of transitions. Formally, Eq.~\eqref{eqn:intro:cp_master_equation} is a first-order equation for a probability distribution defined over a set of discrete states, i.e., the probability $p_n(t)$ of having $n$ active sites in the contact process. Each transition enters the master equation via its rate.

However, we are often interested in a more coarse-grained description of our system, such as the evolution of the density of active states described by Eq. \eqref{eqn:intro:cp_mf}. Crucially, Eq. \eqref{eqn:intro:cp_mf} is a deterministic equation and lacks a noise term that accounts for the stochasticity of our typically finite systems. This level of description can be equivalently achieved by either Langevin or Fokker-Planck equations. The former is a stochastic differential equation, whereas the latter is a partial differential equation that described the evolution of a probability density function (pdf). As we will extensively use them throughout this Thesis, in the last part of this introductory Chapter we will briefly recall the formalism of Langevin and Fokker-Planck equation, and introduce some key concepts from Information Theory.

\subsection{Fokker-Planck and Langevin equations}
We consider a generic Markov process described by the transition probability $P(\vb{x}_n, t_n | \vb{x}_{n-1}, t_{n-1})$, which is the probability of moving from a possibly continuous state $\vb{x}_{n-1} \in \mathbb{R}^N$ at time $t_{n-1}$ to state $\vb{x}_n$ at the next timestep, $t_n$. For this stochastic process, one can expand the Chapman-Kolmogorov equation,
\begin{equation}
    \label{eqn:intro:chapman_kolmogorov}
    P(\vb{x}_{n+1}, t_{n+1} | \vb{x}_{n-1}, t_{n-1}) = \int d\vb{x}_n P(\vb{x}_{n+1}, t_{n+1} | \vb{x}_{n}, t_{n})P(\vb{x}_{n}, t_{n} | \vb{x}_{n-1}, t_{n-1}),
\end{equation}
to derive the master equation as
\begin{equation}
    \label{eqn:intro:master_equation}
    \pdv{P(\vb{x},t | \vb{x}_0, t_0)}{t} = \int d\vb{x}' \left[W(\vb{x}|\vb{x}',t)P(\vb{x}', t | \vb{x}_0, t_0) - W(\vb{x}'|\vb{x},t)P(\vb{x}, t | \vb{x}_0, t_0)\right]
\end{equation}
where we introduced the transition rate $W(\vb{x} | \vb{x}') = \lim_{\Delta t \to 0}p(\vb{x},t+\Delta t|\vb{x}',t)/\Delta t$ \cite{gardiner2004handbook}. Eq.~\eqref{eqn:intro:master_equation} is an equation for the probability that the system is in the state $\vb{x}$ at time $t$, given the initial condition $\vb{x}_0$ at the initial time $t_0$. That is, the stochastic process at a time $t$ is described by the probability density function $p(\vb{x}, t) = \int d\vb{x}_0 P(\vb{x},t | \vb{x}_0, t_0) p(\vb{x}_0)$, where $p(\vb{x}_0)$ is the initial distribution of states.

Eq.~\eqref{eqn:intro:master_equation} is an integro-differential equation, and thus not particularly easy to handle. Under suitable conditions \cite{gardiner2004handbook, risken1996fokker, vankampen1992stochastic}, the master equation can be expanded in the form
\begin{equation}
    \label{eqn:intro:kramers_moyal}
    \pdv{P(\vb{x},t | \vb{x}_0, t_0)}{t} = \sum_{k = 1}^\infty \frac{(-1)^k}{k!} \sum_{i_1, \dots, i_k}\frac{\partial^k}{\partial_{i_1} \cdots \partial_{i_k}} \left[a^{(k)}_{i_1, \dots, i_k}(\vb{x},t)P(\vb{x},t | \vb{x}_0, t_0)\right]
\end{equation}
where $\{i_1, \dots, i_k\}$ is a subset of $\{1, \dots, N\}$, and
\begin{equation*}
    a^{(k)}_{i_1, \dots, i_k}(\vb{x},t) = \int d\vb{x}' \prod_{j = i_1}^{i_k}(x_j'-x_j) W(\vb{x}'|\vb{x}).
\end{equation*}
Eq.~\eqref{eqn:intro:kramers_moyal} is called the Kramers-Moyal expansion of the master equation. If we truncate it at the second order, we obtain
\begin{align}
    \label{eqn:intro:fokker_planck}
    \pdv{P(\vb{x},t | \vb{x}_0, t_0)}{t} = & -\sum_{i= 1}^N \pdv{}{x_i}\left[a^{(1)}_i(\vb{x},t)P(\vb{x},t | \vb{x}_0, t_0)\right] + \nonumber \\
    & + \frac{1}{2} \sum_{ij} \frac{\partial^2}{\partial x_i\partial x_j}\left[a^{(2)}_{ij}(\vb{x},t)P(\vb{x},t | \vb{x}_0, t_0)\right]
\end{align}
which is known as the Fokker-Planck equation. Eq.~\eqref{eqn:intro:fokker_planck} describes the evolution for the transition probability in terms of the first two jump moments $a^{(1)}$ and $a^{(2)}$. Typically, we will refer to $P(\vb{x},t | \vb{x}_0, t_0)$ as the ``propagator'' of the system. For example, in the case of the master equation of the mean-field contact process, Eq.~\eqref{eqn:intro:cp_master_equation}, we have
\begin{align*}
\pdv{P(\rho | \rho_0)}{t} = -\pdv{\rho}\left[\left(\rho(\lambda-1)-\lambda\rho^2\right)P(\rho | \rho_0)\right] + \pdv[2]{\rho}\left[\rho\frac{\lambda(1-\rho)+1}{2N}P(\rho | \rho_0)\right]
\end{align*}
where $\rho = n/N$ is the density of active sites. 

An equivalent description of a Markov process is obtained through stochastic differential equations, which we write as
\begin{equation}
    \label{eqn:intro:langevin}
    \dot{x}_i = f_i(\vb{x},t) + \sum_j g_{ij}(\vb{x},t) \xi_j(t)
\end{equation}
where $\xi(t)$ is a stochastic process. We choose $\xi(t)$ to be a white noise, that is, a Gaussian process satisfying
\begin{equation*}
    \begin{gathered}
        \ev{\xi_i(t)} = 0 \\
        \ev{\xi_i(t_1)\xi_j(t_2)} = 2 D \delta_{ij} \delta (t_2 - t_1)
    \end{gathered}
\end{equation*}
where $D$ is the strength of the noise correlation. With this choice, Eq.~\eqref{eqn:intro:langevin} is called a Langevin equation. Its solution is a Markov process, and thus can be cast in terms of a Kramers-Moyal expansion whose jump moments are given by
\begin{equation*}
\begin{gathered}
    a^{(1)}_i(\vb{x}, t) = f_i(\vb{x},t) + D \sum_{jk} g_{jk}(\vb{x}, t) \pdv{g_{ik}(\vb{x}, t)}{x_j} \\
    a^{(2)}_{ij}(\vb{x}, t) = 2D\sum_{k} g_{ik}(\vb{x}, t)g_{kj}(\vb{x}, t)
    \end{gathered}
\end{equation*}
whereas all other vanish \cite{gardiner2004handbook, vankampen1992stochastic}. Thus, we can write a Fokker-Planck equation for a given Langevin equation, and vice-versa. In particular, the term $g_{ij}$ appearing in Eq.~\eqref{eqn:intro:langevin} plays the role of a - possibly time-dependent and inhomogeneous - diffusion coefficient, whereas the first term $f_i$ describes the deterministic evolution of the system in the absence of noise. In the case of the mean-field contact process, for example, we have
\begin{align*}
\dot{\rho}(t) = \rho(t)(\lambda-1)-\lambda\rho^2(t) + \sqrt{\frac{1}{N}\rho(t)\left[\lambda(1-\rho(t))+1\right]} \xi(t)
\end{align*}
and notice that the deterministic part, which survives in the $N\to\infty$ limit, is exactly the mean-field equation, Eq.~\eqref{eqn:intro:cp_mf}, as can be formally understood in terms of a Van Kampen expansion \cite{gardiner2004handbook, vankampen1992stochastic}. In this Thesis, we will deal with both Langevin and Fokker-Planck equations, depending on the context. For instance, trajectories obtained from Langevin equations can be simulated easily (see Appendix \ref{app:computational}), whereas Fokker-Planck equations are more suitable to analytical treatments. Both, however, are crucially involving stochastic quantities and thus probability distributions. In the next section, we will briefly review how in this setting one can introduce concepts from Information Theory, which we will extensively use in what will follow.

\subsection{Entropy, information, and probability}
How is entropy, one of the fundamental concepts of Statistical Physics, related to probability? Roughly speaking, entropy quantifies the disorder of a system, and thus our uncertainty about its (microscopic) configuration. In a microcanonical setting, where an isolated system occupies each of its $\Omega$ microstates with equal probability, the entropy $S$ is given by the expression first derived by Ludwig Boltzmann,
\begin{equation*}
    S = k_B \log \Omega
\end{equation*}
where $k_B$ is the Boltzmann constant. This formula follows from the assumption that the system finds itself in each microstate with a uniform probability $p_i = 1/\Omega$, for all microstates $i$. In thermodynamic systems where microstates are not occupied with equal probability, we have
\begin{equation}
    \label{eqn:intro:gibbs_entropy}
    S = -k_B \sum_i p_i \log p_i
\end{equation}
which is known as Gibbs entropy. In this context $p_i$ may be, for instance in a canonical ensemble, a Boltzmann distribution $p_i \propto \exp(-\beta E_i)$ where $E_i$ is the energy of the $i$-th state.

Eq.~\eqref{eqn:intro:gibbs_entropy} shows us that the concept of entropy is also connected to probability distributions, or, in other words, to uncertainty regarding the state of a system \cite{jaynes1957information, jaynes2003probability}. Decades later, in his pioneering work on communication Shannon \cite{shannon1948mathematical} defined the entropy $H$ of a random variable $X$ taking values in the discrete set $\{x_1, \dots, x_N\}$ with probabilities $\{p_1, \dots, p_N\}$ as
\begin{equation}
    \label{eqn:intro:shannon_entropy}
    H = - \sum_{i = 1}^N p_i \log_2 p_i
\end{equation}
which is formally identical to Eq.~\eqref{eqn:intro:gibbs_entropy}, only missing Boltzmann constant. $H$ is measured in bits - or nats, if the logarithm is in the natural base -, it is always positive, and quantifies the amount of information contained in the random variable $X$, where information here is measuring our uncertainty about the values of $X$ itself. Keeping in mind that $0 \log 0 = 0$, If all values are equally probable, for instance, the entropy is maximal, whereas if only one outcome is possible it vanishes. In this context, one can introduce the information of an outcome $x_i$ as
\begin{equation*}
    \mathcal{I}_i = -\log_2 p_i
\end{equation*}
so that the less likely the event is, the higher information it yields about $X$. Then, $H = \ev{\mathcal{I}}$, where the expectation value $\ev{\cdot}$ is taken over all possible outcomes. In other words, Shannon entropy is a measure of the uncertainty or randomness of a system in terms of the amount of information required to describe its state - and, precisely, it is the expected amount of information associated with a given probability distribution. The more random the outcomes are, the higher the entropy.

It is crucial to note that Eq.~\eqref{eqn:intro:shannon_entropy} is formally defined only for discrete variables. If $X$ is a continuous random variable described by a probability density function $p(x)$, one can write the differential entropy
\begin{equation}
\label{eqn:intro:differential_entropy}
    H_X = -\int dx \, p(x) \log p(x) = \mathbb{E}_{p}\left[-\log p\right]
\end{equation}
where $\mathbb{E}_{p}[\cdot]$ is the expected value operator with respect to the pdf $p$. Differential entropy, however, can be negative, and it is not invariant under reparametrizations of $p$ \cite{ThomasCover2006, jaynes2003probability}. A more suitable quantity for continuous probability distributions is the relative entropy of a distribution $p(x)$ with respect to a distribution $q(x)$, namely
\begin{equation}
\label{eqn:intro:relative_entropy}
    D_{KL}(p || q) = \int dx \, p(x) \log \frac{p(x)}{q(x)} = - H_X + \mathbb{E}_p\left[-\log q\right]
\end{equation}
which quantifies the information we lose when using $q$ to approximate $p$. Eq.~\eqref{eqn:intro:relative_entropy} is always positive due to Jansen's inequality, and it is known as the Kullback-Leibler divergence between $p$ and $q$ \cite{ThomasCover2006, jaynes2003probability, amari2016information}. Informally, one can obtain once more the differential entropy of $p(x)$ by using an unnormalized uniform distribution in place of $q$ - giving us a heuristic interpretation of the differential entropy, which is otherwise ill-defined. 

A key interpretation of the Kullback-Leibler divergence is that of a statistical distance between the probability distributions $p$ and $q$. Although it is not a metric - it does not satisfy the triangle inequality, nor it is symmetric - one can interpret Eq.~\eqref{eqn:intro:relative_entropy} as a measure of the difference between two points in the space of probability distributions\footnote{Notably, a proper Riemannian metric in probability space, the Fisher information metric, can be obtained from an expansion of the Kullback-Leibler divergence \cite{amari2016information}.}. In particular, the Kullback-Leibler divergence measures the amount of additional information that is needed to encode events from one probability distribution using the other. Several such quantities exist - as we will see in more detail in Chapter \ref{ch:PRE} - and all of them measure, in different ways and with different meanings, how ``far'' a probability distribution is from another. Indeed, Eq.~\eqref{eqn:intro:relative_entropy} vanishes if and only if $p$ and $q$ are equal. For instance, a symmetric version can be obtained by symmetrization,
\begin{equation*}
    D_{KL}^\mathrm{sym}(p || q) = \frac{1}{2}\left[D_{KL}(p || q) + D_{KL}(q || p)\right]
\end{equation*}
which is known as the Jensen-Shannon divergence. It is worth noting that finding a closed expression for the integral in Eq.~\eqref{eqn:intro:relative_entropy} is often non-trivial. One notable case in which such expression is known is that of $d$-dimensional multivariate Gaussian distributions $\mathcal{N}(\bm\mu, \bm\Sigma)$,
\begin{align*}
    D_\mathrm{KL}\left(\mathcal{N}(\bm\mu_1, \bm\Sigma_2) \,||\, \mathcal{N}(\bm\mu_2, \bm\Sigma_2)\right) = \frac{1}{2}\biggl[& \log\frac{\det\bm\Sigma_2}{\det\bm\Sigma_1} + \Tr\bm\Sigma_2^{-1}\bm\Sigma_1 - d + \\
    & + \left(\bm\mu_1 - \mu_2\right)^T\bm\Sigma_2^{-1}\left(\bm\mu_1 - \mu_2\right)\biggl],
\end{align*}
which we will extensively use in this Thesis.

The Kullback-Leibler divergence plays a crucial role in Information Theory, as well as Statistical Physics. For instance, one might compare a joint probability distribution between two random variables $X$ and $Y$, $p_{XY}(x,y)$, and its factorization, $p_X(x)p_Y(y)$, where $p_X$ and $p_Y$ are the corresponding marginal distributions. This quantity is known as the mutual information between $X$ and $Y$,
\begin{equation}
    \label{eqn:intro:mutual_information}
    I_{XY} = D_{KL}(p_{XY} || p_Xp_Y) = \int dxdy \, p_{XY}(x,y)\log\frac{p_{XY}(x,y)}{p_X(x)p_Y(y)}.
\end{equation}

\begin{figure}
    \centering

    \begin{tikzpicture}
    \def\radius{2cm}
    \def\mycolorbox#1{\textcolor{#1}{\rule{2ex}{2ex}}}
    \colorlet{colori}{red!50!black}
    \colorlet{colorii}{blue!50!black}
    
    \coordinate (ceni);
    \coordinate[xshift=\radius] (cenii);
    
    \draw[fill=colori,fill opacity=0.15] (ceni) circle (\radius);
    \draw[fill=colorii,fill opacity=0.15] (cenii) circle (\radius);

    \node[xshift=-.9\radius, scale = 1.5] at (ceni) {$H_{X|Y}$};
    \node[xshift=.9\radius, scale = 1.5] at (cenii) {$H_{Y|X}$};
    \node[xshift=0.99\radius, scale = 1.5] at (ceni) {$I_{XY}$};
    \node[xshift=0pt,yshift=-10pt, scale = 1.5] at (current bounding box.south) {$H_{XY}$};
    \node[xshift=-5pt,yshift=55pt, scale = 1.5] at (current bounding box.west) {\textcolor{colori}{$H_{X}$}};
    \node[xshift=5pt,yshift=55pt, scale = 1.5] at (current bounding box.east) {\textcolor{colorii}{$H_{Y}$}};
    \end{tikzpicture}

    \caption{Depiction of the relation between the mutual information between two random variables $X$ and $Y$, $I_{XY}$, their joint entropy $H_{XY}$, the entropies of the marginal distributions, $H_X$ and $H_Y$, and the corresponding conditional entropies, $H_{X|Y}$ and $H_{Y|X}$}
    \label{fig:mutual_information}
\end{figure}
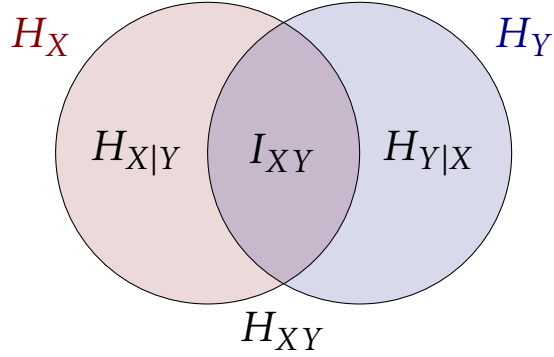

\noindent Mutual information has a straightforward interpretation \cite{ThomasCover2006, jaynes2003probability}. In fact, if $X$ and $Y$ were independent, their probability distribution would be exactly the product $p_X(x)p_Y(y)$. Thus, since we can think of Eq.~\eqref{eqn:intro:mutual_information} as the distance between a joint probability distribution and its factorization, the mutual information quantifies how dependent $X$ and $Y$ are on one another, and vanishes if and only if they are independent. Being always positive and symmetric, $I_{XY}$ measures how much information the two variables share, and this information stems from their overall dependency - e.g., correlations or general higher-order and non-linear relations. More precisely, we can write
\begin{equation*}
    I_{XY} = H_X + H_Y - H_{XY} = H_{XY} - H_{X|Y} - H_{Y|X}
\end{equation*}
where $H_{X|Y}$ is the entropy of the conditional distribution $p_{X|Y}$, as sketched in Figure \ref{fig:mutual_information}. This rewriting tells us that mutual information quantifies the change in information content, as measured by entropy, due to the dependencies between $X$ and $Y$. Indeed, for independent variables, both Shannon entropy and differential entropy are additive, as one would expect from Statistical Physics. As we will see in Chapter \ref{ch:PRL_PRE_1} and Chapter \ref{ch:PRL_PRE_2}, a non-zero mutual information may arise both from observed dependencies - e.g., ``internal'' pairwise interactions between the degrees of freedom - or unobserved ones - e.g., effective dependencies arising from shared environmental changes. One fundamental question, however, is how to generalize Eq.~\eqref{eqn:intro:mutual_information} beyond the two-variable case, where there is more than one way to partition the system. Straightforward extensions to arbitrary numbers of variables suffer from the problem of sometimes being negative \cite{yeung2008information}. Most notably, a generalization first proposed in \cite{williams2010nonnegative}, and known as partial information decomposition, partitions the information between multiple sources and a target into a unique, synergistic, and redundant component. A more recent extension, named O-information, was proposed in \cite{rosas2019quantifying} and takes into account higher-order dependencies by balancing redundancy and synergy. Although we will not make use of such quantities, some results presented in this Thesis may have interesting extensions in this direction \cite{stramaglia2014synergy, faes2017multiscale, camino-pontes2018interaction, marinazzo2019synergy, nuzzi2020synergistic, stramaglia2021quantifying, ehrlich2022partial}.

Finally, as mentioned above, another useful quantity that can be derived from the Kullback-Leibler divergence is the Fisher information matrix $\mathcal{I}_F$ \cite{amari2016information}. Let us write explicitly the parametric dependence of the probability density function $p(\bm x;\bm \vartheta)$ on its parameters $\bm \vartheta = (\vartheta_1, \dots, \vartheta_M)$. If we compute the Kullback-Leibler divergence between $p(\bm x; \vartheta)$ and the pdf resulting from an infinitesimal change of parameters, $p(\bm x;\bm\vartheta + d\bm\vartheta)$, we obtain
\begin{equation*}
    D_{KL} \left[p(\bm x; \bm\vartheta) \,||\, p(\bm x; \bm\vartheta + d\bm\vartheta)\right] = \frac{1}{2} \sum_{ij} d\bm\vartheta_i d\bm\vartheta_j g_{ij}(\bm\vartheta) + \mathcal{O}\left(||d \bm\vartheta ||^3\right)
\end{equation*}
where the Hessian matrix,
\begin{equation}
    \label{eqn:intro:fisher_metric}
    g_{ij}(\bm\vartheta) = \int d\bm{x} \pdv{\log p(\bm x;\bm \vartheta)}{\vartheta_i} \pdv{\log p(\bm x;\bm \vartheta)},{\vartheta_j}
\end{equation}
is known as Fisher information metric for the statistical manifold defined by the family of probability distributions $p(\bm x; \vartheta)$, with coordinates $(\vartheta_1, \dots, \vartheta_M)$. We can rewrite Eq.~\eqref{eqn:intro:fisher_metric} as
\begin{equation}
    \label{eqn:intro:fisher_information}
    \left(\mathcal{I}_F\right)_{ij}(\bm\vartheta) = \int d\bm{x} p(\bm x;\bm \vartheta) \frac{\partial^2 \log p(\bm x;\bm \vartheta)}{\partial \vartheta_i \partial \vartheta_j}
\end{equation}
which we will use in Chapter \ref{ch:PRE}. Heuristically, the Fisher information quantifies how sensible $p(\bm x; \vartheta)$ is to changes in its parameters $\bm\vartheta$ in terms of how much information results from an infinitesimal variation $d\bm\vartheta$.

We will use all of these ideas in different contexts. In particular, computing mutual information will allow us to probe the (pairwise) dependency structure of our stochastic models, understanding how different processes affect the observed degrees of freedom. The Kullback-Leibler divergence and other statistical distances can be used to project one family of models into a different one, as characterized by their possibly time-dependent probability distributions that solve the correspondent Fokker-Planck equation. And, finally, ideas stemming from criticality and phase transitions are powerful tools to understand complex systems with otherwise simple and paradigmatic models, as we will see.

\extraPartText{\lettrine{T}{he} results described in the following part of this Thesis are based upon the published works ``Mutual information disentangles interactions from changing environments'' (\emph{G. Nicoletti, D. M. Busiello, Phys. Rev. Lett. 127, 228301, 2021}) \cite{nicoletti2021mutual}, ``Information-driven transitions in projections of underdamped dynamics'' (\emph{G. Nicoletti, A. Maritan, D. M. Busiello, Phys. Rev. E 106, 0141118, 2022}) \cite{nicoletti2022information}, and ``Mutual information in changing environments: Nonlinear interactions, out-of-equilibrium systems, and continuously-varying diffusivities'' (\emph{G. Nicoletti, D. M. Busiello, Phys. Rev. E 106, 014153, 2022}) \cite{nicoletti2022mutual}.

Parts of the contents presented, including displayed figures, are taken with permission from the published papers \cite{nicoletti2021mutual, nicoletti2022information, nicoletti2022mutual}, copyright (2021-2022) by the American Physical Society.}
\part{Information in stochastic processes and complex systems}

\chapter{Information from unobserved environments}
\label{ch:PRL_PRE_1}
\lettrine{R}{eal-world systems} are usually coupled with noisy, ever-changing environments. In the last twenty years, it was realized that such environments play a crucial role in shaping the properties and dynamics of complex interacting systems. Biological systems \cite{hilfinger2011separating, tsimring2014noise}, biochemical \cite{dass2021furanose} and gene regulatory networks \cite{swain2002extrinsic, thomas2014phenotypic, bowsher2012biochemical}, swarming and oscillatory systems \cite{nosuke2004synchronization, pimentel2008swarming} are only a few examples. Likewise, observed properties believed to be distinctive of neural interactions may be solely explained by an environmental-like dynamics that affects all neurons \cite{touboul2017absence, ferrari2018separating, nicoletti2020scaling, mariani2022disentangling}. Environmental randomness has also been shown to deeply affect the evolution of species in an ecosystem, altering their fixation probabilities and stationary distributions \cite{zhu2009lotkavolterra, wienand2017fluctuating}. Similarly, eco-evolutionary dynamics are deeply affected by possibly sudden and random changes in environmental conditions \cite{wienand2018eco}. Crucially, such randomness may be intrinsically different from periodic, predictable environments \cite{taitelbaum2020periodic}. From a different perspective, crucial non-equilibrium features in chemical systems, such as thermophoresis \cite{piazza2008thermophoresis, liang2021thermophoresis}, and pattern formation \cite{falasco2018turing}, have been recently shown to be sheer consequences of the interplay between environmental and internal interactions acting on different timescales \cite{busiello2020coarsegrained}.

To make things more interesting, an ever-growing wealth of data is populating the realm of biological, chemical, and neural systems, thus fueling the possibility of a direct extrapolation of some properties belonging to the underlying dynamics. In fact, when dealing with experimental data, it is not unusual to solve a given inverse problem, for example using a maximum entropy principle \cite{schneidman2006correlations, mora2010maxent, bialek2012flocks}, to reconstruct the interactions between the internal degrees of freedom that shape the observed behavior. However, one might ask whether these reconstructed couplings may arise from nothing but our ignorance about the unobserved environment in which the system lives. This question is often particularly hard to assess, as effective interactions arise even in non-interacting systems under the influence of a correlated noise \cite{lise1999correlated}.

Hence, describing the coupling to a stochastic environment may be as fundamental as describing those between the internal degrees of freedom. Yet, modeling together these contributions is often a problem too hard to be tackled. Indeed, environmental changes are usually unknown, and the only observed degrees of freedom are the internal ones. Hence, environmental effects are commonly neglected.

\begin{figure*}[t]
    \centering
    \includegraphics[width=\textwidth]{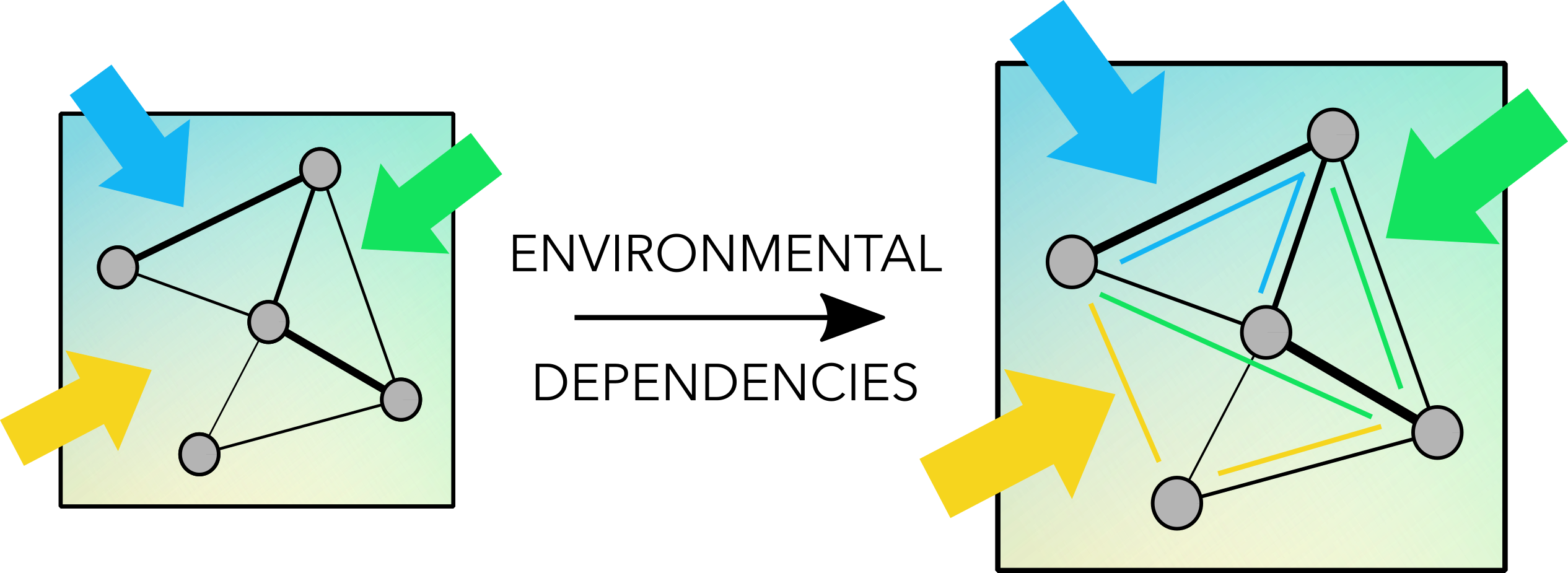}
    \caption{Sketch of a system in a stochastic and typically unobserved environment, e.g., Eqs.~\eqref{eqn:p2c1:internal_langevin_with_env}-\eqref{eqn:p2c1:fokker_planck_with_env}. The internal degrees of freedom (gray dots) with their internal dependencies (black lines) undergo shared environmental changes, affecting their dynamical evolution (colored arrows). As a result, the environment induces new effective dependencies between the internal degrees of freedom (colored lines)}
    \label{fig:p2c1:shared_env_sketch}
\end{figure*}

In this Chapter, we introduce a paradigmatic yet general example of a possibly interacting system under the influence of a shared, but unobserved, stochastic environment. To address the question of understanding how such environments affect the system, we first focus on the case in which internal interactions are not present - i.e., the system's evolution is solely shaped by the environmental dynamics. By computing the mutual information between the internal degrees of freedom, we will explicitly describe the effective dependencies arising from the shared environment in different cases - from discrete environments to continuous ones, in equilibrium and non-equilibrium settings. We will show how a stochastic environment in suitable limits may give rise to non-trivial dependencies in the system, associated with a corresponding environmental information. In particular, in systems placed out-of-equilibrium by the presence of multiplicative noise (e.g., a thermal gradient) such information increases with the magnitude of the non-equilibrium term. Finally, in the presence of continuously varying environments, we will show that the effects of the environment at stationarity can be mapped into a heterogeneous diffusion coefficient, i.e., an effective inhomogeneous medium. The description of these emerging dependencies will help us understand, in the following Chapter, the complex interplay between internal interactions and changing environments.

\section{General formalism}
Consider a system of $N$ particles diffusing in a one-dimensional space. Their positions $\vb{x} = (x_1, \dots, x_N)$ are described by the Langevin equations
\begin{align}
    \label{eqn:p2c1:internal_langevin}
    \frac{d x_\mu}{dt} = F_\mu(\vb{x}; \{\zeta\}) + \sqrt{2 G_\mu\left(\vb{x}; \{\kappa\}\right)} \xi_\mu \qquad\qquad \mu = 1, \dots, N
\end{align}
where $\{\xi_\mu\}$ is a set of independent white noises with zero mean. $F_\mu(\vb{x};\{\zeta\})$ is the $\mu$-th component of a generic force field, which is specified by the set of parameters $\{\zeta\}$. For instance, the case of linear interactions reads $F_\mu = - \sum_\nu A_{\mu \nu} x_\nu$, so that the parameters $\{\zeta\}$ are the elements of the interaction matrix $A_{\mu\nu}$. Similarly, $G_\mu(\vb{x};\{\kappa\})$ is the diffusion coefficient of the $\mu$-th particle, that depends on the set of parameters $\{\kappa\}$.

Eq.~\eqref{eqn:p2c1:internal_langevin} describes the stochastic dynamics of the internal degrees of freedom, $\vb{x}$, of an interacting system. We now assume that this system is coupled to a changing environment, which is shared by all its particles. We model the environment as an independent stochastic process that affects the parameters of the system shared by the $N$ particles. In Eq.~\eqref{eqn:p2c1:internal_langevin}, this corresponds to a process that affects the set of parameters $\{\{\zeta\}, \{\kappa\}\}$, i.e., the environment changes the properties of the model through its parameters, whereas the parametric form itself does not change. Notably, this description corresponds to a large number of compelling scenarios - e.g., a changing carrying capacity in ecosystems \cite{wienand2017fluctuating, wienand2018eco, taitelbaum2020periodic}, a switching environment in a chemical system \cite{dass2021furanose, liang2021thermophoresis}, different strategies in microbial communities \cite{kussell2005phenotypic, visco2010catastrophic} and different regimes of neural activity \cite{touboul2017absence, ferrari2018separating, mariani2022disentangling}, or diffusion in disordered or inhomogeneous media \cite{chechkin2017diffusivities, wang2020randomdiffusivities}. Generalizing the formalism in the case in which different environments correspond to, e.g., different force fields is immediate. A sketch of these ideas is shown in Figure~\ref{fig:p2c1:shared_env_sketch}.

To fix the ideas, consider the case of a finite number $M$ of environmental states indexed by $i = 1, \dots, M$. In this scenario, our system is described by
\begin{align}
    \label{eqn:p2c1:internal_langevin_with_env}
    \frac{d x_\mu}{dt} = F_\mu(\vb{x}; \{\zeta\}_{i(t)}) + \sqrt{2 G_\mu\left(\vb{x}; \{\kappa\}_{i(t)}\right)} \xi_\mu
\end{align}
where $i(t)$ is a realization of the process describing the switch between the $M$ environmental states. Thus, the probability $\pi_i(t) = \mathbb{P}[i(t) = i]$ of being in the $i$-th environmental state at time $t$ is described by the master equation
\begin{equation}
    \label{eqn:p2c1:jump_process}
    \partial_t \pi_i(t) = \sum_{j = 1}^M \left[W_{j \to i} \pi_j(t) - W_{i \to j} \pi_i(t)\right]
\end{equation}
where $W_{i \to j}$ is the transition rate from the $i$-th to the $j$-th environmental state. Notice that this process is independent of all $x_\mu$, hence it indeed plays the role of an environment that affects, but is not affected by, the system. The system itself will jump between states identified by the parameters $\{\{\zeta\}_i, \{\kappa\}_i\}$, so that environmental changes affect all degrees of freedom in the same way. On the other hand, we assume that the internal processes are those determined by the force field $\vb{F}$, which models the interactions between the $N$ degrees of freedom.

The system is described by a joint p.d.f. $p_i(\vb{x}, t)$ to have values $\vb{x}$ at time $t$ and to be in the environmental state $i$. This probability is governed by the Fokker-Planck equation
\begin{align}
    \label{eqn:p2c1:fokker_planck_with_env}
    \partial_t p_i(\vb{x}, t) = & - \sum_{\mu = 1}^N\partial_\mu\left[F_\mu\left(\vb{x};\{\zeta\}_i\right)p_i(\vb{x}, t)\right] + \sum_{\mu= 1}^N\,\partial_\mu^2 \left[G_\mu\left(\vb{x}; \{\kappa\}_i\right)p_i(\vb{x}, t)\right] +\nonumber\\
    & + \sum_{j = 1}^M \left[W_{j \to i} p_j(\vb{x}, t) - W_{i \to j} p_i(\vb{x}, t)\right]
\end{align}
where we used the shorthand notation $\partial_{x_\mu} := \partial_\mu$. Despite its simplicity, general solutions to this model are elusive. Indeed, we are interested in the case in which we regard the environmental changes as unobserved degrees of freedom acting on $\vb{x}$ in the same way. That is, we seek the marginalization over the states of the environment,
\begin{equation}
    \label{eqn:p2c1:marginalization_environment}
    p(\vb x, t) = \sum_i p_i(\vb{x},t).
\end{equation}
Yet, finding even the stationary solution of Eq.~\eqref{eqn:p2c1:fokker_planck_with_env} is a particularly challenging task. Therefore, we resort to a timescale separation approach in which the environment can be either much faster or much slower than all timescales at which the internal dynamics operates.

In what follows, we will often focus on the simple case in which the environment only acts on the diffusion coefficient, i.e., $\vb{F}(\vb{x};\{\zeta\}_i) = \vb{F}(\vb{x};\{\zeta\})$, $\forall i$. For the sake of brevity, we will not write the parametric dependence on $\{\zeta\}$ explicitly when not needed. Furthermore, without loss of generality, we assume that $G_\mu(\vb{x},\{\kappa\}_i) = D_i$, where $D_i$ is a constant diffusion coefficient of the $i$-th environmental state. The calculations can be easily carried out in more general cases. We will eventually relax these conditions and consider systems in the presence of multiplicative noise and continuously varying diffusivities.

\subsection{Fast and slow environments}
\begin{figure*}[t]
    \centering
    \includegraphics[width=1\textwidth]{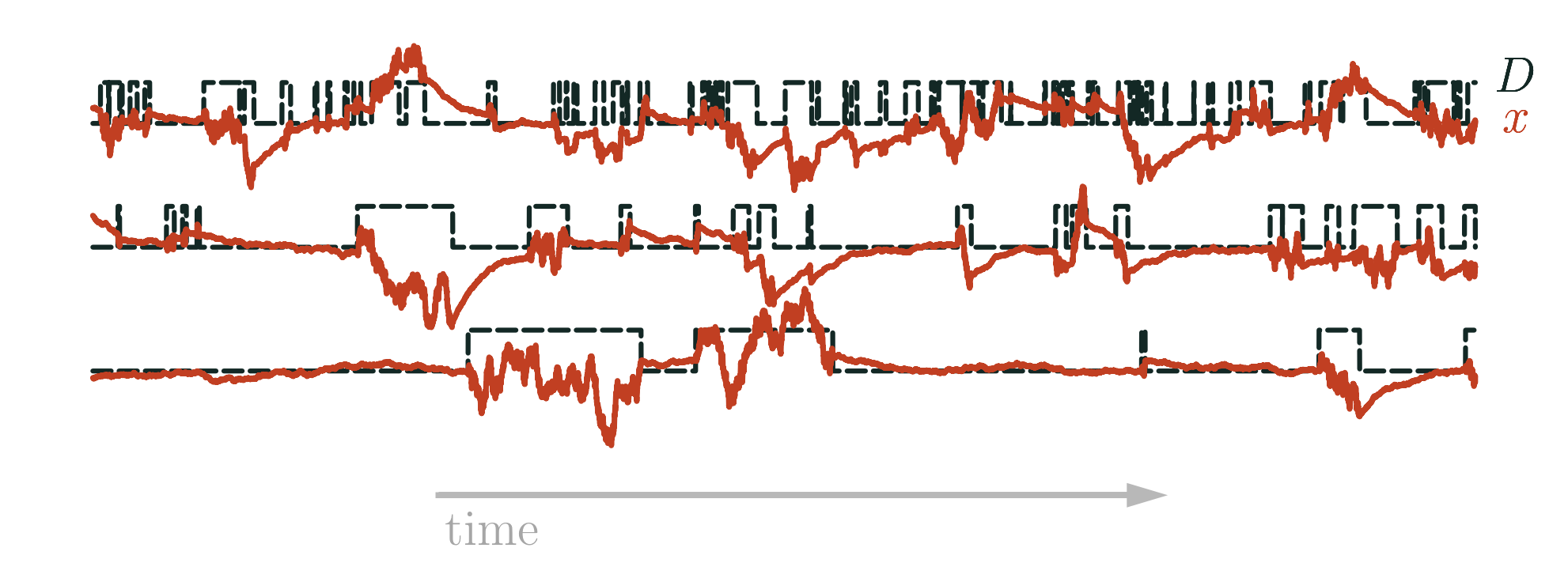}
    \caption{Trajectories of a non-interacting system under harmonic confinement with a switching diffusion coefficient. From top to bottom, the ratio between the typical timescale of the environment and the internal one increases, i.e., the switching rate of the diffusion coefficient becomes slower with respect to the internal relaxation}
    \label{fig:p2c1:trajectories}
\end{figure*}

Let us assume that $\tau_\mathrm{int}$ is the fastest timescale associated with Eq.~\eqref{eqn:p2c1:internal_langevin}, whereas the jump process between the environmental states in Eq.~\eqref{eqn:p2c1:jump_process} occurs on a typical timescale $\tau_\mathrm{env}$. For example, if $M = 2$ we would have $\tau_\mathrm{env} = (W_{1\to2} + W_{2\to1})^{-1}$. In Figure~\ref{fig:p2c1:trajectories} we show the typical trajectories one obtains from a system described by Eq.~\eqref{eqn:p2c1:fokker_planck_with_env} in the absence of interactions and under harmonic confinement, a case which we will study explicitly in Section \ref{sec:p2c1:harmonic}. In particular, notice how such trajectories change as we change the ratio for different ratios $\tau_\mathrm{env}/\tau_\mathrm{int}$.

We first consider the limit $\tau_{\mathrm{env}}/\tau_\mathrm{int} := \epsilon \ll 1$, i.e., the limit in which the environment is much faster than the internal dynamics. We call this case the ``fast-jumps'' limit. We seek a formal solution to Eq.~\eqref{eqn:p2c1:fokker_planck_with_env} of the form
\begin{equation}
    \label{eqn:p2c1:expansion_fast_jumps}
    p_i(\vb{x}, t) = p_i^{(0)}(\vb{x}, t) + \epsilon \, p_i^{(1)}(\vb{x}, t) + \mathcal{O}(\epsilon^2)
\end{equation}
and, in particular, we are interested in the zero-th order steady state marginalized over the environmental states, i.e., $p_\mathrm{fast}(\vb{x}) = \sum_i \lim_{t\to\infty} p_i^{(0)}(\vb{x}, t)$, where the subscript denotes the fast-jumps limit.

To find such a solution, we first rewrite Eq.~\eqref{eqn:p2c1:fokker_planck_with_env} as
\begin{align}
    \label{eqn:p2c1:fast_fokker_planck_rescaled_parameters}
    \partial_t p_i(\vb{x}, t) = & \frac{1}{\tau_\mathrm{int}} \sum_{\mu = 1}^N \biggl[-\partial_\mu\left[\tilde{F}_\mu(\vb{x})p_i(\vb{x}, t)\right] + \partial_\mu^2 \, \left(\tilde{D}_i p_i(\vb{x}, t) \right)  \biggl]+ \nonumber \\
    & + \frac{1}{\tau_\mathrm{env}}\sum_{j = 1}^M \left[\tilde W_{j \to i} p_j(\vb{x}, t) - \tilde W_{i \to j}  p_i(\vb{x}, t)\right]
\end{align}
where $\tilde F_\mu := \tau_\mathrm{int} F_\mu$, $\tilde D_i := \tau_\mathrm{int} D_i$, and $\tilde W_{i \to j} := \tau_{\mathrm{env}} W_{i \to j}$. Then, we rescale the time by the slowest timescale, that is, $t \to t/\tau_\mathrm{int}$. We end up with the rescaled equation
\begin{align}
\label{eqn:p2c1:fast_jumps_expansions_FP}
    \partial_t p_i^{(0)} = \, & \frac{1}{\epsilon}\sum_{j = 1}^M \, \left[\tilde W_{j \to i} p_j^{(0)} - \tilde W_{i \to j} p_i^{(0)}\right] + \nonumber \\
    & + \sum_{\mu = 1}^N \, \biggl[-\partial_\mu\left(\tilde{F}_\mu(\vb x)p_i^{(0)}\right) + \partial_\mu^2 \left(\tilde{D}_i p_i^{(0)} \right) \biggr] + \nonumber\\ 
    & + \sum_{j = 1}^M \, \left[\tilde W_{j \to i} p_j^{(1)} - \tilde W_{i \to j} p_i^{(1)}\right] + \mathcal{O}(\epsilon).
\end{align}
The leading $\epsilon^{-1}$ order in the first row of Eq.~\eqref{eqn:p2c1:fast_jumps_expansions_FP} gives
\begin{equation*}
    0 = \sum_{j = 1}^M \left[\tilde W_{j \to i}  p_j^{(0)}(\vb{x}, t) - \tilde W_{i \to j} p_i^{(0)}(\vb{x}, t)\right]
\end{equation*}
which is the stationary condition of the environmental process alone, Eq.~\eqref{eqn:p2c1:jump_process}. Hence, we can always write the zero-th order solution in the factorized form $p_i^{(0)}(\vb{x}, t) = \pi_i^\mathrm{st} P(\vb{x}, t)$, where
\begin{equation}
    \label{eqn:p2c1:fast_order_0}
    0 = \sum_{j = 1}^M \left[\tilde W_{j \to i} \pi_j^\mathrm{st} - \tilde W_{i \to j} \pi_i^\mathrm{st}\right]
\end{equation}
defines the dependence on the $i$-th index.

Since $\pi_i^\mathrm{st}$ is nothing but the stationary distribution of the environmental states, we immediately find that $p_\mathrm{fast}(\vb{x}) = \sum_i \lim_{t\to\infty} p_i^{(0)}(\vb{x}, t) = \lim_{t\to\infty} P(\vb{x}, t)$. The order $\mathcal{O}(1)$ of Eq.~\eqref{eqn:p2c1:fast_jumps_expansions_FP} reads
\begin{align}
    \label{eqn:p2c1:fast_order_1}
    \partial_t P(\vb{x}, t) = & \sum_{\mu = 1}^N \biggl[-\partial_\mu\left(\tilde{F}_\mu(\vb{x})P(\vb{x}, t)\right) + \left(\sum_i\pi_i\tilde{D}_i\right) \,\partial_\mu^2 \, P(\vb{x}, t) \biggl]
\end{align}
after a summation over the index $i$. Therefore, the stationary joint probability distribution $p_{\mathrm{fast}}(\vb{x})$ that solves Eq.~\eqref{eqn:p2c1:fokker_planck_with_env} is given by the solution of
\begin{align}
    \label{eqn:p2c1:fast_jumps_solution}
    0 = & \sum_{\mu = 1}^N \left[-\partial_\mu\left(F_\mu(\vb x) p_{\mathrm{fast}}(\vb{x})\right) + \left(\sum_i\pi_i D_i\right) \partial_\mu^2 p_{\mathrm{fast}}(\vb{x})\right].
\end{align}
In this fast-jumps limit, as a consequence of the environment, the system feels an effective diffusion coefficient,
\begin{equation}
    \label{eqn:p2c1:fast_diffusion_coefficient}
    D_\mathrm{fast} = \ev{D}_\pi = \sum_i\pi_i D_i,
\end{equation}
which is the stationary ensemble average of all environmental states.

We now consider the opposite limit $\tau_\mathrm{int}/\tau_{\mathrm{env}} = \epsilon^{-1} \ll 1$, i.e., the limit in which the environment is much slower than the internal dynamics. We refer to this case as the ``slow-jumps'' limit. Once again, we seek the formal solution
\begin{equation}
    \label{eqn:p2c1:expansion_slow_jumps}
    p_i(\vb{x}, t) = p_i^{(0)}(\vb{x}, t) + \epsilon^{-1} \, p_i^{(1)}(\vb{x}, t) + \mathcal{O}(\epsilon^{-2})
\end{equation}
and we want to find the zero-th order steady state marginalized over the environmental states, i.e., $p_\mathrm{slow}(\vb{x}) = \sum_i \lim_{t\to\infty} p_i^{(0)}(\vb{x}, t)$. As in the previous case, we rescale the time by the slowest timescale, which is now the environmental characteristic time $\tau_{\mathrm{env}}$. After inserting Eq.~\eqref{eqn:p2c1:expansion_slow_jumps} into Eq.~\eqref{eqn:p2c1:fast_fokker_planck_rescaled_parameters}, we end up with
\begin{align}
\label{eqn:p2c1:slow_jumps_expansions_FP}
    \partial_t p_i^{(0)} = & \, \epsilon \sum_{\mu = 1}^N \, \biggl[-\partial_\mu\left(\tilde{F}_\mu(\vb x)p_i^{(0)}\right) + \partial_\mu^2 \left(\tilde{D}_i p_i^{(0)} \right) \biggr]+ \nonumber\\ 
    & + \sum_{\mu = 1}^N \, \biggl[-\partial_\mu\left(\tilde{F}_\mu(\vb x)p_i^{(1)}\right) + \partial_\mu^2 \left(\tilde{D}_i p_i^{(1)} \right)\biggr]+ \nonumber \\ 
    & + \sum_{j = 1}^M \, \left[\tilde W_{j \to i} p_j^{(0)} - \tilde W_{i \to j} p_i^{(0)}\right] + \mathcal{O}(\epsilon^{-1}).
\end{align}
Once more, the leading order $\mathcal{O}(\epsilon)$ in the first row of Eq.~\eqref{eqn:p2c1:slow_jumps_expansions_FP} corresponds to the stationary solution of the Fokker-Planck equation associated with the fastest dynamics alone. Here, this is equal to the distribution $P_i^{\mathrm{st}}(\vb{x})$ that solves
\begin{equation}
\label{eqn:p2c1:slow_order_0}
    0 = \sum_{\mu = 1}^N \biggl[-\partial_\mu\left(F_\mu(\vb x)P_i^{\mathrm{st}}(\vb{x})\right) + \partial_\mu^2 \left(D_i P_i^{\mathrm{st}}(\vb{x}) \right) \biggr]
\end{equation}
which is nothing but the dynamics at a fixed environmental state $D_i$. Notice that here we are making a slight abuse of notation, writing $P_i^{\mathrm{st}}(\vb{x})$ in place of $P^{\mathrm{st}}(\vb{x}; D_i)$, which explicitly shows that the dependence on the environmental state $i$ only comes through the $i$-th diffusion coefficient $D_i$. Indeed, the functional form of the force field $\vb{F}$ and consequently of $P^{\mathrm{st}}$ is the same for all environmental states. Whenever needed, we will explicitly write the parametric dependence.

Hence, we can write the zero-th order solution of Eq.~\eqref{eqn:p2c1:fast_fokker_planck_rescaled_parameters} in the factorized form $p_i^{(0)}(\vb{x}, t) = \Pi_i(t)P_i^{\mathrm{st}}(\vb{x})$, where the dependence on time only comes through the function $\Pi_i(t)$. Then, after an integration over $\vb{x}$, the order $\mathcal{O}(1)$ gives
\begin{align*}
    \partial_t \Pi_i(t) = \sum_{j = 1}^M \left[\tilde W_{j \to i} \Pi_j(t) - \tilde W_{i \to j} \Pi_i(t)\right]
\end{align*}
which is exactly the master equation governing the environmental dynamics, Eq.~\eqref{eqn:p2c1:jump_process}. Since we are mainly interested in the steady state of $p_i^{(0)}(\vb{x}, t)$, we find that the solution of Eq.~\eqref{eqn:p2c1:fokker_planck_with_env} in the slow-jumps limit is given by
\begin{align}
    \label{eqn:p2c1:solution_slow_jumps}
    p_{\mathrm{slow}}(\vb{x}) := \sum_{i = 1}^M p^{(0)}_i(\vb{x})|_{\epsilon \gg 1}  = \sum_{i = 1}^M \left[\pi_i^\mathrm{st} P_i^{\mathrm{st}}(\vb{x})\right]
\end{align}
where $\pi_i^\mathrm{st}$ are, as before, the stationary probabilities of the jump process alone. Notice that Eq.~\eqref{eqn:p2c1:solution_slow_jumps} is a mixture distribution, where the mixture components are the stationary solutions obtained with a fixed environmental state $i$, $P_i^{\mathrm{st}}(\vb{x})$. As we will see, this is often the most interesting case - hence, we will often work in this limit.

\section{Properties of the mutual information in the slow-jumps limit}
\label{sec:p2c1:mutual_properties}
As we have seen, in the fast jumps limit the stationary distribution corresponds to the one of the interacting system in an average diffusion coefficient, Eq.~\eqref{eqn:p2c1:fast_jumps_solution}. Hence, since the form of the solution does not change, we do not expect the environment to induce any additional dependency between the internal degrees of freedom - as we will see explicitly. This scenario is quite different in the slow-jumps limit, where we find instead the mixture distribution in Eq.~\eqref{eqn:p2c1:solution_slow_jumps}.

For the sake of simplicity, we focus on the case of two particles, so that $\vb{x} = (x_1,x_2)$. A possible generalization to more than two particles can be found in Appendix \ref{app:generalization_jumps}. To quantify the dependency between these internal degrees of freedom, we seek to compute the mutual information
\begin{align}
\label{eqn:p2c1:mutual_information}
    I = \int dx_1 dx_2 \, p_{12}(x_1, x_2) \log \frac{p_{12}(x_1, x_2)}{p_1(x_1) p_2(x_2)}
\end{align}
where $p_{12}$ is the joint p.d.f. of the two particles - e.g. $p_\mathrm{slow}(x_1, x_2)$ - and $p_1$ and $p_2$ are the corresponding marginalizations. Recall that Eq.~\eqref{eqn:p2c1:mutual_information} is nothing but the Kullback-Leibler divergence between the joint probability distribution and its factorization. Therefore, $I$ quantifies how much $x_1$ and $x_2$ depend on one another and vanishes if and only if they are independent. In the case of the probability distribution describing our system, Eq.~\eqref{eqn:p2c1:marginalization_environment}, the mutual information will receive contributions not only from the dependencies due to internal interactions, but also from the ones arising from the marginalization over the unobserved environment.

As a general remark for what follows, let us note that the mutual information $I$ can only depend on dimensionless quantities, since it is itself dimensionless. These, in turn, may depend on environmental features, internal parameters, or combinations of both. To illustrate this crucial detail, consider the case of the slow-jumps limit, Eq.~\eqref{eqn:p2c1:solution_slow_jumps}. Here, the jump rates describing the discrete environment appear only through their dimensionless ratios in $\pi_i^\mathrm{st}$, e.g., $W_{i\to j}/W_{k\to l}$. Hence, the mutual information can only depend on the set of such dimensionless ratios, which we denote with $\{W_\mathrm{jumps}\}$. The remaining parametric dependence appears in the stationary solution of the dynamics at a fixed environment, given by $P_i^\mathrm{st}$. Thus, in principle, $I$ may depend on all dimensionless combinations of the parameters $\{\zeta\}_i$ and $\{\kappa\}_i$ appearing in Eqs.~\eqref{eqn:p2c1:internal_langevin_with_env}-\eqref{eqn:p2c1:fokker_planck_with_env}, for any given environmental state $i$. We call the set of such dimensionless combinations $\{\psi\}_i$. Furthermore, dimensionless combinations may arise by combining parameters from different environmental states, e.g., $D_i/D_j$ in the case of switching diffusion coefficients. We denote the set of these parameters with $\{\varphi\}$. Overall, we end up with
\begin{equation}
    \label{eqn:p2c1:mutual_parametric_dependence}
    I_\mathrm{slow} = I_\mathrm{slow}\left(\{W_\mathrm{jumps}\}, \{\varphi\}, \{\psi\}_{i=1}^M\right)
\end{equation}
where $\{\psi\}_{i=1}^M = \{\{\psi\}_1, \dots, \{\psi\}_M\}$ denotes all the sets of dimensionless combinations at fixed environmental state, for all environmental states.

Crucially, Eq.~\eqref{eqn:p2c1:mutual_parametric_dependence} is the mutual information of a mixture distribution, but it is not necessarily related to that of its components\footnote{Here and further on, for the sake of a simpler notation $P_i^\mathrm{st}$ denotes both the $i$-th component of the joint distribution, $P_i^\mathrm{st}(x_1, x_2)$, and the corresponding marginalizations, $P_i^\mathrm{st}(x_\mu)$ for $\mu = 1,2$.},
\begin{align}
\label{eqn:p2c1:mutual_information_components}
    I_\mathrm{int}(\{\psi\}_i) = \int dx_1 dx_2 \, & P^\mathrm{st}(x_1, x_2;\{\zeta\}_i,\{\kappa\}_i) \times \nonumber \\
    & \times \log \frac{P^\mathrm{st}(x_1, x_2;\{\zeta\}_i,\{\kappa\}_i)}{P^\mathrm{st}(x_1;\{\zeta\}_i,\{\kappa\}_i) P^\mathrm{st}(x_2;\{\zeta\}_i,\{\kappa\}_i)},
\end{align}
where we explicitly wrote the parametric dependence of $P^\mathrm{st}(x_1, x_2;\{\zeta\}_i,\{\kappa\}_i) = P_i^\mathrm{st}(x_1, x_2)$ and its marginalizations.
Yet, $I_\mathrm{int}(\{\psi\}_i)$ is an interesting quantity - it gauges the dependencies due to the internal dynamics, Eq.~\eqref{eqn:p2c1:internal_langevin_with_env}, at a fixed environmental state\footnote{As with $P^{\mathrm{st}}$, since neither $\vb{F}$ nor $\vb{G}$ in Eq.~\eqref{eqn:p2c2:langevin_with_env} change their functional form in different environmental states, we expect $I_\mathrm{int}$ to have the same functional form as well. The $i$-th dependence only comes through the relevant adimensional parameters appearing in the $i$-th set $\{\psi\}_i$. Generalizing to cases in which the internal dynamics changes with $i$ is trivial, as it would result in the explicit dependence $I^i_\mathrm{int}(\{\psi\}_i)$.}. Indeed, it will be useful to introduce its average over the environmental states,
\begin{equation}
    \label{eqn:p2c1:mutual_information_interactions}
    \ev{I_\mathrm{int}}_\pi := \sum_{i = 1}^M \pi_i^\mathrm{st} I_\mathrm{int}(\{\psi\}_i).
\end{equation}
If no internal interactions are present - i.e., if $\vb{F}(\vb{x}) = \sum_\mu F_\mu(x_\mu)$ - $P_i^\mathrm{st}(x_1, x_2)$ is factorizable and thus both $I_\mathrm{int}(\{\psi\}_i)$ and $\ev{I_\mathrm{int}}_\pi$ vanish. As a shorthand notation, we will often write $I_\mathrm{int}^i := I_\mathrm{int}(\{\psi\}_i)$ when the explicit parametric dependence is not needed.

To understand if and how the mutual information of the system living in the stochastic environment, $I$, is related to the internal dependencies, quantified by $I_\mathrm{int}$, let us consider some simple bounds on the Shannon entropy of mixture distributions. Let $H_{12}$ be the joint entropy associated with the mixture distribution $p_{\mathrm{slow}}(x_1,x_2)$, 
\begin{equation}
\label{eqn:p2c1:mixture_entropy_12}
    H_{12} = -\int dx_1dx_2 \, p_\mathrm{slow}(x_1, x_2)\log p_\mathrm{slow}(x_1,x_2),
\end{equation}
and let $H_{12}^i$ be the entropy of its $i$-th component $P_i^{\mathrm{st}}(x_1, x_2)$,
\begin{equation}
\label{eqn:p2c1:mixture_components_entropy_12}
    H^i_{12} = -\int dx_1dx_2 \, P_i^\mathrm{st}(x_1,x_2)\log P_i^\mathrm{st}(x_1,x_2).
\end{equation}
An analytical expression for Eq.~\eqref{eqn:p2c1:mixture_entropy_12} is often elusive, given the mixture nature of $p_\mathrm{slow}$. However, this may not be the case for the entropy in Eq.~\eqref{eqn:p2c1:mixture_components_entropy_12}. In order to bound the former with the latter, let us introduce the joint entropy between the particles and the environment $H_{12, E}$, where the environment is represented by the set of probabilities $\{\pi_i^\mathrm{st}\}$ appearing in Eq.~\eqref{eqn:p2c1:solution_slow_jumps}. Then, the properties of the conditional entropy \cite{ThomasCover2006} allow us to write
\begin{equation*}
    H_{12} \ge H_{12|E}
\end{equation*}
where
\begin{equation*}
    H_{12|E} = \sum_i \pi_i H^i_{12}.
\end{equation*}
Similarly, we can write
\begin{equation*}
    H_{12} \le H_{12,E} = H_{12|E} + H_{E}
\end{equation*}
where
\begin{equation}
\label{eqn:p2c1:Hjumps}
    H_E = -\sum_i \pi_i^\mathrm{st} \log \pi_i^\mathrm{st} := H_\mathrm{jumps}
\end{equation}
is the entropy associated with the steady state of the environmental jump process.

Therefore, we can bound the joint entropy of $p_\mathrm{slow}(x_1,x_2)$ as
\begin{equation}
\label{eqn:p2c1:bound_entropy_12}
    \sum_i \pi_i^\mathrm{st} H_{12}^i \le H_{12} \le \sum_i \pi_i^\mathrm{st} \left[H_{12}^i - \log \pi_i^\mathrm{st}\right].
\end{equation}
Analogous bounds can be cast for $H_{1}$ and $H_2$, i.e., the entropies of the marginal distributions $p_{\mathrm{slow}}(x_{1})$ and $p_{\mathrm{slow}}(x_{2})$, respectively. Then, a lower (upper) bound on the mutual information, $I = H_1 + H_2 - H_{12}$, can be found by taking these lower (upper) bounds on $H_1$ and $H_2$ and the upper (lower) one on $H_{12}$, Eq.~\eqref{eqn:p2c1:bound_entropy_12}. Thus, the mutual information of the mixture distribution is bounded by
\begin{equation}
\label{eqn:p2c1:bound_mutual}
    \ev{I_\mathrm{int}}_\pi - H_{\mathrm{jumps}} \le I \le \ev{I_\mathrm{int}}_\pi + 2 H_{\mathrm{jumps}}.
\end{equation}
Eq.~\eqref{eqn:p2c1:bound_mutual} shows that, at least in principle, the mutual information $I$ between $x_1$ and $x_2$ can be larger than the average mutual information of the internal dynamics, $\ev{I_\mathrm{int}}_\pi$. Furthermore, in a non-interacting system where $I^i_\mathrm{int} = 0$, the two particles may still be dependent - and this dependency is induced by the environment, while being directly connected to its entropy $H_\mathrm{jumps}$.

These bounds can be greatly improved \cite{Kolchinsky2017, nicoletti2021mutual}, provided our ability to compute some suitable information distances both between the components of the mixture distribution in Eq.~\eqref{eqn:p2c1:solution_slow_jumps} and the components of the corresponding marginalizations. Indeed, from \cite{Kolchinsky2017} we can write the estimator
\begin{equation}
\label{eqn:p2c1:pairwise_entropy_estimator}
    \hat{H}_{12} = \sum_{i=1}^M \pi_i^\mathrm{st} H_{12}^i - \sum_{i=1}^M \pi_i^\mathrm{st} \log \left[\sum_{j=1}^M\pi_j^\mathrm{st} e^{-d(P_{i, 12}^{\mathrm{st}} \,||\, P_{j,12}^{\mathrm{st}})}\right]
\end{equation}
where $P_{i,12}^{\mathrm{st}}$ denotes the joint mixture component $P_i^{\mathrm{st}}(x_1, x_2)$ and $d(P_{i, 12}^{\mathrm{st}} \,||\, P_{j,12}^{\mathrm{st}})$ is any generalized distance function in the corresponding probability distributions space. An analogous estimator can be written for $H_1$ and $H_2$, leading to 
\begin{align}
\label{eqn:p2c1:pairwise_mutual_estimator}
    \hat{I} & = \sum_{i=1}^M \pi_i^\mathrm{st} \left[I^i_\mathrm{int} - \log \frac{\left(\sum_{j=1}^M\pi_j^\mathrm{st} e^{-d(P_{i,1}^{\mathrm{st}} \,||\, P_{j,1}^{\mathrm{st}})}\right)\left(\sum_{j=1}^M\pi_j^\mathrm{st} e^{-d(P_{i,2}^{\mathrm{st}} \,||\, P_{j,2}^{\mathrm{st}})}\right)}{\sum_{j=1}^M\pi_j^\mathrm{st} e^{-d(P_{i, 12}^{\mathrm{st}} \,||\, P_{j,12}^{\mathrm{st}})}}\right]
\end{align}
which is our estimator for the mutual information $I$. In particular, Eq.~\eqref{eqn:p2c1:pairwise_entropy_estimator} is a lower bound for the entropy $H_{12}$ when we choose as a distance function the Chernoff-$\alpha$ divergence, defined as
\begin{equation*}
    C_\alpha(p || q) = -\log \int d\vb{x} \,  p^\alpha(\vb{x}) q^{1-\alpha}(\vb{x})
\end{equation*}
for any $\alpha \in [0,1]$ and with $p$ and $q$ probability density functions. Similarly, an upper bound is instead achieved with a Kullback-Leibler divergence
\begin{equation*}
    D_\mathrm{KL}(p || q) = \int d\vb{x}\, p(\vb{x}) \log\frac{p(\vb{x})}{q(\vb{x})}.
\end{equation*}
Therefore, Eq. \ref{eqn:p2c1:pairwise_mutual_estimator} is a lower bound for the mutual information $I$ if we choose the Chernoff-$\alpha$ divergence for the one-variable distances and the Kullback-Leibler divergence for the two-variable distances, and it is an upper bound if we make the opposite choice.

We are now able to bound the mutual information between $x_1$ and $x_2$, at least in the slow-jumps limit. These bounds show explicitly that the dependencies induced by a stochastic environment may go beyond the naive environmental average of the internal ones, i.e., $\ev{I_\mathrm{int}}_\pi$. To find $I$, in general, we rely on a numerical estimation of the mutual information integral, Eq.~\eqref{eqn:p2c1:mutual_information}. Such an estimate can be obtained via a Monte Carlo integration by sampling from the joint distribution, as detailed in Appendix \ref{app:computational}. 

\section{Information from shared environments}
Using the tools introduced in the previous sections, we now address the question of whether and how a changing environment impacts the properties of a system evolving under its influence. In particular, we aim to quantify the dependencies between $x_1$ and $x_2$ through the mutual information $I$. With this in mind, we will first focus on the case in which $x_1$ and $x_2$ are not interacting, i.e., $F_\mu(\vb{x}) = F_\mu(x_\mu)$ for $\mu = 1, 2$. In this scenario, any dependence among the two particles, if any, must stem from the environmental changes they share. We denote with $I_\mathrm{env}$ the related environment-induced mutual information.

For the sake of simplicity, we consider the case in which there are only two environmental states indexed by $i \in \{-, +\}$. The diffusion coefficient follows then a dichotomous process jumping between two states $D_-$ and $D_+ > D_-$, with transition rates $W_{- \to +} = w_+$ and $W_{+ \to -} = w_-$. The state $D_+$ is the state with high diffusion, whereas the state $D_-$ pushes the system in a low-diffusion regime. Generalizing to an arbitrary number of environmental states leads to qualitatively similar results, shown in Appendix \ref{app:generalization_jumps}.

\subsection{Harmonic confinement}
\label{sec:p2c1:harmonic}
Let us begin with a non-interacting system that is confined in a harmonic potential
\begin{equation}
\label{eqn:p2c1:harmonic_potential}
    U(x_\mu) = \frac{x_\mu^2}{2\tau}
\end{equation}
so that the force field appearing in Eqs.~\eqref{eqn:p2c1:internal_langevin_with_env}-\eqref{eqn:p2c1:fokker_planck_with_env} is given by
\begin{align*}
    F_{\mu}(x_\mu) = -\partial_\mu U(x_\mu) = - \frac{x_\mu}{\tau}.
\end{align*}
Here, the internal timescale is simply given by $\tau_\mathrm{int} = \tau$, whereas the environmental one is $\tau_\mathrm{env} = w_\mathrm{sum}^{-1}$ with $w_\mathrm{sum} = w_- + w_+$. Therefore, the relevant dimensionless parameters of this model are: (i) $\tau w_\mathrm{sum}$, which governs the timescale separation between the internal degrees of freedom and the jump process of the environmental states; (ii) $w_-/w_+$, which determines the relative persistence of the environmental states; (iii) $D_-/D_+$, which describes the separation between the environmental states. In particular, the fast- and slow-jumps limits are identified by $\tau w_\mathrm{sum} \gg 1$ and $\tau w_\mathrm{sum} \ll 1$, respectively. Thus, using the notation introduced in Section \ref{sec:p2c1:mutual_properties} we expect the mutual information in these two limits to depend on $\{W_\mathrm{jumps}\} = \{w_-/w_+\}$ and $\{\varphi\} = \{D_-/D_+\}$ only.

We now need to solve the Fokker-Planck equation
\begin{align*}
    \partial_t p_i(\vb{x}, t) = & \sum_{\mu = 1}^2\left[\partial_\mu\left(\frac{x_\mu}{\tau}p_i(\vb{x}, t)\right) + D_i \,\partial_\mu^2 \, p_i(\vb{x}, t)\right] + \sum_{j \in \{-,+\}} \left[w_i  p_j(\vb{x}, t) - w_j  p_i(\vb{x}, t)\right].
\end{align*}
In a fast environment, following Eq.~\eqref{eqn:p2c1:fast_jumps_solution}, we only need to solve
\begin{align}
    \label{eqn:p2c1:harmonic_fokker_planck_fast}
    0 = & \sum_{\mu = 1}^2\left[\partial_\mu\left(\frac{x_\mu}{\tau}p^\mathrm{fast}_{12}(x_1, x_2)\right) + \ev{D}_\pi \,\partial_\mu^2 \, p^\mathrm{fast}_{12}(x_1, x_2)\right]
\end{align}
where $\ev{D}_\pi = (D_+w_+ + D_-w_-)/w_\mathrm{sum}$, and $p^\mathrm{fast}_{12}$ denotes the joint probability between $x_1$ and $x_2$ in the fast-jumps limit. This is nothing but a simple Ornstein-Uhlenbeck process \cite{gardiner2004handbook}, hence we immediately find that 
\begin{align}
\label{eqn:p2c1:harmonic_fast_jumps_pdf}
    p^\mathrm{fast}_{12}(x_1, x_2) = & \frac{1}{2\pi \tau\ev{D}_\pi} \exp[-\frac{x_1^2 + x_2^2}{2\tau \ev{D}_\pi}] \equiv p^\mathrm{fast}_1(x_1)p^\mathrm{fast}_2(x_2)
\end{align}
where $p^\mathrm{fast}_1(x_1) = p^\mathrm{fast}_2(x_2) = \mathcal{N}(0, \tau \ev{D}_\pi)$ is a Gaussian distribution with zero mean and variance $\tau \ev{D}_\pi$. Loosely speaking, this limit describes environmental changes affecting the internal degrees of freedom only on average. Crucially, this leaves the two processes independent of each other, since Eq.~\eqref{eqn:p2c1:harmonic_fast_jumps_pdf} factorizes. Hence, we immediately find that the mutual information induced by the environment, $I_\mathrm{env}$, vanishes in the fast jumps limit,
\begin{equation}
    \label{eqn:p2c1:harmonic_mutual_fast}
    I_\mathrm{env}\bigr|_{\tau w_\mathrm{sum} \gg 1} = 0
\end{equation}
since $D_{KL}(p^\mathrm{fast}_{12} \, || \, p^\mathrm{fast}_1 p^\mathrm{fast}_2) = 0$.

The picture is markedly different in the slow-jumps limit, when $\tau w_\mathrm{sum} \ll 1$. The stationary probability distribution in Eq.~\eqref{eqn:p2c1:solution_slow_jumps} is now a mixture distribution, whose components solve 
\begin{align*}
    0 = \sum_{\mu = 1}^2 \left[\partial_\mu\left(\frac{x_\mu}{\tau}P_i^{\mathrm{st}}(x_1,x_2)\right) + D_i \partial_\mu^2 P_i^{\mathrm{st}}(x_1,x_2)\right]
\end{align*}
from Eq.~\eqref{eqn:p2c1:slow_order_0}. Hence, we immediately the Gaussian mixture 
\begin{align}
\label{eqn:p2c1:harmonic_slow_jumps_pdf_joint}
    p^\mathrm{slow}_{12}(x_1, x_2) & = \frac{1}{2\pi \tau}\left[\frac{\pi^\mathrm{st}_-}{D_-} e^{-\frac{1}{2\tau D_-}\left(x_1^2 + x_2^2\right)} + \frac{\pi^\mathrm{st}_+}{D_+} e^{-\frac{1}{2\tau D_+}\left(x_1^2 + x_2^2\right)}\right] \nonumber \\
    & = \pi^\mathrm{st}_- \mathcal{N}(0, \Sigma_-) + \pi^\mathrm{st}_+\mathcal{N}(0, \Sigma_+)
\end{align}
where $\Sigma_{-(+)} = D_{-(+)}\text{diag}\left(\tau, \tau\right)$. Similarly,
\begin{align}
\label{eqn:p2c1:harmonic_slow_jumps_pdf_marg}
    p^\mathrm{slow}_1(x_1) & = \frac{1}{\sqrt{2\pi\tau}}\left[\frac{\pi^\mathrm{st}_-}{\sqrt{D_-}} e^{-\frac{x_1^2}{2\tau D_-}} + \frac{\pi^\mathrm{st}_+}{\sqrt{D_+}} e^{-\frac{x_1^2}{2\tau D_+}}\right] \nonumber \\
    & = \pi^\mathrm{st}_- \mathcal{N}(0, \tau D_-) + \pi^\mathrm{st}_+\mathcal{N}(0, \tau D_+)
\end{align}
and $p^\mathrm{slow}_2 = p^\mathrm{slow}_1$. Let us note that the joint probabilities we have found do not depend on the dimensionless combinations of parameters only.

\begin{figure*}[t]
    \centering
    \includegraphics[width=1\textwidth]{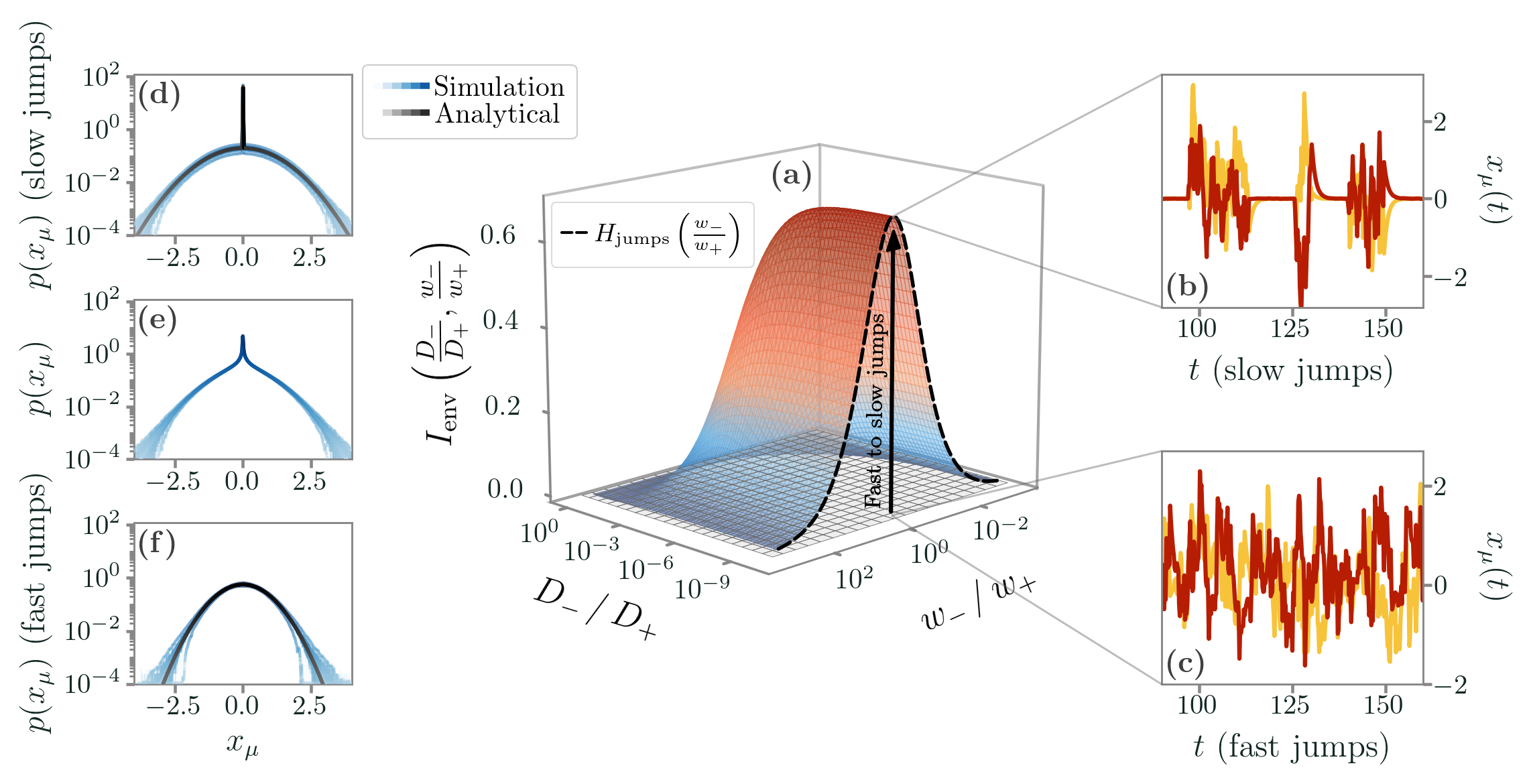}
    \caption{The environmental contribution to the mutual information as a function of $D_-/D_+$ and $w_-/w_+$ in a non-interacting harmonic potential, Eq.~\eqref{eqn:p2c1:harmonic_potential}. (a) The colored surface is the result of a Monte Carlo integration of the mutual information in the slow-jumps limit with $\tau w_\mathrm{sum}= 10^{-3}$. The gray surface is, instead, the vanishing mutual information in the fast-jumps limit $\tau w_\mathrm{sum}= 10^{3}$. In the $D_-/D_+ \to 0$ limit, $I_\mathrm{env}$ becomes exactly $H_\mathrm{jumps}$, the black dashed line, which is also its maximum value. The gray plane is instead the mutual information in the fast-jumps limit, which always vanishes. (b) A realization of $x_1(t)$ and $x_2(t)$ (red and yellow curves) in the slow-jumps limit, at $w_-/w_+ = 1$ and $D_-/D_+ = 10^{-10}$. A bursty, coordinated behavior emerges due to environmental changes. (c) Same, but in the fast-jumps limit, where both variables show a Brownian-like behavior. (d-f) Comparison between the marginalized probability distribution $p(x_\mu)$ from the simulated Langevin dynamics and the analytical distribution ((d) and (f) cases) for $D_- = 10^{-5}$, $D_+ = 1$, $\tau = 1$ in (d) the slow-jumps limit at $w_- = w_+ = 5 \cdot 10^{-4}$,  (f) the fast-jumps limit at $w_- = w_+ = 50$ and (e) in between at $w_- = w_+ = 0.5$}
    \label{fig:p2c1:mutual_extrinsic}
\end{figure*}

It is clear that in this slow environment limit the two processes are not always independent, since in general $p^\mathrm{slow}_{12} \ne p^\mathrm{slow}_1p^\mathrm{slow}_2$. Thus, the unobserved environment is inducing an effective dependency between the two variables $x_1$ and $x_2$, which are otherwise independent. An example of a realization and its corresponding probability distribution is shown in Figure~\ref{fig:p2c1:mutual_extrinsic}b and \ref{fig:p2c1:mutual_extrinsic}d, respectively. In the intermediate regime between the fast- and slow-jumps limits we cannot solve the Fokker-Planck equation explicitly, but a direct simulation of the Langevin equations \cite{gillespie1996OU} shows that the resulting probability interpolates between Eq.~\eqref{eqn:p2c1:harmonic_slow_jumps_pdf_marg} and Eq.~\eqref{eqn:p2c1:harmonic_fast_jumps_pdf} in a smooth fashion, as we see in Figure~\ref{fig:p2c1:mutual_extrinsic}d-f. Therefore, we will now focus on the slow-jumps limit, where we can tackle the problem analytically, and the mutual information takes non-zero values.

Even though no closed form exists for the mutual information of a Gaussian mixture, we can compute the estimator in Eq.~\eqref{eqn:p2c1:pairwise_mutual_estimator} and the associated bounds. Therefore, we are now interested in both the Chernoff-$\alpha$ divergence and the Kullback-Leibler divergence between the components of these Gaussian mixtures. For the two one-dimensional components of Eq.~\eqref{eqn:p2c1:harmonic_slow_jumps_pdf_marg} we have the Chernoff-$\alpha$ divergence
\begin{equation*}
    C_\alpha(\mathcal{N}(0, \tau D_+) || \mathcal{N}(0, \tau D_-)) = \frac{1}{2}\log\frac{(1-\alpha) + \alpha (D_-/D_+)}{(D_-/D_+)^\alpha}
\end{equation*}
which depends only on the ratio $D_-/D_+ := \varepsilon_\pm$. Since we are free to choose $\alpha \in [0,1]$, we take $\partial_\alpha C_\alpha = 0$ so that the Chernoff divergence is minimum. We find
\begin{align*}
    \alpha = \frac{1- \varepsilon_\pm - \varepsilon_\pm\log\varepsilon_\pm}{(\varepsilon_\pm-1)\log\varepsilon_\pm}
\end{align*}
so that
\begin{align}
\label{eqn:p2c1:harmonic_chernoff_1D}
    C(\mathcal{N}(0, \tau D_+) || \mathcal{N}(0, \tau D_-)) & = \frac{1}{2}\left[-1 + \log \frac{(\varepsilon_\pm-1) \varepsilon_\pm^\frac{1}{\varepsilon_\pm -1}}{\log \varepsilon_\pm}\right] \nonumber \\
    & := \frac{1}{2}z(\varepsilon_\pm).
\end{align}
Similarly, for the components of Eq. \ref{eqn:p2c1:harmonic_slow_jumps_pdf_joint} we have 
\begin{equation*}
    C_\alpha(\mathcal{N}(0, \Sigma_+) || \mathcal{N}(0, \Sigma_-)) = \log\frac{(1-\alpha) + \alpha (D_+/D_-)}{(D_+/D_-)^\alpha}
\end{equation*}
and upon optimization over $\alpha$ we find the same result as before up to a factor $1/2$, 
\begin{equation}
\label{eqn:p2c1:harmonic_chernoff_2D}
    C(\mathcal{N}(0, \Sigma_+) || \mathcal{N}(0, \Sigma_-))) = -1 + \log \frac{(\varepsilon_\pm-1) \varepsilon_\pm^\frac{1}{\varepsilon_\pm -1}}{\log \varepsilon_\pm} = z(\varepsilon_\pm).
\end{equation}
Notice that $z(\varepsilon_\pm) = z(1/\varepsilon_\pm)$ implies that the divergences in Eqs.~\eqref{eqn:p2c1:harmonic_chernoff_1D}-\eqref{eqn:p2c1:harmonic_chernoff_2D} are symmetric. We also note that the function $z(\varepsilon_\pm)$ has the following properties:
\begin{equation}
\label{eqn:p2c1:harmonic_z_limits}
    \begin{gathered}
    \lim_{\varepsilon_\pm \to 0} z(\varepsilon_\pm) = +\infty = \lim_{\varepsilon_\pm \to 0} z(1/\varepsilon_\pm) \\
    \lim_{\varepsilon_\pm \to 1} z(\varepsilon_\pm) = 0
    \end{gathered}
\end{equation}
which means that if $D_- \ll D_+$ both Chernoff divergences between the components of the Gaussian mixtures diverge.

We also need to write down explicitly the Kullback-Leibler divergences between the mixture components, which are
\begin{align}
\label{eqn:p2c1:harmonic_kl_1D}
    D_\mathrm{KL}(\mathcal{N}(0, \tau D_+) \,||\, \mathcal{N}(0, \tau D_-)) = \frac{1}{2}\left[\frac{1-\varepsilon_\pm}{\varepsilon_\pm}+\log\varepsilon_\pm\right] := \frac{1}{2} \, h(\varepsilon_\pm)
\end{align}
and
\begin{align}
\label{eqn:p2c1:harmonic_kl_2D}
    D_\mathrm{KL}(\mathcal{N}(0, \Sigma_+) \,||\, \mathcal{N}(0, \Sigma_-)) = \frac{1-\varepsilon_\pm}{\varepsilon_\pm}+\log\varepsilon_\pm = h(\varepsilon_\pm).
\end{align}
These distances are not symmetric anymore, but the function $h(\varepsilon_\pm)$ is such that
\begin{equation}
\label{eqn:p2c1:harmonic_h_limits}
    \begin{gathered}
    \lim_{\varepsilon_\pm \to 0} h(\varepsilon_\pm) = +\infty = \lim_{\varepsilon_\pm \to 0} h(1/\varepsilon_\pm) \\
    \lim_{\varepsilon_\pm \to 1} h(\varepsilon_\pm) = 0
    \end{gathered}
\end{equation}
so the limit $D_- \ll D_+$ is, perhaps unsurprisingly, the limit in which the distances between the mixture components diverge.

Overall, the upper and lower bounds on the environmental mutual information are respectively given by\vspace*{-0.3cm}
\begin{align}
\label{eqn:p2c1:harmonic_mutual_bounds}
    I_\mathrm{env}^\mathrm{slow,up}\left(\frac{D_-}{D_+}, \frac{w_-}{w_+}\right) = & -\pi^\mathrm{st}_+\log\frac{\left[\pi^\mathrm{st}_+ + \pi^\mathrm{st}_- e^{-\frac{h(D_-/D_+)}{2}}\right]^2}{\pi^\mathrm{st}_++\pi^\mathrm{st}_- e^{-z(D_-/D_+)}} + \nonumber \\
    & -\pi^\mathrm{st}_-\log\frac{\left[\pi^\mathrm{st}_+ e^{-\frac{h(D_+/D_-)}{2}} + \pi^\mathrm{st}_-\right]^2}{\pi^\mathrm{st}_+ e^{-z(D_+/D_-)} + \pi^\mathrm{st}_-} \nonumber \\
    I_\mathrm{env}^\mathrm{slow,low}\left(\frac{D_-}{D_+}, \frac{w_-}{w_+}\right) = & -\pi^\mathrm{st}_+\log\frac{\left[\pi^\mathrm{st}_+ + \pi^\mathrm{st}_- e^{-\frac{z(D_-/D_+)}{2}}\right]^2}{\pi^\mathrm{st}_++\pi^\mathrm{st}_- e^{-h(D_-/D_+)}} + \nonumber \\
    & -\pi^\mathrm{st}_-\log\frac{\left[\pi^\mathrm{st}_+ e^{-\frac{z(D_+/D_-)}{2}} + \pi^\mathrm{st}_-\right]^2}{\pi^\mathrm{st}_+ e^{-h(D_+/D_-)} + \pi^\mathrm{st}_-}
\end{align}
and they only depend on the dimensionless ratios $D_-/D_+$ and $w_+/w_-$. Notice that, since the particles are not interacting, the first term of Eq.~\eqref{eqn:p2c1:pairwise_mutual_estimator}, which depends on $I^\pm$, vanishes due to the fact that the mixture components are always factorizable.

Crucially, although both bounds in Eq.~\eqref{eqn:p2c1:harmonic_mutual_bounds} are not tight in general, from Eqs.~\eqref{eqn:p2c1:harmonic_z_limits}-\eqref{eqn:p2c1:harmonic_h_limits} we see that they do saturate in the limits $D_-/D_+ \to 0$ or $D_-/D_+ \to 1$. These limits are particularly significant. The former corresponds to drastic environmental changes, which lead to markedly different dynamics \textcolor{black}{and give rise to a bursty, seemingly coordinated behavior of the internal degrees of freedom. The latter, on the other hand,} describes the trivial case in which $D_-$ and $D_+$ are very similar and thus {\color{black}environmental} changes are effectively negligible. We end up with 
\begin{align}
\label{eqn:p2c1:harmonic_environment_mutual_limits}
    I^\mathrm{slow}_\text{env}\left(\frac{w_-}{w_+}\right) =
    \begin{cases}
        H_\mathrm{jumps}  &\, \text{if} \quad D_+ \gg D_- \\
        \,0 &\, \text{if} \quad D_+ \approx D_-
    \end{cases}
\end{align}
where $H_\mathrm{jumps} = -\pi^\mathrm{st}_+ \log \pi^\mathrm{st}_+ -\pi^\mathrm{st}_- \log \pi^\mathrm{st}_-$ is the environmental entropy, defined in Eq.~\eqref{eqn:p2c1:Hjumps}. Eq.~\eqref{eqn:p2c1:harmonic_environment_mutual_limits} gives us the exact limits of the mutual information in the slow-jumps regime. Clearly, when $D_-/D_+ \to 1$, the dynamics is insensitive to the environment, thus $x_1$ and $x_2$ are independent processes. Instead, and interestingly, the first line is nothing but the Shannon entropy of the jump distribution, $H_\mathrm{jumps}(w_-/w_+)$. A Monte Carlo integration of the mutual information integral, Eq.~\eqref{eqn:p2c1:mutual_information}, shows that $H_\mathrm{jumps}$ is also the maximum value of the mutual information that emerges due to the environment, see Figure~\ref{fig:p2c1:mutual_extrinsic}a.

This result has a quite clear intuitive interpretation, even though the bounds in Eq.~\eqref{eqn:p2c1:bound_mutual} would allow the mutual information to be larger. In fact, from an information-theoretic point of view, $H_\mathrm{jumps}$ quantifies precisely the information lost once we integrate out the stochastic environment, i.e. our ignorance about the system as a whole. Hence, the dependency that the unobserved environment is inducing between $x_1$ and $x_2$ is at most equal to the entropy associated with the jump process describing the environmental changes.

\subsection{Non-linear confining potentials}
\label{sec:p2c1:nonlinear}
Are the results shown in the previous section strictly related to the presence of harmonic confinement? We can now generalize them to a non-interacting system in a non-linear confinement. We consider a generic non-linear relaxation stemming from the potential
\begin{equation}
    \label{eqn:p2c1:nonlinear_potential}
    U(x_\mu) = \frac{x_\mu^{2n}}{2 n \tau},
\end{equation}
where $n$ is a positive integer. In this scenario, the internal timescale is proportional to $\tau$, which has the dimension of a  - that is, we need to solve the rescaled Fokker-Planck equation
\begin{align*}
    \partial_t p_i(\vb{x}, t) = & \frac{1}{\tau_\mathrm{int}}\sum_{\mu = 1}^2\left[\partial_\mu\left(\frac{x_\mu^{2n - 1}}{\rho^n}p_i(\vb{x}, t)\right) + \tau_\mathrm{int} D_i \,\partial_\mu^2 \, p_i(\vb{x}, t)\right] \\
    & + \frac{1}{\tau_\mathrm{env}}\sum_{j \in \{-,+\}} \left[\frac{w_i}{w_\mathrm{sum}}  p_j(\vb{x}, t) - \frac{w_j}{w_\mathrm{sum}}  p_i(\vb{x}, t)\right]
\end{align*}
where $\rho = \sqrt[n]{\tau/\tau_\mathrm{int}}$ is a characteristic length-scale associated with the potential in Eq.~\eqref{eqn:p2c1:nonlinear_potential}. As before, the timescale of the environment is $\tau_\mathrm{env} = w_\mathrm{sum}^{-1}$ and, in the fast- and slow-jumps limit, the only dimensionless parameters are given by $w_-/w_+$ and $D_-/D_+ = \varepsilon_\pm$.

Similarly to the harmonic case, the fast-jumps limit leads to the Fokker-Planck equation
\begin{equation*}
    0 = \sum_{\mu = 1}^2\left[\partial_\mu\left(\frac{x_\mu^{2n-1}}{\tau}p^\mathrm{fast}_{12}(x_1, x_2)\right) + \ev{D}_\pi \,\partial_\mu^2 \, p^\mathrm{fast}_{12}(x_1, x_2)\right]
\end{equation*}
whose solution is trivially factorizable. Yet again, the environmental changes affect the internal degrees of freedom only on average - leaving the two processes independent of each other. Therefore, the mutual information induced by the environment vanishes in the fast-jumps limit even in the presence of non-linear confinements.

In the slow-jumps limit, for the mixture distribution in Eq.~\eqref{eqn:p2c1:solution_slow_jumps}, the picture is clearly different. The components of the joint distribution that solve Eq.~\eqref{eqn:p2c1:slow_order_0} can be written as
\begin{equation}
\label{eqn:p2c1:nonlinear_components_joint}
    P_i^{\mathrm{st}}(x_1, x_2) = \frac{(2 D_i n \tau)^{-1/n}}{4\Gamma^2\left(1 + \frac{1}{2n}\right)}\exp \left[-\frac{x_1^{2n}+x_2^{2n}}{2 D_i n \tau}\right],
\end{equation}
and their marginalization reads
\begin{equation}
\label{eqn:p2c1:nonlinear_components_marg}
    P_i^{\mathrm{st}}(x_\mu) = \frac{n}{\Gamma\left(\frac{1}{2n}\right)(2 D_i n \tau)^{1/2n}}\exp\left[-\frac{x_\mu^{2n}}{2 D_i n \tau}\right]
\end{equation}
with $\Gamma(\cdot)$ the Gamma function. From these expressions, we can write the mixture distributions $p^{\mathrm{slow}}_{12}(x_1,x_2) = \sum_{i}\pi^\mathrm{st}_iP_i^{\mathrm{st}}(x_1, x_2)$ and $p^{\mathrm{slow}}_{12}(x_\mu) = \sum_{i}\pi^\mathrm{st}_i P_i^{\mathrm{st}}(x_\mu)$, for $i \in \{-,+\}$.

In order to derive the bounds from the estimator in Eq.~\eqref{eqn:p2c1:pairwise_mutual_estimator}, we first compute the Chernoff-$\alpha$ divergence between the components of the mixture distribution. For the sake of simplicity and without loss of generality, being free to choose $\alpha \in [0,1]$, we set $\alpha = 1/2$, so that the expression
\begin{equation*}
    C_{1/2}(P_-^{\mathrm{st}}, P_+^{\mathrm{st}}) = -\log \int d\vb{x} \sqrt{P_-^{\mathrm{st}}(\vb{x})P_+^{\mathrm{st}}(\vb{x})}
\end{equation*}
can be easily computed analytically. We find
\begin{equation}
\begin{gathered}
    C^{\mathrm{joint}}_{1/2}(\varepsilon_\pm) = -\frac{1}{n}\log\left[\frac{2\sqrt{\varepsilon_\pm}}{1 + \varepsilon_\pm}\right] \\
    C^{\mathrm{marg}}_{1/2}(\varepsilon_\pm) = \frac{1}{2}\,C^{\mathrm{joint}}_{1/2}(P_-^{\mathrm{st}}, P_+^{\mathrm{st}})
\end{gathered}
\end{equation}
for Eq.~\eqref{eqn:p2c1:nonlinear_components_joint} and Eq.~\eqref{eqn:p2c1:nonlinear_components_marg}, respectively. Similarly,
\begin{equation}
\begin{gathered}
    D^{\mathrm{joint}}_{\mathrm{KL}}(\varepsilon_\pm) = \frac{\varepsilon_\pm - 1 - \log\varepsilon_\pm}{n} \\
    D^{\mathrm{marg}}_{\mathrm{KL}}(\varepsilon_\pm) = \frac{1}{2}\,D^{\mathrm{joint}}_{\mathrm{KL}}(P_-^{\mathrm{st}} || P_+^{\mathrm{st}})
\end{gathered}
\end{equation}
are the Kullback-Leibler divergences between the $-$ components and the $+$ components.
Then, the lower bound reads
\begin{align}
    I_{\mathrm{env}}^{\mathrm{low}} = &  -\pi^\mathrm{st}_+ \log\frac{\left(\pi^\mathrm{st}_+ + e^{-C^{\mathrm{marg}}_{1/2}(\varepsilon_\pm)}\pi^\mathrm{st}_-\right)^2}{\pi^\mathrm{st}_+ + e^{-D^{\mathrm{joint}}_{\mathrm{KL}}(\varepsilon_\pm)}\pi^\mathrm{st}_-} -\pi^\mathrm{st}_- \log\frac{\left(e^{-C^{\mathrm{marg}}_{1/2}(1/\varepsilon_\pm)}\pi^\mathrm{st}_+ + \pi^\mathrm{st}_-\right)^2}{e^{-D^{\mathrm{joint}}_{\mathrm{KL}}(1/\varepsilon_\pm)}\pi^\mathrm{st}_+ + \pi^\mathrm{st}_-}
\end{align}
and the upper bound is
\begin{align}
    I_{\mathrm{env}}^{\mathrm{up}} = &  -\pi^\mathrm{st}_+ \log\frac{\left(\pi^\mathrm{st}_+ + e^{-D^{\mathrm{marg}}_{\mathrm{KL}}(\varepsilon_\pm)}\pi^\mathrm{st}_-\right)^2}{\pi^\mathrm{st}_+ + e^{-C^{\mathrm{joint}}_{1/2}(\varepsilon_\pm)}\pi^\mathrm{st}_-} -\pi^\mathrm{st}_- \log\frac{\left(e^{-D^{\mathrm{marg}}_{\mathrm{KL}}(1/\varepsilon_\pm)}\pi^\mathrm{st}_+ + \pi^\mathrm{st}_-\right)^2}{e^{-C^{\mathrm{joint}}_{1/2}(1/\varepsilon_\pm)}\pi^\mathrm{st}_+ + \pi^\mathrm{st}_-}.
\end{align}
Crucially, notice that both $C^{\mathrm{joint}}_{1/2}(\varepsilon_\pm)$ and $D^{\mathrm{joint}}_{\mathrm{KL}}(\varepsilon_\pm)$ diverge as $-\log\varepsilon_\pm$ when $\varepsilon_\pm \to 0$. Thus, as before, the upper and lower bounds on $I_{\mathrm{env}}$ converge to $H_{\mathrm{jumps}}$ in this limit, implying that
\begin{align}
\label{eqn:p2c1:nonlinear_mutual_env_limits}
    I_\text{env}\left(\frac{D_-}{D_+}, \frac{w_-}{w_+}\right) =
    \begin{cases}
        H_{\mathrm{jumps}} &\, \text{if} \quad D_-/D_+ \ll 1 \\
        \,0 &\, \text{if} \quad D_-/D_+ \approx 1
    \end{cases},
\end{align}
exactly as in the harmonic case. That is, the non-linearities of the potential do not change the fact that, when the two environmental states are infinitely separated, the dependency induced by the unobserved environment is exactly equal to its entropy, $H_\mathrm{jumps}$.

\begin{figure*}[t]
    \centering
    \includegraphics[width=\textwidth]{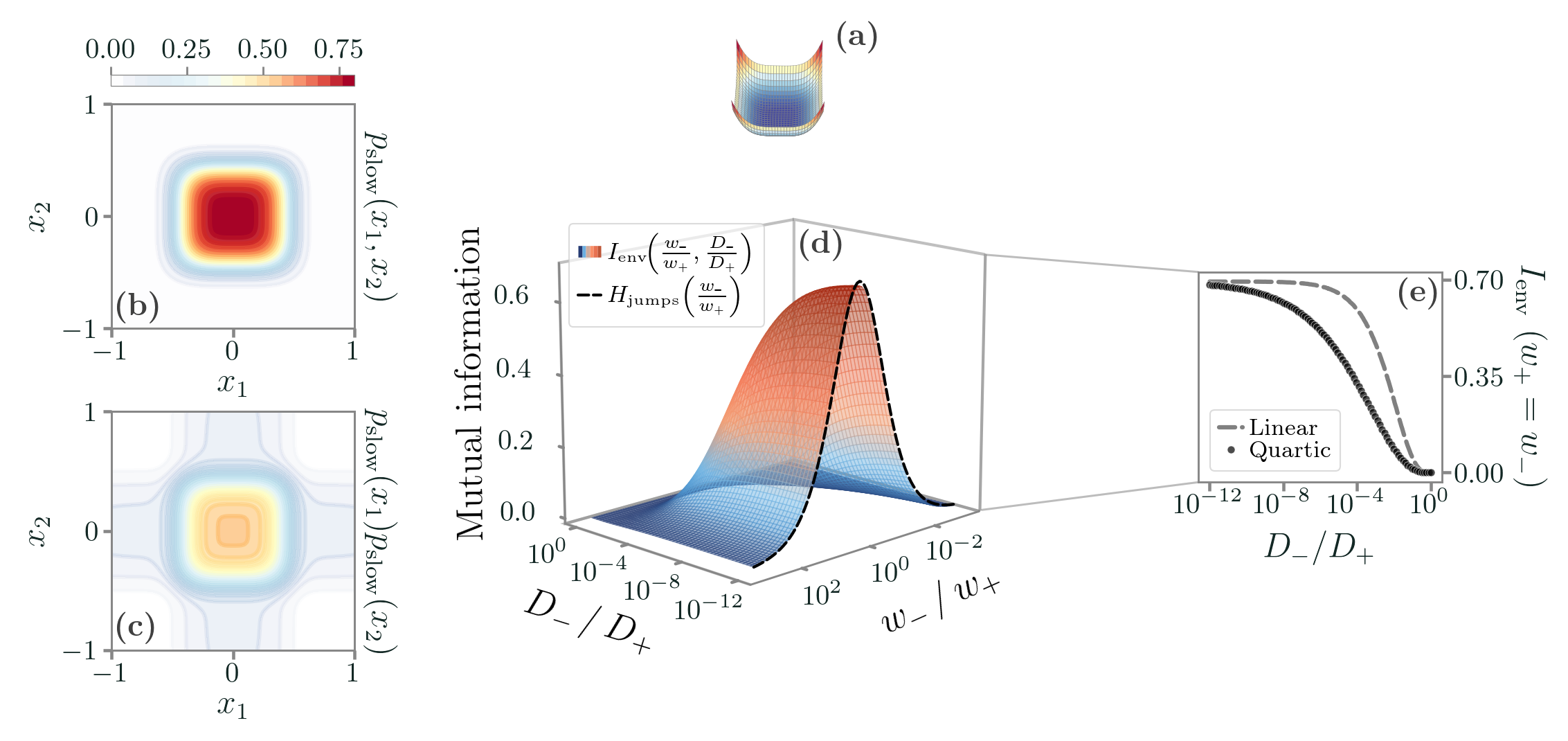}
    \caption{The environmental contribution to the mutual information as a function of $D_-/D_+$ and $w_-/w_+$ in a quartic potential and in the slow-jumps limit. For all plots, $\tau = 1$. (a) The quartic potential considered here. (b-c) Contour plots of the joint probability distribution and its factorization, respectively, for $D_+ = 10$, $D_- = 10^{-2}$, $w_- = w_+$. Notice that the marginalized probability has much longer tails along the axis $x = 0$ and $y = 0$. (d) The colored surface is the result of a Monte Carlo integration of the mutual information. In the $D_-/D_+ \to 0$ limit, $I_\mathrm{env}$ approaches $H_\mathrm{jumps}$, the black dashed line, which is also its maximum value. (e) Compared with the linear case, the non-linear relaxation reflects into a considerably slower convergence towards $H_\mathrm{jumps}$ of the mutual information}
    \label{fig:p2c1:mutual_quartic}
\end{figure*}

Let us show explicitly the results with $n=2$ in Eq.~\eqref{eqn:p2c1:nonlinear_potential}. In the slow-jumps limit, the mixture components read
\begin{gather*}
\label{eqn:p2c1:quartic_mixture_joint}
    P_i^{\mathrm{st}}(x_1, x_2) = \sqrt{\frac{D_i \tau \pi^2}{2}}\frac{\Gamma\left(\frac{3}{4}\right)}{\Gamma\left(\frac{1}{4}\right)} \exp\left(-\frac{x_1^4+x_2^4}{4 D_i \tau}\right), \qquad P_i^{\mathrm{st}}(x_\mu) = \frac{\Gamma\left(\frac{3}{4}\right)}{\pi (D_i \tau)^{1/4}} e^{-\frac{x_\mu^4}{4 D_i \tau}}.
\end{gather*}
Let us remark that $P_i^{\mathrm{st}}(x_1,x_2) = P_i^{\mathrm{st}}(x_1)P_i^{\mathrm{st}}(x_2)$ since the particles are not interacting. In Figure~\ref{fig:p2c1:mutual_quartic}b-c we plot the joint mixture distribution, defined in Eq.~\eqref{eqn:p2c1:solution_slow_jumps}, and its factorization,
\begin{equation*}
     \prod_{\mu=1}^2 p_{\mathrm{slow}}(x_\mu) = \prod_{\mu=1}^2 \Bigg( \sum_{i = \{+,-\}} \left[\pi_i P_i^{\mathrm{st}}(x_\mu)\right] \Bigg).
\end{equation*}
Notably, the effects of the environment on the joint distribution, with respect to the factorized one, reflect into a suppression of the tails along the axes. Indeed, the shared diffusion coefficient implies that both $x_1$ and $x_2$ experience either a high- or low-diffusion regime at the same time. Thus, the probability that one particle has diffused away from the potential minima at the origin, while the other remains close to it, is significantly lower. This is exactly the dependency induced by the environment - although the particles are not directly interacting, they are not exploring the $(x_1, x_2)$ space independently due to the shared changes they experience.

In Figure~\ref{fig:p2c1:mutual_quartic}d, we plot the mutual information in this slow-jumps limit, computed via Monte Carlo sampling \cite{landau2021guide}. In particular, we sample the components of the joint distribution starting from the potential in Eq.~\eqref{eqn:p2c1:nonlinear_potential} via Hamiltonian Monte Carlo \cite{neal2011mcmc, betancourt2017conceptual} (see Appendix \ref{app:computational} for more details). Then, each component is weighted according to the stationary distribution of the environment, for any given $w_-/w_+$, to obtain samples of Eq.~\eqref{eqn:p2c1:solution_slow_jumps}. In Figure~\ref{fig:p2c1:mutual_quartic}e we see that the convergence rate of the mutual information to $H_\mathrm{jumps}$ in the limit $D_-/D_+\to0$ is slower than the one obtained in the linear regime. This holds in general - at a given value of $D_-/D_+$, the mutual information due to the environment is typically smaller than in the case of a linear relaxation. This is perhaps unsurprising since the non-linear relaxation increases the typical auto-correlation timescale and thus reduces the impact of environmental changes.

Overall, we have shown that, for any potential of the form $U(x_\mu) \propto x_\mu^{2n}$, the bounds obtained from Eq.~\eqref{eqn:p2c1:pairwise_mutual_estimator} always saturate to $H_\mathrm{jumps}$ when $D_-/D_+ \to 0$ and vanish when $D_- \to D_+$. This result remarks that, when the variability of the environment is maximal, any two non-interacting degrees of freedom share exactly the information described by Shannon entropy associated with the external jump process, $H_{\mathrm{jumps}}$. Importantly, in all these non-interacting cases, the only dimensionless parameters we can build are $w_-/w_+$, which determines the persistence of the two environmental states, and $D_-/D_+$, describing how similar the environmental states are. Although the probability distributions in Eq.~\eqref{eqn:p2c1:nonlinear_components_marg} and Eq.~\eqref{eqn:p2c1:nonlinear_components_joint} do not depend only on such combinations, the mutual information does. We will see how this scenario is changed in the presence of internal interactions in the next Chapter.

\section{Out-of-equilibrium systems}
So far, we have been investigating non-interacting systems coupled with an ever-changing environment that eventually relaxes to equilibrium. However, nature usually operates out-of-equilibrium, and most of the environments of biochemical, neural, and ecological systems are in non-equilibrium conditions. Hence, before introducing internal interactions to study their interplay with changing environments, we focus yet again on a non-interacting case that is however out-of-equilibrium. We consider the dynamics 
\begin{equation}
    \dot{x}_\mu = - \frac{1}{\tau} x_\mu +  \sqrt{2 \gamma_{i(t)} T(x_\mu)} \xi_\mu
\end{equation}
where $i(t)$ is a realization of the stochastic process governing the environment, $\mu = 1, 2$, and $T(x_\mu) = T_0 + x_\mu \Delta T$ is a linear temperature gradient. This picture captures key features of a diffusing molecule that can live in two conformational states \cite{barducci2015non, gaspari2007aggregation, busiello2021dissipation, liang2021thermophoresis}. Alternatively, it can be a simple way to describe proteins in an environment with patches of different densities - e.g., liquid condensates \cite{weber2019physics, hyman2014liquid} - subject to an external gradient. Since diffusion and temperature are connected by the Einstein relation,
\begin{equation*}
    D_i(x_\mu) \propto \gamma T_i(x_\mu),
\end{equation*}
the environment may act as a modification of the viscosity, in the case of patches of different densities, or the motility, when the switching describes two different conformational states. These diffusive properties are encoded into $\gamma_{i(t)}$ that can take two values, $\gamma_-$ and $\gamma_+$, replacing the role of $D_-$ and $D_+$ of the previous models.

\begin{figure}[t]
    \centering
    \includegraphics[width=\textwidth]{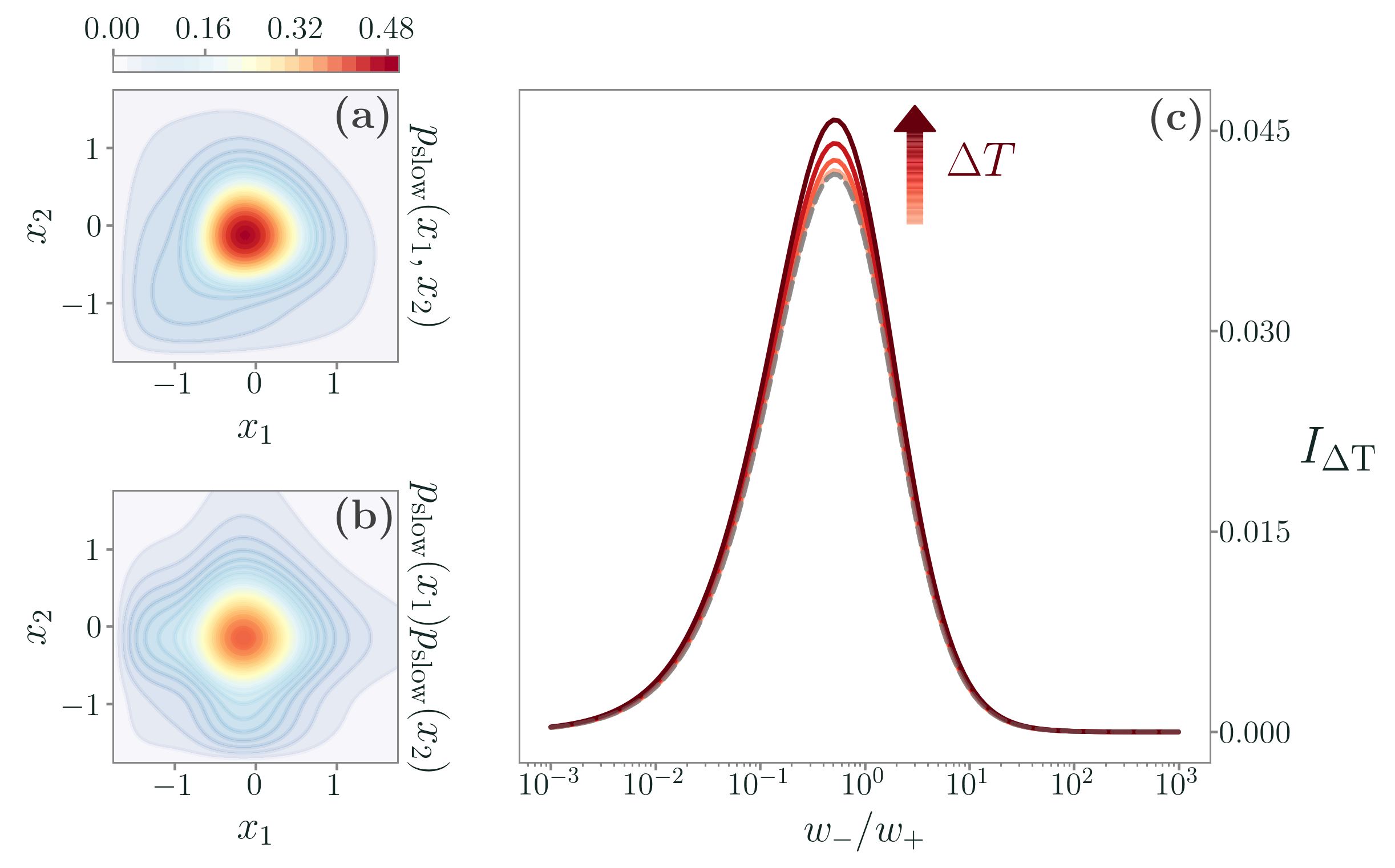}
    \caption{Mutual information in the presence of a linear temperature gradient. (a-b) The joint and the marginal probability distributions in the slow-jumps limit. (c) As we increase the strength of the temperature gradient $\Delta T$, the system is farther from equilibrium, and the mutual information induced by the environment increases (the gray dashed line is the baseline equilibrium value). Here, the different curves are for $\Delta T = \{0.15, 0.29, 0.43, 0.57\}$, $T_0 = \tau = 1$, $\gamma_+ = 1.5$ and $\gamma_- = 0.5$}
    \label{fig:p2c1:multiplicative}
\end{figure}

The peculiarity of this model is the presence of a multiplicative noise proportional to $x_\mu$ rather than the additive noise considered in the previous cases. Once more, the limit of a slow environment is identified by $\tau w_\mathrm{sum} \ll 1$, following the notation of the previous sections. The components of the mixture distribution defined in Eq.~\eqref{eqn:p2c1:solution_slow_jumps} are the solution of the Fokker-Planck equation
\begin{equation*}
    0 = \sum_{\mu = 1}^2\left[\partial_\mu\left(\frac{x_\mu}{\tau}P^\mathrm{st}_i(x_1, x_2)\right) + \partial_\mu^2 \left(\gamma_i T(x_\mu)P^\mathrm{st}_i(x_1, x_2)\right)\right]
\end{equation*}
where for consistency we need to have $x_\mu \in [-T_0/\Delta T, +\infty]$, so that $T(x_\mu)$ is always positive. We find
\begin{equation}
    P^\mathrm{st}_i(x_\mu) = \frac{N_i\left(\frac{T_0/\Delta T}{\Delta T \gamma_i \tau}\right)}{\Gamma\left(\frac{T_0/\Delta T}{\Delta T \gamma_i \tau}\right)} \left( 1 + \frac{\Delta T}{T_0} x_\mu \right)^{\frac{T_0/\Delta T}{\Delta T \gamma_i \tau} - 1}e^{-\frac{1}{\tau} \frac{x_\mu}{\gamma_i \Delta T}}
\end{equation}
with $N_i(z) = \Delta T^{1-2z} \, T_0^{z - 1}e^{-z}(\gamma_i \tau)^{-z}$, and $P^\mathrm{st}_i(x_1, x_2) = P^\mathrm{st}_i(x_1)P^\mathrm{st}_i(x_2)$. In particular, the boundary condition 
\begin{equation*}
    \frac{P^\mathrm{st}_i(x_\mu)}{\partial x_\mu} \biggr|_{x_\mu = -T_0/\Delta T} = 0
\end{equation*}
ensures that no particle can escape the system, and it is always satisfied provided that the temperature gradient is such that $\Delta T < \sqrt{T_0/\gamma_i \tau}$ for all environmental states $i$. In Figure~\ref{fig:p2c1:multiplicative}a-b, we show the joint and factorized distributions of this system, respectively. Notice that the dependency induced by the environment translates into a suppression of the probability of finding a particle diffusing away from the origin while the other is close to it, much like in the additive noise case.

We are now interested in the mutual information associated with the joint distribution $p^{\mathrm{slow}}_{12}$ and its dependence on the strength of the temperature gradient $\Delta T$ - that is, on how far the system is from equilibrium. We compute it as a standard numerical integral and plot the results in Figure~\ref{fig:p2c1:multiplicative}c as a function of $w_-/w_+$ for increasing $\Delta T$. As we can see, the strongest the temperature gradient, the higher the mutual information, suggesting that environmental-induced dependencies may be stronger in out-of-equilibrium systems.

\section{Continuously-varying environments}
Finally, as a last example of two particles that are not interacting but share the same environment, we consider the case in which the environment changes in a continuous fashion \cite{chechkin2017diffusivities, wang2020randomdiffusivities}. To fix the ideas, let us consider the paradigmatic example of two Ornstein-Uhlenbeck processes,
\begin{equation}
\label{eqn:p2c1:continuous_env_langevin}
    \begin{cases}
        \dot{x}_\mu = -x_\mu/\tau_X + \sqrt{2} D \,\xi_\mu \\
        \dot{D} = -D/\tau_D + \sqrt{2 \theta} \,\xi_D
    \end{cases}
\end{equation}
where the only dimensionless parameter of the system is now $\tau_X/\tau_D$, which governs the timescale separation of the two dynamics. Hence, contrary to the case of a discrete-state environment, we cannot define the separation between environmental states - previously quantified by $D_-/D_+$ - nor their relative persistence - which was given by $w_-/w_+$.

\begin{figure}[t]
    \centering
    \includegraphics[width=\textwidth]{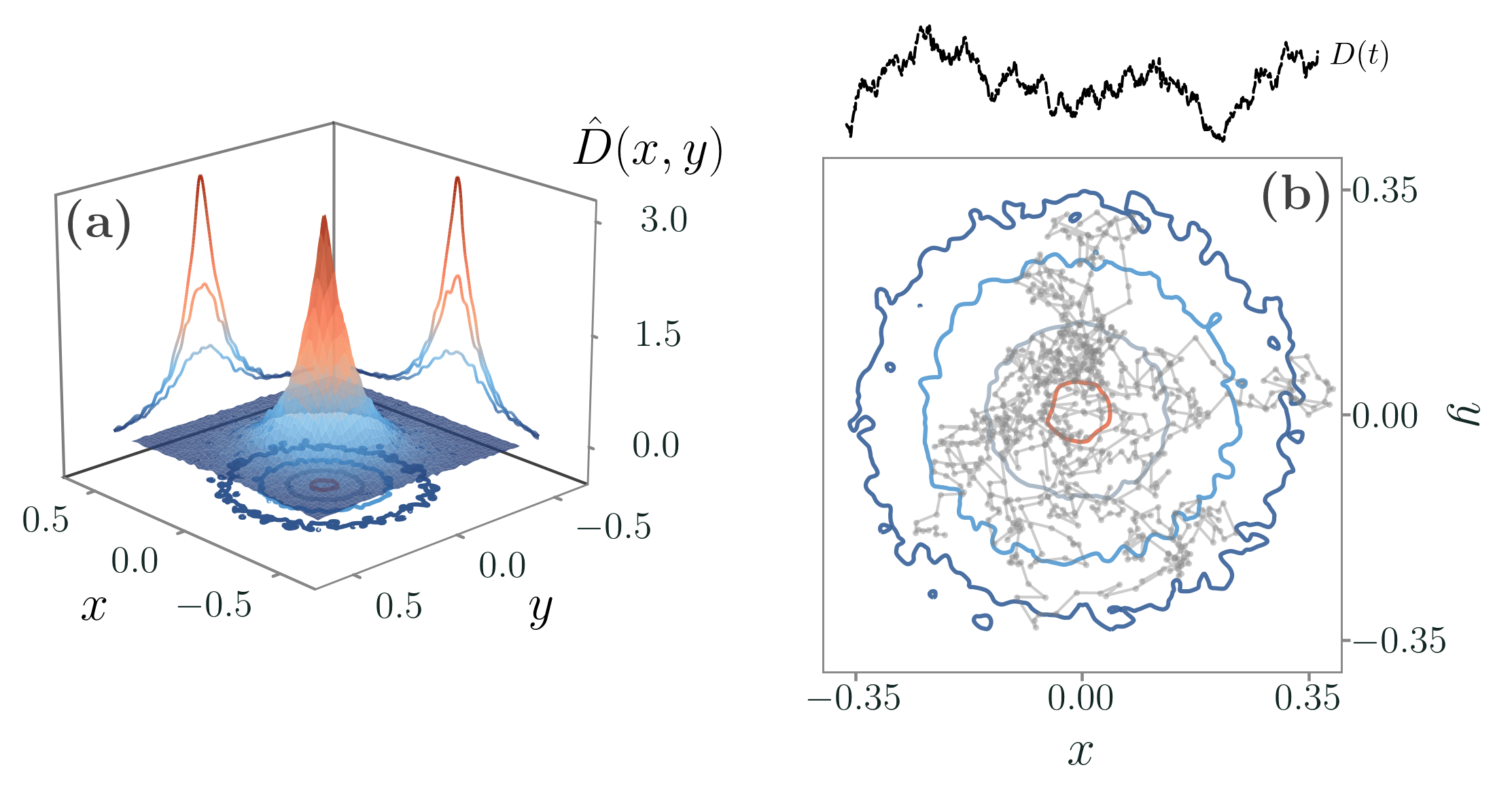}
    \caption{A time-varying diffusion coefficient is equivalent to a constant yet spatially-varying one, Eq.~\eqref{eqn:p2c1:continuous_env_D_eff}. (a) Plot of $\hat{D}(x, y)$, obtained self-consistently from the numerical solution $p(D | x, y)$, estimated from simulations of the Langevin equations in Eq.~\eqref{eqn:p2c1:continuous_env_langevin}. (b) Comparison of the contour lines of $\hat{D}(x, y)$ and a trajectory of the system (gray) in a time-varying diffusion (black). The system experiences a higher diffusion close to the origin, as expected}
    \label{fig:p2c1:effective_diffusion}
\end{figure}

The corresponding stationary Fokker-Planck equation is given by
\begin{align}
    \label{eqn:p2c1:continuous_env_FP}
    0 = & \sum_{\mu = 1}^2\left[\partial_\mu\left(\frac{x_\mu}{\tau_X} p(\vb{x}, D)\right) + D^2\partial_\mu^2 p(\vb{x}, D)\right] + \partial_D \left[\frac{D}{\tau_D} p(\vb{x}, D)\right] + \theta \partial_D^2 p(\vb{x}, D)
\end{align}
and, as before, we are interested in the marginalization over the environment
\begin{equation*}
    p(\vb{x},t) = \int dD p(\vb{x}, D, t).
\end{equation*}
Notably, we can try and explicitly marginalize Eq.~\eqref{eqn:p2c1:continuous_env_FP} before considering any timescale separation limit. If we do so, at stationarity we obtain
\begin{align}
    \label{eqn:p2c1:continuous_env_FP_marg}
    0 = & \sum_{\mu = 1}^2\left[\partial_\mu\left(\frac{x_\mu}{\tau_X} p(\vb{x})\right) + \partial_\mu^2\left(\hat{D}^2(\vb x) p(\vb{x})\right)\right] 
\end{align}
where
\begin{equation}
    \label{eqn:p2c1:continuous_env_D_eff}
    \hat{D}^2(\vb x) = \int_{-\infty}^{+\infty} dD\,D^2 p(D|\vb{x})
\end{equation}
is an effective spatial diffusion coefficient. An explicit example of such mapping is shown in Figure~\ref{fig:p2c1:effective_diffusion}.

Clearly, Eq.~\eqref{eqn:p2c1:continuous_env_D_eff} lacks a closed form since we cannot compute $p(D|\vb{x})$ explicitly. Nevertheless, Eq.~\eqref{eqn:p2c1:continuous_env_FP_marg} shows that the effective dependencies induced by the environment are equivalent to those that arise from an inhomogeneous medium. Surprisingly, such dependencies are fundamentally different from an effective coupling between $x_1$ and $x_2$, as one might naively expect. It is also worth noting that, in principle, space-dependent diffusion coefficients interpreted in the Ito sense might always emerge from the variations of an external stochastic environment, which is also the sole responsible for a non-zero mutual information here. This result might shed some light on the controversial topic of the Ito-Stratonovich dilemma in diffusing chemical systems. A similar perspective, where the internal states play a role analogous to a changing environment, is presented in \cite{liang2021thermophoresis}.

In the limit in which the environment is either much faster or much slower than the internal relaxation, i.e., respectively $\tau_X/\tau_D \gg 1$ and $\tau_X/\tau_D \ll 1$, we can repeat the calculations of the previous sections. In the presence of a slower environment, we find the following stationary joint probability distribution
\begin{align}
\label{eqn:p2c1:continuous_env_joint}
    p^{\mathrm{slow}}_{12}(x_1, x_2) & = \int_{-\infty}^{+\infty} dD\, p^{\mathrm{st}}(D) p^{\mathrm{st}}(x_1, x_2|D) \nonumber \\
    & = \frac{1}{2\pi\sqrt{(x_1^2+x_2^2)\theta \tau_D \tau_X}} \exp\left[-\sqrt{\frac{x^2 + y^2}{\theta \tau_D \tau_X}}\right]
\end{align}
where $p^{\mathrm{st}}(D) \sim \mathcal{N}(0,\tau_D\theta)$ is the stationary distribution of the process governing the evolution of the diffusion coefficient, and $p^{\mathrm{st}}(x_1, x_2|D) \sim \mathcal{N}(0, \tau_X D^2)$ is the stationary distribution of $(x_1, x_2)$ at fixed $D$.

\begin{figure}[t]
    \centering
    \includegraphics[width=\textwidth]{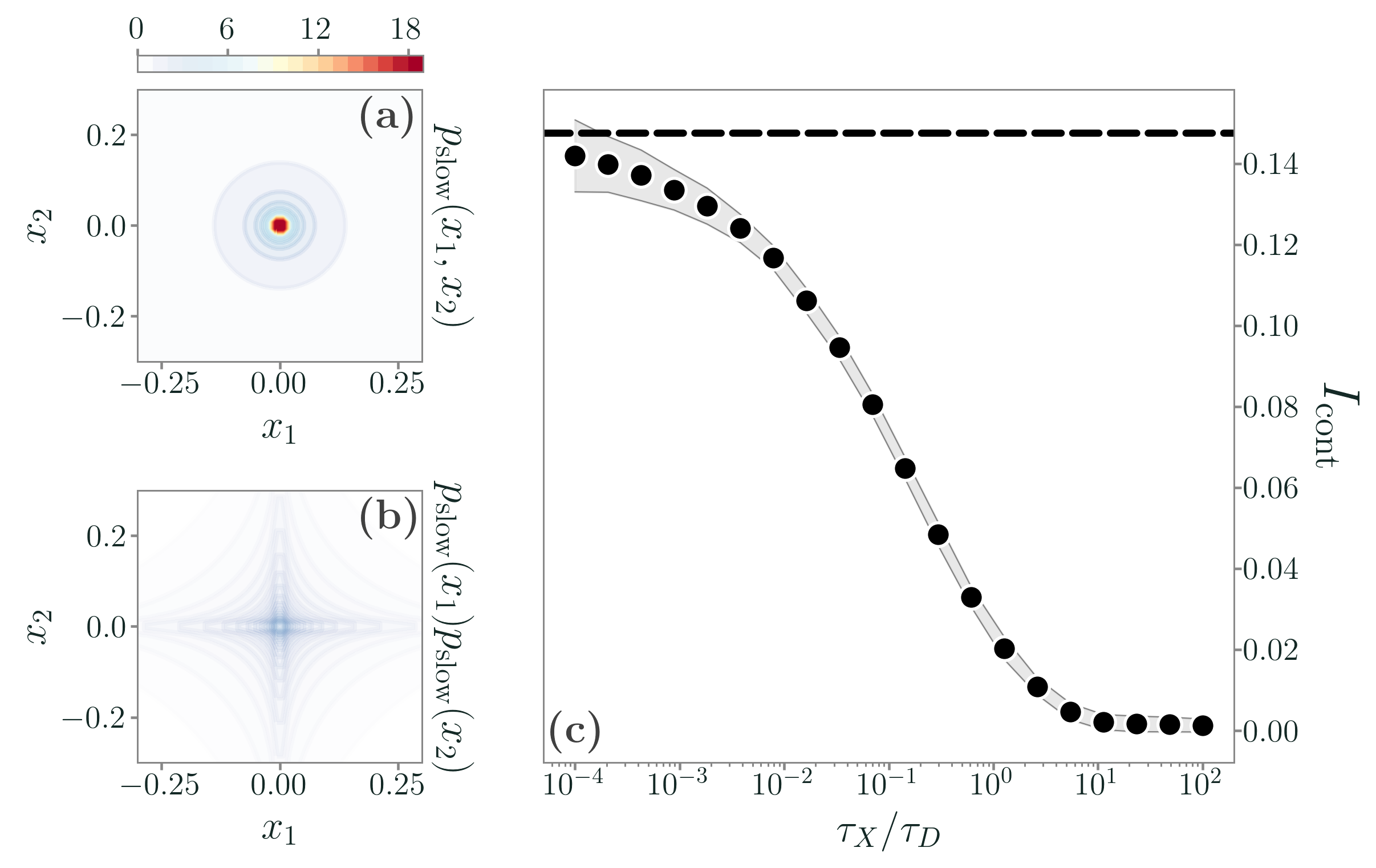}
    \caption{The mutual information in the presence of a continuously varying environment, described by the Langevin equations in Eq.~\eqref{eqn:p2c1:continuous_env_langevin}. (a-b) Plots of the joint and the factorized distribution in the limit of a slow environment, $\tau_X/\tau_D \ll 1$. (c) Mutual information at different values of $\tau_X/\tau_D$ estimated through a $k$-nearest neighbors estimator. The gray shaded area represents $3$ standard deviations from the mean value obtained from different simulations of the Langevin equations. As expected, in the limit $\tau_X/\tau_D \ll 1$ the mutual information converges to Eq.~\eqref{eqn:p2c1:continuous_env_integral}, whereas it vanishes in the opposite limit}
    \label{fig:p2c1:continuous_env}
\end{figure}

Eq.~\eqref{eqn:p2c1:continuous_env_joint} can be marginalized exactly over one of the two degrees of freedom, in order to evaluate the mutual information. The marginalization leads to
\begin{align}
\label{eqn:p2c1:continuous_env_marg}
    p^{\mathrm{slow}}_\mu(x_\mu)  = \frac{1}{\pi\sqrt{\theta \tau_D \tau_X}}K_0\left(\frac{|x_\mu|}{\sqrt{\theta \tau_D \tau_X}}\right)
\end{align}
where $K_n(\cdot)$ is the modified Bessel function of the second kind.
These probability distributions are plotted in Figure~\ref{fig:p2c1:continuous_env}a-b. As in the previous case, we see that the shared environment favors regions in which both particles are close to the potential minima. Notably, as we have seen, this can be interpreted as a spatially-varying diffusion coefficient that peaks at the origin (see Figure~\ref{fig:p2c1:effective_diffusion}).

The joint probability in Eq.~\eqref{eqn:p2c1:continuous_env_joint} is not factorizable, and thus we expect that in this limit the mutual information, $I_{\mathrm{cont}}^{\mathrm{slow}}$, will be different from zero due to the shared environment. Since in this case there are no dimensionless parameters left, we also expect no parametric dependence. Indeed, let us rewrite the mutual information integral after the change of variables $(x_1, x_2) = \sqrt{\theta\tau_D\tau_X}(s \cos\phi, s\sin\phi)$ as
\begin{align*}
    I_{\mathrm{cont}}^{\mathrm{slow}} & = \int_0^\infty ds \int_0^{2\pi} d\phi \, \frac{e^{-s}}{2\pi}\, \log\frac{e^{-s}\pi}{2s K_0\left(s|\cos\phi|\right)K_0\left(s|\sin\phi|\right)} \\
    & = - 1 + \frac{1}{2\pi}\int_0^\infty ds \, e^{-s} \int_0^{2\pi} d\phi \, \log\frac{\pi}{2s K_0\left(s|\cos\phi|\right)K_0\left(s|\sin\phi|\right)}
\end{align*}
where no free parameters are left. Using the integral definition of the Euler's constant, $\gamma_E = - \int_0^\infty e^{-s} \log s$, we end up with
\begin{align}
\label{eqn:p2c1:continuous_env_integral}
    I_{\mathrm{cont}}^{\mathrm{slow}} = & \, \gamma_E + \log\frac{\pi}{2} - 1 - \frac{1}{\pi} \int_0^\infty ds\,e^{-s} \int_{0}^{2\pi} d\phi \log K_0\left(s|\cos\phi|\right)
\end{align}
whose numerical value is $I_{\mathrm{cont}}^{\mathrm{slow}} \approx 0.148$.

In the opposite limit, $\tau_X/\tau_D \gg 1$, the effective diffusion coefficient averaged over the environmental states is simply $\ev{D}_{p^{\mathrm{st}}(D)} = 0$. Thus, we trivially find that 
\begin{equation*}
    p^{\mathrm{fast}}_{12}(x_1, x_2) = \delta(x_1)\delta(x_2),
\end{equation*}
which is a factorized distribution. Therefore, and as expected from the results of the previous sections, no mutual information between the internal degrees of freedom emerges from a fast continuously-varying environment.

At intermediate values of $\tau_X/\tau_D$, we cannot solve Eq.~\eqref{eqn:p2c1:continuous_env_FP} exactly. Therefore, in order to obtain samples from the stationary joint distribution, we simulate the Langevin equations in Eq.~\eqref{eqn:p2c1:continuous_env_langevin}. Then, from these samples, we estimate the mutual information through the $k$-nearest neighbors estimator proposed in \cite{kraskov2004estimating, holmes2019estimation}. The results are plotted in Figure~\ref{fig:p2c1:continuous_env}c. The mutual information changes smoothly and monotonically with $\tau_X/\tau_D$ and, in the limit $\tau_X/\tau_D \to 0$, approaches Eq.~\eqref{eqn:p2c1:continuous_env_integral}. Notably, we explicitly see that slower environments lead to a higher dependency between the internal degrees of freedom, and thus to a larger information induced by the environmental changes.

Overall, this example of a continuously-varying diffusion coefficient and the previous ones of linear, non-linear, and out-of-equilibrium models have helped us understand how a stochastic environment shapes the dependencies in a system made of otherwise independent particles. With these results in mind, in the next Chapter we will reintroduce internal interactions and explore their complex interplay with environmental changes.

\chapter{Information of interacting systems in unobserved environments}
\begin{figure*}[t]
    \centering
    \includegraphics[width=\textwidth]{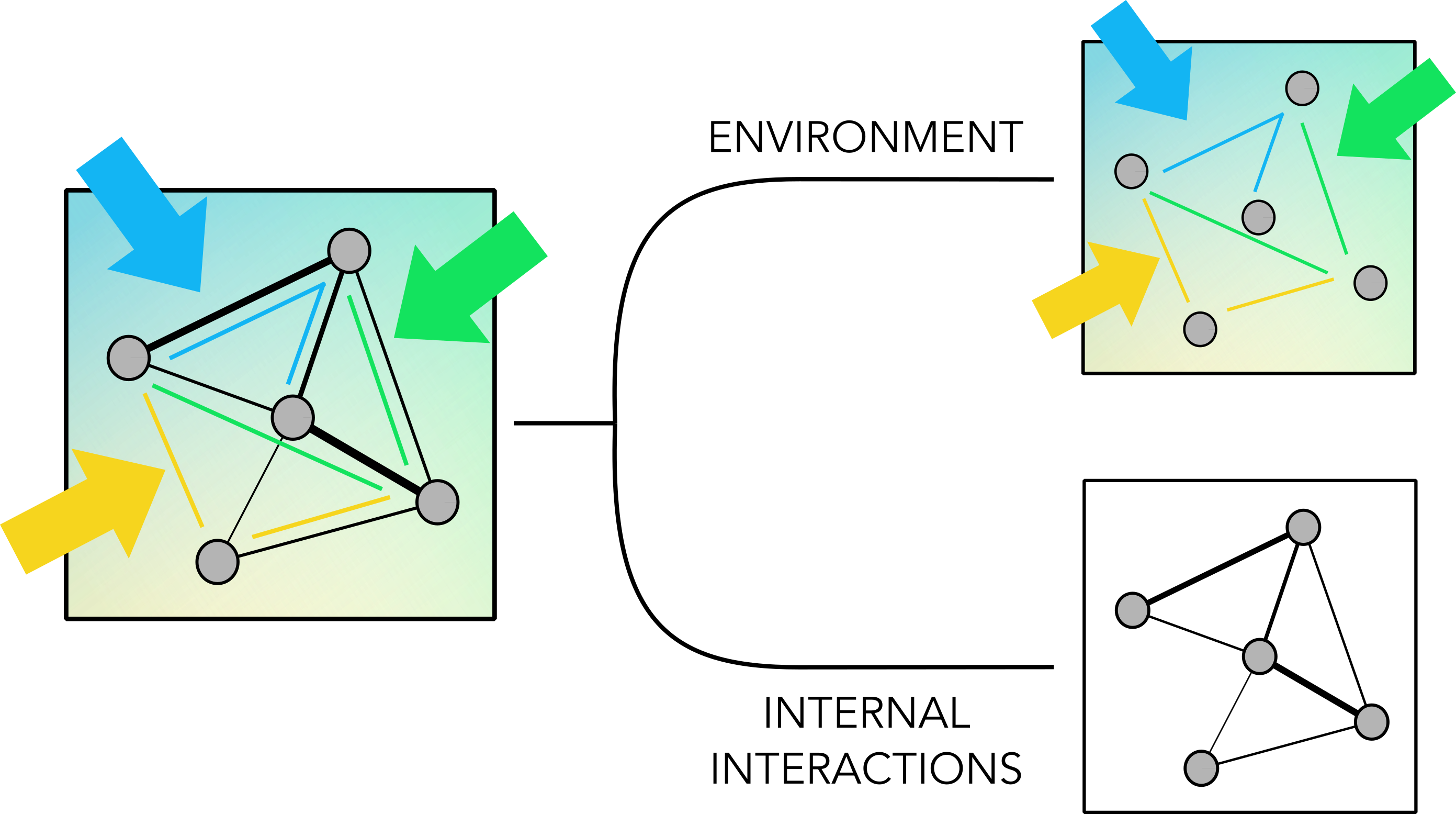}
    \caption{The degrees of freedom of a complex system (gray dots) living in a changing environment (colored arrows) will experience different sources of dependencies, both coming from internal interactions (black lines) and induced by the environment (colored lines). Can these dependencies be disentangled, and if so, in which cases? As we will see, they are shaped by the complex interplay between interactions and environmental changes}
    \label{fig:p2c2:disentangling_sketch}
\end{figure*}

\chaptermark{Interacting systems in unobserved environments}
\label{ch:PRL_PRE_2}
\lettrine{I}{n this Chapter} we will study the complete dynamical model introduced in Eq.~\eqref{eqn:p2c1:internal_langevin_with_env}, working with systems in which the force field $\vb{F}(\vb{x}; \{\zeta\})$ describes interactions between the internal degrees of freedom $\vb{x}$. Without loss of generality, we will focus once more on the case in which environmental changes describe a diffusion coefficient that switches between $M$ discrete states. Namely, the dynamics of the particles is described by
\begin{align}
    \label{eqn:p2c2:langevin_with_env}
    \frac{d x_\mu}{dt} = F_\mu(\vb{x}; \{\zeta\}) + \sqrt{2 D_{i(t)}} \xi_\mu
\end{align}
where $i(t)$ is a realization of the jumps between the environmental states. Our goal is to describe how the different processes - the one stemming from internal interactions, and the stochastic environment - shape the dependencies of the system, and thus the information between its internal degrees of freedom. These ideas are sketched in Figure~\ref{fig:p2c2:disentangling_sketch}.

Recently, this problem has gained momentum from a theoretical perspective in different contexts \cite{bressloff2016randenv, hufton2016intrinsic, bressloff2017dichotomous, grebenkov2019switching}, but the general question of how we can characterize and possibly disentangle the effects of internal interactions from those of a stochastic environment is very much open and elusive. In this Chapter, we will show that in the presence of linearized interactions the mutual information of the whole system encodes the internal and environmental processes as distinct contributions. Furthermore, such contributions can always be fully disentangled in suitable limits. Non-linear interactions, on the other hand, lead to a much richer structure of the mutual information, giving rise to a new term that models the interference between the dependencies induced by internal interactions and those stemming from the environment. This interference can be either constructive - with the dependency between the particles amplified by the presence of both interactions and environment at once - or destructive, depending on the shape of the internal interactions. Overall, although characterizing the specific nature of internal interactions through mutual information remains a challenge, we will show how a fast-varying environment might reveal the presence of underlying internal couplings in any general system.

\section{Exact disentangling with linear interactions}
Let us begin with the paradigmatic example of linear interactions, i.e., of two interacting Ornstein-Uhlenbeck processes \cite{gardiner2004handbook}. This particular choice is twofold. First, an Ornstein-Uhlenbeck process is one of the simplest multidimensional stochastic processes with a non-trivial stationary distribution. Second, Ornstein-Uhlenbeck processes can often be seen as a linearization of more complex, non-linear interacting models.

\begin{figure*}[t]
    \centering
    \includegraphics[width=0.6\textwidth]{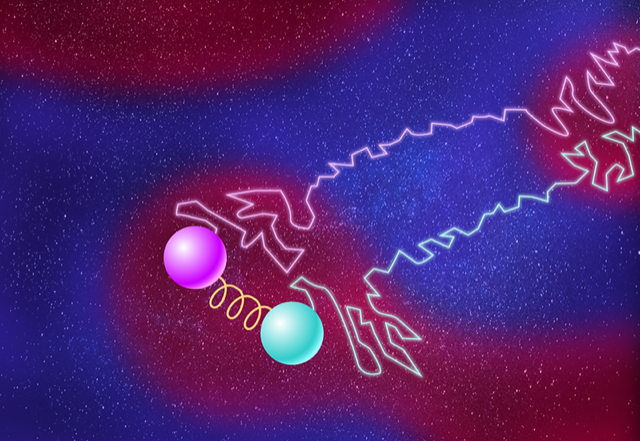}
    \caption{A cartoon representing a possible interpretation of the system described by Eq.~\eqref{eqn:p2c2:linear_langevin}. Two particles coupled by a spring move experience a change in, e.g., temperature affecting their diffusion. The wiggly lines show the dependencies arising from the intrinsic force of the spring and from the extrinsic temperature changes, fueled by the environment through which the particles move. This cartoon was taken from the commentary on our paper \cite{nicoletti2021mutual} that appeared in \emph{Physics 14, 162} (see Appendix \ref{app:viewpoint}) (copyright APS/Carin Cain)}
    \label{fig:p2c2:cartoon}
\end{figure*}

As in the previous Chapter, for the sake of simplicity we consider the case of two particles, whose position is described by the internal degrees of freedom $x_1$ and $x_2$. Environmental changes are modeled through a dichotomous process of the diffusion coefficient $D_{i(t)}$ between the states $i \in \{-,+\}$, with $D_-$ and $D_+ > D_-$ and transition rates $W_{- \to +} = w_+$ and $W_{+ \to -} = w_-$. A general formalism for $N$ particles and $M$ jumps can be found in Appendix \ref{app:generalization_jumps}. Given the linear nature of this system, the internal dynamics is described by an interaction matrix $\vb{A}$, whose off-diagonal elements describe the internal couplings between $x_1$ and $x_2$. Our model can be written as the set of Langevin equations
\begin{align}
\label{eqn:p2c2:linear_langevin}
    \dv{x_\mu}{t} = -\sum_{\nu}A_{\mu\nu}\frac{x_\nu}{\tau} + \sqrt{2 D_{i(t)}} \xi_\mu(t)
\end{align}
where $i(t)$ is a realization of the jump process between $\{-, +\}$ and $\xi_1$ and $\xi_2$ are independent white noises with zero mean. Albeit simple, this model is related to ``diffusing diffusivity'' processes, often used to describe spatially disordered or inhomogeneous environments \cite{chechkin2017diffusivities, wang2020randomdiffusivities}, represented by the cartoon in Figure~\ref{fig:p2c2:cartoon}. Our goal is to understand whether these two distinct contributions to the dynamics can be disentangled, and, if so, under which conditions.

Notice that in the case $\vb{A} = \mathbb{1}$ no interactions are present, and we recover the harmonic case described in the previous Chapter, Section \ref{sec:p2c1:harmonic}. Therefore, we expect the mutual information stemming from the environment, $I_\mathrm{env}$, to obey the limiting behaviors
\begin{align}
\label{eqn:p2c2:harmonic_environment_mutual_limits}
    I_\text{env}\left(\frac{w_-}{w_+}\right) =
    \begin{cases}
        H_\mathrm{jumps} &\, \text{if} \quad D_+ \gg D_- \\
        \,0 &\, \text{if} \quad D_+ \approx D_-
    \end{cases}
\end{align}
where $H_\mathrm{jumps} = - \sum_i \pi^\mathrm{st}_i \log \pi^\mathrm{st}_i$. How does the mutual information $I$ between $x_1$ and $x_2$ change when we add back linear interactions on top of the shared environment?

Let us consider the matrix
\begin{equation*}
    \vb A = \begin{pmatrix}
                1 & - g_1 \\
                - g_2 & 1
            \end{pmatrix}
\end{equation*}
and assume that its eigenvalues have positive real parts, so that a stationary state exists. For instance, this is always satisfied if $g_1$ and $g_2$ have opposite signs, or if $1 \geq \sqrt{g_1g_2}$. We can now write down the expected parametric dependence of the mutual information, Eq.~\eqref{eqn:p2c1:mutual_parametric_dependence}. The set of parameters $\{\zeta\}$ appearing in Eq.~\eqref{eqn:p2c2:langevin_with_env} is $\{\tau, g_1/\tau, g_2/\tau\}$, where $\tau = \tau_\mathrm{int}$ plays the role of the internal timescale. As before, the environmental one is again $\tau_\mathrm{env} = w_\mathrm{sum}^{-1}$ with $w_\mathrm{sum} = w_- + w_+$. Therefore, assuming that $\tau w_\mathrm{sum}$ is fixed by the timescale separation, we can rewrite Eq.~\eqref{eqn:p2c1:mutual_parametric_dependence} as
\begin{equation}
    I = I\left(\frac{w_-}{w_+}, \frac{D_-}{D_+}, g_1, g_2\right) 
\end{equation}
where, using the notation of Section \ref{sec:p2c1:mutual_properties}, $\{W_\mathrm{jumps}\} = \{w_-/w_+\}$, $\{\varphi\} = \{D_-/D_+\}$ and $\{\psi\}_i = \{g_1, g_2\}$. The fact that the set of dimensionless parameters of the internal dynamics, $\{\psi\}_i$, does not depend on the environmental states has crucial implications, as we will see further on. In particular, we have that the mutual information of the internal dynamics $I_\mathrm{int}(\{\psi\}_i)$ is equal to its environmental average, $\ev{I_\mathrm{int}}_\pi$.

Let us assume, for the time being, that we are in the slow-jumps limit $\tau_\mathrm{env} \gg \tau_\mathrm{int}$, so that we can solve the Langevin equations in Eq.~\eqref{eqn:p2c2:linear_langevin} separately for $D_+$ and $D_-$ and then average them over $\pi^\mathrm{st}_\pm$ as in Eq.~\eqref{eqn:p2c1:solution_slow_jumps}. The components of the resulting mixture distribution $p^\mathrm{slow}_{12} = \sum_i \pi^\mathrm{st}_i P^\mathrm{st}_i$ are multivariate Gaussian distributions,
\begin{equation*}
    P^\mathrm{st}_i(x_1,x_2) \sim \mathcal{N}(0,\bm{\Sigma}_i),
\end{equation*}
where $\bm{\Sigma}_i$ are symmetric matrices obeying the Lyapunov equation \cite{gardiner2004handbook}
\begin{equation}
    \label{eqn:p2c2:lyapunov_equation}
    \bm{A}\bm{\Sigma}_i + \bm\Sigma_i \bm{A}^T = 2 \tau D_i \mathbb{1}
\end{equation}
with $\mathbb{1}$ the identity matrix. We now introduce the rescaled covariance matrix $\tilde{\bm\Sigma}_i = \bm\Sigma_i / (D_i \tau)$, which is determined by the equation $\bm{A}\tilde{\bm{\Sigma}}_i + \tilde{\bm{\Sigma}}_i \bm{A}^T = 2 \mathbb{1}$. Since no dependence on the environmental state $i$ is present, we have that $\tilde{\bm{\Sigma}}_i = \tilde{\bm{\Sigma}}$ $\forall i$. Thus we can rewrite Eq.~\eqref{eqn:p2c2:lyapunov_equation} as 
\begin{equation}
\label{eqn:p2c2:lyapunov_equation_tilde}
    \frac{1}{2}\left[\vb{A}\tilde{\bm\Sigma} + \tilde{\bm\Sigma} \vb{A}^T\right] = \mathbb{1}
\end{equation}
where
\begin{equation}
\label{eqn:p2c2:rescaled_covmat}
    \bm\Sigma_i = D_i \tau \tilde{\bm\Sigma}.
\end{equation}
Eq.~\eqref{eqn:p2c2:rescaled_covmat} shows that the covariance matrix appearing in the mixture components of $p^\mathrm{slow}_{12}$ receives separate contributions from the environmental diffusion coefficient $D_i$ and the internal interactions $\bm A$.

We can immediately solve Eq.~\eqref{eqn:p2c2:lyapunov_equation_tilde}, finding
\begin{equation}
\label{eqn:p2c2:rescaled_covmat_sol}
    \tilde{\bm\Sigma} = \frac{1}{g_1 g_2 - 1}\begin{pmatrix}
                        \frac{g_1(g_2-g_1)}{2} - 1 & -\frac{g_1+g_2}{2} \\
                        -\frac{g_1+g_2}{2} & \frac{g_2(g_1-g_2)}{2} - 1
                        \end{pmatrix}.
\end{equation}
In order to compute the bounds on the mutual information from the estimator in Eq.~\eqref{eqn:p2c1:pairwise_mutual_estimator}, we need to find the mutual information $I_\mathrm{int}(\{\psi\}_i) = I^i$ in a fixed environment, i.e.,
\begin{equation*}
    I^i = \int dx_1\,dx_2 P^\mathrm{st}_i(x_1,x_2) \log\frac{P^\mathrm{st}_i(x_1,x_2)}{P^\mathrm{st}_i(x_1)P^\mathrm{st}_i(x_2)}.
\end{equation*}
In this case of our Gaussian distributions $\mathcal{N}(0,\tau D_i \tilde{\Sigma})$, we can compute $I^i$ exactly as
\begin{align}
    \label{eqn:p2c2:mutual_interactions_linear}
    I^i & = \frac{1}{2}\log \frac{\Sigma_{11} \Sigma_{22}}{\mathrm{det} \Sigma} = \frac{1}{2}\log \frac{\tilde\Sigma_{11} \tilde\Sigma_{22}}{\mathrm{det} \tilde\Sigma} \nonumber \\
    & = \frac{1}{2}\log\left[1 - \frac{4}{4 + (g_1 - g_2)^2}+ \frac{1}{1 - g_1 g_2} \right] \nonumber \\
    & := I_\mathrm{int}(\{g_\mu\})
\end{align}
where $\{g_\mu\} = \{g_1, g_2\}$. Since $\sum_i\pi^\mathrm{st}_i = 1$, this also implies $\ev{I_\mathrm{int}}_\pi = I_\mathrm{int}$. That is, as expected from the dimensionless parameters of this system, the dependency between $x_1$ and $x_2$ induced by the internal interactions does not change with the environmental state.

Then, to find the estimator in Eq.~\eqref{eqn:p2c1:pairwise_mutual_estimator} we also need the Chernoff and the Kullback-Leibler divergences between the components of the slow-jumps distribution and its factorization. Since we are dealing with Gaussian distributions, such divergences can be computed analytically \cite{Kolchinsky2017, amari2016information, ThomasCover2006}. We find
\begin{align*}
    C_\alpha\left(\mathcal{N}(0, \bm{\Sigma}_+) \,||\, \mathcal{N}(0, \bm{\Sigma}_-)\right) & = \frac{1}{2}\log\frac{\det \left[(1-\alpha)\bm\Sigma_++ \alpha\bm\Sigma_-)\right]}{\det^{1-\alpha} \bm\Sigma_+ \det^{\alpha} \bm\Sigma_-} \nonumber \\
    & = \frac{1}{2}\log\frac{\det \tau \tilde{\bm\Sigma}\left[(1-\alpha)D_++ \alpha D_-\right]}{\det^{1-\alpha} \tau D_+\tilde{\bm\Sigma} \det^{\alpha} \tau D_-\tilde{\bm\Sigma}} \nonumber \\
    & = \log \frac{\left[(1-\alpha)D_++ \alpha D_-\right]}{D_+^{1-\alpha}D_-^\alpha}
\end{align*}
and
\begin{align*}
    D_\mathrm{KL}\left(\mathcal{N}(0, \bm{\Sigma}_+) \,||\, \mathcal{N}(0, \bm{\Sigma}_-)\right) & = \frac{1}{2}\left[\log\frac{\det\bm\Sigma_-}{\det\bm\Sigma_+} + \Tr\bm\Sigma_-^{-1}\bm\Sigma_+ - 2\right] \nonumber \\
    & = \frac{1}{2}\left[\log\frac{\det\tau D_-\tilde{\bm\Sigma}}{\det\tau D_+\tilde{\bm\Sigma}} + \Tr \frac{1}{\tau D_-}\tilde{\bm\Sigma}^{-1}\tau D_+\tilde{\bm\Sigma} - 2\right] \nonumber \\
    & = \log\frac{D_-}{D_+} + \frac{D_+}{D_-} - 1.
\end{align*}
As we can see, due to the factorization of the covariance matrix, these distances are the same as the ones of the non-interacting case in Eqs.~\eqref{eqn:p2c1:harmonic_chernoff_2D}-\eqref{eqn:p2c1:harmonic_kl_2D}. As such, they only depend on the ratio $D_-/D_+$. Hence, we can bound the mutual information with
\begin{equation}
\label{eqn:p2c2:linear_mutual_bounds}
    I^\mathrm{slow,up/low}\left(\frac{w_-}{w_+}, \frac{D_-}{D_+}, \{g_\mu\}\right) = I_\mathrm{int}\left(\{g_\mu\}\right) + I_\mathrm{env}^\mathrm{slow,up/low}\left(\frac{w_-}{w_+}, \frac{D_-}{D_+}\right)
\end{equation}
where $I_\mathrm{env}^\mathrm{slow,up/low}$ are the bounds on the mutual information in the absence of interactions, Eqs.~\eqref{eqn:p2c1:harmonic_mutual_bounds}.

\begin{figure}[t]
    \centering
    \includegraphics[width=0.9\textwidth]{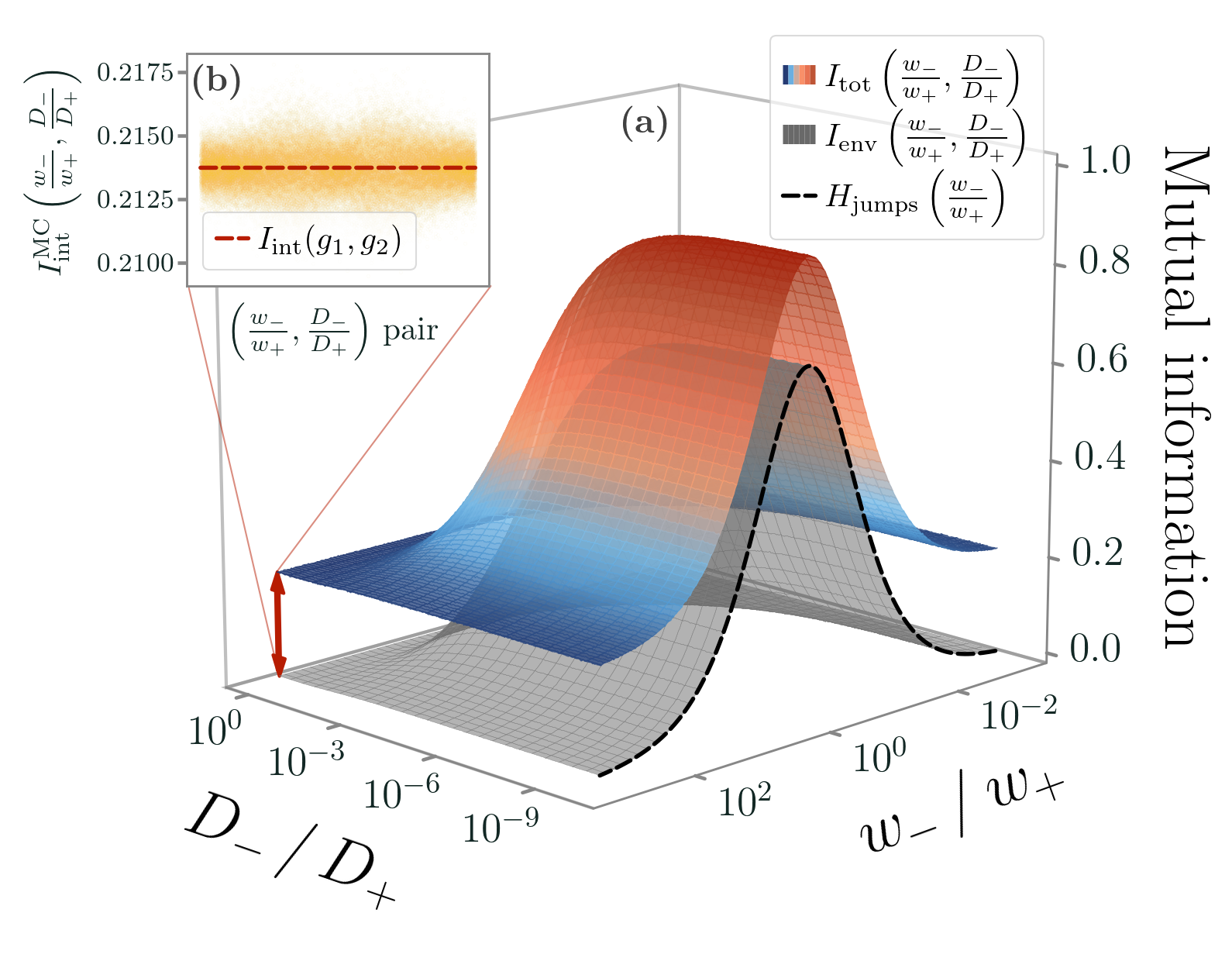}
    \caption{The total mutual information as a function of $D_-/D_+$ and $w_-/w_+$ at fixed $\tau w_{\mathrm{sum}}= 10^{-3}$, i.e., in the slow-jumps limit. (a) The colored surface is the result of a Monte Carlo integration of $I_\text{tot}$, in the slow-jumps limit, for the interacting model with $g_1 = 5$, $g_2 = -0.1$, and $\tau = 1$. The gray surface is instead the non-interacting case, $I_{\mathrm{env}}$. The two contributions to the mutual information disentangle and the interactions simply result in a constant shift. (b) A comparison between the predicted shift $I_\text{int}(g_1, g_2)$, Eq.~\eqref{eqn:p2c2:mutual_interactions_linear}, and the difference of the Monte Carlo estimates of the two surfaces for every sampling point $(w_-/w_+, D_-/D_+)$, namely $I_\text{int}^{MC}(w_-/w_+, D_-/D_+)$}
    \label{fig:p2c2:mutual_linear}
\end{figure}

Eq.~\eqref{eqn:p2c2:linear_mutual_bounds} tells us that, in the slow-jumps regime, the bounds on the mutual information are fully disentangled in form - i.e., interactions and environment are not mixing. This further implies the limiting behaviors
\begin{align}
\label{eqn:p2c2:linear_mutual_limits}
    I^\mathrm{slow}\left(\{g_\mu\}, \frac{w_-}{w_+}\right) =
    \begin{cases}
        H_\mathrm{jumps} + I_\text{int}\left(\{g_\mu\}\right) & \text{if} \, D_+ \gg D_- \\
        I_\text{int}\left(\{g_\mu\}\right) & \text{if} \, D_+ \approx D_-
    \end{cases}
\end{align}
since the environmental bounds saturate. When $D_-/D_+ \ll 1$ and thus the two environmental states are markedly different from one another the mutual information is simply the sum of the one stemming from the interactions, $I_\text{int}$, and the entropy of the environment, $H_\mathrm{jumps}$. Crucially, a Monte Carlo integration in the slow-jumps limit shows that this is true not only for these limits - but rather holds in general, see Figure~\ref{fig:p2c2:mutual_linear}.

On the other hand, in the fast-jumps limit, once we solve the Lyapunov equation in Eq.~\eqref{eqn:p2c2:lyapunov_equation_tilde}, the stationary probability distribution is the multivariate Gaussian $\mathcal{N}(0, \ev{D}_\pi \tau \tilde{\bm \Sigma})$. Hence, as shown in the previous Chapter, it only depends on the single effective diffusion coefficient $\ev{D}_\pi$. In this limit we can compute the mutual information exactly as in Eq.~\ref{eqn:p2c2:mutual_interactions_linear}, since it only depends on $\tilde{\bm \Sigma}$. That is, we find
\begin{equation}
    I^\mathrm{fast} = I_\text{int}\left(\{g_\mu\}\right)
\end{equation}
which is perhaps unsurprising, since we know from the previous results that no information arises from the environment in the fast-jumps limit. Finally, in the intermediate regime between the fast- and slow-jumps limits - i.e., for intermediate values of $\tau w_\mathrm{sum}$ - we can estimate the stationary p.d.f. from simulations of the Langevin equations in Eq.~\eqref{eqn:p2c2:langevin_with_env}. Then, through either numerical integration or k-nearest neighbors estimators \cite{kraskov2004estimating, holmes2019estimation}, we always find that the presence of linear internal interactions simply shifts the mutual information with respect to the non-interacting case, $I_\mathrm{env}$.

Therefore, our results suggest that the mutual information receives two distinct contributions - one from the environment, $I_\mathrm{env}$, and one from the internal linearized interactions, $I_\mathrm{int}$ - and that they are disentangled in form:
\begin{equation}
    I\left(\{g_\mu\}, \frac{D_-}{D_+}, \frac{w_-}{w_+},  \tau w_\mathrm{sum}\right) = I_\mathrm{int}(\left\{g_\mu\}\right) + I_\mathrm{env}\left(\frac{D_-}{D_+}, \frac{w_-}{w_+},  \tau w_\mathrm{sum}\right).
\end{equation}
That is, in systems linearized close to equilibrium the internal dependencies are not altered by the environment and, vice versa, the information induced by the unobserved environmental changes is not affected by interactions between the observed degrees of freedom. In principle, as we will see in the rest of this Chapter, this result needs not to hold in general - but it is rather a fundamental property of linear interactions. In Appendix \ref{app:viewpoint}, we report the non-technical commentary that appeared in \emph{Physics 14, 162} about this result \cite{nicoletti2021mutual}, which further elucidates its relevance in different contexts.

Let us also note that, from the previous Chapter, we know that even with non-linear interactions in the fast-jumps limit the environmental contribution vanishes exactly. Hence, and independently of the underlying interactions, any non-zero mutual information in the fast-jumps limit acts as a fingerprint of the presence of internal couplings. This result is extremely interesting. In fact, although the environmental states - identified by $D_-$ and $D_+$ in our model - are usually not experimentally accessible, it might be possible to characterize the frequency of such environmental changes. Neural activity originating from external stimuli \cite{mcdonnel2011noise, temereanca2008whisker, mariani2021avalanches}, stirring in chemical conglomerates \cite{viedma2011homochirality}, temperature-activated chemical reactions in solutions \cite{astumian2019kinetic,dass2021furanose}, and population growth \cite{wienand2017fluctuating, kussell2005phenotypic, visco2010catastrophic}, are only a few examples in which this framework might apply. Even if fast-varying environments have been shown to be informative, our approach might provide hints about the presence of interactions even away from the fast-jumps limit, by bounding the environmental contribution to the mutual information. Yet, to fully understand how internal interactions and changing environments shape the dependencies of a system, we need to introduce non-linear interactions - often at the price of analytical tractability, as we will see in the next Section.

\section{Non-linear potentials and information interference}
Solving Fokker-Planck equations in the presence of non-linear interactions is often a hard task. Hence, to keep things as analytically tractable as possible, we assume that the force field $\vb{F}$ appearing in Eq.~\eqref{eqn:p2c2:langevin_with_env} can be written as the gradient of a potential of the form
\begin{equation}
\label{eqn:p2c2:interacting_potentials}
    V(x_1, x_2) = \sum_{\mu = 1}^2 U(x_\mu) + V_{\mathrm{int}} (x_1, x_2)
\end{equation}
so that we can immediately solve the Fokker-Planck equation in a fixed environment as
\begin{equation}
\label{eqn:p2c2:potential_components}
    P_i^{\mathrm{st}}(x_1, x_2) \propto {\mathrm{exp}}\left[-\frac{V(x_1,x_2)}{D_i}\right]
\end{equation}
up to a normalization factor. Without loss of generality, we will often consider the case of $U(x_\mu) = x_\mu^4/4\tau$, so we can leverage the results of the previous Chapter. Hence, we are focusing on equilibrium systems with non-linear relaxation and non-linear interactions.

\subsection{General decomposition of the mutual information}
From Section \ref{sec:p2c1:mutual_properties} we know that the mutual information can only depend on dimensionless parameters. Besides the ratio of the rates of the environmental jump process, we denoted the dimensionless parameters for a given environmental state $i$ with $\{\psi\}_i$. In particular, recall that the average over the environmental states of the mutual information associated with the joint distribution in Eq.~\eqref{eqn:p2c2:potential_components} is given by
\begin{equation*}
    \ev{I_\mathrm{int}}_\pi := \sum_{i = 1}^M \pi_i^\mathrm{st} I_\mathrm{int}(\{\psi\}_i).
\end{equation*}
This term stems from the presence of internal interactions, since in the absence of $V_{\mathrm{int}}$ such components are factorizable and their mutual information is zero. If the dimensionless parameters do not change with $i$ - such as in the case of linear interactions - it reduces to $\ev{I_\mathrm{int}}_\pi = I_\mathrm{int}(\{\psi\})$.

With this in mind, and inspired by the results obtained with linearized interactions, it will be useful to identify all the different contributions to the mutual information of the overall system. As before, we call $I_{\mathrm{env}}$ the mutual information stemming from the shared environment alone, which we obtained in Section \ref{sec:p2c1:nonlinear} for $U(x_\mu) = x_\mu^4/4\tau$. Using the notation of Section \ref{sec:p2c1:mutual_properties}, we then choose to decompose the mutual information of the overall system as
\begin{align}
    \label{eqn:p2c2:mutual_decomposition}
    I\left(\{W_\mathrm{jumps}\}, \{\varphi\}, \{\psi\}_{i=1}^M \right) = & \,\, I_{\mathrm{env}}\left(\{W_\mathrm{jumps}\}, \{\varphi\}\right) + \nonumber \\
    & + \ev{I_{\mathrm{int}}}_\pi\left(\{\psi\}_{i=1}^M \right) + \nonumber \\
    & + \Xi \left(\{W_\mathrm{jumps}\}, \{\varphi\}, \{\psi\}_{i=1}^M \right)
\end{align}
where $\{\psi\}_{i=1}^M = \{\{\psi\}_1, \dots, \{\psi\}_M\}$. For our system with two environmental states, we have $\{W_\mathrm{jumps}\} = \{w_-/w_+\}$ and $\{\varphi\} = \{D_-/D_+\}$. The term $\Xi$ in Eq.~\eqref{eqn:p2c2:mutual_decomposition} quantifies the contributions due to the presence of both the environment and the interactions at once. In general, this is not a mutual information, i.e., it needs not be positive, and may depend on all dimensionless parameters. For these reasons, we name this term as \emph{information interference}. Clearly, when interactions can be linearized this interference term is always zero.

Since we know that in fast environments the system is solved by
\begin{equation*}
    p_\mathrm{fast}(x_1, x_2) \propto {\mathrm{exp}}\left[-\frac{V(x_1,x_2}{\ev{D}_\pi}\right]
\end{equation*}
and that there will be no dependency induced by the environment, we will focus on the slow-jumps limit throughout the rest of this Chapter. In this limit, finding analytical expressions for all the terms in Eq.~\eqref{eqn:p2c2:mutual_decomposition} is often challenging, and we have to resort to numerical integration once more. Hence, unless otherwise specified, the mutual information in Eq.~\eqref{eqn:p2c2:mutual_decomposition} is obtained as outlined in the previous Chapter. We employ Hamiltonian Monte Carlo to sample the joint distribution in Eq.~\eqref{eqn:p2c2:potential_components}, and we weigh these samples according to the corresponding mixture distribution, $\pi_i^\mathrm{st}$. Then, the mutual information integral is evaluated by Monte Carlo integration. Crucially, Monte Carlo sampling requires the knowledge of the analytical expressions of both the joint and the marginal mixture components \cite{landau2021guide}, which we need to compute for every choice of potential.

\subsection{Destructive information interference}
\begin{figure*}[t]
    \centering
    \includegraphics[width=\textwidth]{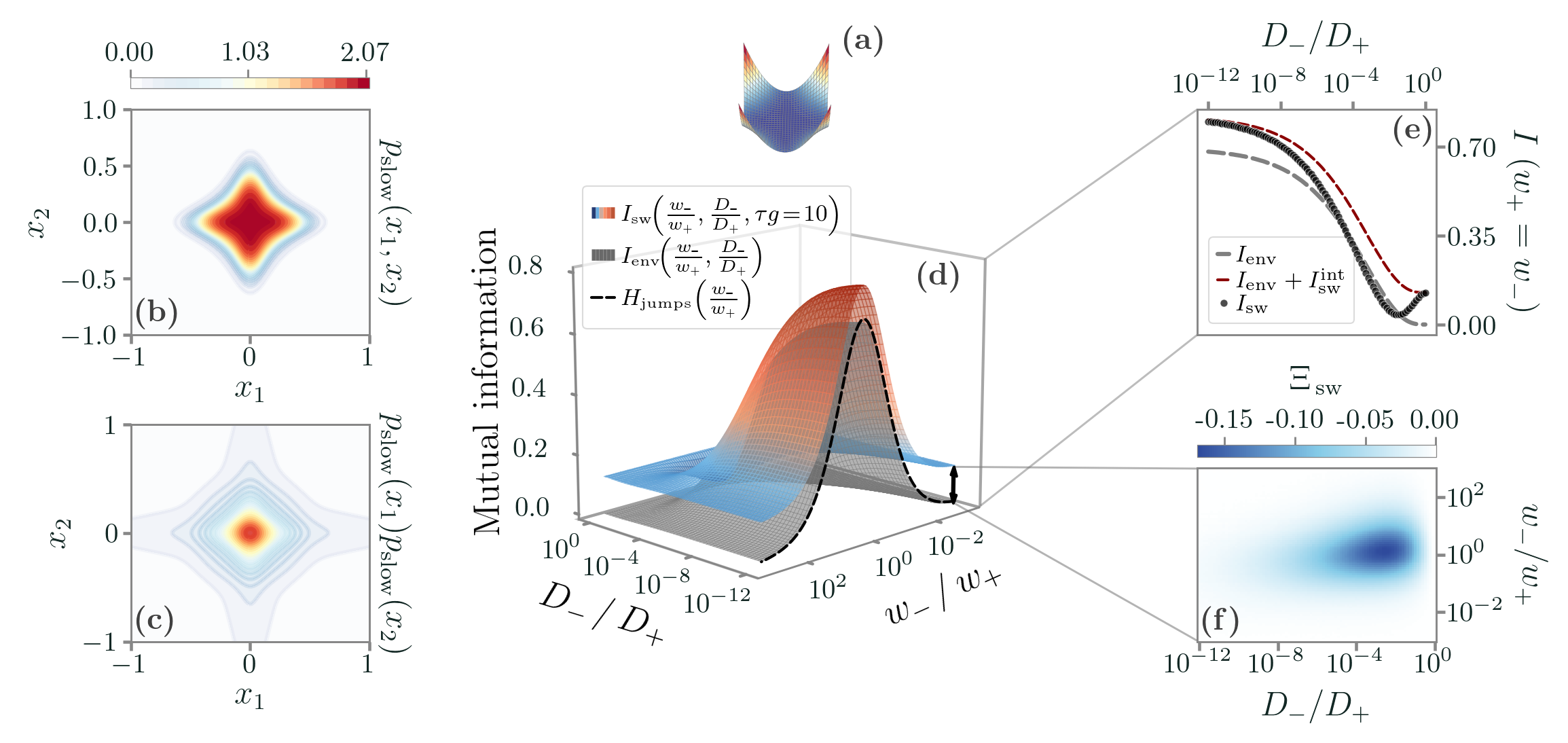}
    \caption{The total mutual information $I_{\mathrm{sw}}$ in the single-well case, Eq.~\eqref{eqn:p2c2:single_well_potential}, as a function of $D_-/D_+$ and $w_-/w_+$, in the slow-jumps limit. For all plots, $\tau g = 10$. (a) The single-well potential considered here. (b-c) Contour plots of the joint probability distribution and its factorization, respectively, for $D_+ = 10$, $D_- = 10^{-2}$, $w_- = w_+$. The marginalized probability has much longer tails along the axis $x= 0$ and $y= 0$, which tend to be suppressed by the environment. However, these tails are still present in the joint probability as a consequence of the interactions. (d) The colored surface is the result of a Monte Carlo integration of $I_{\mathrm{sw}}$, whereas the gray surface represents the environmental contribution alone $I_{\mathrm{env}}$. (e) $I_{\mathrm{sw}}$ (black dots) can be smaller than $I_{\mathrm{env}}$ (gray dashed line) and, in general, it is lower than the sum of $I^{\mathrm{int}}_{\mathrm{sw}}$ and $I_{\mathrm{env}}$ (red dashed line). (f) In fact, the term $\Xi_{\mathrm{sw}}$ is always negative, showing that the effects of the environment and of the interactions are reciprocally masked at low-enough values of $D_-/D_+$. For $D_-/D_+ \to 0, 1$ we find $\Xi_{\mathrm{sw}} = 0$, hence the environmental and the interactions contributions are disentangled}
    \label{fig:p2c2:mutual_single_well}
\end{figure*}

Let us begin with the potential
\begin{equation}
    \label{eqn:p2c2:single_well_potential}
    V(x_1, x_2) = \frac{x_1^4 + x_2^4}{4\tau} - g \frac{x_1^2 x_2^2}{2} := V_{\mathrm{sw}} (x_1, x_2)
\end{equation}
where, for stability, $g > 0$. The subscript ``sw'' refers to the fact that this is a single-well potential, depicted in Figure~\ref{fig:p2c2:mutual_single_well}a, with one stable minimum at $(x_1, x_2) = (0,0)$. In the slow-jumps limit, the mixture components of the joint distributions follow the Boltzmann-like distribution
\begin{equation}
    \label{eqn:p2c2:single_well_joint}
    P_i^{\mathrm{st}}(x_1, x_2) = \frac{1}{\mathcal{N}_\mathrm{sw}(\tau, g, D_i)}e^{-V_{\mathrm{sw}}(x_1, x_2)/D_i}
\end{equation}
where the normalization $\mathcal{N}_{\mathrm{sw}}$ can be computed analytically in terms of hypergeometric functions. The corresponding marginal components are
\begin{equation}
    \label{eqn:p2c2:single_well_marginal}
    P_i^{\mathrm{st}}(x_\mu) = \frac{|x_\mu|}{\mathcal{N}_{\mathrm{sw}(\tau, g, D_i)}} \sqrt{\frac{g \tau}{2}} K_\frac{1}{4}\left(\frac{g^2x_\mu^4\tau}{8 D_i}\right)e^{(-2+g^2\tau^2)\frac{x_\mu^4}{8 D_i \tau}}
\end{equation}
where $K_{n}(x)$ is the modified Bessel function of the second kind. We plot the corresponding mixture distributions in Figure~\ref{fig:p2c2:mutual_single_well}b-c.

With this choice of the potential, the sets dimensionless parameters are given by $\{W_\mathrm{jumps}\} = \{w_-/w_+\}$, $\{\varphi\} = \{D_-/D_+\}$ and $\{\psi\}_i = \{g \tau\}$. Once more, as with the case of linear interactions, we have $\{\psi\}_i = \{\psi\}$, leading to
\begin{equation*}
    \ev{I_\mathrm{int}}_\pi := I_{\mathrm{sw}}^{\mathrm{int}}\left(\tau g\right),
\end{equation*}
which is the mutual information arising from the presence of the non-linear interacting potential. In other words, and as for the linear case, the mutual information of the joint distribution, $I_{\mathrm{sw}}$, cannot depend separately on $D_-$ and $D_+$, thus being independent of the environmental state. Consequently, we write it as
\begin{equation}
    \label{eqn:p2c2:single_well_mutual_decomposition}
    I_{\mathrm{sw}}\left(\frac{w_-}{w_+}, \frac{D_-}{D_+}, \tau g \right) = I_{\mathrm{env}}\left(\frac{w_-}{w_+}, \frac{D_-}{D_+}\right) + I^{\mathrm{sw}}_{\mathrm{int}}\left(\tau g\right) + \Xi_{\mathrm{sw}} \left(\frac{w_-}{w_+}, \frac{D_-}{D_+}, \tau g \right).
\end{equation}
Unfortunately, no analytical expression can be found for these terms.

Nevertheless, let us investigate separately the impact of interactions and environmental changes in this example. First, the effect of the interactions in the joint probability distribution reflects into the appearance of tails along the axes $x_1 = 0$ and $x_2 = 0$, as we can see in Figure~\ref{fig:p2c2:mutual_single_well}b-c. The higher $g$, the longer the tails - indeed, due to the shape of the potential and the flat regions around the axes (Figure~\ref{fig:p2c2:mutual_single_well}a), the dependency induced by the presence of interactions leads to a high probability of finding one particle close to the minimum at the origin, and the other particle farther from it. Conversely, as shown in Section \ref{sec:p2c1:nonlinear}, the environment affects the joint distribution by suppressing such tails as the difference between $D_-$ and $D_+$ becomes more pronounced. Indeed, the shared environmental changes lead to the particles experiencing either a high- or low-diffusion regime at the same time - suppressing the probability of one particle diffusing away from the potential minima at the origin while the other is close to it.

As the dependencies induced by the environment and interactions operate in opposite ways, the mutual information takes contrasting contributions. In Figure~\ref{fig:p2c2:mutual_single_well}d-e-f, we see that the mutual information of the overall system, $I_{\mathrm{sw}}$, is always smaller than the sum of $I_{\mathrm{env}}$ and $I_{\mathrm{sw}}^{\mathrm{int}}$. Furthermore, it can also be smaller than $I_{\mathrm{env}}$ for some values of $(w_-/w_+, D_-/D_+)$. This means that
\begin{equation}
\label{eqn:p2c2:destructive}
     \Xi_{\mathrm{sw}} \left(\frac{w_-}{w_+}, \frac{D_-}{D_+}, \tau g \right) \leq 0
\end{equation}
in the entire space, and $\Xi_{\mathrm{sw}} < -I_{\mathrm{sw}}^{\mathrm{int}}$ in some regions of the parameter space (see Figure~\ref{fig:p2c2:mutual_single_well}f). Naively speaking, non-linear interactions can mask environmental information by counteracting the dependency induced by a switching environment and effectively reducing the information that $x_1$ and $x_2$ share. We name the phenomenon described by Eq.~\eqref{eqn:p2c2:destructive} \textit{destructive information interference}.

However, the limiting behaviors of $I_{\mathrm{sw}}$ can still be understood. When $D_-/D_+ \to 1$, the only contribution to the mutual information comes from the interactions alone, $I_{\mathrm{sw}}^{\mathrm{int}}(\tau g)$. Similarly, in the opposite limit $D_-/D_+ \to 0$, the numerical integration shows that the two contributions to the mutual information are exactly disentangled, i.e., 
\begin{equation}
    \label{eqn:p2c2:single_well_limits}
    I_{\mathrm{sw}}\left(\frac{w_-}{w_+}, \frac{D_-}{D_+}, \tau g \right) = 
    \begin{cases}
        H_{\mathrm{jumps}} + I_{\mathrm{sw}}^{\mathrm{int}}(\tau g) & \text{if} \quad D_-/D_+ \ll 1 \\
        \,I_{\mathrm{sw}}^{\mathrm{int}}(\tau g) & \text{if} \quad D_-/D_+ \approx 1
    \end{cases}.
\end{equation}
Eq.~\eqref{eqn:p2c2:single_well_limits} implies that in both limits $\Xi_{\mathrm{sw}} \to 0$. In Figure~\ref{fig:p2c2:mutual_single_well}e, we compare this fully disentangled form (in red) with the mutual information at fixed $w_-/w_+$ and for different values of $D_-/D_+$ (black dots). We see that, indeed, this disentangling is only achieved in the limits in Eq.~\eqref{eqn:p2c2:single_well_limits} - whereas at intermediate values of $D_-/D_+$ destructive interference reduces $I_{\mathrm{sw}}$. This suggests that, away from dynamical fixed points where internal interactions cannot be linearized, the interplay between them and environmental changes can create non-trivial mixed dependencies in the system.

\subsection{Constructive information interference}
In the previous section, we argued that the destructive information interference stems from the fact that interactions and environment tend to generate opposite dependencies in the system. If this phenomenon is at the root of the destructive information interference we observed, what happens when we rotate the interaction term in Eq.~\eqref{eqn:p2c2:single_well_potential} of an angle $\pi/4$?

\begin{figure*}[t]
    \centering
    \includegraphics[width=\textwidth]{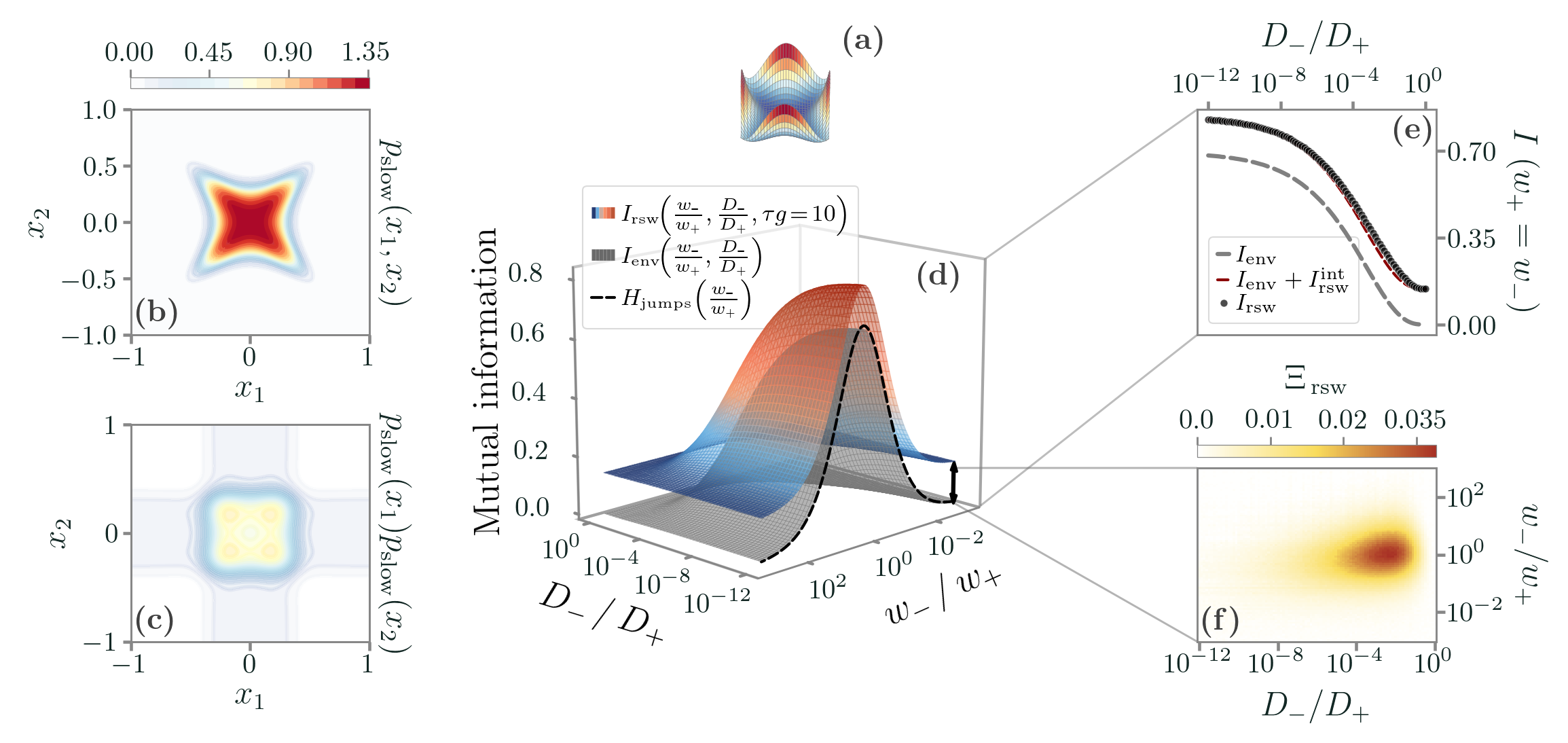}
    \caption{The total mutual information $I_{\mathrm{rsw}}$ in the rotated single-well case, Eq.~\eqref{eqn:p2c2:rotated_single_well_potential}, as a function of $D_-/D_+$ and $w_-/w_+$, in the slow-jumps limit. For all plots, $\tau g= 10$. (a) The rotated single-well potential considered here. (b-c) Contour plots of the joint probability distribution and its factorization, respectively, for $D_+ = 10$, $D_- = 10^{-2}$, $w_- = w_+$. The marginalized probability has much longer tails along the axis $x= 0$ and $y= 0$, which are suppressed in the joint probability as a consequence of the environment. Contrary to the single-well case, the interactions here trigger the presence of tails along the bisectors of the $(x,y)$ plane. (d) The colored surface is the result of a Monte Carlo integration of $I_{\mathrm{rsw}}$, whereas the gray surface represents the environmental contribution alone $I_{\mathrm{env}}$. (e) $I_{\mathrm{rsw}}$ (black dots) is always greater than $I_{\mathrm{env}}$ (gray dashed line) and, in general, it is greater than the sum of $I^{\mathrm{int}}_{\mathrm{rsw}}$ and $I_{\mathrm{env}}$ (red dashed line). (f) The term $\Xi_{\mathrm{rsw}}$ is always positive, and in particular it is different from zero at high enough values of $D_-/D_+$. For $D_-/D_+ \to 0, 1$ we end up with  $\Xi_{\mathrm{rsw}} = 0$ as expected}
    \label{fig:p2c2:mutual_single_well_rotated}
\end{figure*}

The potential governing the system is now
\begin{equation}
    \label{eqn:p2c2:rotated_single_well_potential}
    V_{\mathrm{rsw}}(x_1, x_2) = \frac{x_1^4 + x_2^4}{4\tau} - g \frac{||R_{\pi/4}(x, y)||^2}{2}
\end{equation}
where $R_\theta$ is the rotation matrix of angle $\theta$, $||\cdot||^2$ is the $L_2$ norm and $g>0$. The ``rsw'' subscript stands for rotated single-well, in analogy with the previous case. Such potential is shown in Figure~\ref{fig:p2c2:mutual_single_well_rotated}a, and as expected from the rotation the flat region around the minimum at $(x_1, x_2) = (0,0)$ now expands along the bisectors of the $(x_1, x_2)$ plane, rather than its axes.

Since the rotation does not change the parametric dependence, the sets dimensionless parameters are still $\{W_\mathrm{jumps}\} = \{w_-/w_+\}$, $\{\varphi\} = \{D_-/D_+\}$ and $\{\psi\} = \{g \tau\}$. Hence, we must have that
\begin{equation*}
    \ev{I_\mathrm{int}}_\pi := I_{\mathrm{rsw}}^{\mathrm{int}}\left(\tau g\right),
\end{equation*}
similarly to the previous case. The joint and marginal mixture components in the slow-jumps limit can be again found analytically, and are given by
\begin{equation}
    \label{eqn:p2c2:rotated_single_well_joint}
    P_i^{\mathrm{st}}(x_1, x_2) = \frac{1}{\mathcal{N}_\mathrm{rsw}(\tau, g, D_i)}e^{-V_{\mathrm{rsw}}(x_1, x_2)/D_i}
\end{equation}
and 
\begin{align}
    \label{eqn:p2c2:rotated_single_well_marginal}
    P_i^{\mathrm{st}}(x_\mu) = & \sqrt{\frac{g\pi^2}{2 \alpha}} \frac{|x_\mu| \left[I_{-\frac{1}{4}}\left(\beta_i x_\mu^4\right) + I_{\frac{1}{4}}\left(\beta_i x_\mu^4\right)\right]}{2\mathcal{N}_\mathrm{rsw}(\tau, g, D_i)} \exp\left[\frac{x_\mu^4(g^2 - 128 \alpha)}{128\alpha D_i}\right]
\end{align}
where $I_n(x)$ is the modified Bessel function of the first kind, $\alpha = \tau^{-1}/4 + g/8$, $\beta_i = g^2/(128 \alpha D_i)$, and $\mathcal{N}_\mathrm{rsw}$ has an exact expression in terms of hypergeometric functions.

Overall, the decomposition of the mutual information
\begin{equation}
    \label{eqn:p2c2:rotated_single_well_mutual_decomposition}
    I_{\mathrm{rsw}}\left(\frac{w_-}{w_+}, \frac{D_-}{D_+}, \tau g \right) = I_{\mathrm{env}}\left(\frac{w_-}{w_+}, \frac{D_-}{D_+}\right) + I^{\mathrm{rsw}}_{\mathrm{int}}\left(\tau g\right) + \Xi_{\mathrm{rsw}} \left(\frac{w_-}{w_+}, \frac{D_-}{D_+}, \tau g \right)
\end{equation}
is formally identical to Eq.~\eqref{eqn:p2c2:single_well_mutual_decomposition}. However, in this case, the role of the interactions is to introduce tails along the bisectors of the $(x_1, x_2)$ plane, whereas the environment keeps acting on the $x_1 = 0$ and $x_2 = 0$ axes. Hence, non-linear interactions do not counteract the dependency induced by the environment anymore. As a consequence, as shown in Figure~\ref{fig:p2c2:mutual_single_well_rotated}d-e, the mutual information of the overall system is almost always close to the sum of the environmental and the interaction terms, i.e., $\Xi_{\mathrm{rsw}} \approx 0$. However, in general we find
\begin{equation}
\label{eqn:p2c2:constructive}
     \Xi_{\mathrm{rsw}} \left(\frac{w_-}{w_+}, \frac{D_-}{D_+}, \tau g \right) \geq 0
\end{equation}
as we see from Figure~\ref{fig:p2c2:mutual_single_well_rotated}f. This implies that, in some regions of the parameter space, $x_1$ and $x_2$ share more information than the one coming from the changing environment and their sheer couplings. In analogy with the previous case, we name this phenomenon \textit{constructive information interference} - naively speaking, this particular choice of the potential leads to a cooperation of the environmental term and the interaction term, boosting the overall mutual information.

We remark that, yet again, the limiting behaviors of the mutual information exhibit an exact disentangling, as before. Thus, when $D_-/D_+ \to 0$, $\Xi_{\mathrm{rsw}} \to 0$ and $I_{\mathrm{env}} \to H_{\mathrm{jumps}}$, while for $D_- \to D_+$ only $I_{\mathrm{rsw}}^{\mathrm{int}}$ survives.

\subsection{Bistable systems and information peaks}
\begin{figure*}[t]
    \centering
    \includegraphics[width=\textwidth]{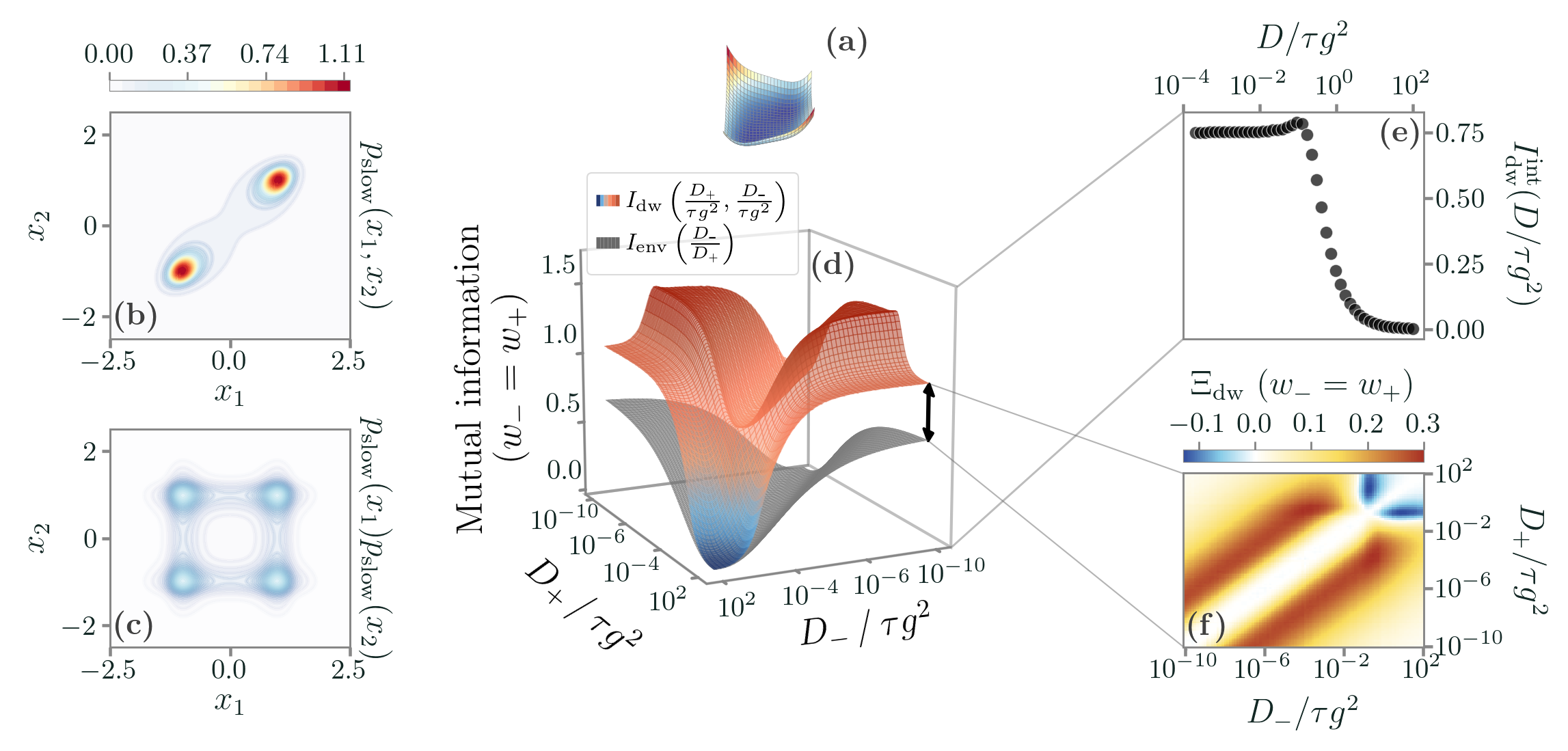}
    \caption{The total mutual information $I_{\mathrm{dw}}$ in the double-well case, Eq.~\eqref{eqn:p2c2:double_well_potential}, as a function of the adimensional parameters $D_\pm/(\tau g^2)$, in the slow-jumps limit. For all plots, $w_-/w_+ = 1$, $\tau = g = 1$. (a) The double-well potential considered here. (b-c) Contour plots of the joint probability distribution and its factorization, respectively, for $D_+ = 10$, $D_- = 10^{-2}$. The joint probability has two peaks, corresponding to the two minima of the potential. On the contrary, the marginalized probability is markedly different, with four peaks. (d) The colored surface is the result of a Monte Carlo integration of $I_{\mathrm{dw}}$, whereas the gray surface represents the environmental contribution alone $I_{\mathrm{env}}$, in the plane $(D_-/\tau g^2, D_+/\tau g^2)$. (e) The mutual information $I_{\mathrm{dw}}^{\mathrm{int}}$ of the interactions only. At large $D$, we expect the two minima to be less relevant, and indeed the mutual information vanishes. At small $D$, instead, $I_{\mathrm{dw}}^{\mathrm{int}}$ is markedly different from zero since the particles are typically trapped in one of the two minima. At intermediate values the mutual information peaks due to an interplay between trapping and diffusion. (e) The term $\Xi_{\mathrm{dw}}$ can be either positive or negative, meaning that, at different values of $D_\pm/(\tau g^2)$, we find both constructive and destructive interference. Clearly, when $D_- \approx D_+$, we have $\Xi_{\mathrm{dw}} \approx 0$}
    \label{fig:p2c2:mutual_double_well}
\end{figure*}

As a last example, here we study the slightly more complicated case of a bistable system. In particular, we consider the potential
\begin{equation}
    \label{eqn:p2c2:double_well_potential}
    V_{\mathrm{dw}}(x_1, x_2) = \frac{x_1^4 + x_2^4}{4\tau} - g x y
\end{equation}
where $g>0$. The ``dw'' subscript stands for double-well. In fact, this potential, depicted in Figure~\ref{fig:p2c2:mutual_double_well}a, has two stable minima at $(x_1,x_2) = (\pm \sqrt{g\tau}, \pm \sqrt{g\tau})$. The joint and the marginal mixture components in the slow-jumps limit are:
\begin{equation}
    \label{eqn:p2c2:double_well_joint}
    P_i^{\mathrm{st}}(x_1, x_2) = \frac{1}{\mathcal{N}_{\mathrm{dw}}(\tau, g, D_i)}e^{-V_{\mathrm{dw}}(x_1, x_2)/D_i}
\end{equation}
and 
\begin{align}
    \label{eqn:p2c2:double_well_marginal}
    P_i^{\mathrm{st}}(x_\mu) = & \,\biggl[(D_i^9\tau)^\frac{1}{4} \Gamma\left(\frac{1}{4}\right) {_0}F_2\left(\frac{1}{2},\frac{3}{4}; \alpha_i x_\mu^4\right) + \nonumber \\
    & + g^2x_\mu^2(D_i\tau)^\frac{3}{4} \Gamma\left(\frac{3}{4}\right) {_0}F_2\left(\frac{5}{4},\frac{3}{2}; \alpha_i x_\mu^4\right)\biggl] \frac{e^{-\frac{x_\mu^4}{4 D_i \tau}}}{\sqrt{2}D_i^2 \mathcal{N}_\mathrm{dw}(\tau, g, D_i)}\nonumber
\end{align}
where ${_p}F_q(a_1, \dots, a_p;b_1, \dots, b_q;x)$ is the generalized hypergeometric function, $\alpha_i = g^4 \tau/(64D_i^3)$, and $\mathcal{N}_\mathrm{dw}$ can be found analytically. As we can see in Figure~\ref{fig:p2c2:mutual_double_well}b-c, the joint probability distribution has two peaks corresponding to the two minima of the potential, whereas the factorized distribution presents four peaks - that is, the dependency induced by the internal interactions reflects into the fact that both particles are likely found in either $(\sqrt{g\tau}, \sqrt{g\tau})$ or $(-\sqrt{g\tau}, -\sqrt{g\tau})$. On the other hand, the shared environment here induces more complex dependencies. In fact, the switching between high- and low-diffusion regimes favors the transition from one minimum to the other, and of both particles at the same time. This can be readily understood by the presence, in the factorized probability, of connections among the four peaks, which are not present in the joint probability distribution that accounts for the environmental-induced dependencies.

The crucial difference between this case and the previous ones is that the dimensionless parameters appearing in the mutual information mix environmental and interaction features. Indeed, we have $\{W_\mathrm{jumps}\} = \{w_-/w_+\}$, $\{\varphi\} = \{D_-/D_+\}$ and $\{\psi\}_i = \{D_i/\tau g^2\}$. Hence, now the mutual information can be decomposed as
\begin{equation*}
    I_{\mathrm{dw}}\left(\frac{w_-}{w_+}, \frac{D_-}{\tau g^2}, \frac{D_+}{\tau g^2}\right) = I_{\mathrm{env}}\left(\frac{w_-}{w_+}, \frac{D_-}{D_+}\right) + \sum_{i \in \pm} \pi_i I_{\mathrm{dw}}^{\mathrm{int}}\left(\frac{D_i}{\tau g^2}\right) + \Xi_{\mathrm{dw}}\left(\frac{w_-}{w_+}, \frac{D_-}{\tau g^2}, \frac{D_+}{\tau g^2}\right)
\end{equation*}
where $I_{\mathrm{dw}}^{\mathrm{int}}$ is the mutual information associated to Eq.~\eqref{eqn:p2c2:double_well_joint}, which depends explicitly on the environmental state $D_i$. Notice also that the interference term, $\Xi_\mathrm{dw}$, does not depend explicitly on $D_-/D_+$ alone since this parameter can be obtained by combining $D_+/\tau g^2$ and $D_-/\tau g^2$.

To explore such dependence of the mutual information on the single environmental states, we plot it in Figure~\ref{fig:p2c2:mutual_double_well}d in the case $w_- = w_+$, whereas in Figure~\ref{fig:p2c2:mutual_double_well}e we plot the dependence of the interaction term, $I_{\mathrm{dw}}^{\mathrm{int}}$, on the diffusion coefficient. We see that, at intermediate values of $D$, this term peaks - meaning that the particles are more dependent. This fact can be understood at least heuristically. Indeed, the distance between $P_i^{\mathrm{st}}(x_1, x_2)$ and its factorization receives the most contributions from the fact that the latter has four peaks, due to the implicit assumption of independence between $x_1$ and $x_2$ in the factorized distribution. However, when $D$ is large, the system can easily escape the potential minima, and thus they will not contribute to $I_{\mathrm{dw}}^{\mathrm{int}}$, which vanishes as $D$ grows. Conversely, small values of $D$ weigh more the potential minima, since the system is substantially trapped in them. In this limit, $I_{\mathrm{dw}}^{\mathrm{int}}$ converges to a non-zero value due to the fact that only two of the peaks of the factorized distribution are present in the joint distribution. Finally, we observe an emerging peak of $I_{\mathrm{dw}}^{\mathrm{int}}$ at a finite value of $D$. This optimal diffusion naively allows the system to explore both minima from time to time, still being trapped for a consistent amount of time during each stochastic realization.

In Figure~\ref{fig:p2c2:mutual_double_well}f, we show the interference term $\Xi_{\mathrm{dw}}$ for the specific case $w_- = w_+$. All other choices do not qualitatively change the picture. In this scenario, we have
\begin{equation}
\label{eqn:p2c2:interference_double_well}
     \Xi_{\mathrm{dw}} \left(\frac{w_-}{w_+}, \frac{D_-}{\tau g^2}, \frac{D_+}{\tau g^2} \right) \gtreqless 0
\end{equation}
leading to a non-trivial pattern of constructive and destructive information interference. This pattern, although hard to understand analytically, is intuitively a consequence of the system switching from a state in which it is trapped in one single minimum, to a state in which it can freely explore larger regions of the $(x_1,x_2)$ plane. When both environmental states are represented by large diffusion coefficients, we find destructive interference - the diffusion is too large for the system to be trapped into one of the two minima and thus the environment is masking the internal dependencies. On the other hand, we find a significant constructive interference when one environmental state displays a small diffusion coefficient, and the other an intermediate one. As outlined above, in this scenario the system can explore both minima thanks to the switching environment, achieving a larger information between the two particles than with interactions alone. In this case, it is difficult to define the usual limiting behaviors of the mutual information, in which the disentangling is recovered. Indeed, $D_-/D_+$ is not the only relevant parameter of the system and the limit $D_-/D_+ \to 0$ is not particularly informative anymore.

Finally, let us remark that, although we focused on paradigmatic but rather comprehensive physical models, these ideas have a much larger scope. In fact, disentangling the different dependencies of a system is a far-reaching question. Techniques such as Bayesian networks and other probabilistic graphical models have been successfully used in biological data, for instance, to disentangle different sources of interactions and dependencies in general \cite{burger2008bayesian, layeghifard2017disentangling, burger2010disentangling}. Connections may be also drawn to machine learning and artificial neural networks, particularly in the context of learning disentangled representations of the data, i.e. representations in which the informative latent factors are described by a factorized distribution \cite{kim2018disentangling, chen2018vae, locatello2019disentangled, raban2020nn}, or in generative models with latent variables, such as switching state-space models \cite{fox2009bayesian, linderman2017rslds}. The environment in our model, in fact, can be seen as a latent variable, i.e. unobserved and independent of the observed degrees of freedom, while affecting the observed dynamics. Unlike the one presented in this thesis, these approaches are often harder to interpret and are less prone to the derivation of exact results, even though they remain extremely powerful in dealing with experimental data.

\chapter{Information-driven transitions in optimal effective models}
\chaptermark{Information-driven transitions}
\label{ch:PRE}
\lettrine{T}{the dependencies} between the many degrees of freedom of complex systems, as we have seen in the previous Chapters, are shaped by the interplay between internal interactions and shared, changing environments. However, it is often the case that we are not able to describe them as a whole, but rather seek an effective representation of their dynamics. Indeed, we may not have experimental access to all of the degrees of freedom - e.g., as we will see, we only observe the position subspace rather than the whole velocity-position phase space - or it may be more useful to study the behavior of coarse-grained variables. In this Chapter, we will show how Information Theory can help us define these effective models in such a way that they resemble the original dynamical evolution as closely as possible - and what this choice implies for the relation between the underlying system and our effective description.

\begin{figure*}[t]
    \centering
    \includegraphics[width=\textwidth]{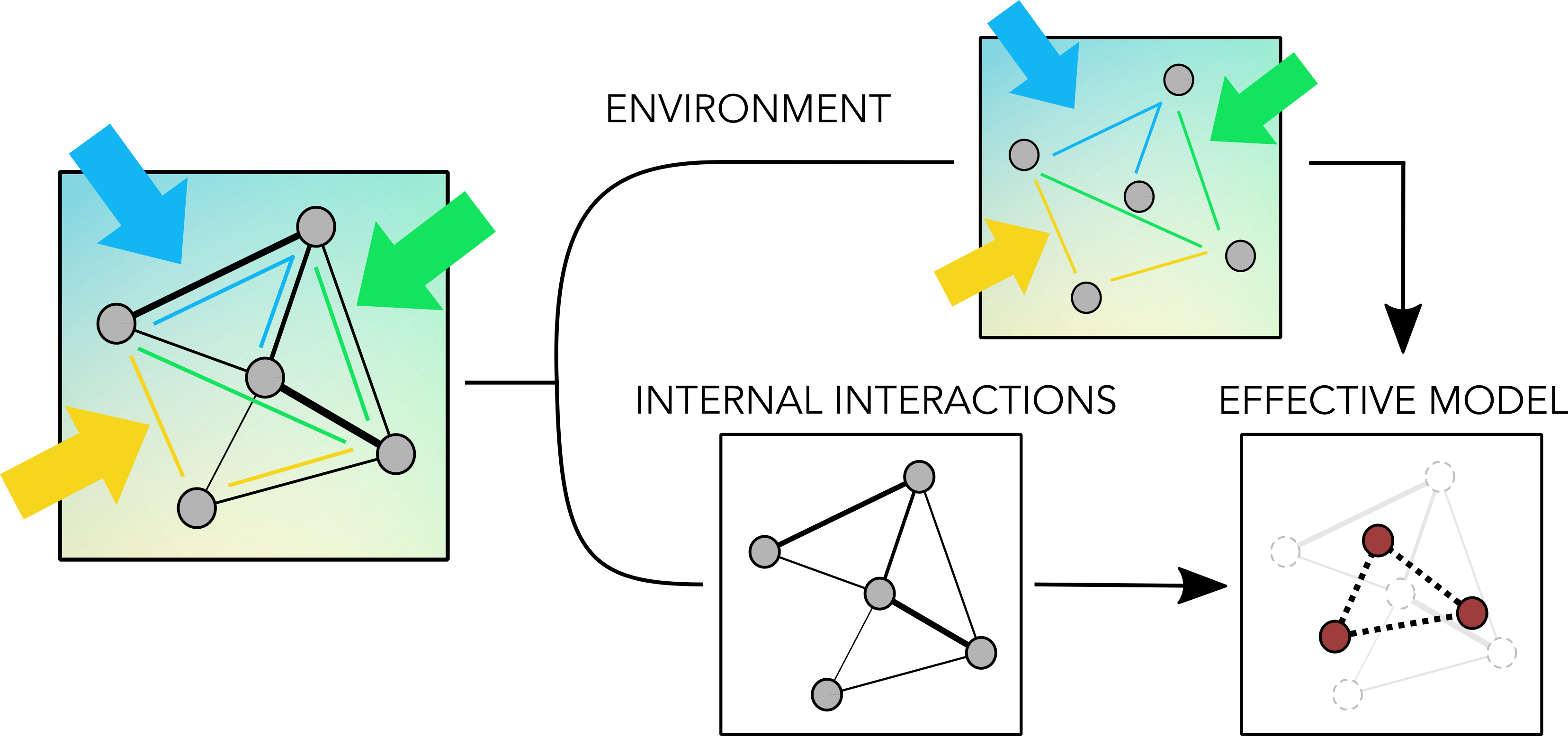}
    \caption{In the previous chapters, we have seen how the dependencies of the degrees of freedom of complex systems (gray dots) are shaped by the interplay between changing environments (colored arrows and lines) and internal interactions (black lines). Yet, often we do not seek to describe all these internal degrees of freedom and their dependencies, but rather a subset or a coarse-grained version of them. Thus, we usually deal with effective models to tackle the underlying complexity. In this Chapter, we will explore if and how such effective models can retain information about the original dynamics}
    \label{fig:p2c3:sketch_effective_models}
\end{figure*}

In Figure~\ref{fig:p2c3:sketch_effective_models} we sketch these ideas that, in general, are deeply related to the question of how to infer such effective models from experimental data. For example, higher-order discretization schemes have been recently proposed to build effective Langevin models from discrete datasets \cite{ferretti2020building}. Different Bayesian techniques have also been proposed, leveraging artificial neural networks for density estimations and the field of simulation-based inference to obtain posterior distributions of an effective model given the data \cite{papamakarios2016inference, lueckmann2017inference, wilkinson2018stochastic, cranmer2020frontier, tejero-cantero2020sbi}. Furthermore, data-driven approaches to infer dynamical features from single trajectories are solidly taking hold, especially as the experimental spatiotemporal resolution rapidly increases. This inference problem is well understood for deterministic systems \cite{crutchfield1987equations, daniels2015automated, brunton2016discovering}, while criticalities arise in stochastic systems, where fast variables have to be treated as external noise \cite{risken1996fokker}. Fundamental advances have recently been made for, e.g., stationary underdamped \cite{bruckner2020inferring} and overdamped stochastic processes \cite{el2015inferencemap, garcia2018high, frishman2020learning, ragwitz2001indispensable,gnesotto2020learning}, where the fast equilibration of velocities is employed\footnote{Crucially, it has been shown that employing this simplification \textit{ab initio} might lead to erroneous results \cite{liang2021thermophoresis,lau2007state}, whose origin dates back to the Ito-Stratonovich dilemma \cite{kupferman2004ito}. Nevertheless, the overdamped framework remains a paramount tool to gain analytical insights.}.

The idea of building projections from complex to simpler models is also at the heart of coarse-graining procedures, which can be implemented to obtain informative coarse-grained models that preserve relevant features \cite{katsoulakis2003coarse, gfeller2007spectral, altaner2012fluctuation}, and dimensionality reduction techniques \cite{coifman2008diffusion, wehmeyer2018time, otto2019linearly, swischuk2019projection, wright2022high}. However, the properties of such low-dimensional representations and how they relate to the original, high-dimensional space are often non-trivial to address.

In this Chapter, we frame the definition of an optimal model in the context of information projections between the probability space of the observed dynamics and the one of the effective model. An optimal model is such that it captures the maximum amount of information on possibly short-time trajectories of the observed degrees of freedom. We then focus on the physically relevant case of reducing the dynamics of an underdamped system in the full position-velocity phase space, $(\bm{x},\bm{v})$, to an effective overdamped dynamics in the $\bm{x}$-space. This question is, even by itself, challenging and far-reaching and might impact several other fields, ranging from the estimation of dissipation in biological systems \cite{busiello2019hyperaccurate,manikandan2020inferring,skinner2021estimating}. In doing so, we relax the stationarity assumption usually employed, and include the effect of the initial conditions. We show that the information-preserving feature of our approach is associated with an unforeseen discontinuity in the parameter space of the optimal model, associated with two minima of the information loss that exchange stability. This information-driven transition is present both in simple harmonic confinements - where we can solve the model analytically - as well as in anharmonic potentials, where we employ a Gaussian ansatz to infer the relevant quantities directly from simulated trajectories. This result poses a significant and unforeseen limitation on the efficacy of effective models in predicting underlying dynamics, as slight changes in underdamped parameters may give rise to large variations in the optimal prediction.

\section{Effective models through information theory}
\begin{figure}[t]
    \centering
    \includegraphics[width=0.8\textwidth]{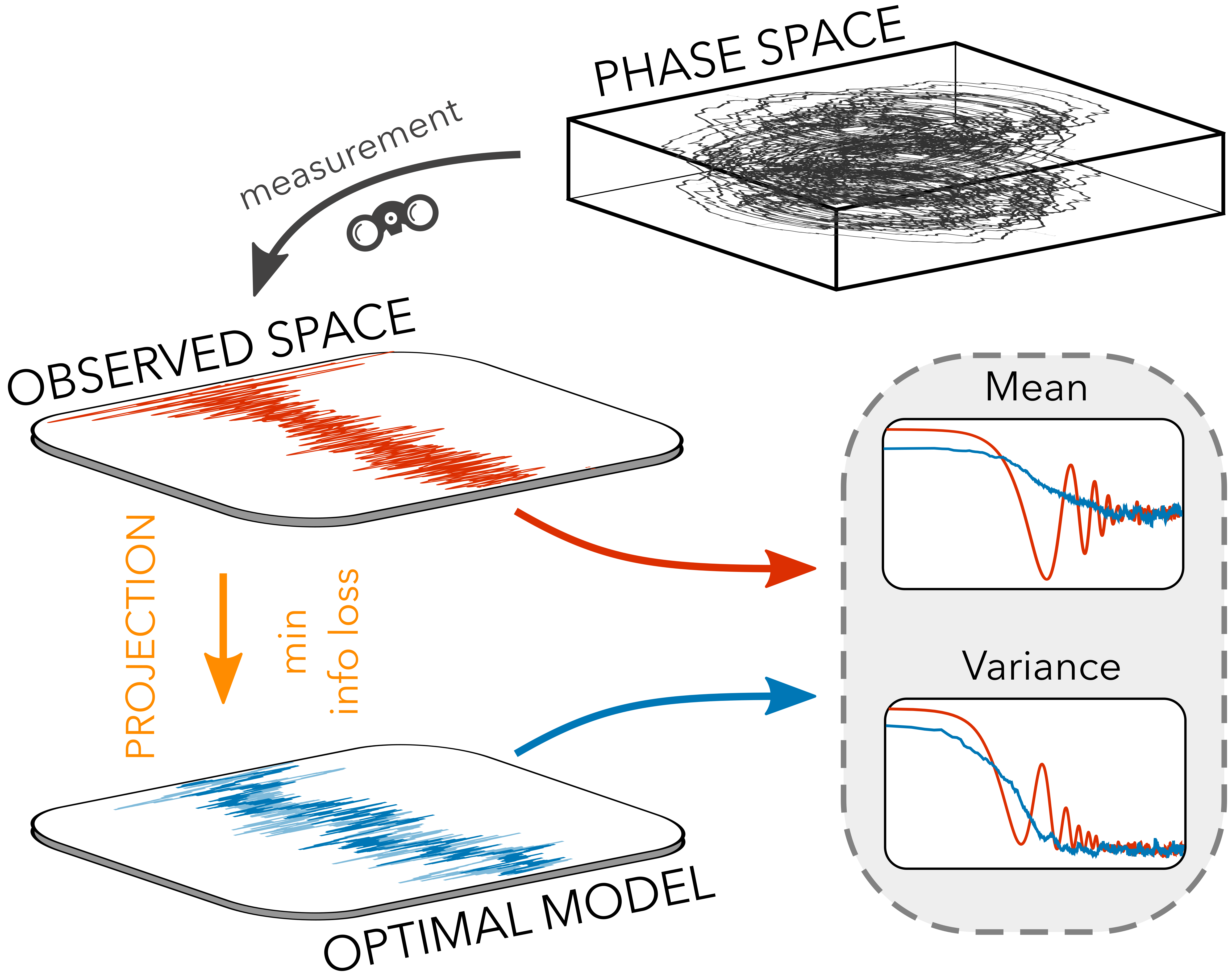}
    \caption{Sketch of the definition of optimal effective models used in this Chapter. We observe trajectories that reflect a subspace or a coarse-grained version of an underlying dynamics in a higher-dimensional phase space. We build an optimal effective model that captures as much information as possible on the observed evolution via a projection that minimizes the information loss integrated in time, Eq.~\eqref{eqn:p2c3:infoloss}. Such an optimal model is, by definition, the best effective description we can find and thus does not depend on a specific inference procedure. To make this approach computationally feasible, we often assume that the observed and effective trajectories are fully characterized by their mean and variance, which corresponds to a Gaussian ansatz}
    \label{fig:p2c3:figure1}
\end{figure}

Let us assume that our system lives in an $N$ dimensional phase space, and that it is described by the probability distribution
\begin{equation}
\label{eqn:c2p3:pcomplex}
    P_\mathrm{complex} = P_\mathrm{complex}(x_1, \dots, x_N, t).
\end{equation}
In principle, Eq.~\ref{eqn:c2p3:pcomplex} depends on time as well, as the underlying process needs not to be stationary. We further assume that we do not have access to the evolution of all degrees of freedom $\{x_n(t)\}_{n=1, \dots, N}$, but rather observe only the set $\{y_k(t)\}_{k=1, \dots, K}$ with $y_k = y_k(x_1, \dots, x_N)$ - e.g., $y_k$ may be a coarse-grained version of the system, or we may only partially observe its evolution, so that $y_k = x_k$ with $K < N$. Typically, such observed space - see the sketch in Figure~\ref{fig:p2c3:figure1} - is lower-dimensional with respect to the full phase space. Hence, we can only experimentally measure the probability distribution
\begin{equation}
\label{eqn:c2p3:pobs}
    P_\mathrm{obs} = P_\mathrm{obs}(y_1, \dots, y_K, t)
\end{equation}
and not Eq.~\ref{eqn:c2p3:pcomplex}, which is defined on the whole phase space. Determining the dynamics that generates Eq.~\ref{eqn:c2p3:pobs} is, in general, a hard task. For instance, as we will see later on, $P_\mathrm{obs}$ may be the result of a non-Markovian evolution even if $P_\mathrm{complex}$ is the solution of a Markovian dynamics. Furthermore, describing our complex system as a whole might be not possible, especially in experimental settings. A physically relevant example, which we will study throughout this chapter, is the case in which we only observe spatial trajectories while the complete dynamics is defined in the full position-velocity phase space.

Hence, one often resorts to an effective model described by
\begin{equation}
\label{eqn:c2p3:peff}
    P_\mathrm{eff} = P_\mathrm{eff}(y_1, \dots, y_K, t ; \bm{\theta}_p)
\end{equation}
where $\bm{\theta}_p$ is the vector of parameters that specify, e.g., the Fokker-Planck equation whose solution is $P_\mathrm{eff}$. In principle, we can choose a simple enough effective model that we are able to solve it - however, tuning its parameters $\bm{\theta}_p$ is a non-trivial task, and depending on the specific setting different inference procedures have been proposed \cite{ferretti2020building, papamakarios2016inference, lueckmann2017inference, wilkinson2018stochastic, cranmer2020frontier, tejero-cantero2020sbi, crutchfield1987equations, daniels2015automated, brunton2016discovering, el2015inferencemap, garcia2018high, frishman2020learning, ragwitz2001indispensable,gnesotto2020learning, bruckner2020inferring}. Yet, the goal of any inference procedure can be thought of as tuning our effective model in such a way that $P_\mathrm{eff}$ is as close as possible to our observations, $P_\mathrm{obs}$. Thus, rather than focusing on the features of a specific inference problem, here we assume that we are able to find the best possible description of the original dynamics in terms of a chosen, most likely simpler, effective model.

To this end, we build an information projection that maps $P_{\mathrm{obs}}(\bm{y},t)$ into the solution of the effective model minimizing the overall information loss during the dynamical evolution, defined as
\begin{equation}
    \label{eqn:p2c3:infoloss}
    \mathcal{F}_\mathcal{M}(\bm{\theta}_p) = \int_0^{t_{\mathrm{obs}}} dt~\mathcal{M}(P_{\mathrm{obs}}, P_{\mathrm{eff}})
\end{equation}
where $t_\mathrm{obs}$ is the duration of our observations and $\mathcal{M}$ is any information metric - e.g., a Kullback-Leibler divergence or a Chernoff divergence. That is, we choose the parameters of the effective model $\bm{\theta}_p$ using the information projection
\begin{equation}
    {\bm \theta_{\mathrm{opt}}} = \mathrm{argmin}_{\bm \theta_p} \mathcal{F}_\mathcal{M}(\bm{\theta}_p),
    \label{eqn:p2c3:minM}
\end{equation}
so that the global minimum of $\mathcal{F}_\mathcal{M}$ determines the optimal effective model. The corresponding pdf then is
\begin{equation}
    \label{eqn:p2c3:popt}
    P_{\mathrm{opt}}(\bm{y}, t) = P_{\mathrm{eff}}(\bm{x}, t \, | \,\bm{\theta}_{\mathrm{opt}}).
\end{equation}
By definition, once the effective model is specified, this is the best description an inference problem can determine in terms of information loss over the entire dynamical evolution.

Crucially, this approach is often computationally hard to solve. Even if we were able to explicitly obtain the dependence of $P_\mathrm{eff}$ on $\bm\theta_p$, finding a reliable estimate of $P_{\mathrm{obs}}$ from the data may be particularly challenging. Thus, computing exactly the metric $\mathcal{M}$ appearing in the information loss is often unfeasible. To avoid this issue, we introduce a Gaussian ansatz and assume that we can describe our probability distributions with
\begin{equation*}
    P_\mathrm{obs} \sim \mathcal{N}\left(\bm{\mu}_{\mathrm{obs}}(t), \hat{\sigma}_{\mathrm{obs}}(t)\right), \quad P_\mathrm{eff} \sim \mathcal{N}\left(\bm{\mu}_{\mathrm{eff}}(t), \hat{\sigma}_{\mathrm{eff}}(t)\right)
\end{equation*}
where $\bm{\mu}_{\mathrm{obs}}(t)$ and $\hat{\sigma}_{\mathrm{obs}}(t)$ are the evolution of the mean and covariance matrix in the observed space, and similarly $\bm{\mu}_{\mathrm{eff}}(t)$ and $\hat{\sigma}_{\mathrm{eff}}(t)$ are the analogous evolution of the effective model. Clearly, we can let go of the Gaussian assumption for the effective model if we can solve explicitly for $P_\mathrm{eff}(\bm{y}, t ; \bm\theta_p)$. Otherwise, this assumption corresponds to an effective model described by an Ornstein-Uhlenbeck (OU) process of the form
\begin{equation}
    \dot{\bm{x}}(t) = - w_p^2 \left(\bm{x}(t) - \bm{\mu}_p \right) + \sqrt{2} \hat{\Delta}_p \bm{\xi}_p(t),
    \label{eqn:p2c3:OU}
\end{equation}
where $\bm{\xi}_p$ is a white noise, and $\{w_p$, $\bm{\mu}_p$, $\hat{\Delta}_p \} := {\bm \theta_p}$. The solution of Eq.~\eqref{eqn:p2c3:OU} is $P_{\mathrm{eff}}(\bm{x},t) \equiv P_{\mathrm{OU}}(\bm{x},t) = \mathcal{N}(\bm{\mu}_{\mathrm{OU}}(t),\hat{\sigma}_{\mathrm{OU}}(t))$. Since with this assumption the pdfs appearing in Eq.~\eqref{eqn:p2c3:infoloss} are Gaussian, we can always find an exact expression for the information metric $\mathcal{M}$, greatly simplifying the task of solving Eq.~\eqref{eqn:p2c3:minM}. In practice, this also corresponds to a maximum entropy ansatz at each time if we assume that the observed trajectories are fully characterized by their mean and covariance \cite{jaynes2003probability}. This proposed definition of optimal effective models is sketched in Figure~\ref{fig:p2c3:figure1}.

Importantly, one must be careful that, in principle, $P_{\mathrm{obs}}$ and $P_{\mathrm{eff}}$ may admit different steady states for some parameters $\bm\theta_p$ - if they admit one in the first place. In fact, $\mathcal{M}(P_{\mathrm{obs}}, P_{\mathrm{eff}})$ vanishes if and only if $P_{\mathrm{eff}} = P_{\mathrm{obs}}$. Clearly, since in this case the integral in Eq.~\eqref{eqn:p2c3:minM} would diverge in the long-time limit, these parameters will not be the ones that minimize the information loss. Hence, we are free to impose that $\mathcal{M} \to 0$ when $t_{\mathrm{obs}} \to +\infty$, resulting in a constraint on the stationary mean and variance of $P_{\mathrm{eff}}$ and thus reducing the number of free parameters to optimize. As we will see explicitly in a concrete example, we can check a posteriori that the optimal solution is indeed the one in which the steady state of the effective model matches the observed one.

\section{Information-driven transitions in overdamped approximations of underdamped dynamics}
\sectionmark{Overdamped approximations of underdamped dynamics}
From now on, we will leverage the definitions introduced in the previous section and focus on the physically relevant case of a system described by an underdamped model,
\begin{equation}
    \label{eqn:p2c3:under}
    \begin{gathered}
    \dot{\bm{x}} = \bm{v} \\
    \dot{\bm{v}} = - \gamma \bm{v} + \bm{F}(\bm{x}) + \sqrt{2} \hat{\Delta} \bm{\xi}(t)
    \end{gathered}
\end{equation}
where $\gamma$ is the friction coefficient, $\hat{\Delta}^T \hat{\Delta} = \hat{D}$ is the diffusion matrix, $\bm{F}$ a generic non-linear position-dependent force, and $\bm{xi}$ uncorrelated noise with zero mean. Masses are set to unity for simplicity. This general model also includes chiral diffusion \cite{hargus2021odd}. However, it is often the case that the details of Eq.~\eqref{eqn:p2c3:under} are not known, or the model cannot be solved analytically. Furthermore, we usually have access only to measurements of short-time trajectories in the $\bm x$-space. These trajectories are those described by the probability distribution $P_{\mathrm{obs}}({\bm x},t)$, which ideally coincides with the exact solution of Eq.~\eqref{eqn:p2c3:under} marginalized over the $\bm v$-space.

Then, we need to specify the effective model, whose parametric form will shape the information loss in Eq.~\eqref{eqn:p2c3:infoloss}. In full generality, we prescribe a generic overdamped model,
\begin{equation}
    \label{eqn:p2c3:over}
    \dot{\bm{x}} = \bm{F}_p(\bm{x};\bm\vartheta_p) + \sqrt{2} \hat{\Delta}_p \bm{\xi}_p(t)
\end{equation}
where $\bm{F}_p(\bm{x};\bm\vartheta_p)$ may be a new force-field that depends on the parameters $\bm\vartheta_p$, $\hat{\Delta}_p^T \hat{\Delta}_p$ a new diffusion matrix, and $\bm{\xi}_p$ uncorrelated white noise with zero mean. The parameters of this effective model that need to be optimized by solving Eq.~\eqref{eqn:p2c3:minM} are $\bm\theta_p = \{\bm\vartheta_p, \hat\Delta_p\}$. Let us stress that, in principle, this is an arbitrary choice - our effective model could have been, e.g., a delayed process with memory. Indeed, in general, there are no Markov processes whose pdf coincides with $P_{\mathrm{obs}}(\bm{x},t)$, which is obtained from the marginalization of the Markovian dynamics described by $P_\mathrm{complex}(\bm x, \bm v, t)$. This implies that our optimal model will be simpler and possibly tractable, at the price of not being able to match perfectly the observed dynamics. That is, the overall information loss along the dynamical evolution will never be exactly zero.

As a proof of concept, we will first apply the method to systems with one spatial dimension. The underdamped dynamics lives in a $2D$ phase space and the optimal model is a $1D$ OU process. The multidimensional extension is conceptually straightforward but deserves proper attention in dealing with non-diagonal diffusivities. We will present a simple example of a $2D$ process at the end of this section. Furthermore, since the focus of this Chapter is to show that information-preserving projections might lead to emerging singularities even in simple scenarios, we will first consider the $t_{\mathrm{obs}} \to +\infty$ limit, so we have access to the whole dynamical evolution and we can effectively match the steady state distribution. Hence, no further inference error emerges from small observation times. In the last part of the Chapter, we will discuss a trajectory-based approach where this assumption is relaxed.

\subsection{One-dimensional harmonic potentials}
We begin by considering the paradigmatic case of a one-dimensional harmonically bounded particle, i.e., we write Eq.~\eqref{eqn:p2c3:under} as
\begin{equation}
    \label{eqn:p2c3:under_harmonic}
    \begin{gathered}
    \dot{{x}} = {v} \\
    \dot{{v}} = - \gamma v - \omega^2 x + \sqrt{2 k_B T\gamma^2} \xi(t)
    \end{gathered}
\end{equation}
where $\hat{\Delta} = \sqrt{k_B T\gamma^2}$ for thermodynamic consistency \cite{risken1996fokker}. We will call $v_{\mathrm{th}} = \sqrt{k_B T}$ the thermal velocity. With this choice, Eq.~\eqref{eqn:p2c3:under_harmonic} can be solved exactly. These Langevin equations correspond to the Kramers equation
\begin{align*}
        \partial_ t p + v \partial_x p = \partial_v\left[(\omega^2 x + \gamma v) p\right] + v_{\mathrm{th}}^2 \gamma \partial_v^2 p,
        \label{eqn:p2c3:kramersHBP}
    \end{align*}
with $p = p(x, v, t| x_0, v_0, 0)$. We can solve this equation, finding the Gaussian propagator
\begin{equation*}
p(x, v, t | x_0, v_0, 0) \sim \mathcal{N}(\vb{M}, \hat S)
\end{equation*}
with mean and covariance given by
\begin{gather*}
    \vb M = e^{-\Gamma t} \begin{pmatrix}x_0 \\ v_0\end{pmatrix} \\
    S_{xx} = S_0 \left(\frac{\lambda_1+\lambda_2}{\lambda_1 \lambda_2}+\frac{4 E}{\lambda_1+\lambda_2}-\frac{e^{-2 \lambda_1 t}}{\lambda_1}-\frac{e^{-2 \lambda_2 t}}{\lambda_2}\right) \\
    S_{vv} = S_0 \left(\lambda_1+\lambda_2+\frac{4\lambda_1\lambda_2 E}{\lambda_1+\lambda_2}-e^{-2 \lambda_1 t}\lambda_1-e^{-2 \lambda_2 t}\lambda_2\right) \\
    S_{xv} = S_{vx} = \frac{\gamma  v_{\mathrm{th}}^2}{(\lambda_1-\lambda_2)^2}\left(e^{-\lambda_1 t} + e^{-\lambda_2 t}\right)^2
\end{gather*}
where $S_0 = \frac{\gamma  v_{\mathrm{th}}^2}{(\lambda_1-\lambda_2)^2}$, $E = \left( e^{t (-\lambda_1-\lambda_2)}-1 \right)$, and
\begin{align*}
    \Gamma = \begin{pmatrix}
    0 & -1 \\
    \omega^2 & \gamma \\
    \end{pmatrix}, \quad\quad \lambda_{1, 2} = \frac{\gamma \pm \sqrt{\gamma^2-4\omega^2}}{2}.
\end{align*}
In order to find a Gaussian pdf, we further assume that the initial conditions $(x_0, v_0)$ are described by two independent Gaussian distributions $p(x_0) \sim \mathcal{N}(\mu_{x_0}, \sigma_{x_0}^2)$ and $p(v_0) \sim \mathcal{N}(\mu_{v_0}, \sigma_{v_0}^2)$, so that
\begin{align}
\label{eqn:p2c3:solution_kramers_1D}
    p(x, v, t) \sim \mathcal{N}\left(e^{-\Gamma t} \begin{pmatrix}\mu_{x_0} \\ \mu_{v_0}\end{pmatrix}, S + e^{-\Gamma t} \begin{pmatrix}\sigma_{x_0}^2 & 0 \\ 0 & \sigma_{v_0}^2\end{pmatrix}\left(e^{-\Gamma t}\right)^T\right)
\end{align}
is the solution to the Kramers equation we are seeking.

Eq~\eqref{eqn:p2c3:solution_kramers_1D} describes the dynamics of the system in the complete phase space. Yet, we are interested only in the position space, which we may measure from experimental spatial trajectories. Since we know the full analytical solution of the system, this assumption amounts to computing the marginal probability distribution over the $x$-space,
\begin{equation*}
    p_{\mathrm{har}}(x, t) = \int dv \, p(x, v, t).
\end{equation*}
Clearly, this is still a Gaussian distribution, with mean and variance
\begin{equation*}
\begin{gathered}
    \mu_{\mathrm{har}}(t) = e^{-\frac{\gamma t}{2}} \left[\frac{\gamma\mu_{x_0}+2 \mu_{v_0} \sinh\left(\frac{\lambda t}{2}\right)}{\lambda}+\mu_{x_0} \cosh\left(\frac{\lambda t}{2}\right)\right] \\
    \sigma^2_{\mathrm{har}}(t) = \frac{e^{-\gamma t}}{\omega^2\lambda^2} \biggl[\gamma \sigma_{x_0}^2 \omega ^2\lambda \sinh \left(\lambda t\right) + C \cosh \left(\lambda t\right) + v_{\mathrm{th}}^2\lambda\left[\lambda e^{\gamma t}-\gamma\sinh(\lambda  t)\right]  - D \biggl]
\end{gathered}
\end{equation*}
where $C = \omega^2 \left(\sigma_{x_0}^2(\gamma^2 - 2\omega^2)+2 \sigma_{v_0}^2\right)-\gamma ^2 v_{\mathrm{th}}^2$, $D = 2\omega ^2\left(\sigma_{v_0}^2+\sigma_{x_0}^2 \omega ^2-2 v_{\mathrm{th}}^2\right)$, and $\lambda = \sqrt{\gamma ^2-4 \omega ^2}$. Hence, in the harmonic case, the Gaussian ansatz proposed in the previous Section is exact, and we only need to know $\mu_{\mathrm{obs}}(t) = \mu_{\mathrm{har}}(t)$ and $\sigma_{\mathrm{obs}}(t) = \sigma_{\mathrm{har}}(t)$.

We now need to specify the effective model. We consider an Ornstein-Uhlenbeck (OU) process of the form
\begin{equation}
    \dot{x}(t) = - w_p^2 \left(x(t) - \mu_p\right) + \sqrt{2} \hat{\Delta}_p \bm{\xi}_p(t),
    \label{eqn:p2c3:OUharm}
\end{equation}
where $\{ w_p$, $\mu_p$, $\hat{\Delta}_p \} := {\bm \theta_p}$. The solution of Eq.~\eqref{eqn:p2c3:OUharm} is again a Gaussian propagator that can be marginalized over the initial condition $p(x_0)$ introduced above. Thus, we end up with the distribution $P_{\mathrm{eff}}(x,t) = P_{\mathrm{OU}}(x,t) = \mathcal{N}(\mu_{\mathrm{OU}}(t),\sigma_{\mathrm{OU}}(t))$, whose mean and variance are
\begin{gather*}
    \mu_{\mathrm{OU}}(t) = \mu_p (1 - e^{-w_p^2 t}) + \mu_{x_0}e^{-w_p^2 t} \\
    \sigma^2_{\mathrm{OU}}(t) = \frac{\Delta_p^2}{w_p^2} (1 - e^{-2 w_p^2 t}) + \sigma_{x_0}^2 e^{-2 w_p^2 t}.
\end{gather*}
Once more, in this simple case of harmonic confinement, we only need to compute the mean and the variance of the distribution to obtain the exact information loss. That is, we are not introducing any approximation in the information projection, as we are matching two Gaussian distributions through their means and variances. This allows us to study exactly the properties of the optimal model defined by Eq.~\eqref{eqn:p2c3:minM}.

\begin{figure*}[t]
    \centering
    \includegraphics[width=\textwidth]{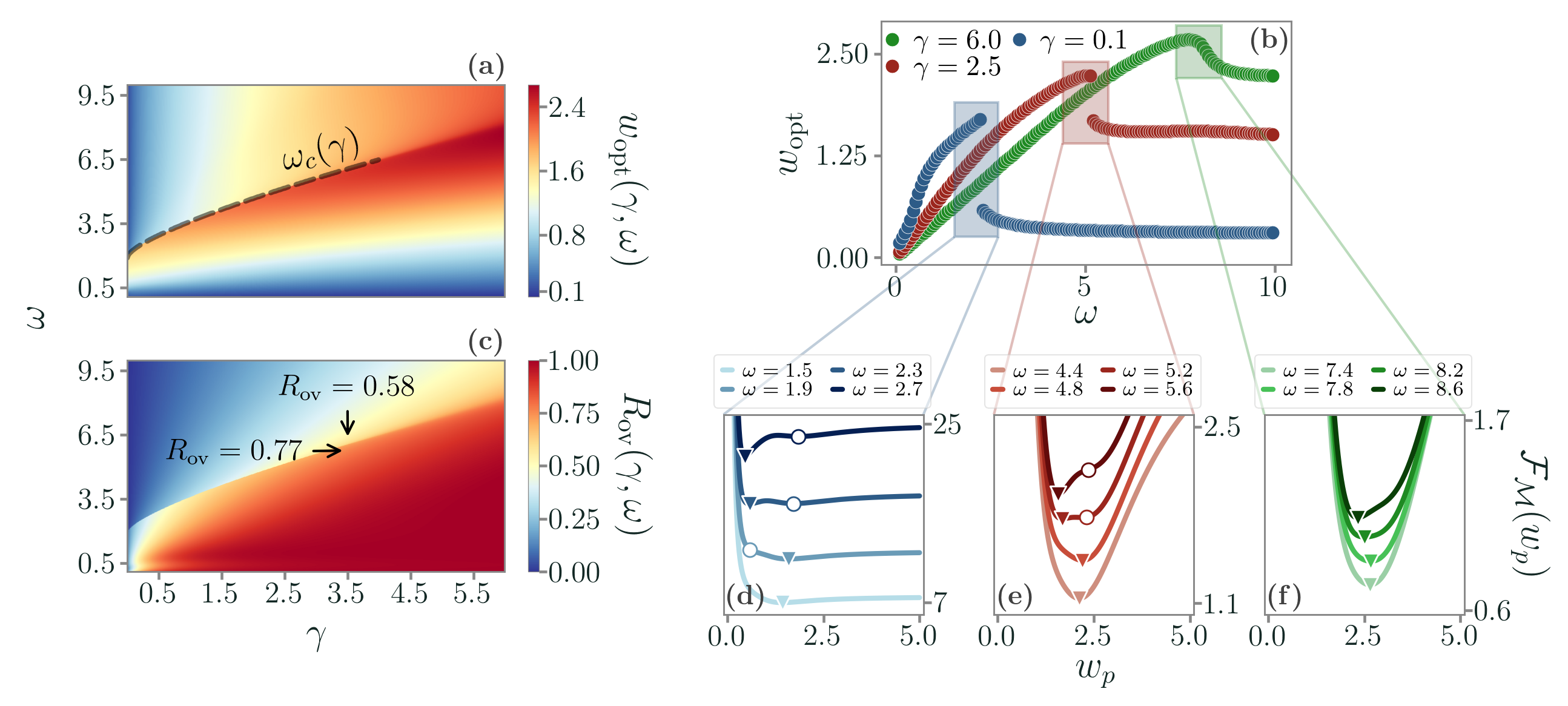}
    \caption{(a) Contour plot of $w_{\mathrm{opt}}(\gamma, \omega)$ in the harmonic case, showing a discontinuity line $\omega_c(\gamma)$. (b) By plotting $w_{\mathrm{opt}}(\omega)$ for selected values of $\gamma$, we see that the discontinuity gradually decreases as $\gamma$ increases (plotted from bottom to top) and eventually disappears. At large $\omega$, $w_{\mathrm{opt}}$ displays a plateau. (c) The ratio $R_{\mathrm{ov}}(\gamma, \omega)$ shows that for high values of $\gamma$ our method coincides with the overdamped limit. When the transition line is crossed, we find drastically different values of $w_{\mathrm{opt}}$ even for relatively large $\gamma$. (d-f) The transition is driven by the presence of two minima of $\mathcal{F}_\mathcal{M}$ that exchange stability. The global minimum is highlighted by a triangle and the unstable minimum (if present) by a circle. At low enough values of $\gamma$, as we increase $\omega$ a second minimum appears at low $w$. These two minima eventually coalesce at larger $\gamma$, smoothing the transition}
    \label{fig:p2c3:figure2}
\end{figure*}

Importantly, as argued in the previous Section, we are free to constrain the steady state of $P_{\mathrm{eff}}$ to be equal to the one of $P_{\mathrm{obs}}$, since otherwise the information loss diverges in the $t_\mathrm{obs} \to \infty$ limit. The steady state of the former defines
\begin{align*}
    \mu^{\mathrm{stat}}_{\mathrm{obs}} = 0, \quad \sigma^{\mathrm{stat}}_{\mathrm{obs}} = \frac{v_{\mathrm{th}}}{\omega},
\end{align*}
and that of the effective Ornstein-Uhlenbeck gives
\begin{align*}
    \mu_{\mathrm{eff}}^{\mathrm{stat}} = \mu_p, \quad \sigma_{\mathrm{eff}}^{\mathrm{stat}} = \frac{\Delta_p}{w_p}.
\end{align*}
This constraint allows us to fully specify the effective dynamics, Eq.~\eqref{eqn:p2c3:OUharm}, in terms of one single parameter to optimize. We choose to set $\mu_p = 0$ and $\Delta_p = v_{\mathrm{th}} w_p /\omega$, so that ${\bm \theta}_p = \{w_p\}$. The optimal information projection is then solely determined by $w_p$. Notice that in the multi-dimensional case, the constraint above will reduce the number of model parameters, but they are generally more than one. We will later check that, as $t_{\mathrm{obs}} \to +\infty$, the multi-parameter optimization leads to the same minimum as the one obtained imposing the steady state, and that even for short observation time the results we find are qualitatively unchanged.

For the sake of simplicity, as a metric we choose the symmetrized Kullback-Leibler divergence $\mathcal{M} = D^{\mathrm{sym}}_{\mathrm{KL}}$, so that the information loss
\begin{align*}
    \mathcal{F}_{D_{\mathrm{KL}}^{\mathrm{sym}}} & = \int_{0}^\infty D_{\mathrm{KL}}^{\mathrm{sym}}(P_{\mathrm{obs}}, P_{\mathrm{eff}})dt
\end{align*}
can be obtained from
\begin{align}
    D_{\mathrm{KL}}^{\mathrm{sym}}(P_{\mathrm{obs}}, P_{\mathrm{eff}}) & = \frac{D_{\mathrm{KL}}(P_{\mathrm{obs}} || P_{\mathrm{eff}}) + D_{\mathrm{KL}}(P_{\mathrm{eff}} || P_{\mathrm{obs}})}{2} \nonumber \\
    & = \frac{\sigma^2_{\mathrm{eff}}(t) + \left[\mu_{\mathrm{eff}}(t) - \mu_{\mathrm{obs}}(t)\right]^2}{4 \sigma^2_{\mathrm{obs}}(t)} + \frac{\sigma^2_{\mathrm{obs}}(t) + \left[\mu_{\mathrm{obs}}(t) - \mu_{\mathrm{eff}}(t)\right]^2}{4 \sigma^2_{\mathrm{eff}}(t)}.
\end{align}
In Figs.~\ref{fig:p2c3:figure2}a-b we plot the space of the $w_p$ that minimizes $\mathcal{F}_{D_{\mathrm{KL}}^{\mathrm{sym}}}$ as a function of the parameters of the original model,
\begin{equation}
    w_{\mathrm{opt}} = w_{\mathrm{opt}}(\gamma, \omega).
\end{equation}
Surprisingly, we find a line of discontinuities at $\omega = \omega_c(\gamma)$.  We can also compare the values of $w_{\mathrm{opt}}$ with those predicted by a standard overdamped limit, $w_{\mathrm{ov}} = \omega/\sqrt{\gamma}$, a projection in the $x$-space usually employed for strong friction regimes. By plotting the ratio $R_{\mathrm{ov}} = w_{\mathrm{opt}}/w_{\mathrm{ov}}$, it is evident that the optimal model is markedly different from the overdamped model, even at relatively large values of $\gamma$ (see Figure~\ref{fig:p2c3:figure2}c). We can think of $\mathcal{F}_\mathcal{M}$ as a quantity analogous to a free energy, whose global minimum defines the optimal model. Then, the discontinuity can then be seen as a first-order phase transition due to the presence of two minima that exchange stability (see Figs.~\ref{fig:p2c3:figure2}d-f).

\begin{figure}
    \centering
    \includegraphics[width = \textwidth]{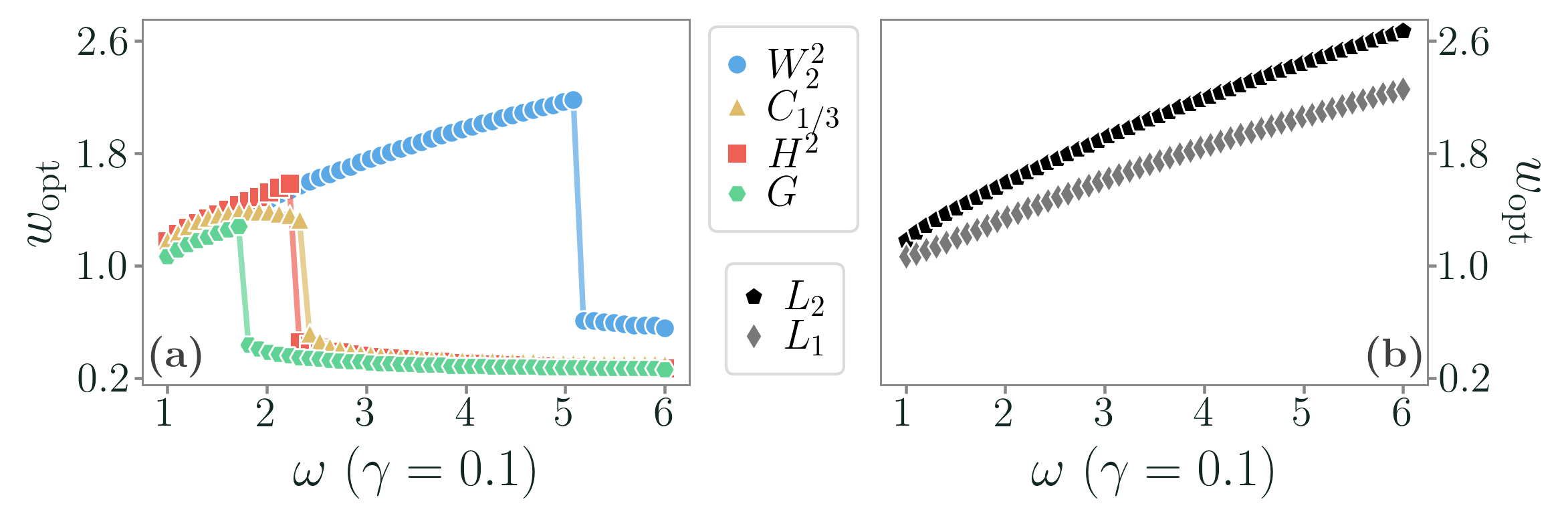}
    \caption{$w_{\mathrm{opt}}(\omega)$ for different metrics, with $\gamma = 0.1$. (a) Different colors correspond to different choices of the information metric. They all exhibit a discontinuous transition, for different $\omega_c(\gamma)$. (b) Non-information metrics result in a continuous $w_{\mathrm{opt}}(\omega)$, which monotonously increase with $\omega$}
    \label{fig:p2c3:figS1}
\end{figure}

Crucially, this discontinuity is independent of the specific choice of $\mathcal{M}$. To illustrate this, we choose as alternative options the Hellinger distance $H^2$, 
\begin{equation*}
    H^2(P_{\mathrm{obs}}, P_{\mathrm{eff}}) = 1 - \sqrt{2\frac{\sigma_{\mathrm{obs}}\sigma_{\mathrm{eff}}}{\sigma_{\mathrm{obs}}^2 + \sigma_{\mathrm{eff}}^2}}e^{-\frac{1}{4}\frac{\Delta\mu^2}{\sigma_{\mathrm{obs}}^2 + \sigma_{\mathrm{eff}}^2}}
\end{equation*}
with $\Delta\mu = \mu_{\mathrm{obs}} - \mu_{\mathrm{eff}}$, the geodesic distance $G$,
\begin{equation*}
    G(P_{\mathrm{obs}}, P_{\mathrm{eff}}) = 2\sqrt{2} \tanh^{-1}\left[\sqrt{\frac{\Delta\mu^2 + 2(\sigma_{\mathrm{obs}} - \sigma_{\mathrm{eff}})^2}{\Delta\mu^2 + 2(\sigma_{\mathrm{obs}} + \sigma_{\mathrm{eff}})^2}}\right],
\end{equation*}
the Chernoff-alpha divergence $C_\alpha$,
\begin{equation*}
    C_\alpha(P_{\mathrm{obs}} || P_{\mathrm{eff}}) = \frac{\alpha(1-\alpha)\Delta\mu^2}{2 \psi_\alpha} + \frac{1}{2}\log\left[\frac{\psi_\alpha}{\sigma_{\mathrm{obs}}^{2(1-\alpha)} + \sigma_{\mathrm{eff}}^{2\alpha}}\right]
\end{equation*}
with $\psi_\alpha = (1-\alpha)\sigma_{\mathrm{obs}}^2 + \alpha\sigma_{\mathrm{eff}}^2$, and the Wasserstein distance $W_2^2$,
\begin{equation*}
    W_2^2(P_{\mathrm{obs}}, P_{\mathrm{eff}}) = \sqrt{\Delta\mu^2} + \sigma_{\mathrm{obs}}^2 + \sigma_{\mathrm{eff}}^2 + \sqrt{\sigma_{\mathrm{obs}}^2 \sigma_{\mathrm{eff}}^2},
\end{equation*}
which are the expressions of such distances between two $1D$ Gaussian distributions. Clearly, each distance has a different information-geometric meaning \cite{ThomasCover2006, amari2016information}. In Figure~\ref{fig:p2c3:figS1}a we show that the discontinuous transition is present for all these choices, although their functional forms are vastly different - which implies that the exact transition line $\omega_c(\gamma)$ is metric-dependent, as expected. Then, we minimize the $L_\beta$ norm between the first two moments of the two Gaussian distributions $P_{\mathrm{obs}}$ and $P_{\mathrm{eff}}$,
\begin{equation*}
L_\beta(\{\mu\},\{\sigma\}) = \left|\mu_{\mathrm{obs}} - \mu_{\mathrm{OU}}\right|^\beta + \left|\sigma^2_{\mathrm{obs}} - \sigma^2_{\mathrm{OU}}\right|^\beta.
\end{equation*}
This, in general, is not a metric in the probability space. Remarkably, in Figure~\ref{fig:p2c3:figS1}b we show that with this choice the transition disappears, and $w_{\mathrm{opt}}$ monotonously increases with $\omega$. This result strongly suggests that the transition is an intrinsic feature deeply related to the minimization of the information loss, and not just a byproduct of our specific choice of the metric.

\begin{figure}[t]
    \centering
    \includegraphics[width=\textwidth]{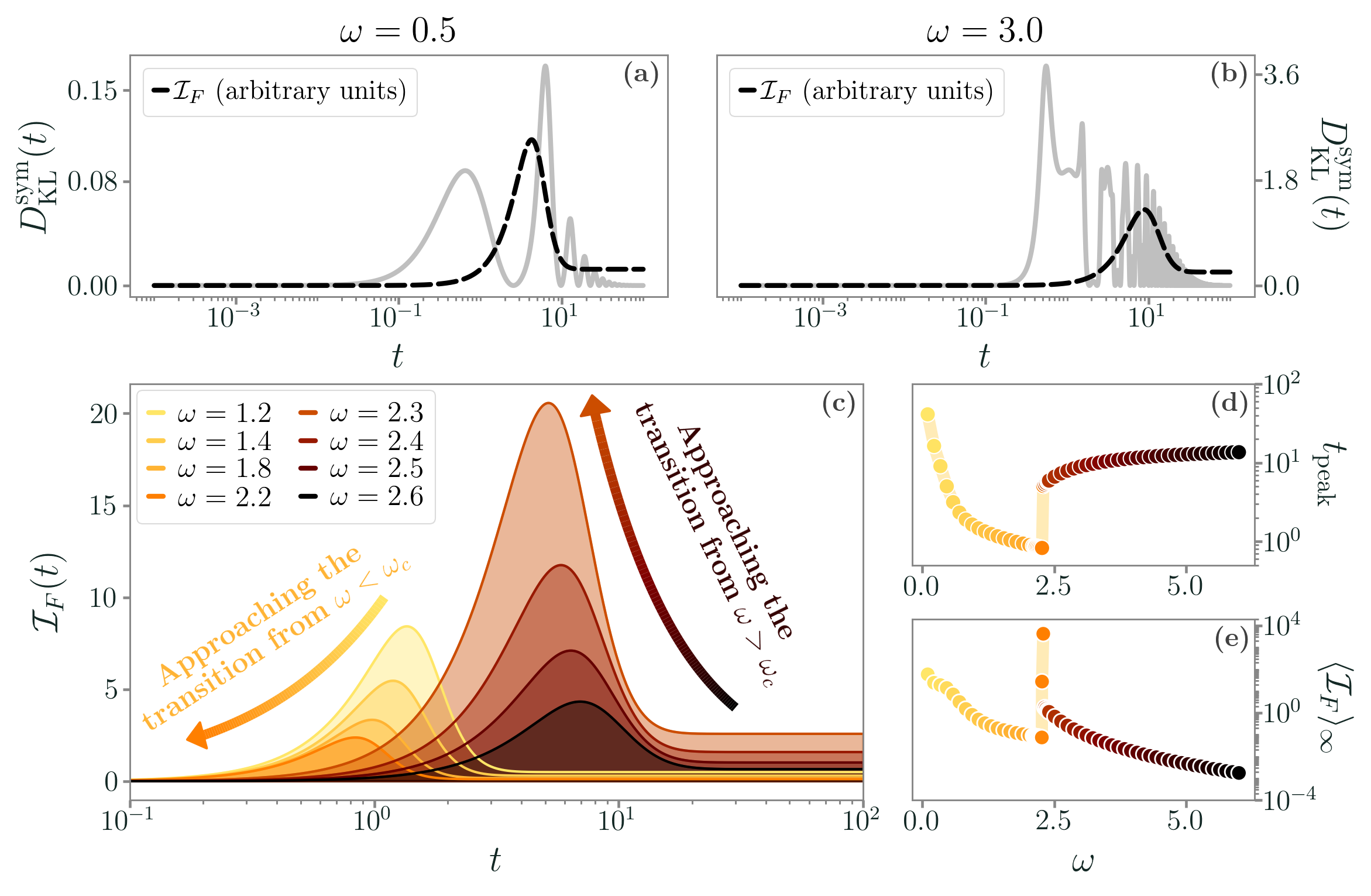}
    \caption{(a-b) Plot of $D^{\mathrm{sym}}_{KL}(t)$ (gray) and $\mathcal{I}_F(t)$ (black dashed line) in the harmonic case before and after the transition, respectively. (c) $\mathcal{I}_F(t)$ exhibits opposing trends before and after the transition. (d) These trends are well characterized by the peak time of $\mathcal{I}_F$, $t_{\mathrm{peak}}$, which captures the transient dynamics for $\omega < \omega_c$, and the oscillatory dynamics for $\omega > \omega_c$. Notice that at small $\omega$ we are close to the free diffusion regime with a longer transient. (e) At the transition, the integral mean of the Fisher information diverges. In all these plots, $\gamma = 0.1$, $\sigma_{x_0}^2 =  \sigma_{v_0}^2 = \mu_{x_0} = v_{\mathrm{th}} = 1$ and $\mu_{v_0} = 0$}
    \label{fig:p2c3:figure3}
\end{figure}

To investigate the meaning of the phases associated with the two minima of $\mathcal{F}_\mathcal{M}$, we introduce the Fisher information \cite{ThomasCover2006, amari2016information} at fixed $\gamma$,
\begin{align}
    \mathcal{I}_F(\omega,t \,|\,\gamma) & = \int_{-\infty}^{+\infty} P_{\mathrm{opt}}(x, t \, | \, w_{\mathrm{opt}}(\omega,\gamma)) \frac{\partial^2 \log(P_{\mathrm{opt}}(x, t \, | \, w_{\mathrm{opt}}(\omega,\gamma)))}{\partial^2 \omega} \bigg|_\gamma dx \nonumber \\
    & = \frac{(\partial_\omega w_{\mathrm{opt}})^2\left[2 \sigma_{\mathrm{opt}} (\partial_{w_{\mathrm{opt}}} \mu_{\mathrm{opt}})^2 + (\partial_{w_{\mathrm{opt}}} \sigma_{\mathrm{opt}})^2\right]}{2 \sigma_{\mathrm{opt}}^2}\bigg|_\gamma
    \label{eqn:p2c3:fisher}
\end{align}
where $\mu_{\mathrm{opt}}$ and $\sigma^2_{\mathrm{opt}}$ are the mean and variance of the optimal pdf, respectively, and $\partial_{w_{\mathrm{opt}}} = \partial / \partial w_{\mathrm{opt}}$. Eq.~\eqref{eqn:p2c3:fisher} quantifies the sensitivity of $P_{\mathrm{opt}}$ to changes in $\omega$, at a fixed value of $\gamma$. In Figure~\ref{fig:p2c3:figure3}a-b we show the temporal evolution of $\mathcal{I}_F$. Approaching the transition from below ($\omega \lesssim \omega_c$), $\mathcal{I}_F$ peaks at short times, indicating that the information projection weighs more earlier stages of the dynamics, i.e., the transient regime. On the other hand, for $\omega \gtrsim \omega_c$, the peak of $\mathcal{I}_F$ appears at longer times, capturing the persistent oscillating behavior. Notice that, for very small values of $\omega$, the system is close to the free-diffusion regime. This reflects into longer transients and, in turn, an increase of the peak time (see Figs.~\ref{fig:p2c3:figure3}c-d). The integral mean
\begin{equation*}
    \ev{\mathcal{I}_F}_T = \frac{1}{T} \int_0^T dt \, \mathcal{I}_F
\end{equation*}
in the limit $T\to+\infty$ quantifies the total susceptibility of the optimal model to changes in $\omega$. This quantity diverges at the transition point, as expected. Moreover, we observe that $\ev{\mathcal{I}_F}_\infty$ considerably decreases at large values of $\omega$ since $w_{\mathrm{opt}}(\omega)$ saturates, indicating increasing robustness of the information projection (see Figure~\ref{fig:p2c3:figure3}e).

These results suggest that the two phases of the optimal model, represented by the minima of $\mathcal{F}_\mathcal{M}$, are characterized by the dynamical regimes they capture the most. The transition appears at low values of $\gamma$ and high enough values of $\omega$, when the strong harmonic confinement and the low damping generate long-lasting oscillations around the steady state. Conversely, the optimal model relaxes exponentially to the steady state with a characteristic time $w_{\mathrm{opt}}^{-2}$, thus being unable to optimally capture both the transient and the oscillatory behavior along the entire dynamical evolution.

\begin{figure}
    \centering
    \includegraphics[width =\textwidth]{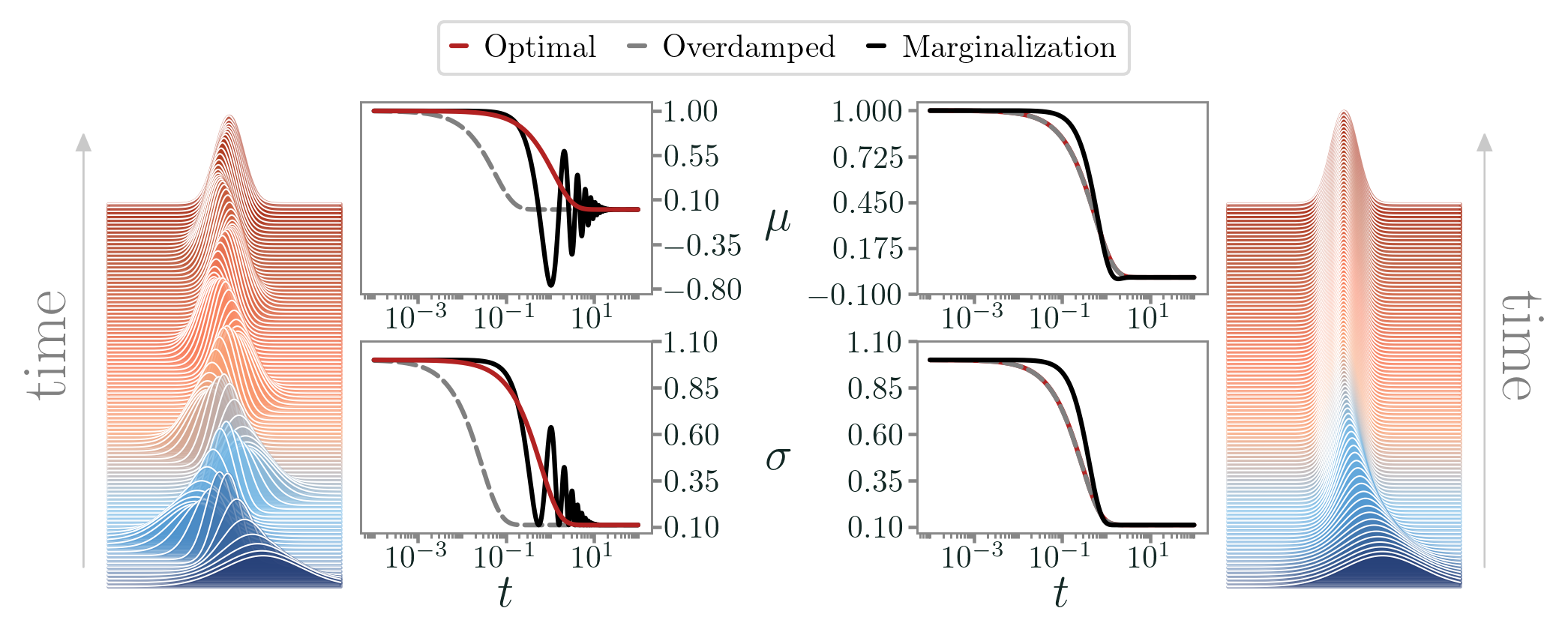}
    \caption{An example of the evolution of $P_\mathrm{obs}$, its mean and its variance at low values of damping (left) and high values of damping (right). As both the optimal model and the standard overdamped approximation are exponential models, they are able to match closely the behavior of $P_\mathrm{obs}$ at large damping. Instead, at low $\gamma$, we are approximating an oscillatory evolution with an exponential one. The constraint of keeping as much information as possible on this complex dynamics with a simpler one is at the heart of the observed information-driven transition}
    \label{fig:p2c3:complex_to_simple}
\end{figure}

To provide a heuristic interpretation of the transition, let us assume that we are at a low enough $\gamma$, so that the transition is present. If $\omega$ is low as well, the particle is loosely confined and $P_{\mathrm{obs}}$ relaxes with a few oscillations to the steady state. Hence, the optimal model is the one that better matches the initial transient dynamics, as this weighs the most in the time-integrated Kullback-Leibler divergence. This means that the characteristic time is small, i.e., $w_{\mathrm{opt}}$ is large and the information loss tends to be small, as we see in Figure~\ref{fig:p2c3:figure2}d-e. As we increase $\omega$, however, the confinement grows stronger, and more persistent oscillations appear. At $\omega_c(\gamma)$, the optimal model becomes the one that captures these long-lasting oscillations rather than the initial transient, hence its characteristic time suddenly increases. Therefore, the two minima of the information loss $\mathcal{F}_{\mathcal{M}}(w)$ represent these two optimal models that capture at best either the initial transient dynamics or the persistent oscillations. These ideas are illustrated in Figure~\ref{fig:p2c3:complex_to_simple}.

\begin{figure}
    \centering
    \includegraphics[width =\textwidth]{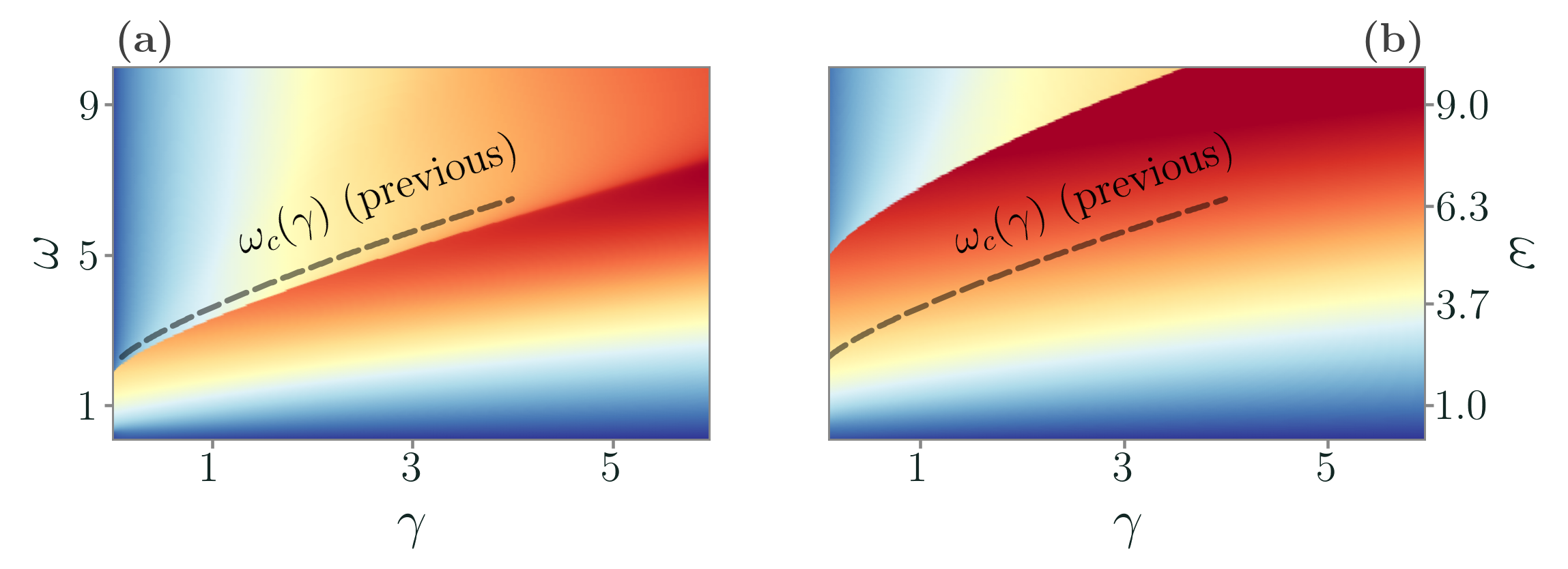}
    \caption{Contour plot of $w_{\mathrm{opt}}$ as a function of $\omega$ and $\gamma$ for (a) $v_{\mathrm{th}}^2 = 0.75$, $\mu_{x_0} = 1.25$, $\sigma_{x_0} = 1$, $\mu_{v_0} = 0$, $\sigma_{v_0}^2 = 0.5$, and (b) $\mu_{x_0} = 1$, $\sigma_{x_0} = 0.5$, $\mu_{v_0} = 0$, and $\sigma_{v_0}^2 = 1$. The dashed lines indicate the transition line $\omega_c(\gamma)$ for the initial conditions of the previous case}
    \label{fig:p2c3:figS2}
\end{figure}

Crucially, in the underdamped model, the dynamics changes smoothly across the transition line, highlighting that the information-preserving feature of the projection is at the root of the discontinuous transition. Remarkably, as shown before, this transition does not appear if $\mathcal{M}$ is not an information metric, e.g., the $L^2$-distance between the mean and variance of $P_{\mathrm{obs}}$ and $P_{\mathrm{OU}}$, while being a robust property for different appropriate choices of $\mathcal{M}$. In Figure~\ref{fig:p2c3:figS2} we also show that the transition line changes as we change the initial conditions, but the results are qualitatively identical. We also remark that a variation in $\sigma_{x_0}^2$ (Figure~\ref{fig:p2c3:figS2}b) has a greater impact than a variation in all the other initial conditions (Figure~\ref{fig:p2c3:figS2}a). This is compatible with the heuristic interpretation that the transition appears due to the effective model capturing either the long-time oscillations or the initial transient dynamics, which is affected by the initial conditions.

Interestingly, although in general there is no analytical expression for $w_{\mathrm{opt}}$, by tuning the initial conditions it is possible to gain an analytical grasp of its form in some regions of model parameters. We choose the initial distributions in such a way that the variance stays constant at all times, i.e., $\sigma_{\mathrm{har}}^2(t) = \sigma_{\mathrm{OU}}^2(t) = \sigma_{x_0}^2$. To this end, we need $\sigma_{x_0}^2 = k_B T/\omega^2$ and $\sigma_{v_0}^2 = k_B T$. Hence, the Kullback-Leibler divergence depends only on the mean and greatly simplifies. We have to minimize
\begin{align*}
    \mathcal{F}_\mathcal{M}(w_p^2) & = \int_{0}^\infty \frac{(\mu_{\mathrm{OU}}(t) - \mu_{\mathrm{har}}(t))^2}{2v^2_{\mathrm{th}}/\omega^2} dt
\end{align*}
which gives
\begin{align*}
    \mathcal{F}_\mathcal{M}(w_p^2) & = \frac{\omega^2}{2 v_{\mathrm{th}}^2 \lambda^2}\int_0^{\infty} e^{-t \left(\gamma +2 w_p^2\right)} \left[\mu_{x_0} \lambda e^{\frac{\gamma t}{2}}-e^{t w_p^2}f(t) \right]^2 = \\
    & = \frac{1}{4v_{\mathrm{th}}^2}\left[\frac{\mu_{x_0}^2\omega^2 + (\mu_{v_0} + \gamma\mu_{x_0})^2}{\gamma} + \frac{\mu_{x_0}^2\omega^2}{w_p^2} - \frac{4 \mu_{x_0}\omega^2[\mu_{v_0} + \mu_{x_0}(\gamma + w_p^2)]}{\omega^2 + \gamma w_p^2 + w_p^4}\right].
\end{align*}
where $\lambda = \sqrt{\gamma ^2-4 \omega^2}$ and $f(t) = (\gamma \mu_{x_0}+2 \mu_{v_0}) \sinh {\lambda t}/{2} +\mu_{x_0} \lambda \cosh{\lambda t}/{2}$. Thus, we have to solve
\begin{align*}
     0 = \frac{4 \mu_{v_0} w_p^4 \left(\gamma +2 w_p^2\right)-\mu_{x_0} K(w_p)}{w_p^3 \left(\gamma  w_p^2+\omega^2+w_p^4\right)^2} = 4 \mu_{v_0} w_p^4 \left(\gamma +2 w_p^2\right)-\mu_{x_0} K(w_p)
\end{align*}
where $K(w_p) = 2 \omega ^2 w_p^2 \left(\gamma +3 w_p^2\right)-3 w_p^4 \left(\gamma +w_p^2\right)^2+\omega^4$. This equation has a solution that is always positive and analytical, although particularly cumbersome and not reported here. Notably, when $\mu_{v_0} = 0$, we only need to find the positive and real solution of the equation
\begin{align*}
    \omega^4 + 2\gamma\omega^2w_p^2-3(\gamma^2-2\omega^2)w_p^4-6\gamma w_p^6-3w_p^8 = 0
\end{align*}
which does not depend on $\mu_{x_0}$. We find
\begin{gather*}
    w_{\mathrm{opt}} = \sqrt{\frac{\sqrt{-\frac{8 \sqrt{3} \gamma  \omega ^2}{\sqrt{A(\gamma, \omega)+B(\gamma, \omega)}}-A(\gamma, \omega)+2 B(\gamma, \omega)}+\sqrt{A(\gamma, \omega)+B(\gamma, \omega)}}{2 \sqrt{3}}-\frac{\gamma}{2}}
\end{gather*}
with
\begin{gather*}
    A(\gamma, \omega) = \frac{\gamma^4}{\sqrt[3]{8 \omega ^3 \sqrt{16 \omega ^6-\gamma ^6}+\gamma ^6-32 \omega ^6}} + \sqrt[3]{8 \omega ^3 \sqrt{16 \omega ^6-\gamma ^6}+\gamma ^6-32 \omega ^6} \\
    B(\gamma, \omega) = \gamma^2 + 4\omega^4.
\end{gather*}
Remarkably, if we expand this solution for $\gamma \to +\infty$ we find
\begin{equation*}
    w_{\mathrm{opt}} = \frac{\omega}{\sqrt{\gamma}}\left[1 - \frac{3}{8}\left(\frac{\omega}{\gamma}\right)^4\right] + \mathcal{O}\left(\frac{1}{\gamma^{11/2}}\right)
\end{equation*}
which is the correction to the overdamped solution, $w_\mathrm{ov} = {\omega}/{\sqrt{\gamma}}$. Instead, when $\gamma \to 0$, we have
\begin{equation*}
    w_{\mathrm{opt}} = \sqrt{\omega} \left[\left(1+\frac{2}{\sqrt{3}}\right)^{1/4} - \frac{(3+2\sqrt{3})^{1/4}}{2\sqrt{6}\omega}\gamma\right] + \mathcal{O}\left(\gamma^{4/3}\right)
\end{equation*}
so the behavior in the small-$\gamma$ regime is drastically different. In particular, the zero-th order approximation of $w_{\mathrm{opt}}$ does not depend on $\gamma$ anymore.

\begin{figure}[t]
    \centering
    \includegraphics[width=\textwidth]{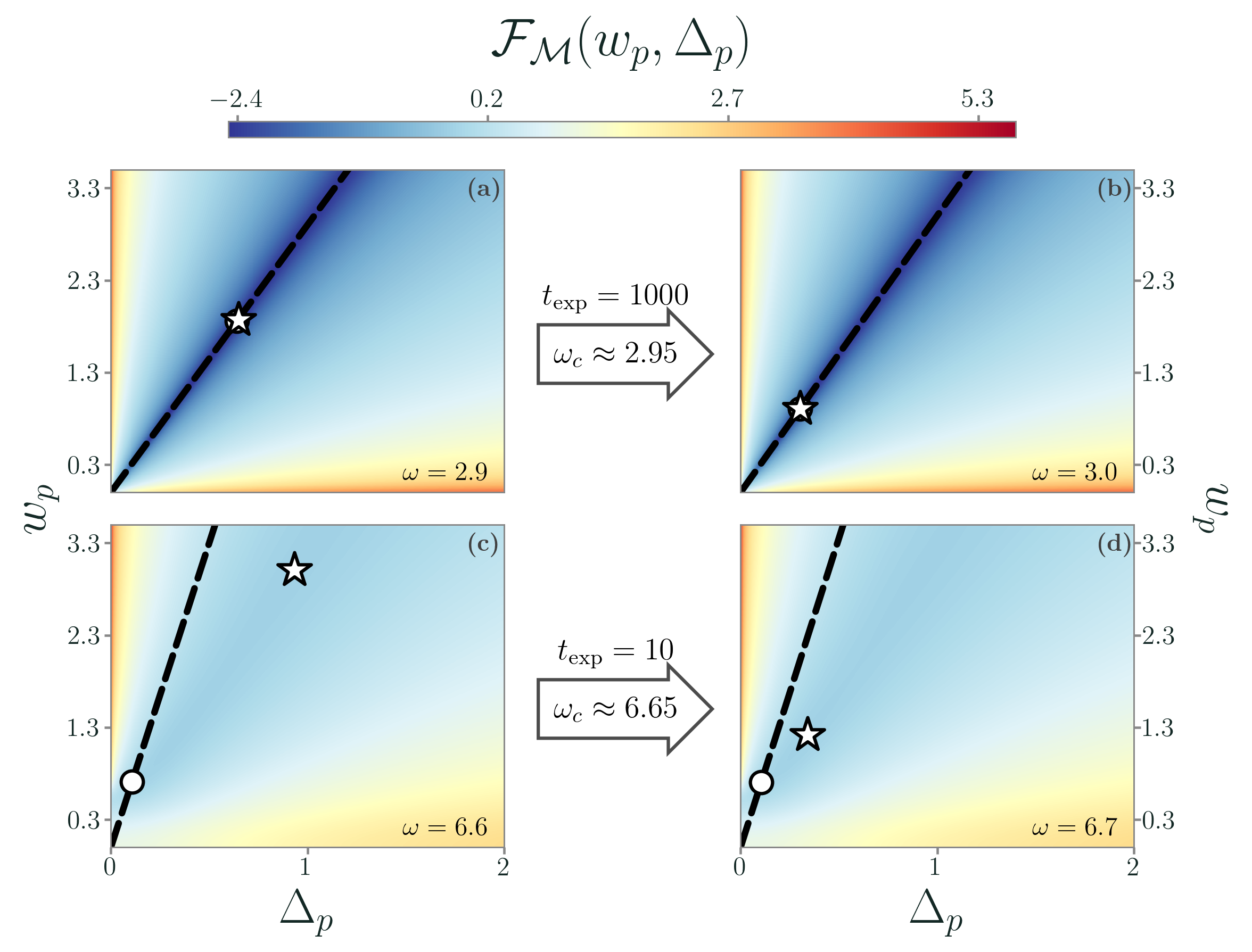}
    \caption{Optimization over both $w_p$ and $\Delta_p$ at different observation times $t_{\mathrm{obs}}$ in the harmonic case, for $\gamma = 0.5$. The black dashed line represents the stationary constraint $\Delta_p = v_{\mathrm{th}}w_p/\omega$, the white star represents the minima of the information loss $\mathcal{F}_{\mathcal{M}}(w, D)$, whereas the white dot represents the minima obtained from the constrained optimization. Results are obtained with the symmetric Kullback-Leibler divergence, but do not change for other information metrics. (a-b) The large $t_{\mathrm{obs}}$ limit gives the same results as the constrained case. (c-d) If $t_{\mathrm{obs}}$ is small, the minimum is different since the observed trajectories do not reach stationarity. Yet, the transition is still present for larger values of $\omega$, when a large number of oscillations in the mean and variance of $P_{\mathrm{obs}}(x,t)$ are present even at small times}
    \label{fig:p2c3:new_figure_S1}
\end{figure}

Clearly, in this case there is no discontinuous transition, since $\mathcal{F}_\mathcal{M}(w_p)$ has always a unique minimum. This implies that long-lasting oscillations in both the mean and the variance of $P_{\mathrm{obs}}$ are needed to observe the transition - since here we are forcing the variance to stay constant in the first place. This observation, together with the necessity highlighted above for $\mathcal{M}$ to be an information metric and thus a generalized distance in probability space, suggests that the two phases of the effective model emerge from the interplay between the non-exponential temporal evolution of the system and the minimization of the information loss, which involves both mean and variance at once.

\subsection{Optimal model at short times and stationary constraints}
We now check for consistency that, as $t_{\mathrm{obs}} \to +\infty$, the multi-parameter optimization leads to the same minimum as the one obtained by imposing the steady state. Furthermore, we show that the optimal model exhibits the same qualitative properties even without this constraint. In particular, to keep the information loss comparable at different observation times, we introduce an integral mean in Eq.~\eqref{eqn:p2c3:minM} to determine the optimal parameters,
\begin{equation*}
    {\bm \theta_{\mathrm{opt}}} = \underset{{\bm \theta_p}}{\mathrm{argmin}} \frac{1}{t_{\mathrm{obs}}} \int_0^{t_{\mathrm{obs}}} dt~\mathcal{M}(P_{\mathrm{obs}}, P_{\mathrm{eff}}).
    \label{eqn:p2c3:SM:minM}
\end{equation*}
With this choice, we still avoid divergences in the long-time limit if the stationary distributions are different. For simplicity, and in order to visualize the results, we still set $\mu_p = 0$ and we optimize over the $(\Delta_p, w_p)$ plane. In Figure~\ref{fig:p2c3:new_figure_S1}a-b we plot the information loss $\mathcal{F}_\mathcal{M}(w_p, \Delta_p)$ with $t_{\mathrm{obs}} \gg 1$ - i.e., when the observation time is long enough to reach the steady state. The results are identical to the ones presented in the main text.

On the other hand, in Figure~\ref{fig:p2c3:new_figure_S1}c-d we show the information loss for a shorter observation time $t_{\mathrm{obs}}$. Crucially, the discontinuous transition is still present, but for a much larger value of $\omega$. Intuitively, this transition appears when enough oscillations are present in the given observation time, strengthening the observation that it originates from an interplay between transient and oscillatory regimes. Hence, for small $t_{\mathrm{obs}}$, the discontinuous jump of $w_{\mathrm{opt}}$ appears when the potential is more confining, i.e., larger $\omega$, so that strong oscillations arise even at short times.

\subsection{Two-dimensional harmonic potentials}
Here, we extend the previous results by considering a simple generalization to a two-dimensional harmonic potential. In particular, we set
\begin{align}
\label{eqn:p2c3:F2D}
\bm{F}(\bm{x}) = \begin{pmatrix}\omega_1^2 & 0 \\ 0 & \omega_2^2\end{pmatrix} \begin{pmatrix}x_1 \\ x_2\end{pmatrix}
\end{align}
while keeping $\hat{D} = v_{\mathrm{th}} \gamma \, \mathbb{1}$, i.e., thermal noise acts independently on each degree of freedom. As an effective dynamics, we choose the immediate two-dimensional extension of the previous OU process, that is:
\begin{equation}
    \dot{\bm{x}}(t) = - w_p^2 \bm{x}(t) + \sqrt{2 v_{\mathrm{th}}^2 w_p^2} \begin{pmatrix}\omega_1^{-1} & 0 \\ 0 & \omega_2^{-1}\end{pmatrix} \bm{\xi}_p(t),
\end{equation}
where the steady-state constraints have been incorporated explicitly. In this scenario, the optimal model is still defined by one single parameter to optimize, $w_p = w_p(\omega_1, \omega_2, \gamma)$. In other words, the effective dynamics needs to approximate the different evolution along each of the two spatial dimensions, $x_1$ and $x_2$, with a single characteristic time $w_{\mathrm{opt}}^{-2}$.

\begin{figure}[t]
    \centering
    \includegraphics[width=\textwidth]{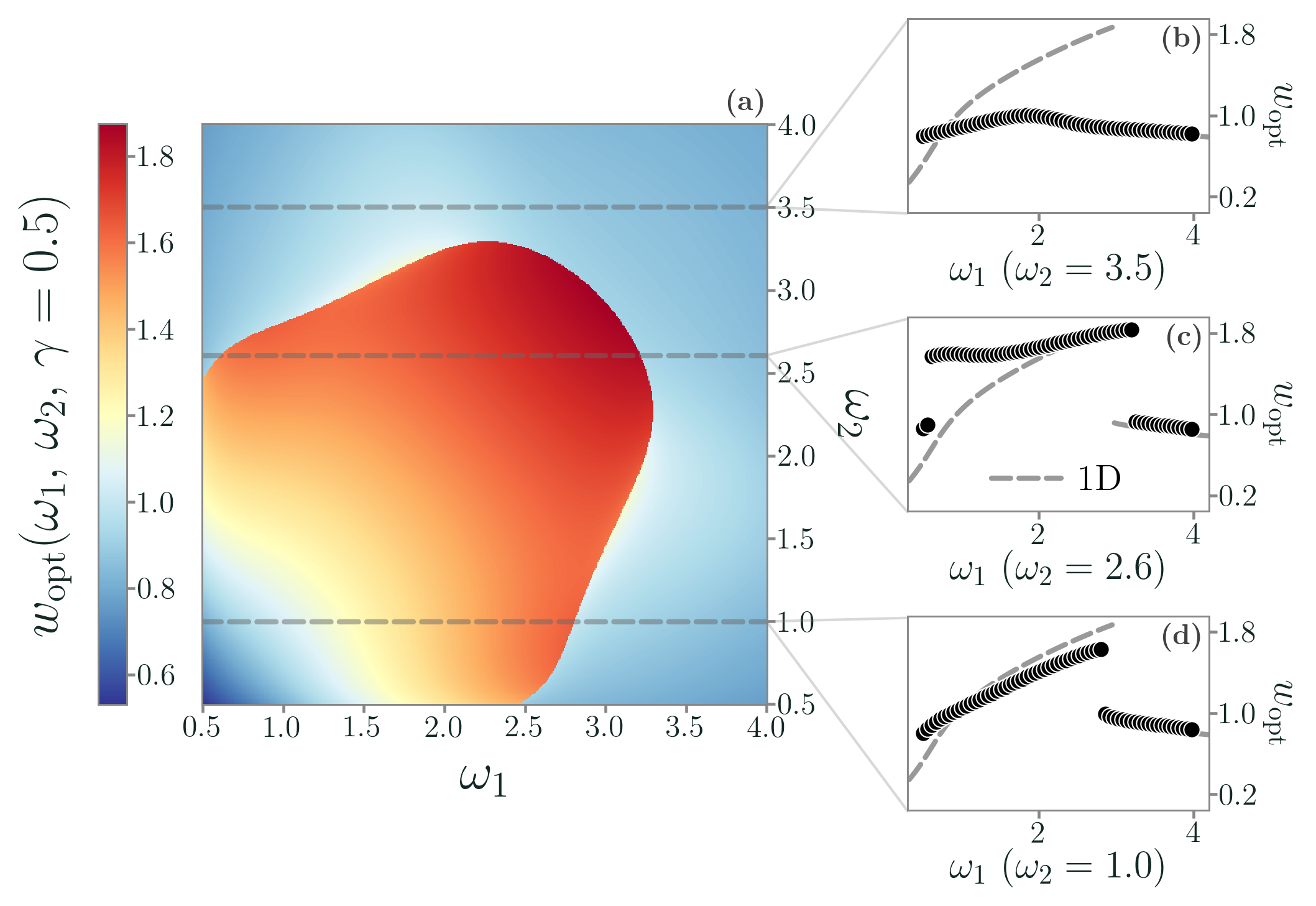}
    \caption{(a) Contour plot of $w_{\mathrm{opt}}(\omega_1, \omega_2)$ at fixed $\gamma = 0.1$ in the $2D$ harmonic case. As before, we find that $w_{\mathrm{opt}}$ undergoes a discontinuous transition. (b-d) If $\omega_2$ (or $\omega_1$) is large, no discontinuity is present since the overall information loss is dominated by its corresponding minima (see Figure~\ref{fig:p2c3:figure2}d), regardless of the value of $\omega_1$. On the contrary, if $\omega_2$ is small, its contribution to the information loss is negligible and the optimal parameters as a function of $\omega_1$ are similar to the $1D$ case (gray dashed line). Intermediate regimes show two discontinuous transitions. In all these plots, $\sigma_{x_0}^2 =  \sigma_{v_0}^2 = \mu_{x_0} = v_{\mathrm{th}} = 1$ and $\mu_{v_0} = 0$}
    \label{fig:p2c3:new_figure4}
\end{figure}

The underlying $2D$ dynamics determined by the force in Eq.~\eqref{eqn:p2c3:F2D} entails two independent confinements along the two spatial directions, each one analogous to the one-dimensional case. Therefore, as before, we expect a discontinuous transition to appear at large values of $\omega_{1,2}$ at a small $\gamma$. Loose confinements in both directions are then associated with higher values of $w_{\mathrm{opt}}$, while a strong oscillatory behavior results in lower values of the optimal parameter. In Figure~\ref{fig:p2c3:new_figure4}a we plot $w_{\mathrm{opt}}(\omega_1, \omega_2)$, at fixed $\gamma$. As expected, along the diagonal, we recover the results of the one-dimensional case. However, the interplay between $\omega_1$ and $\omega_2$ reflects into a complex, yet still discontinuous, front.

Let us describe the behavior of $w_{\mathrm{opt}}(\omega_1)$ for a given value of $\omega_2$. First, we notice that the total information loss $\mathcal{F}_{\mathcal{M}}$, evaluated using the symmetric Kullback-Leibler divergence, is the sum of two $1D$ contributions, $\mathcal{F}_{\mathcal{M}_1}$ and $\mathcal{F}_{\mathcal{M}_2}$, respectively associated with $\omega_1$ and $\omega_2$. If $\omega_2$ is large (Figure~\ref{fig:p2c3:new_figure4}b), persistent oscillations along the $x_2$ directions are present. From the $1D$ optimization problem (Figure~\ref{fig:p2c3:figure2}d-f), we know that $\mathcal{F}_{\mathcal{M}_2} \gg \mathcal{F}_{\mathcal{M}_1}$ in this regime, hence $\mathcal{F}_{\mathcal{M}} \approx \mathcal{F}_{\mathcal{M}_2}$, independently of the value of $\omega_1$. Since $\mathcal{F}_{\mathcal{M}_2}$ is minimized for low values of $\omega_2$, $\omega_{\mathrm{opt}} \approx \omega_2$, showing a post-transition behavior. As we decrease $\omega_2$, when $\mathcal{F}_{\mathcal{M}_1} \approx \mathcal{F}_{\mathcal{M}_2}$, $\mathcal{F}_{\mathcal{M}}$ takes both contributions, exhibiting two discontinuities that are associated with the crossings $\mathcal{F}_{\mathcal{M}_2} \gtrsim \mathcal{F}_{\mathcal{M}_1}$ and $\mathcal{F}_{\mathcal{M}_1} \gtrsim \mathcal{F}_{\mathcal{M}_2}$ (see Figure~\ref{fig:p2c3:new_figure4}c). Finally, at small $\omega_2$, $\mathcal{F}_{\mathcal{M}} \approx \mathcal{F}_{\mathcal{M}_1}$, and we see small deviation from the $1D$ optimal solution as a function of $\omega_1$ (see Figure~\ref{fig:p2c3:new_figure4}d).

This simple example shows that information-driven transitions are still present in higher dimensions, although for general systems the parameter space quickly grows and the features of the transition become harder to interpret and predict.

\section{Anharmonic potentials and trajectory-based approximations}
So far, we have studied the case of a harmonically bounded particle starting from the exact expression of $P_{\mathrm{obs}}$ without relying on trajectory estimations. Now, we consider the more realistic case in which we only have access to a few, and possibly short-time, trajectories. We now explicitly employ the Gaussian ansatz,
\begin{equation*}
    P_{\mathrm{obs}}(\bm{x},t) = \mathcal{N}(\bm{\mu}_{\mathrm{obs}}(t), \hat{\sigma}_{\mathrm{obs}}(t)),
\end{equation*}
starting from simulated trajectories of Eq.~\eqref{eqn:p2c3:under}. Notice that $\bm{\mu}_{\mathrm{obs}}(t)$ and $\hat{\sigma}_{\mathrm{obs}}(t)$ depend on $\gamma$, $\bm{F}$ and $\hat{\Delta}$ in non-trivial ways. Similarly, we keep the Gaussian form of the effective model,  
\begin{equation*}
    \dot{\bm{x}}(t) = - w_p^2 \left(\bm{x}(t) - \bm{\mu}_p \right) + \sqrt{2} \hat{\Delta}_p \bm{\xi}_p(t),
\end{equation*}
as before. Thus, we extract from the simulated trajectories: (i) the mean, $\mu_{\mathrm{obs}}$, and the variance, $\sigma^2_{\mathrm{obs}}$, at any time, to obtain the maximum entropy ansatz for $P_{\mathrm{obs}}$; (ii) the initial conditions in the $x$-space, $\mu_{x_0}$ and $\sigma_{x_0}$; (iii) the \textit{observed} steady state, which converges to the analytical one only as we increase number and duration of trajectories. Then, we obtain the optimal parameters from Eq.~\eqref{eqn:p2c3:minM}.

\begin{figure*}[t]
    \centering
    \includegraphics[width=\textwidth]{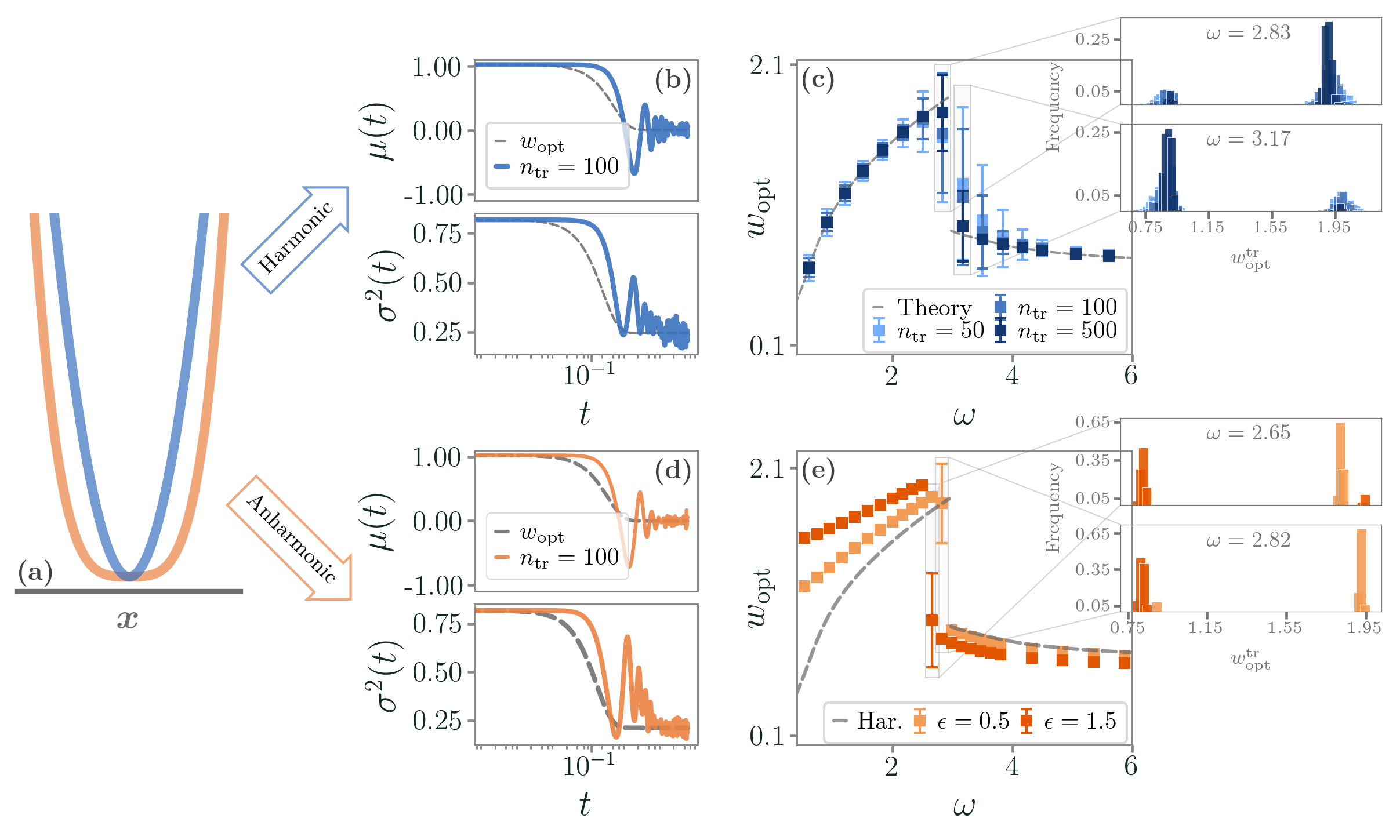}
    \caption{Results obtained from a limited number of spatial trajectories, in the presence of harmonic and anharmonic potentials. For all these plots, $\sigma_{x_0}^2 =  \sigma_{v_0}^2 = \mu_{x_0} = v_{\mathrm{th}} = 1$, $\mu_{v_0} = 0$ and $\gamma = 0.1$. (a) Depiction of the two potentials. (b) Mean and variance in the harmonic case, obtained from $n_{\tr} = 100$ trajectories. The gray dashed line represents the mean and variance of the corresponding optimal model. (c) The optimal value $w_{\mathrm{opt}}$ in a harmonic potential, as a function of $\omega$ and for different numbers of trajectories. The error bars represent one standard deviation over $10^3$ numerical experiments, and the gray dashed line is obtained from the exact solution of the system. Close to the transition the standard deviation increases due to the statistical errors that exchange the depth of the two minima, as we see from the inset histograms. (d) Same as (b), for an anharmonic potential with $\epsilon = 1.5$. (e) Similarly to (c), the optimal value $w_{\mathrm{opt}}$ as a function of $\omega$ obtained from $10^4$ trajectories, for different values of $\epsilon$. The gray dashed line corresponds to the harmonic case, i.e. $\epsilon = 0$. Notice that, the larger $\epsilon$ (darker points), the sooner with respect to $\omega$ the transition happens}
    \label{fig:p2c3:figure5}
\end{figure*}

We first test this trajectory-dependent approach using simulated trajectories for the harmonic case, comparing its results with the analytical ones obtained above. We integrate numerically Eq.~\eqref{eqn:p2c3:under} using a standard Euler-Maruyama method (see Appendix \ref{app:computational}). In Figure~\ref{fig:p2c3:figure5}b-c, we show that the method generally leads to accurate results. However, close to the transition, the number of trajectories plays a crucial role. Indeed, a small sample size induces statistical errors in the estimates of $\mu_{\mathrm{obs}}$ and $\sigma^2_{\mathrm{obs}}$, as well as initial conditions and steady states. These errors, in turn, might lead to an inaccurate estimation of the deepest minimum. In the insets of Figure~\ref{fig:p2c3:figure5}c we show that, at fixed $\omega \approx \omega_c$, for a small number of trajectories one can end up with a value of $w_{\mathrm{opt}}$ corresponding to either of the two minima.

Finally, we apply our method to simulated trajectories generated with an anharmonic force,
\begin{equation}
    F(x) = \omega^2 x + \epsilon x^3
\end{equation}
whose potential is depicted in Figure~\ref{fig:p2c3:figure5}a. In this case, the dynamics is not solvable, and the maximum entropy ansatz for $P_{\mathrm{obs}}$ is not exact, hence we do not have an exact theoretical baseline to compare the results. Remarkably, the information-driven transition is still present, albeit slightly shifted with respect to the case of a harmonic potential. The number of trajectories plays the same role as before, i.e., it introduces uncertainty close to the transition as it decreases. This result highlights that the transition is a robust feature also for not analytically tractable models, short-time trajectories, and small observational sample sizes. As a consequence, the potential appearance of a discontinuity has to be considered when building projections that capture the maximum amount of information of a (relatively small) set of experimental trajectories. Indeed, without knowing the parameters of the underlying system, the presence of an abrupt transition may lead to markedly different behaviors of the effective model.

Overall, in this Chapter we introduced an archetypal method that allowed us to build information-preserving projections of a possibly unknown complex dynamics, which is arguably the solution to a generic inference problem. By focusing on the paradigmatic case of underdamped systems, we have shown that, when approximating complex models with simpler effective ones, the optimal parameter space may be unexpectedly singular - and that this discontinuity in the optimal parameter space is triggered by the minimization of the information loss. This information-driven discontinuous transition induces abrupt changes in the effective model, which switches between qualitatively different phases. Our results pose fundamental challenges to the ambition of inferring underlying parameters from effective low-dimensional models, as the appearance of this transition in paradigmatic systems translates into an alarming warning signal for more general cases. Naively speaking, we expect that discontinuities in the parameter space might emerge when the effective model is unable to simultaneously capture all the dynamical features of the original model (e.g., oscillations and transient in the case presented here).

Notwithstanding, the proposed method to build optimal effective models through information projections has broad applicability - e.g., passive tracers in active media \cite{dabelow2019irreversibility}, species dynamics in ecological communities \cite{vacher2016learning}, effective models to probe neural activity \cite{mastrogiuseppe2018linking, williams2018unsupervised}, or any dynamics with unobserved degrees of freedom. Yet, we remark that our approach did not consider thermodynamic features. Non-equilibrium thermodynamics suffers from coarse-graining procedures \cite{esposito2012stochastic,busiello2019entropyA,busiello2019entropyB,busiello2020coarsegrained}, and building projections that preserve the underlying thermodynamics represents a completely different and far-reaching task. A fascinating future idea will be to simultaneously optimize dynamics and thermodynamics in a Pareto-like multi-optimization problem \cite{seoane2015phase}. A more immediate extension would be to perturbatively include higher moments in the estimation of $P_{\mathrm{obs}}$ from the experimental data to improve the Gaussian ansatz proposed here. In principle, if a large number of trajectories is accessible, one can directly estimate the full marginal distribution numerically. Moreover, more general classes of effective models should be explored, with particular attention to understanding if and how the corresponding optimal model improves upon a classical overdamped limit. Further generalizations of the model might also include other types of noise \cite{jung1987dynamical,di2018non}, although they must rely mostly on extensive numerical simulations.

Ultimately, we believe that this work sheds light on the fundamental properties of effective representations of complex dynamics. Indeed, emerging singularities in low-dimensional models, while crucial in shaping their behavior, might be a sheer consequence of the employed projection method, without reflecting any property of the original system.


\extraPartText{\lettrine{T}{he} results described in the following part of this Thesis are based upon the published works ``Scaling and criticality in a phenomenological renormalization group'' (\emph{G. Nicoletti, S. Suweis, A. Maritan, Phys. Rev. Research 2, 023144, 2020}) \cite{nicoletti2020scaling}, ``Disentangling the critical signatures of neural activity'' (\emph{B. Mariani, G. Nicoletti, M. Bisio, M. Maschietto, S. Vassanelli, S. Suweis, Sci. Rep. 12, 107702022}) \cite{mariani2022disentangling}, and ``Criticality and network structure drive emergent oscillations in a stochastic whole-brain model'' (\emph{G. Barzon, G. Nicoletti, B. Mariani, M. Formentin, S. Suweis, J. Phys. Complexity 3, 025010, 2022}) \cite{barzon2022oscillations}.

Parts of the contents presented, including displayed figures, are taken with permission from the published paper \cite{nicoletti2020scaling}, copyright (2020) by the American Physical Society.}
\part{Criticality and phase transitions in Neuroscience}

\chapter{Disentangling the critical signatures of neural activity}
\label{ch:scirep}
\lettrine{T}{he brain} is one of the most impressively complex systems we are able to study. Although a complete understanding of the properties of neural dynamics is a fundamental, far-reaching, and very much open question in science, the large number of degrees of freedom in the brain has inspired the study of minimal, paradigmatic models - often related to Statistical Physics. In 2003, the seminal work of Beggs and Plenz \cite{beggs2003avalanches} showed that in Local Field Potentials (LFPs) in cortical slices and cultures on chips neural activity occurred in cascades, named ``neuronal avalanches''. Remarkably, they found that both the sizes and the lifetimes of these avalanches were power-law distributed, with exponents surprisingly close to the ones of a critical mean-field branching process. From then on, the idea that the collective behavior of networks of neurons might emerge from a possibly self-organized critical state grew into a widespread and active research field \cite{de2006self, kinouchi2006optimal, dearcangelis2010learning, shew2013criticality, hesse2014self, hidalgo2014information, tkavcik2015thermodynamics, rocha2018homeostatic, rocha2022recovery, munoz2018colloquium, marinazzo2014information, legenstein2007edge, boedecker2012information}. One of the most compelling aspects of this idea is that the large scale collective behavior of critical systems does not depend on their microscopic details. Therefore, from the point of view of Statistical Physics, one might argue that we could describe the emergent properties of the brain without necessarily knowing the fine structure of its $\approx 10^{11}$ neurons and $\approx 10^{15}$ synapses - an impossible task so far.

In this and the following Chapters, we will show how the idea that neural activity might be driven by an underlying critical dynamics shapes experimental and modeling approaches. Here, we will first review some of the most compelling experimental evidence that supports the so-called ``critical brain hypothesis'', as well as the most prominent examples of how such experimental evidence can be explained with non-critical models. We focus on LFPs from the rat's somatosensory barrel cortex, where we find both scale-free avalanches and signatures of criticality beyond avalanches in the form of scale-free spatial correlations. From these data, and inspired by the results obtained so far in this Thesis, we then introduce a general class of stochastic processes describing an archetypal evolution of neural activity driven by another, but unobserved, external process. We show how the properties of this environmental-like latent variable are crucial in producing seemingly power-law neuronal avalanches in the observed degrees of freedom. Further, we exploit the properties of the mutual information derived in the previous Chapters to study the interplay between internal and extrinsic activity, allowing us to understand how the underlying dependencies shape the observed properties of neural activity. In particular, our work suggests that, whereas avalanches may emerge from an external stochastic modulation that affects all degrees of freedom in the same way, interactions between neural populations are the fundamental biological mechanism that gives rise to seemingly scale-free correlations.

\section{The critical brain hypothesis}
It is experimental evidence that the cortex is never silent, but rather rife with spontaneous activity. The critical brain hypothesis suggests that this is the result of the brain operating in the vicinity of the critical point of a phase transition, leading to a rich and variable dynamics at rest. Indeed, most experimental efforts to study criticality in the brain have focused on the spatiotemporal organization of outbursts of spontaneous activity, i.e., neuronal avalanches. Since the work of Beggs and Plenz, power-law neuronal avalanches have been repeatedly observed in experiments \cite{petermann2009spontaneous, yu2011higherorder, hahn2010avalanches, gireesh2008avalanches, mazzoni2007dynamics, pasquale2008self, fontenele2019criticality, dehghani2012avalanche, gireesh2008neuronal, hahn2010neuronal, fontenele2019criticality}, although with varying exponents. In particular, the original observation \cite{beggs2003avalanches} was that neuronal avalanches displayed exponents remarkably close to those of a mean-field branching process - or, equivalently, of the mean-field directed percolation universality class. Furthermore, the idea that the collective behavior of networks of neurons might emerge from a self-organized critical state is undeniably tempting from a modeling perspective. Large-scale correlations - a feature typical of systems close to a phase transition - would allow a coherent global response to external stimuli \cite{mora2011biological, munoz2018colloquium}, as well as an optimal information transmission \cite{marinazzo2014information, dearcangelis2010learning, hidalgo2014information, boedecker2012information}, and at the same time the near-divergent susceptibility would translate into a high sensitivity to sensory stimuli \cite{kinouchi2006optimal, khajehabdollahi2022when}. From a computational point of view, it has been advocated that the trade-off between a disordered phase where perturbations and noise can propagate unboundedly - high information transmission, but low storage capacity - and an ordered phase where changes are rapidly erased - high storage capacity, but low information transmission - might be a naturally optimal solution. In general, it has been argued that criticality provides biological systems with an optimal balance between robustness against perturbations and the flexibility to adapt to changing conditions. For instance, Hidalgo et al. \cite{hidalgo2014information} have shown that complex adaptive systems that have to cope with a great variety of stimuli are much more efficient when operating in the vicinity of a critical point, and thus they benefit from dynamically tuning themselves to such state.

On the other hand, the debate about the nature of the transition - if there is any - is very much open, and thus its hypothetical universality class is poorly understood. Some suggest that the avalanche exponents might be those found in a mean-field branching process \cite{beggs2003avalanches, niebur2014criticality}, and thus connected to directed percolation. Yet, recent works have proposed that the observed transition might be related to a synchronous-asynchronous one \cite{disanto2018landau, dallaporta2019avalanches, poil2012critical, buendia2021hybrid, buendia2020selforganized, buendia2022broad}, a disorder-induced transition \cite{poncealvarez2018crackling}, or a contact process with inhibitory degrees of freedom \cite{corrallopez2022excitatoryinhibitory}. Further, and perhaps unsurprisingly, a number of works showed that the presence of power-law avalanches is not a sufficient condition for criticality, as they might emerge from different mechanisms \cite{touboul2010avalanches, touboul2017absence, martinello2017neutral, priesemann2018can, faqeeh2019emergence}. We will now focus on the description of neuronal avalanches, and then show how in state-of-the-art spatially extended experimental recordings the structure of spatial correlations reveals further signatures of criticality.

\subsection{Scale-free neuronal avalanches}
Historically, the study of neuronal avalanches has mostly dealt with LFPs, i.e., the electric potential recorded in the extracellular space of neuronal tissues. As such, this is a typically mesoscopic measure that samples the activity of neural populations rather than single-neuron activity \cite{brette2012handbook}. Then, standard procedures exist to detect neural activity events from LFPs, obtaining a binary timeseries in discrete time, e.g., a threshold detection based on the standard deviation of the signal \cite{gireesh2008avalanches, petermann2009spontaneous, shriki2013avalanches}. From this discrete timeseries, avalanches are extracted - an avalanche starts when activity is detected from a previous timestep where no activity was present, and lasts until the timestep in which no events are detected. The size of an avalanche $S$ is then the number of such events, and its duration is $T$.

Therefore, to study avalanches' statistics, the data are temporally binned, and avalanches are defined as sequences of bins that present activity. In particular, the temporal bin is often chosen to coincide with the average inter-event interval \cite{beggs2003avalanches}. Then, scale-free neuronal avalanches are characterized by size distributed as the power-law
\begin{equation}
    \label{eqn:p3c1:av_size}
    p_S(S)\sim S^{-\tau}
\end{equation}
and by durations
\begin{equation}
    \label{eqn:p3c1:av_duration}
    p_T(T)\sim T^{-\tau_T}.
\end{equation}
In \cite{beggs2003avalanches} and subsequent works, it was suggested that $\tau \approx 1.5$ and $\tau_t \approx 2$, which are the exponents one would find in, e.g., a mean-field branching or contact process. These distributions are computed and fitted using a corrected maximum likelihood method \cite{gerlach2019testing, mariani2021avalanches}. Following the methods proposed in \cite{clauset2009powerlaw, gerlach2019testing}, avalanche sizes and lifetimes are usually fitted with the discrete power-law
\begin{equation*}
    p(x; \alpha) = \frac{x^{-\alpha}}{\sum_{y = x_{min}}^{y = x_{max}}y^{-\alpha}}
\end{equation*}
where the parameter $x_{max}$ is set to the maximum observed size or duration. $x_{min}$ is selected as the one that minimizes the Kolmogorov-Smirnov distance
\begin{equation*}
\mathrm{KS} = \max_{x\geq x_{min}}|\hat{S}(x) - \hat{P}(x)|
\end{equation*} 
where $\hat{S}(x)$ is the cumulative distribution function (CDF) of the data and $\hat{P}(y)$ is the CDF of the theoretical distribution fitted with the parameter that best fits the data for $x \geq x_{min}$. Goodness-of-fit is typically assessed by comparing against surrogate datasets drawn from the best-fit power-law distribution with the same number of samples as the experimental dataset. We also take into account the fact that while maximum likelihood methods rely on the independence assumption, actual timeseries data are often dependent. As the authors of \cite{gerlach2019testing} suggest, before performing the fit, the avalanche data are undersampled in order to decorrelate them, by estimating the time $\tau^*$ after which two observations are independent of each other.

Then, a robust test of criticality is often to verify whether the so-called ``crackling-noise relation'' between avalanche exponents holds. This scaling relation was first developed in the context of crackling noise \cite{sethna2001crackling}, but nonetheless it is expected to hold in general in all systems close to their critical point \cite{friedman2012universal}, and in particular in systems with absorbing states \cite{disanto2017randomwalks}. The relation predicts that the critical exponent $\delta$, which relates the duration of an avalanche to its mean size through the relation
\begin{equation}
\label{eqn:p3c1:S_average}
    \langle S \rangle(T) \sim T^{\delta_{\mathrm{fit}}},
\end{equation}
obeys the scaling relation
\begin{equation}
\label{eqn:p3c1:crackling_noise}
    \delta_{\mathrm{pred}} = \frac{\tau_T-1}{\tau -1} = \delta_\mathrm{fit}.
\end{equation}
Clearly, both $\delta_{\mathrm{pred}}$ and $\delta_{\mathrm{fit}}$ can be estimated independently. In principle, if these two estimates are compatible, then the system is compatible with criticality. Proving this relation is however challenging. First, it is sensitive to the fitting methods of the distributions of avalanche sizes and lifetimes. Second, in the case of LFPs, the range of avalanche lifetimes typically extends over one order of magnitude only, which undermines the reliability of power-law fitting. Yet, recent findings \cite{fontenele2019criticality, buendia2021hybrid, carvalho2021subsamples} suggest that, while the avalanche exponents found in different experimental settings do vary, they all lay along the scaling line defined by the crackling-noise relation with a seemingly universal exponent $\delta \approx 1.28$. Nevertheless, it was recently suggested that this relation can be fulfilled in different settings \cite{scarpetta2018hysteresis} and even in models of independent spiking units, for a range of choices of the power-law fitting method \cite{destexhe2020criticality}. In fact, one can derive this relation with the sole assumption that avalanches are power-law distributed and that they satisfy $S \sim T^\delta$, i.e., fluctuations in the size of an avalanche given its duration are negligible. Then,
\begin{equation}
    p(S(T)) \left|\frac{dS}{dT}\right|dT = p(T)dT \implies (T^\delta)^{-\tau} \delta T^{\delta-1} = T^{-\tau_T}
\end{equation}
from which it follows immediately that $\delta = (\tau_T -1)/(\tau -1)$ \cite{scarpetta2018hysteresis}. These assumptions are certainly satisfied in critical points, where the exponent $\delta$ is related to other critical exponents by a number of scaling relations \cite{sethna2001crackling}. Yet, the crackling-noise relation may hold also hold in other settings, as we will also find.

\begin{figure*}[t]
    \centering
    \includegraphics[width=\textwidth]{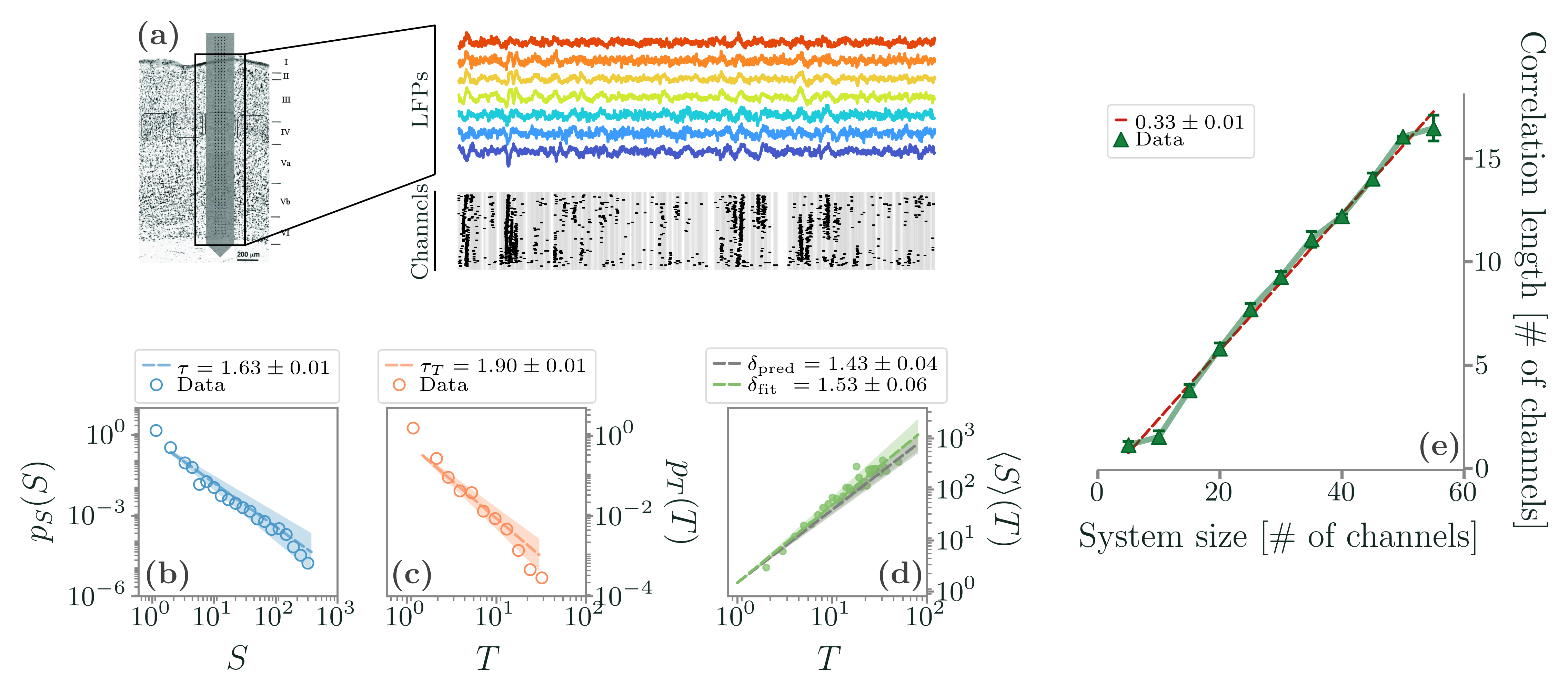}
    \caption{(a) Left: scheme of the array used to obtain the LFPs data from all the cortical layers of the barrel cortex (adapted from \cite{zhang1997intracortical}); right an example of the LFPs signals for different layers and the corresponding discretization. An array of 256 channels organized in a $64 \times 4$ matrix is inserted in a barrel column and the signals from the cortical layers are collected by $55 \times 4$ electrodes. (b-d) Avalanche statistics obtained from the analysis of LFPs data in a rat. Both the distribution of the avalanches (b) sizes and (c) durations are power-laws, and (d) the crackling-noise relation is satisfied. (e) Scaling of the correlation length with the system size in LFPs data, averaging over four different rats. The error bars are shown as $5$ standard deviations from the mean for visual ease. The correlation length scales linearly with the system size with no plateau in sight, a hallmark of criticality}
    \label{fig:p3c1:LFPData}
\end{figure*}

Here, we apply these ideas to LFPs activity from the primary somatosensory cortex of four rats, recorded through spatially extended multi-electrodes arrays\footnote{In order to distinguish real LFP events from noise, a three standard deviation threshold was chosen based on the distribution of the signal amplitudes, which significantly deviated from a Gaussian best fit above that threshold. Both negative and positive LFPs ( nLPFs and pLFPs, respectively) were considered as events in accordance with previous works \cite{shew2013criticality}. One reason is that polarity changes in the LFP signal take place across the depth of the cortex due to compensatory capacitive ionic currents, particularly along the dendrites of pyramidal cells \cite{buzsaki2012extracellular}. Since in our experiments electrodes span multiple cortical layers, both nLFPs and pLFPs were found and detected. For detection, each event was considered terminated only after it crossed the mean of the signal. See also \cite{mariani2021avalanches} for further details and comparison with MUAs, as well as avalanches in non-resting state conditions.}. The cortical activity is recorded through a 256-channels array organized in a $64$ rows $\times$ $4$ columns matrix with an inter-electrode distance of 32 $\mu m$ (see \cite{mariani2021avalanches, mariani2022disentangling} for further details on the experimental procedures). We find that both are statistically compatible with the expected power-laws, as we show in Figure~\ref{fig:p3c1:LFPData}b-c. Averaging over four rats, we find an inter-rat variability with average exponents $\ev{\tau}=1.75 \pm 0.1$ and $\ev{\tau_T} =2.1 \pm 0.3$. In Figure~\ref{fig:p3c1:LFPData}d we show that the crackling-noise relation holds, by comparing $\delta_\mathrm{pred} = \frac{\tau_T-1}{\tau-1}$ with the exponent obtained by fitting the average avalanche sizes as a function of their duration, i.e., $\ev{S}(T) \sim T^{\delta_\mathrm{fit}}$. Averaging over each of our rats, we find
\begin{equation*}
    \langle \delta_{\mathrm{pred}} \rangle =1.47\pm0.18, \quad \ev{\delta_\mathrm{fit}} = 1.46 \pm 0.14
\end{equation*}
Further details on the fitting procedure can be found in \cite{mariani2021avalanches}, where we also use higher frequency data (MUAs) from the same experimental condition, but with a much less dense array. MUAs reproduce more closely the avalanche exponents and in particular the scaling exponent $\delta \approx 1.28$ \cite{fontenele2019criticality, buendia2021hybrid, carvalho2021subsamples}, possibly due to the fact that LFPs are known to be strongly affected by finite-size effects \cite{beggs2003avalanches}.

\subsection{Scale-free correlations}
Although MUAs are better suited to study avalanches, the advantage of our LFPs is that we can leverage the spatially extended structure of the electrodes. Indeed, since the first observation of scale-free neuronal avalanches, many works showed that they might emerge from different mechanisms \cite{touboul2010avalanches, touboul2017absence, martinello2017neutral, priesemann2018can, faqeeh2019emergence}. Arguably, a more fundamental signature of criticality is the presence of power-law correlations in space \cite{binney1992, henkel2009}. In particular, a key feature of both equilibrium and non-equilibrium systems is a correlation length that, in the thermodynamic limit, diverges at criticality. In finite systems, such scale-free correlations manifest themselves in a correlation length that scales linearly with the system size. Yet, the study of these correlations has usually been applied at coarser scales, such as in whole-brain data \cite{haimovici2013connectome, poncealvarez2018crackling}, and only recently in specific cortical areas \cite{ribeiro2020correlations}.

In our data, thanks to the extended spatial resolution of the multi-electrodes array, we are able to obtain the spatial correlations of the fluctuations of the measured LFP activity \cite{fraiman2012noise, haimovici2013connectome}. We study the scaling of the correlation length $\xi$ as a function of the system sizes $L$ by selecting different portions of the array \cite{cavagna2010scalefree, martin2020boxscaling}. The correlation length $\xi$ can be defined as the average distance at which the correlation of the fluctuations around the mean crosses zero \cite{cavagna2010scalefree} at different system sizes. For each timeseries of LFP activity $v_i(t)$, we first compute their fluctuations around the mean activity,
\begin{equation}
    \Tilde{v}_i(t) = v_i(t) - \frac{\sum_{i= 1}^Nv_i(t)}{N}
\end{equation}
with $N$ the number of electrodes. Different sizes of the system, i.e., different portions of the array, are selected and, importantly, the mean activity is computed for each system size, considering the channels inside the portion of the array of linear size $L$ \cite{martin2020boxscaling, ribeiro2020correlations}. For our array, $L$ decreases from the maximum of $55$ channels down to $5$ channels.

For each system's subset, we compute the average correlation function of the fluctuations between all pairs of channels separated by a distance $r$,
\begin{equation}
    {C(r)} = \ev{\frac{\ev{\left(\tilde{v}_{i}-\overline{\tilde{v}}_{i}\right)\left(\tilde{v}_{j}-\overline{\tilde{v}}_{j}\right)}_{t}}{\sigma_{\tilde{v}_{i}} \sigma_{\tilde{v}_{j}}}}_{i, j}
\end{equation}
where $\ev{\cdot}_t$ stands for the average over time, $\ev{\cdot}_{i,j}$ is the average over all pairs of channels separated by a distance $r$ and
\begin{gather*}
    \overline{\tilde{v}}_{i} =\frac{1}{N_T} \sum_{t=1}^{N_T} \tilde{v}_i\left(t\right) \\
\sigma_{\tilde{v}_{i}}^{2} =\frac{1}{N_T} \sum_{t=1}^{N_T}\left(\tilde{v}_i\left(t\right)-\overline{\tilde{v}}_{i}\right)^{2}
\end{gather*}
with $N_T$ is the length of the time series. $\xi$ is computed as the zero of the correlation function, $C(r = \xi) = 0$.

We find that $\xi$ scales linearly with $L$, as shown in Figure~\ref{fig:p3c1:LFPData}. This result can be interpreted as a signature of the presence of underlying long-range correlations that scale with the size of the system. This behavior matches exactly what would happen at a critical point, where the correlation length diverges in the thermodynamic limit and thus grows with the size of a finite system. Hence, we find that the measured neural activity in the barrel cortex at rest displays two different signatures of a possible underlying critical dynamics - power-law avalanches and scale-free spatial correlations. In the rest of this Chapter, and inspired by previous results on stochastic environments, we will show a possible origin of these signatures of criticality.

\section{Extrinsic neural activity as a stochastic environment}
In order to try and unfold the underlying processes from which these collective properties emerge, we assume that neural activity may be decomposed in two parts \cite{priesemann2018can, ferrari2018separating}: (i) \emph{intrinsic activity}, which is the activity driven by interactions between neurons or populations of neurons - in our case, the propagation dynamics across the multi-layer network of the interconnected neurons along the barrel; (ii) \emph{extrinsic activity}, which corresponds to activity modulated by an external or global unit - in our case, the external inputs triggering or modulating the propagation (e.g. synaptic current injection from the thalamic inputs). Taking into account \emph{extrinsic activity} becomes particularly important when neural activity is not analyzed in an isolated context - e.g., from neural slices - but rather from a portion of the animal brain, as in our case.

Let us begin with a biologically sound model. We consider the neural activity described by a Wilson-Cowan model \cite{wilson1972model}, which includes both excitatory and inhibitory synapses, as well as non-linearities in the transfer function. Its derivation is based on arguments on neural dynamics and action potentials, which makes it a general tool to model mesoscopic neural regions. In particular, we study a stochastic version of the Wilson-Cowan model \cite{disanto2018nonnormality, benayoun2010avalanches, decandia2021critical, wallace2011oscillations}, which includes a stochastic term that accounts for the finite size of the populations.

In order to model extrinsic activity alone, i.e., activity purely driven by external modulation, we consider $N$ non-interacting neural populations. Each one is modeled through the activity of two sub-populations, one of excitatory neurons $E_i$ and one of inhibitory neurons $I_i$. $E_i$ and $I_i$ are defined as the densities of active excitatory or inhibitory neurons, and can be interpreted as firing rates. They evolve according to
\begin{equation}
\label{eqn:p3c1:wilson}
    \begin{cases}
        \dot{E}_i = -\alpha E_i + (1-E_i) f(\zeta_i) + \sqrt{(\alpha E_i + (1-E_i) f(\zeta_i))}\eta_{E_i}\\
        \dot{I}_i = -\alpha I_i + (1-I_i) f(\zeta_i) + \sqrt{(\alpha I_i + (1-I_i) f(\zeta_i))}\eta_{I_i}
    \end{cases}
\end{equation}
where
\begin{equation*}
    \zeta_i = \omega_{E} E_i - \omega_{I} I_i + h
\end{equation*}
is the input to the $i$-th population, $\alpha$ is the rate of spontaneous activity decay, $\omega_{E, I}$ are the synaptic efficacies, and $\eta_{E,I}$ are uncorrelated Gaussian white noises with population-size dependent strength $\sigma \propto 1/\sqrt{K}$, with $K$ the number of excitatory and inhibitory neurons \cite{benayoun2010avalanches, wallace2011oscillations}. The response function $f(x)$ is given by
\begin{equation}
    \begin{cases}
        f(x) = \beta \tanh(x) & x \geq 0\\
        f(x) = 0 & x < 0
    \end{cases}
\end{equation}
where $x = \omega_{E} E_i - \omega_{I} I_i + h$ is the average incoming current from the other synaptic inputs and an external input $h$. For each unit, we are interested in the firing rate of the overall population, $\Sigma_i = (E_i + I_i)/2$. $\beta$ will be set to 1 from now on. Importantly, we consider the case in which the units are inhibition dominated, i.e., when $\omega_{I} > \omega_{E}$, with a small noise amplitude $\sigma$, and are non-interacting with each other. Importantly, it was recently shown in \cite{decandia2021critical} that this model admits a critical point at $\omega_{0_C} = \omega_E - \omega_I = \frac{\alpha}{\beta}$, where power-law distributed avalanches will emerge independently of the size of the system. 

\begin{figure*}[t]
    \centering
    \includegraphics[width=0.7\textwidth]{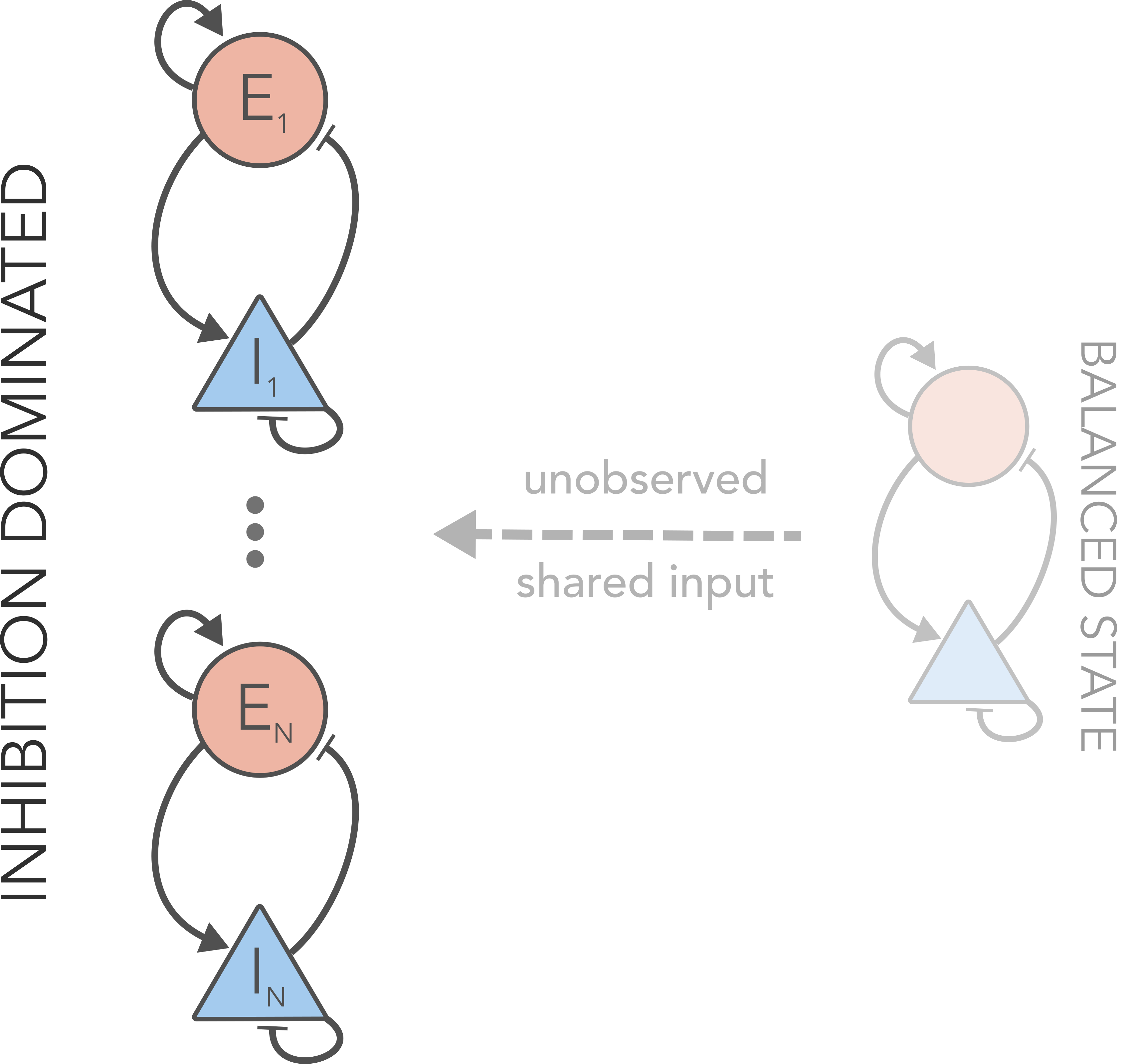}
    \caption{We consider the case of $N$ independent neural populations, each described by a Wilson-Cowan model in the inhibition dominated phase. These populations all share a common unobserved input from another neural population in, e.g., a balanced state. These shared inputs act as a stochastic environment and, depending on their features, may provide strong extrinsic modulation to the observed neural activity}
    \label{fig:p3c1:WC_sketch}
\end{figure*}

We now add external modulation, which comes into this model through the external current $h$. Intuitively, this is equivalent to an unobserved environment that we studied in the previous Chapters - in this scenario, we assume that such a role is played by other, yet unobserved, neural populations, as we sketch in Figure~\ref{fig:p3c1:WC_sketch}. Hence, we model $h$ as the firing rate $h = \left(E^{(h)} + I^{(h)}\right)/2$ of another Wilson-Cowan model, namely
\begin{equation}
\label{eqn:p3c1:wilsoninput}
    \frac{dh}{dt} = \frac{d}{dt}\left[\frac{E^{(h)} + I^{(h)}}{2}\right]
\end{equation}
whose evolution is parametrized by $\omega^{(h)}_E$, $\omega^{(h)}_I$, and $\sigma^{(h)}$. In principle, we may want to choose an external input $h$ that displays bursts of activity separated by periods of silence. Thus, clearly, a possible choice for the external stochastic modulation would be a Wilson-Cowan unit in the critical state. Another and more compelling candidate is a neural population in a balanced state \cite{benayoun2010avalanches, disanto2018nonnormality} defined by $\omega^{(h)}_0 = \omega^{(h)}_E - \omega^{(h)}_I\ll \omega^{(h)}_\mathrm{S} = \omega^{(h)}_E + \omega^{(h)}_I$, and  we set the parameters so that $\omega^{(h)}_0 > \omega^{(h)}_{0_C}$. Crucially, in this scenario, the mechanism giving rise to avalanches is fundamentally different. With these parameters, and in the absence of noise, the dynamics predicts a stable up state. Yet, by increasing the noise amplitude, this up state can be destabilized, leading to large excursions in the down state and thus to avalanches. This phenomenon is a consequence of the non-normality of the matrix describing the linearized dynamics, that can cause a system to be reactive - i.e., its dynamics can exhibit unusually long-lasting transient behaviors even if it asymptotically converges to a stable fixed point, and that coincides with the condition $\omega_0 \ll \omega_S$ \cite{disanto2018nonnormality}.

\begin{figure*}[t]
    \centering
    \includegraphics[width=\textwidth]{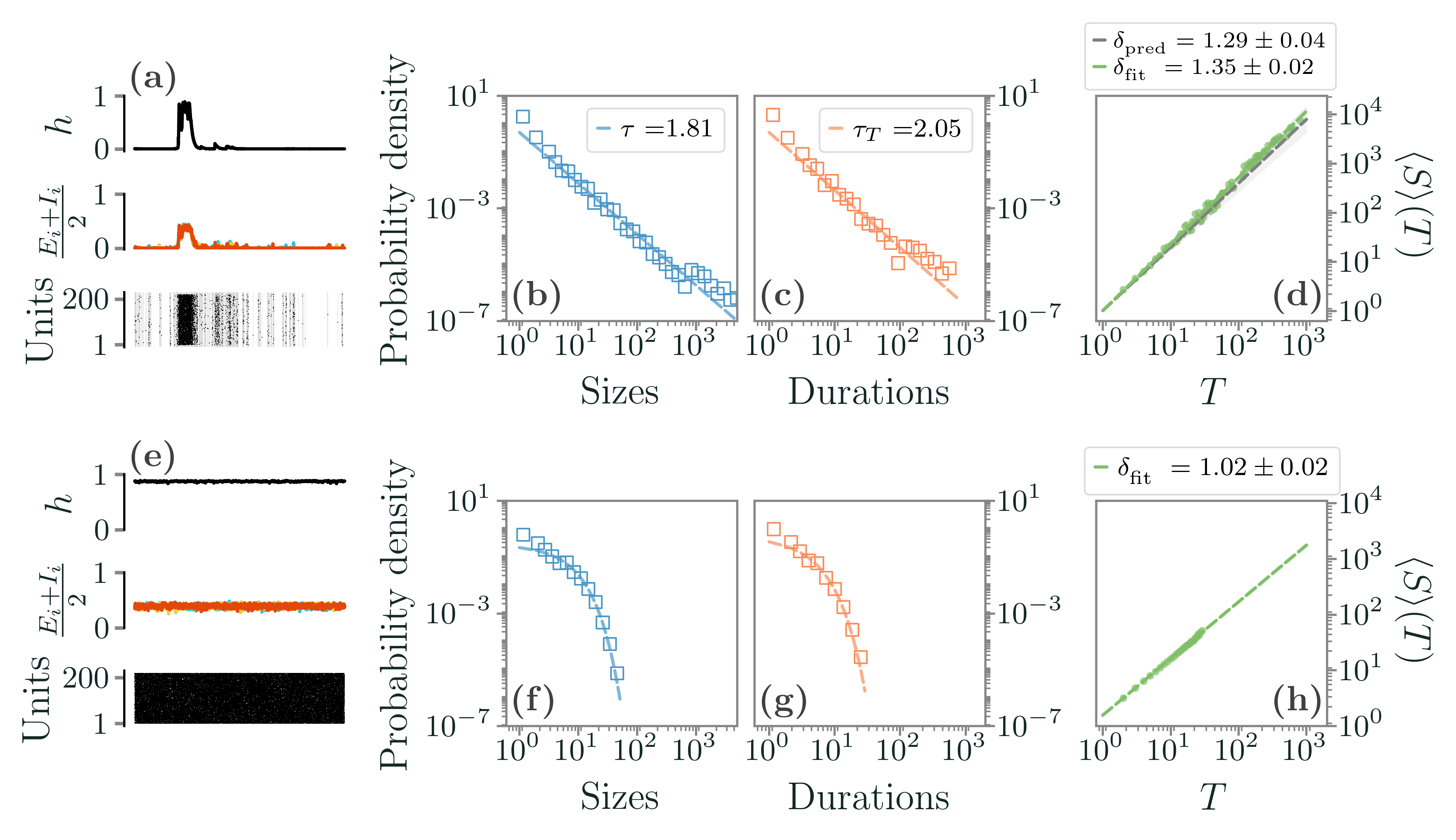}
    \caption{Avalanche statistics generated by the Wilson-Cowan units driven by a balanced unobserved population. Internal units are in an inhibition dominated phase, with $\omega_{I} = 7$, $\omega_{E} = 6.8$, and $\alpha = 1$. The external input $h$ is in a balanced state, with $\omega^{(h)}_{E} = 50.5$, $\omega^{(h)}_{I} = 49.5$, $h^{(h)} = 10^{-3}$, and $\alpha^{(h)} = 0.1$. In Figures (a-d) we set the amplitude of the noise of the unobserved population, $\sigma^{(h)}$, to $2.5 \times 10^{-2}$, so that its up state can be destabilized by the noise. In Figures (e-h), instead, $\sigma^{(h)} = 5 \times 10^{-3}$ and the unobserved population is always in the up state. (a, e) Comparison between the trajectories of $h$, $\frac{E_i + I_i}{2}$ and the corresponding trains of events in the high (a) and low (e) $\sigma^{(h)}$ regime. (b-d) If $\sigma^{(h)}$ is high, avalanches are power-law distributed and the crackling-noise relation is verified. (f-h) Same plots, now in the low $\sigma^{(h)}$ regime. Avalanches are fitted with an exponential distribution. (h) The average avalanche size as a function of the duration scales with an exponent that, as $\sigma^{(h)}$ decreases, becomes closer to the trivial one $\delta_\mathrm{fit} \approx 1$}
    \label{fig:p3c1:avalanches_WC}
\end{figure*}

\begin{figure}[t]
    \centering
    \includegraphics[width=\textwidth]{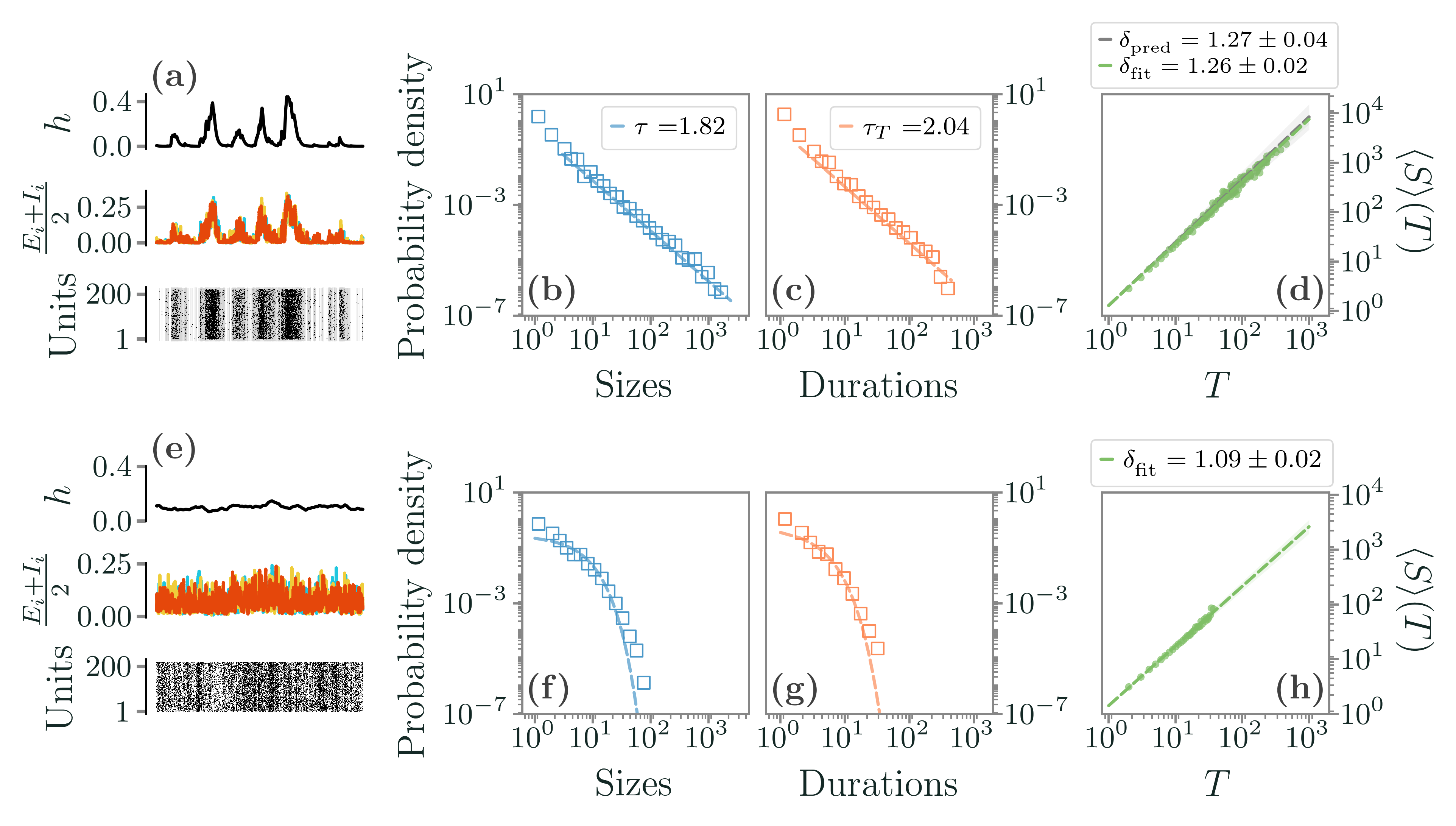}
    \caption{ {Avalanche statistics generated by the Wilson-Cowan units driven by a critical unobserved population. Internal units are once more in an inhibition dominated phase, with the same parameters as in Figure~\ref{fig:p3c1:avalanches_WC}. The unobserved population is now in the critical state $\omega^{(h)}_{E} = 50.05$, $\omega^{(h)}_{I} = 49.95$ and $\alpha^{(h)} = 0.1$, with $h^{(h)} = 10^{-3}$. In panels (a-d) we have $\sigma^{(h)} = 1.2 \times 10^{-3}$, whereas in panels (e-h) $\sigma^{(h)} = 5 \times 10^{-5}$. In panels (a) and panel (e) we compare the trajectories of $h$, $\frac{E_i + I_i}{2}$ and the corresponding trains of events in the high (a) and low (e) $\sigma^{(h)}$ regime. (b-d) If $\sigma^{(h)}$ is high avalanches are power-law distributed and the crackling-noise relation is verified. (f-h) Same plots, now in the low $\sigma^{(h)}$ regime. Avalanches are fitted with an exponential distribution}
    }
    \label{fig:p3c1:criticalwilson}
\end{figure}

For these reasons, we first choose as an effective input $h$ the firing rate coming from a neural population in a balanced state. Although we cannot analytically tackle this model, we simulate the Langevin equations Eqs.~\eqref{eqn:p3c1:wilson}-\eqref{eqn:p3c1:wilsoninput} and from each firing rate $\Sigma_i$ we generate trains of events and analyze avalanches by temporal binning through the average inter-event interval. If the noise strength $\sigma_h$ is high enough, $h$ spends most of the time close to the down state, while showing frequent bursts of activity - this behavior is qualitatively equivalent to the low-$\mathcal{D}^*$ limit previously considered, and is depicted in Figure~\ref{fig:p3c1:avalanches_WC}a. Again, in this regime, we find that avalanche sizes and durations are power-law distributed and satisfy the crackling-noise relation, as we see in Figure~\ref{fig:p3c1:avalanches_WC}b-d. On the other hand, if $\sigma_h$ is lower, the external input $h$ is not modulated as in the previous high $\mathcal{D}^*$ limit. Once more, in this regime avalanches obtained through temporal binning are exponentially suppressed both in their size and in their duration, as we see in Figure~\ref{fig:p3c1:avalanches_WC}e-h.

Let us note that a balanced $h = \left(E^{(h)} + I^{(h)}\right)/2$ is only one of the possible choices that could be considered for the external modulation. Crucially, we believe it is one that achieves a significant biological realism, being linked to E-I balance \cite{poil2012critical}. Moreover, a large number of parameters can satisfy this condition, thus not requiring extreme fine-tuning, as long as the size of the system remains finite and thus driven by noise. We highlight that other choices of external modulation should be able to generate bursts of activity separated by periods of silence - for instance, Brunel's model in the synchronous irregular phase \cite{brunel2000spiking}. Remarkably, another choice is the same Wilson-Cowan model, but in a critical rather than balanced state. In Figure~\ref{fig:p3c1:criticalwilson} we analyze some Wilson-Cowan units that receive the same input from another Wilson-Cowan population that is in a critical state. Indeed, in this case we set $\alpha^{(h)} = \omega_0^{(h)} = \omega_E^{(h)} - \omega_I^{(h)} = 0.1$, thus we are exactly at the critical point $\omega_{0_C} = \omega_E - \omega_I = \frac{\alpha}{\beta}$ \cite{decandia2021critical}. Here, in principle, we expect avalanches to be present for all noise amplitudes (i.e., system sizes) since the external driving experiences power-law distributed cascades. However, note that in Figure~\ref{fig:p3c1:criticalwilson} we set $h^{(h)} = 10^{-3}$, so that the dynamics slightly deviates from the critical point defined for $h = 0$. For this reason, avalanches disappear if we reduce the noise to a sufficiently low value (see Figure~\ref{fig:p3c1:criticalwilson}e-h). Notably, the noise amplitude has to be dramatically reduced in order not to see avalanches, at $\sigma^{(h)} = 5 \times 10^{-5}$.

We further note that these exponents - and those that we will find throughout the Chapter - are different from the ones obtained in LFP data, but this is perhaps not surprising. In fact, besides the simplicity of this paradigmatic model, avalanche exponents have been found to depend on the experimental settings \cite{fosque2021quasicritical} and on individual variability \cite{fontenele2019criticality}. Nevertheless, our framework reproduces the seemingly universal scaling exponent $\delta \approx 1.28$ found in many different experimental settings \cite{fontenele2019criticality, buendia2021hybrid}, including our own in MUAs data \cite{mariani2021avalanches}.

\subsection{Avalanches from extrinsic activity}
Why do we find power-law avalanches in a non-interacting model? Our results suggest that, in principle, any external modulation that alternates periods of silence with periods of varying activity might be able to generate power-law avalanches in the observed degrees of freedom. Thus, we now resort to a simpler model that reproduces these features, where we may have a deeper analytical understanding of the origin of such phenomena. Furthermore, as we will see, in this archetypal model we will have the possibility to add internal interactions and leverage the information-theoretic results we obtained in the previous Chapters.

To this end, we introduce a paradigmatic model of $N$ continuous real variables $(v_1, \dots, v_N)$, denoting the activity of $N$ units (e.g. neurons or, as in our case, distinct populations of neurons as measured by our LFPs). Intrinsic activity will eventually correspond to pairwise interactions among these units, whereas extrinsic activity is still modeled through a common external input that affects all the units in the same way. Inspired by the results of the previous Chapters, we consider the simple case of a multivariate Ornstein-Uhlenbeck (mOU) process, in which the external input corresponds to a common modulation of the noise strength. Although not realistic from a biophysical point of view, this model is simple enough to be treated analytically and to provide a clear physical interpretation, while being complex enough to display non-trivial behaviors \cite{nozari2020brain}. Moreover, mOU processes have been already considered in the literature in the context of fMRI signals \cite{saggio2016structural, gilson2016fmri, gilson2019fmri, arbabyazd2021connectomics}. We have
\begin{equation}
\label{eqn:p3c1:OUprocess_vi}
     \frac{dv_i(t)}{dt} =- \sum_{j} A_{ij}v_j(t) + \sqrt{\mathcal{D}(t)} \eta_i(t),
\end{equation}
where $\eta_i(t)$ are standard white noises{, $A$ is a $N\times N$ symmetric matrix} and {$\mathcal{D}(t)>0$} corresponds to a noise strength modulation from an external input shared among all the units. We also write {$A_{ij} = \frac{1}{\gamma_i} - \mathcal{W}_{ij}$}, where $\gamma_i$ is the characteristic time of the $i$-th unit, and $\mathcal{W}$ is the matrix of the effective synaptic strengths, whose diagonal entries are set to zero. In order to derive analytical results, following \cite{touboul2017absence}, we define the noise modulation $\mathcal{D}(t)$ as
\begin{equation}
\label{eqn:p3c1:OU_Dt}
    \mathcal{D}(t) = \begin{cases}
                     \mathcal{D}^* & \text{if} \quad D(t) \le \mathcal{D}^* \\
                     D(t) & \text{if} \quad D(t) > \mathcal{D}^*
                     \end{cases}
\end{equation}
where $D(t)$ is itself an OU process,
\begin{equation*}
    \dot{D}(t) = -D(t)/\gamma_D + \sqrt{\theta} \eta_D(t)
\end{equation*}
and $\mathcal{D}^*>0$ is a properly chosen threshold. Therefore, the noise modulation $\mathcal{D}(t)$ is described by periods in which it is constant in time and equal to $\mathcal{D}^*$, and periods in which it changes according to an OU process with values $\mathcal{D}(t)>\mathcal{D^*}$.

We first consider the case in which the units are driven only by this extrinsic activity and not by the intrinsic one, i.e., we set the internal interactions to $\mathcal{W}_{ij} = 0$, $\forall i, j$. We refer to this case as the ``extrinsic model''. Then, we add back interactions by reconstructing an effective connectivity \cite{gilson2016fmri} in order to match the correlations matrix of our LFP data - and we will refer to this as the ``interacting model''.

In the absence of internal interactions, at each time the units are conditionally independent given the common external modulation $\mathcal{D}$. However, we typically do not have experimental access to such external modulation. In general, we can only describe the joint stationary probability distribution $p(v_1, \dots, v_N)$ of the units alone. Let us now consider the timescale separation limit $\gamma_{D} \gg \gamma_i$, which corresponds to the assumption that the timescale of the modulation is much slower than the one of the units \cite{priesemann2018can}. In this limit, as shown in the previous Chapters, the process of $v_i$ reaches stationarity much faster than the process of $\mathcal{D}$, thus the joint stationary distribution is given by the marginalization
\begin{equation}
    p(v_1, \dots, v_N) = \int d\mathcal{D} \prod_{i=1}^N p(v_i\,|\, \mathcal{D}) p(\mathcal{D})
\end{equation}
where $p(v_i|\mathcal{D})$ is the stationary solution to the Fokker-Planck equation\cite{gardiner2004handbook} associated to Eq.~\eqref{eqn:p3c1:OUprocess_vi} at a fixed $\mathcal{D}$, and $p(\mathcal{D})$ is the stationary solution associated to Eq.~\eqref{eqn:p3c1:OU_Dt}. Notice that, although the conditional probability distribution is factorizable, in general $p(v_1, \dots, v_N) \ne \prod_{i=1}^Np(v_i)$, i.e., the presence of the unobserved modulation results in an effective dependence between the units.

With our choice of an Ornstein-Uhlenbeck process for $v_i$, $p(v_i | \mathcal{D})$ is Gaussian and we are able to compute these distributions analytically. We find
\begin{align}
\label{eqn:p3c1:pstat_D}
    p(\mathcal{D}) = \frac{1}{2}\left[1+ \text{Erf}\left(\frac{\mathcal{D}^*}{\sqrt{\theta \gamma_D}}\right)\right]\delta(\mathcal{D}-\mathcal{D}^*) + \frac{H(\mathcal{D}-\mathcal{D}^*)}{\sqrt{\pi \theta \gamma_D}}\exp{-\frac{\mathcal{D}^2}{\theta \gamma_D}},
\end{align}
where $H$ is the Heaviside step function, and
\begin{align}
\label{eqn:p3c1:pstat_joint}
    p(v_i, v_j) = & + \frac{1+ \text{Erf}\left(\frac{\mathcal{D}^*}{\sqrt{\theta \gamma_D}}\right)}{2\pi \mathcal{D}^* \sqrt{\gamma_i\gamma_j}}\exp^{-\frac{1}{\mathcal{D}^*}\left(\frac{v_i^2}{\gamma_i}+\frac{v_j^2}{\gamma_i}\right)} + \nonumber \\
    & + \frac{1}{\sqrt{\gamma_i\gamma_j\gamma_D \pi^3\theta}} \int_{\mathcal{D}^*}^\infty \frac{dD}{D} e^{-\frac{1}{D}\left(\frac{v_i^2}{\gamma_i}+\frac{v_j^2}{\gamma_i}\right)}e^{-\frac{D^2}{\theta \gamma_D}}.
\end{align}
It is clear that $p(v_i, v_j) \ne p(v_i)p(v_j)$ - as in the previous Chapters, the marginalization over the environment is introducing effective dependencies between the units. In principle, we are able to compute the joint probability distribution for any number of variables in the same way. Crucially, notice that
\begin{equation}
\label{eqn:p3c1:uncorrelated}
    \langle v_i v_j \rangle - \langle v_i \rangle \langle v_j \rangle = 0 \quad \forall i \ne j,
\end{equation}
which implies that the units, although not independent, are always uncorrelated. This follows immediately from the fact that all the expectation values where a variable $v_i$ appears an odd number of times vanish, e.g.,
\begin{align*}
 \langle v_i v_j \rangle & = \int dv_i \,dv_j\, p(v_i, v_j) v_i\, v_j  \nonumber \\
 & = \int d\mathcal{D} \, p(\mathcal{D}) \left(\int dv_i \, v_i\,p(v_i|\mathcal{D})\right)\left(\int dv_j\, v_j \,p(v_j|\mathcal{D})\right) \nonumber \\
 & = 0 \quad \forall i \ne j
\end{align*}
since 
\begin{equation*}
    \int dv_i \,v_i \,p(v_i | \mathcal{D}) = 0.
\end{equation*}
Therefore, in this extrinsic Ornstein-Uhlenbeck model, the variables are always uncorrelated. This property will be useful when we will consider the case of a non-zero $\mathcal{W}_{ij}$, which will be the sole source of correlations in the model.

\begin{figure}[t]
    \centering
    \includegraphics[width = 1\textwidth]{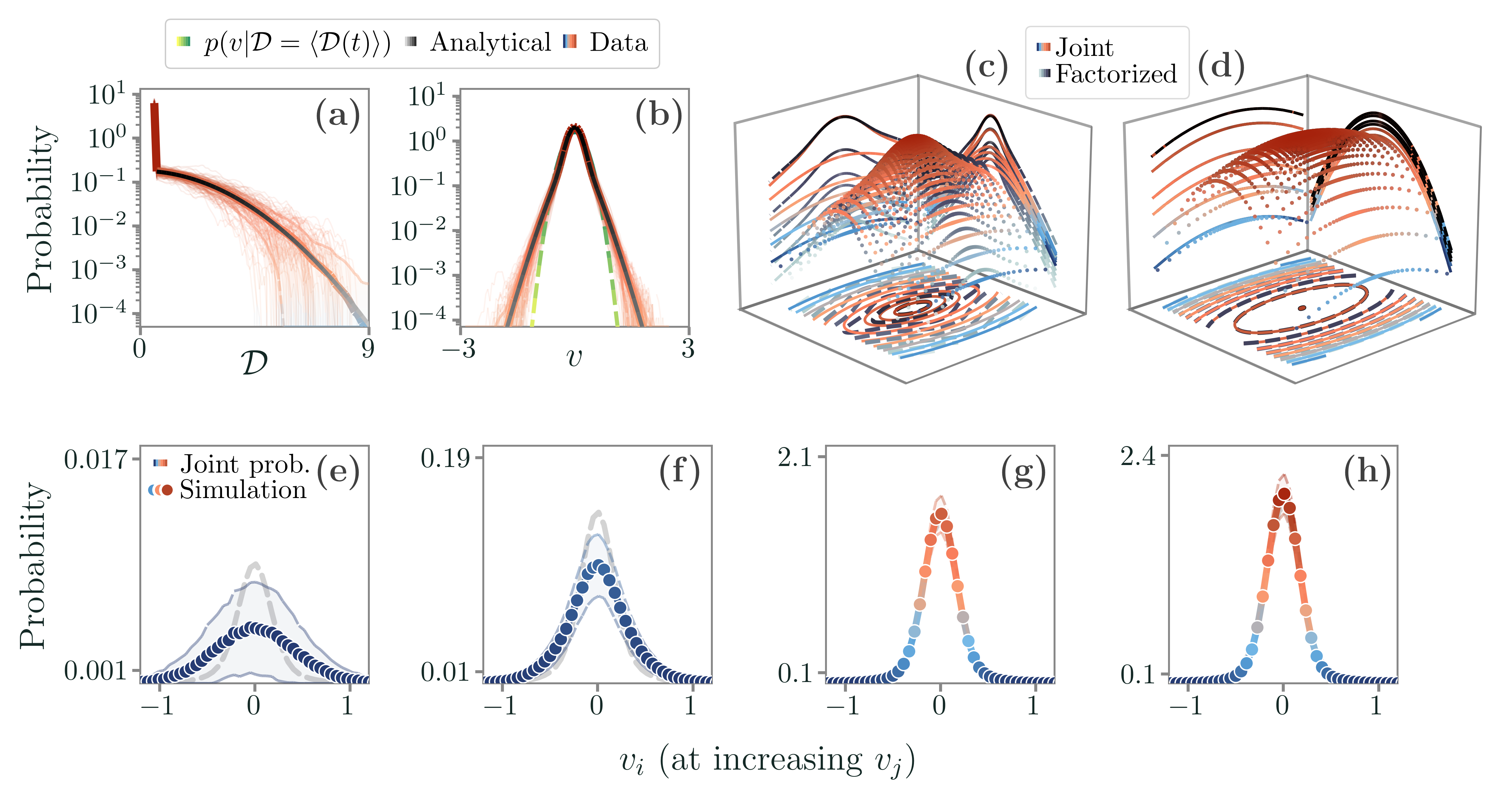}
    \caption{Probability distribution in the extrinsic model. (a-b) The stationary distributions $p(\mathcal{D})$ and $p(v)$. Colored lines represent the results from $10^3$ simulations, and filled areas one standard deviation. The green line in panel (b) represents a Gaussian distribution with average diffusion coefficient. (c-d) Comparison between the (logarithm of the) joint probability distribution in Eq. (\ref{eqn:p3c1:pstat_joint}) (colored lines) and its factorization $p(v_i)p(v_j)$ (black lines), in the (c) low $\mathcal{D}^*$ and (d) high $\mathcal{D}^*$ regimes. (e-h) Comparison between the analytical expressions of the joint probability distribution (colored lines), its factorization (gray lines), and simulations (colored dots). Filled areas represent one standard deviation. Plots are for the low $\mathcal{D}^*$ regime. Notice that most differences between the joint probability and its factorization are found in tails of the distribution, similarly to previous Chapters}
    \label{fig:p3c1:extrinsic_model}
\end{figure}

In Figure~\ref{fig:p3c1:extrinsic_model}a-b we plot the stationary distribution of $\mathcal{D}$ and of $v_i$ obtained from direct simulation of the corresponding Langevin equations. If we compare this distribution to a standard distribution of an Ornstein-Uhlenbeck process with a diffusion coefficient equal to the mean $\ev{\mathcal{D}(t)}$ - the one we would obtain in the opposite timescale separation limit - we immediately see that the distribution of our model is peaked around zero and displays longer tails, as expected. Indeed, due to the fact that $\mathcal{D}^* < \ev{\mathcal{D}(t)}$, the system tends to wander around the potential minimum in zero, particularly during the time windows where the diffusion coefficient is constant and equal to $\mathcal{D}^*$. At the same time, the fact that $\mathcal{D}(t)$ can suddenly increase favors the presence of values of $v$ that are larger in absolute value, suggesting the presence of a bursting behavior. These phenomena can be seen in the joint probability distribution $p(v_i, v_j)$ as well. In Figure~\ref{fig:p3c1:extrinsic_model}(c-h) we compare it with its factorization $p(v_i)p(v_j)$, which is equivalent to ignoring the feedback effects between $v_i$ and $v_j$ due to the shared extrinsic modulation. We see that at low $\mathcal{D}^*$ the most important differences between the two occur in the tails of the two-dimensional distribution, with the joint distribution typically showing dramatically longer tails. This translates to the fact that far-from-zero values of the two variables can occur more easily at the same time. Perhaps unsurprisingly, this scenario is completely reversed when we increase $\mathcal{D}^*$. As expected, for higher values of $\mathcal{D}^*$ the joint probability distribution and the factorized distribution become identical.

\begin{figure*}[t]
    \centering
    \includegraphics[width=\textwidth]{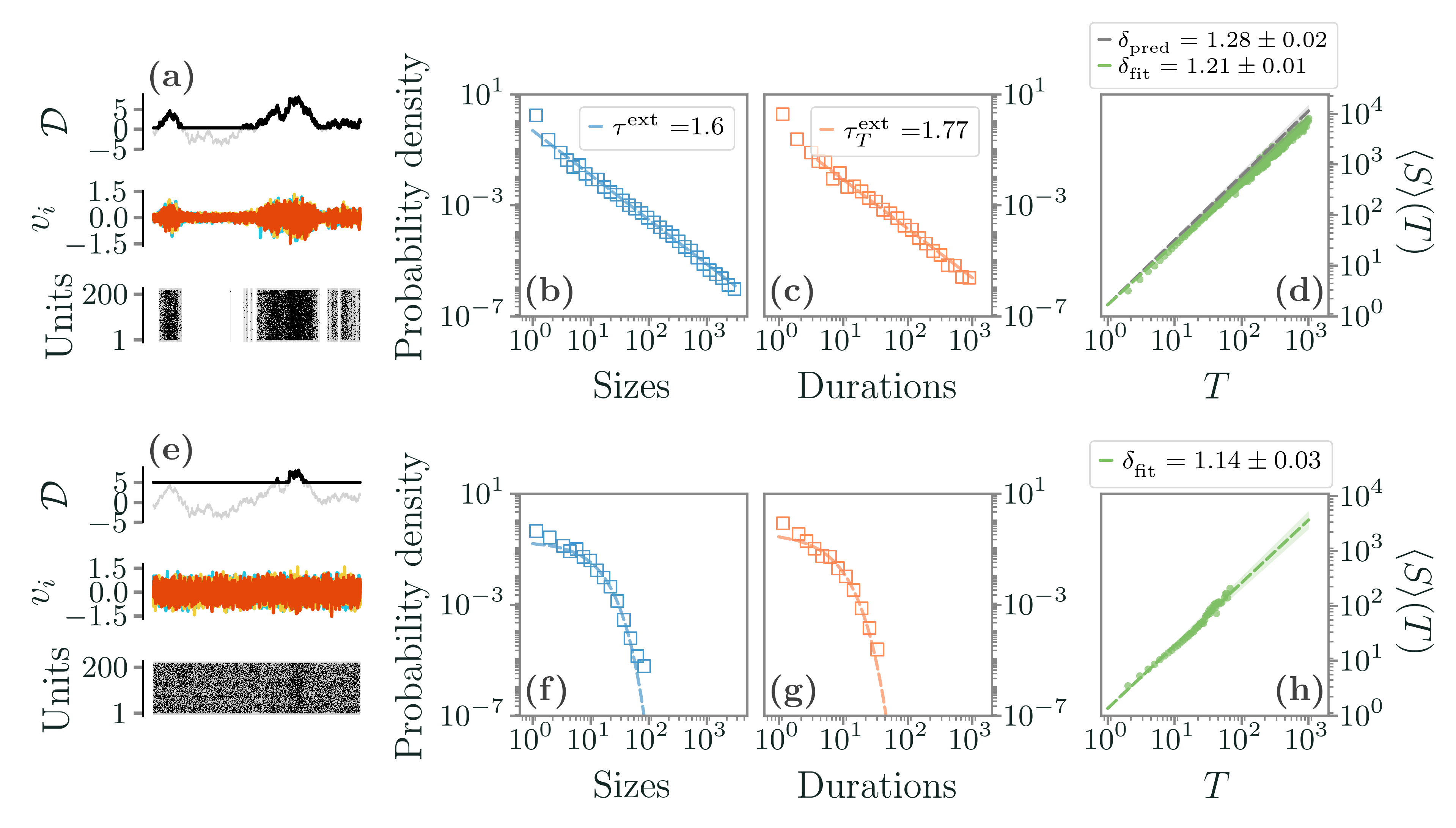}
    \caption{Avalanche statistics generated by the extrinsic model with $\gamma_D = 15$, $\theta = 1$, and $\gamma_i = 0.05$. (a) Comparison between the trajectories of $\mathcal{D}(t)$, $v_i$ and the corresponding discretization with $\mathcal{D}^* = 0.3$. (b-d) If $\mathcal{D}^*$ is low, avalanches are power-law distributed with exponents $\tau^\mathrm{ext} = 1.60 \pm 0.01$ and $\tau^\mathrm{ext}_T = 1.77 \pm 0.01$. The crackling-noise relation is verified in both cases. (e) Trajectories with $\mathcal{D}^* = 5$. (f-h) Avalanches are now fitted with an exponential distribution. The average avalanche size as a function of the duration scales with an exponent that, as $D^*$ increases, becomes closer to the trivial one $\delta_\mathrm{fit}^\mathrm{ext} \approx \delta_\mathrm{fit}^\mathrm{int} \approx 1$}
    \label{fig:p3c1:avalanches_extrinsic}
\end{figure*}

Overall, the net effect of the marginalization is the widening of the tails of both the one-point $p(v)$ and the two-point $p(v_i, v_j)$ probability distributions when $\mathcal{D}^*$ is small enough. As we increase $\mathcal{D}^*$, this effect becomes more and more negligible. In this sense, we can think of $\mathcal{D}^*$ as a control parameter that changes the qualitative behavior of the system - and, in particular, shapes the dependence between the units induced by the modulation. Most importantly, the fact that the tails of the joint probability distribution are wider when $\mathcal{D}^*$ is small reflects dynamically in the emergence of a non-trivial coordination between the variables, leading to a bursty behavior - even without internal interactions between the units. The effect of a low $\mathcal{D}^*$ in the trajectories can be appreciated in Figure~\ref{fig:p3c1:avalanches_extrinsic}a. Whenever $\mathcal{D}(t) = \mathcal{D}^*$, the noise contribution to the units is vanishing and the activity follows an exponential decay. On the other hand, when $\mathcal{D}(t) > \mathcal{D}^*$, noise dominates the system and leads all variables to large excursions from the mean. Therefore, in this regime, each $v_i$ will typically alternate periods of quasi-silence to periods of activity, reproducing a bursty, coordinated behavior. 

Most importantly, the low $\mathcal{D}^*$ regime is also the onset of power-law distributed avalanches, as we see by simulating the model at different $\mathcal{D}^*$ and performing the same analysis as in LFPs. We find that, as $\mathcal{D}^*$ decreases, a transition between exponential avalanches and power-law distributed ones appears. Figures \ref{fig:p3c1:avalanches_extrinsic}b-d show that, if $\mathcal{D}^*$ is small enough, the stochastic modulation produces scale-free avalanches in both size and time with exponents $\tau^\mathrm{ext} = 1.60 \pm 0.01$ and $\tau^\mathrm{ext}_t = 1.77 \pm 0.01$. Crucially, these avalanches satisfy the crackling-noise relation - we fit $\delta_\mathrm{fit}^\mathrm{ext} = 1.21 \pm 0.01$ while expecting $\delta_\mathrm{pred}^\mathrm{ext} = 1.28 \pm 0.02$ from the avalanches exponents. These results are strictly related to the low value of $\mathcal{D}^*$, that is, on the alternating periods of low and high noise strength. In fact, for higher $\mathcal{D}^*$, modulation is rare, as we see in Figure~\ref{fig:p3c1:avalanches_extrinsic}e. This leads to an almost unperturbed Ornstein-Uhlenbeck process, which is only capable of producing exponentially distributed avalanches, which we plot in Figures~\ref{fig:p3c1:avalanches_extrinsic}f-h. Hence, our simple extrinsic model reproduces the results we found in the previous sections with Wilson-Cowan units, and we can understand them in terms of the modulation-induced dependency appearing in the joint probability distribution.

\begin{figure}[t]
    \centering
    \includegraphics[width=0.9\textwidth]{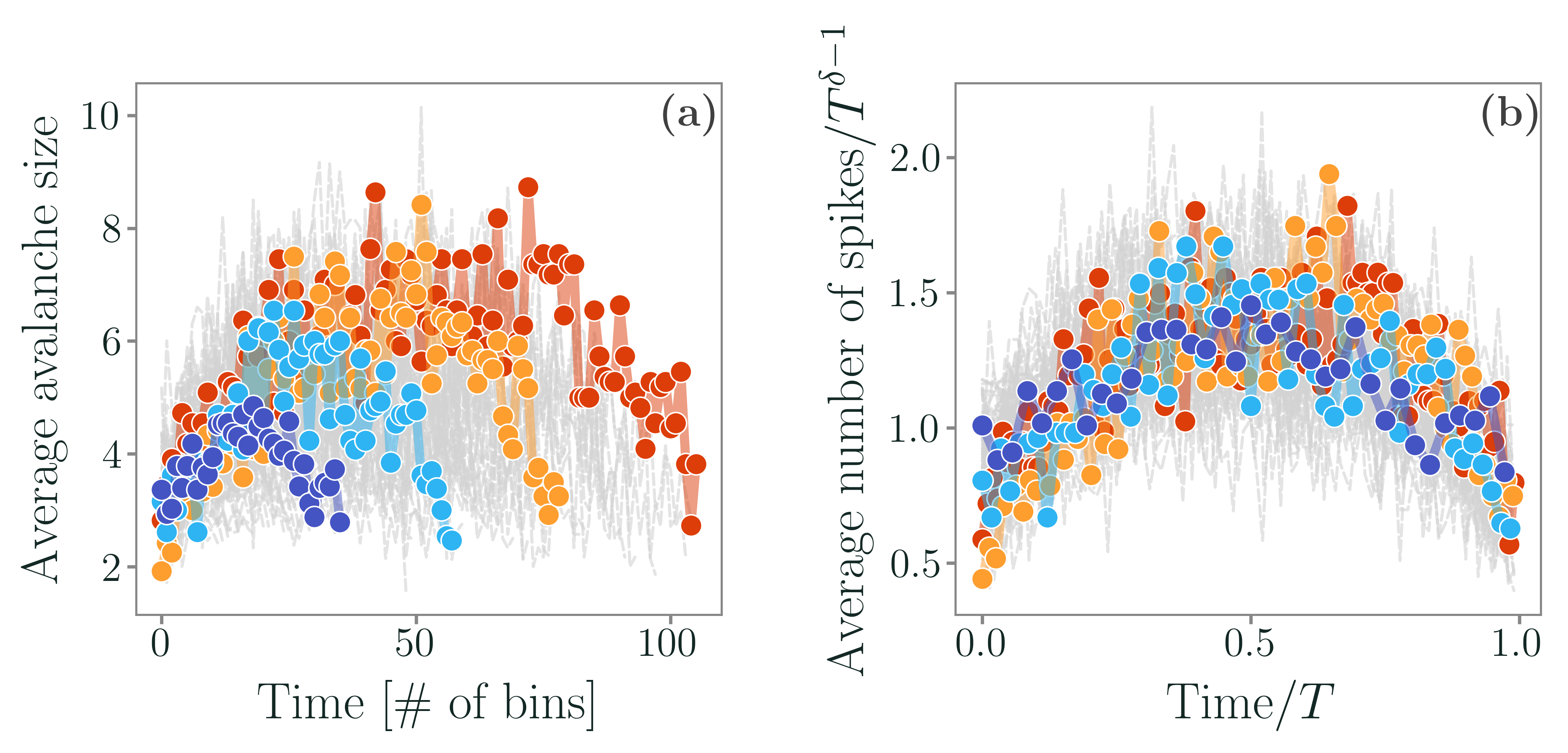}
    \caption{Collapse of the average profile of avalanches of different durations in the extrinsic model, for the low $\mathcal{D}^*$ regime. (a) Profile of the avalanches before the rescaling. (b) If we rescale with an exponent $\delta \approx 1.33$, which is remarkably close to the one found through the crackling-noise relation, we obtain an optimal collapse onto the same scaling function}
    \label{fig:p3c1:collapse}
\end{figure}

\begin{figure}[t]
    \centering
    \includegraphics[width = \textwidth]{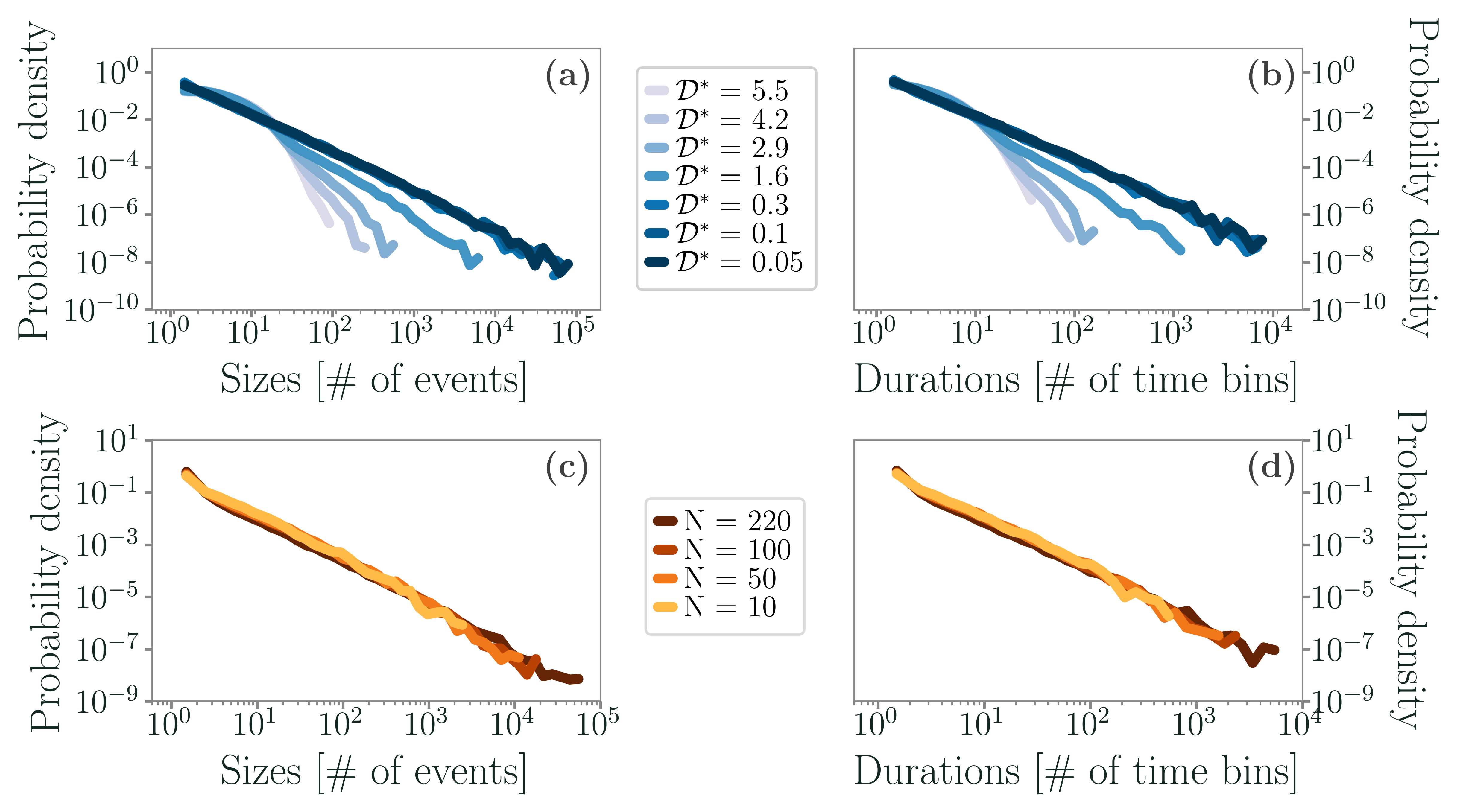}
    \caption{Properties of avalanches in the extrinsic model. (a-b) At increasing values of $\mathcal{D}^*$, the shape of both avalanche sizes and durations smoothly changes from a power-law to an exponential one. Notably, as $\mathcal{D}^*$ approaches zero, the power-law exponents become stable. (c-d) At different system sizes, the power-law distributions of avalanches do not display the expected finite-size cutoff. This is not surprising, since units are conditionally independent and thus no border effects are present}
    \label{fig:p3c1:avalanche_properties}
\end{figure}

Another signature of criticality is the collapse of the average profile of avalanches of different durations onto a single scaling function. For avalanches of duration $T$, we can write down the average number of firing at time $t$ as $S(t,T) = T^{\delta-1} F(t/T)$ where $F$ is a universal scaling function that determines the shape of the average temporal profile. $\ev{S(T)}$ and $S(t,T)$ are related by
\begin{equation}
    \label{eqn:p3c1:shapecollapse}
    \ev{S(T)} = \int_0^TS(t,T)dt.
\end{equation}
At the critical point, we expect that plots of $t/T$ versus $s(t,T)T^{1-\delta}$ for different $T$ will collapse onto the same universal scaling function $F$ \cite{sethna2001crackling}. Thus, finding the exponent for which the goodness of the collapse is higher provides another way to estimate $\delta$. For testing the avalanche shape collapse, we used the methodology introduced in \cite{marshall2016analysis}. To determine the quality of the collapse, the averaged and rescaled avalanche profiles of different lifetimes $F(t/T) = T^{1-\delta}S(t/T,T) $ are first linearly interpolated at $1000$ points along the scaled duration. The variance across the different $F(t/T)$ is calculated at each interpolated point, and the shape collapse error $\epsilon(\delta)$ is then defined as the mean variance divided by the squared span of the avalanche shapes. We plot the results in Figure~\ref{fig:p3c1:collapse}. We find that the exponent that minimizes $\epsilon(\delta)$ is $\delta \approx 1.33$, close to the estimates of $\delta$ we find from the crackling-noise relation. Again, it is also close to the apparently super-universal exponent found in \cite{fontenele2019criticality} and in \cite{friedman2012universal}. 

Importantly, the avalanche exponents are unaffected by the size of the system (see Figure~\ref{fig:p3c1:avalanche_properties}) - that is, the avalanches obtained do not present a clear cutoff. Yet, it is evident that when increasing the number of units the values of the maximum avalanche size and duration increase, while the exponent of the distribution remains the same, displaying a property typical of power-laws. This phenomenon is perhaps unsurprising in our model, where the units are conditionally independent to begin with. On the contrary, in \cite{mariani2021avalanches} we find that avalanches in LFPs from the same set of experiments reproduce the expected finite-size scaling features \cite{beggs2003avalanches}. Finally, Figure~\ref{fig:p3c1:avalanche_properties} shows avalanches' distributions in the extrinsic model at different values of $\mathcal{D}^*$. We can see that, while the exponents converge to a fixed value for small $\mathcal{D}^*$, they change smoothly with $\mathcal{D}^*$ until the distribution becomes exponential for high threshold values. 


\section{Extrinsic and intrinsic neural activity}
We have shown that in the paradigmatic model described by Eq.~\eqref{eqn:p3c1:OUprocess_vi} with $\mathcal{W}_{ij} = 0$ the extrinsic modulation alone generates power-law avalanches in the low-$\mathcal{D}^*$ limit. Yet, this extrinsic activity cannot explain correlations such as the ones observed in Figure~\ref{fig:p3c1:LFPData}e, {as shown by Eq.~\eqref{eqn:p3c1:uncorrelated}}. Hence, we now consider the interacting model, i.e., we consider the case in which both the extrinsic and the intrinsic components of activity are present.

\subsection{Scale-free correlations from internal interactions}
Since in the extrinsic model described in the previous section the units are uncorrelated, we can infer the values of $A_{ij}$ directly from the data \cite{gilson2016fmri}. In particular, we solve the inverse problem in such a way that the correlations of Eq.~\eqref{eqn:p3c1:OUprocess_vi} match the experimentally-measured correlations $\sigma_{ij}$ of our LFPs. For simplicity, we start from a general multivariate Ornstein-Uhlebeck process \cite{gardiner2004handbook} of the form
\begin{equation}
    d \boldsymbol{v}(t)=-A \boldsymbol{v}(t) d t+ B(t)d \boldsymbol{Z}(t),
\end{equation}
where $B(t)$ is a diagonal matrix whose diagonal elements are given by
\begin{equation*}
    B_{ij} = \sqrt{\mathcal{D}(t)} \delta_{ij}
\end{equation*}
and $\boldsymbol{Z}(t)$ denotes a Wiener process. In the case of non-interacting units, which we use to model extrinsic activity, the matrix $A$ is again diagonal with entries $A_{ij} = \delta_{ij}/\gamma_i$. If the matrix $B$ was constant in time, the covariance matrix $\sigma$ of the process would be determined by the continuous Lyapunov equation
\begin{align}
    \label{eqn:p3c1:Lyapu_EQ}
    A \sigma +\sigma A^{\mathrm{T}}= B B^{\mathrm{T}}.
\end{align} 
Since in our case the matrix $B$ is a stochastic variable, we need to marginalize over its stationary distribution $p(B)$. Then, we immediately get
\begin{equation}
    A \sigma +\sigma A^{\mathrm{T}} =  Q, 
    \label{eqn:p3c1:l_eq}
\end{equation} 
where $Q$ is a diagonal matrix, whose elements are given by
\begin{equation}
    Q_{ij} = \delta_{ij}\int_{\mathcal{D}^*}^\infty \mathcal{D} \,p(\mathcal{D})\, d\mathcal{D} := \delta_{ij} f (\mathcal{D}^*, \gamma_D, \theta)
    \label{eqn:p3c1:Q_eq}
\end{equation}
where $f (\mathcal{D}^*, \gamma_D, \theta)$ is nothing but the average value of the modulation. Then, by taking the transpose of Eq.~\eqref{eqn:p3c1:l_eq} and assuming that $A$ is symmetric, we end up with a Lyapunov equation for the matrix $A$, where we can set $\sigma$ to be the covariance matrix of the data. In principle, we could relax the assumption of symmetry of the matrix $A$ by considering the covariance matrix and the time-shifted covariances \cite{gilson2016fmri}. However, this introduces further approximations and in the present work we are only interested in the covariance matrix. Hence, we end up with a model
\begin{equation*}
    \dot{v}_i(t) = -\sum_j A_{ij}v_j(t) + \sqrt{\mathcal{D}(t)} \xi_i(t)
\end{equation*}
where $A_{ij}$ depends on the the parameters of the stochastic modulation $(\mathcal{D}^*, \gamma_D, \theta)$. Notice that, if we write
\begin{equation*}
    \tilde{A}_{ij} = \frac{A_{ij}}{f(\mathcal{D}^*, \gamma_D, \theta)},
\end{equation*}
we need to solve the Lyapunov equation $\sigma \tilde A + \tilde A\sigma = \mathbb{1}$ that only depends on $\sigma$, the covariance matrix of the data. Clearly, if we rescale the experimental timeseries by their standard deviation, $\sigma$ coincides with the correlation matrix. 

\begin{figure}[t]
    \centering
    \includegraphics[width=\textwidth]{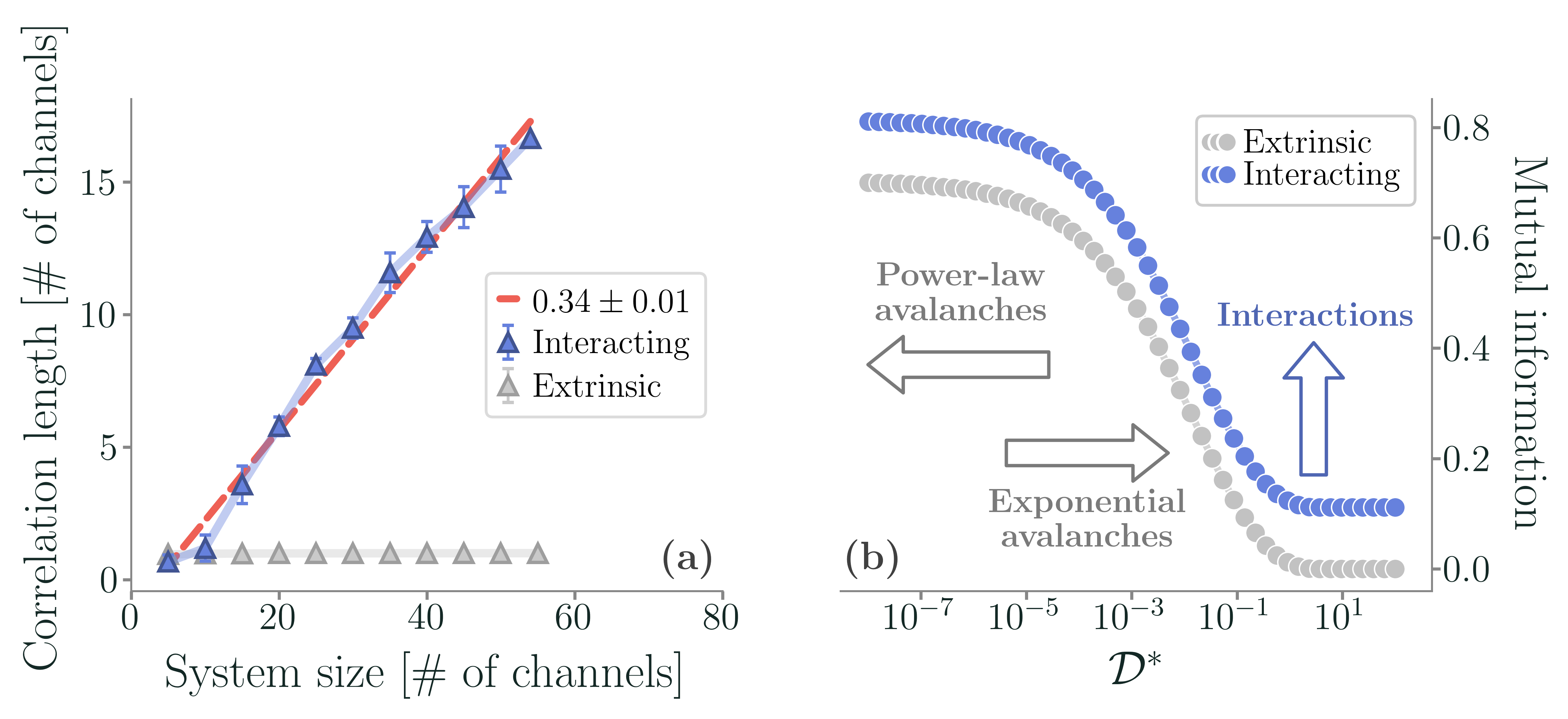}
    \caption{(a) The correlation length of the interacting model scales linearly with the system size, as in the data. In the extrinsic model, as expected, the correlation length of the fluctuations is constant and equal to 1, i.e., the correlation function drops to zero for adjacent electrodes. (b) Comparison between the mutual information in the extrinsic model ($\theta = 1$, $\gamma_D = 10$, $\gamma_1 = 0.1$, $\gamma_2 = 0.5$) and, as an example, in the interacting model with two units. Notice that the onset of a non-vanishing mutual information induced by $\mathcal{D}(t)$ is also the onset of power-law distributed avalanches, whereas the mutual information arising from interactions is independent of $\mathcal{D}^*$}
    \label{fig:p3c1:disentangling_correlations}
\end{figure}

Overall, the effective connectivity $A$ obeys the Lyapunov equation
\begin{equation}
\label{lyapunov}
\sum_{k}\left[\sigma_{ik} A_{kj} +A_{ik} \sigma_{kj} \right] = \delta_{ij}\int_{\mathcal{D}^*}^\infty \mathcal{D} \,p(\mathcal{D})\, d\mathcal{D} := \delta_{ij} f (\mathcal{D}^*, \gamma_D, \theta).
\end{equation}
With these effective interactions, we are now able to study the scaling of the correlation length $\xi$ as a function of the system sizes $L$. We plot the results in Figure~\ref{fig:p3c1:disentangling_correlations}a. As in our data, $\xi$ scales linearly with $L$ in the interacting model - in fact, we can show that even though $A_{ij}$ depends on the modulation parameters $(\mathcal{D}^*, \gamma_D, \theta)$, the scaling of the correlation length does not. In fact, by introducing the rescaled parameters $\tilde{\mathcal{D}} = \mathcal{D}/f$ and $\tilde{v}_i = v_i/\sqrt{f}$, we end up with
\begin{equation*}
    \frac{d \tilde v_i}{dt} = -\sum_j \tilde{A}_{ij} \tilde{v}_j(t) + \sqrt{\tilde{\mathcal{D}}(t)} \xi_i(t)
\end{equation*}
and clearly
\begin{equation*}
    \langle \tilde v_i \tilde v_j \rangle - \langle \tilde v_i \rangle \langle \tilde v_j \rangle = \frac{\langle v_i v_j \rangle - \langle v_i \rangle \langle v_j \rangle}{f}.
\end{equation*}
Therefore, the covariance between $\tilde v_i$ and $\tilde v_j$ is proportional to the covariance between $v_i$ and $v_j$. This implies that at different $(D^*, \theta, \gamma_D)$ we simply find a rescaled interaction matrix $A_{ij}$, but the scaling of correlation length does not change.

\subsection{Disentangling extrinsic and intrinsic activity}
In our model, scale-free correlations are strictly dependent on the interaction network. But what happens to modulation-induced avalanches in the presence of such interactions? Let us remark that our model is formally similar to the ones studied in the previous Chapters \cite{nicoletti2021mutual, nicoletti2022information}. Thus, we can leverage mutual information to understand what kind of dependency structure arises between units, and what such dependencies can tell us about the origin of the critical signatures observed in LFPs.

To this end, we compute the mutual information between pairs of variables
\begin{equation}
    \label{eqn:p3c1:mutual}
    I = \int_{-\infty}^{+\infty} dv_i \int_{-\infty}^{+\infty} dv_j p(v_i, v_j) \log\frac{p(v_i,v_j)}{p(v_i)p(v_j)},
\end{equation}
which captures pairwise dependencies in the system that go well beyond simple correlations. In Figure~\ref{fig:p3c1:disentangling_correlations}b we show that a non-zero mutual information emerges in the extrinsic model in the low $\mathcal{D}^*$ limit. That is, the unobserved modulation - much like the stochastic environments previously considered - induces effective dependencies among the units, even without internal interactions. Remarkably, the onset of this dependency is also the onset of the coordinated behavior between the units and thus of power-law distributed avalanches we have shown in Figure~\ref{fig:p3c1:avalanches_extrinsic}. On the other hand, the mutual information vanishes only in the trivial limit $\mathcal{D}^* \to \infty$, since at large but finite $\mathcal{D}^*$ Eq.~\eqref{eqn:p3c1:pstat_joint} is not exactly factorizable.

When interactions are added back, the stationary probability distribution solution of the interacting model with a generic interaction matrix $A$ is
\begin{align}
    p(v_1, \dots, v_N) = & \frac{1 + \mathrm{Erf}\left[\frac{\mathcal{D}^*}{\sqrt{\theta\gamma_D}}\right]}{2\sqrt{(\pi\mathcal{D}^*)^N \det \Sigma}} e^{-\frac{1}{\mathcal{D}^*}\vb{v}^T {\Sigma}^{-1}\vb{v}} + \nonumber \\
    & + \frac{1}{\sqrt{(\gamma_D\theta)^N \pi^{N+1}\det \Sigma}}G_N\left(\frac{\mathcal{D}^*}{\sqrt{\theta \gamma_D}}, \frac{\vb{v}^T {\Sigma}^{-1}\vb{v}}{\sqrt{\theta \gamma_D}}\right) 
\end{align}
where the matrix $\Sigma$ is determined by $\left({A}{\Sigma} + {\Sigma}{A}^T \right)/2 = \mathbb{1}$ and
\begin{equation}
    G_N(\alpha, \beta) = \int_{\alpha}^{\infty} \frac{dx}{x^{N/2}}e^{-\frac{\beta}{x}-x^2}
\end{equation}
does not have, in general, a closed form. In general, we could define a multivariate information between these $N$ variables. In practice, however, it is very hard to perform the related numerical integration if $N$ is large. Therefore, and without loss of generality, we consider yet again pairwise dependencies, i.e., the case of two variables that interact through the matrix element $A_{12} = \tilde{A}_{12} f(\mathcal{D}^*, \gamma_D, \theta)$ with $\tilde{A}_{12}$ the inferred interaction matrix, and average over all pairs of variables. As previously shown, since this is a Gaussian model we expect the mutual information to receive a distinct - and constant - contribution from the interaction matrix. In fact, in Figure~\ref{fig:p3c1:disentangling_correlations}b we see that the mutual information is simply shifted independently of $\mathcal{D}^*$.

\begin{figure*}[t]
    \centering
    \includegraphics[width=\textwidth]{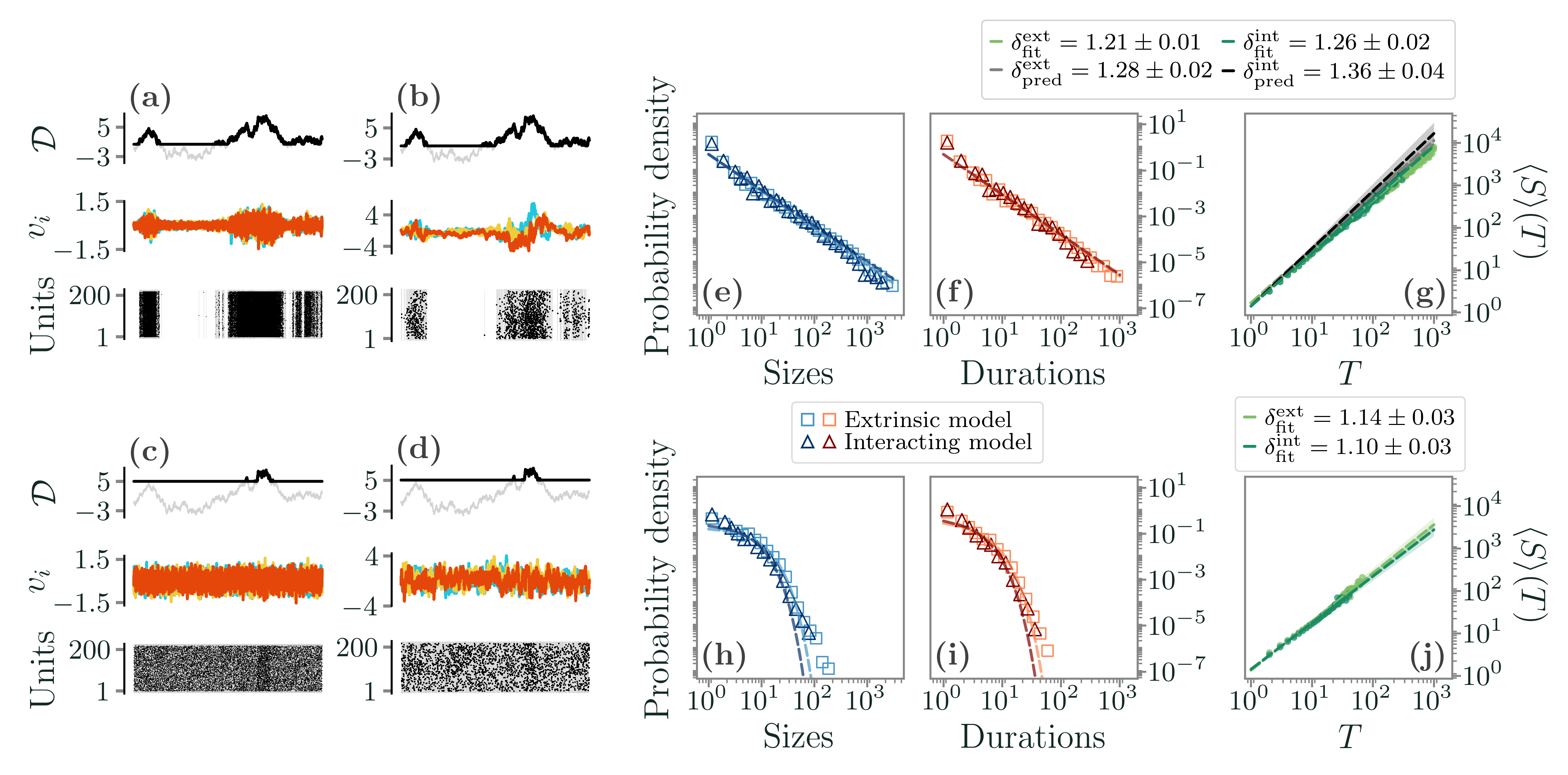}
    \caption{Avalanche statistics generated by the model at $\mathcal{D}^* = 0.3$ (a-b, e-g) and at $\mathcal{D}^* = 5$ (c-d,h-j), with $\gamma_D = 15$ and $\theta = 1$ and $\gamma_i = \gamma = 0.05$ for the extrinsic model. (a-b) Comparison between the trajectories of $\mathcal{D}(t)$, $v_i$ and the corresponding discretization in the low-$\mathcal{D}^*$ regime for (a) the extrinsic model and (b) the interacting one. (c-d) Same, but in the high-$\mathcal{D}^*$ regime. (e-g) If $\mathcal{D}^*$ is low, avalanches are power-law distributed with almost identical exponents in the extrinsic and interacting model, $\tau^\mathrm{ext} = 1.60 \pm 0.01$, $\tau^\mathrm{int} = 1.55 \pm 0.01$ and $\tau^\mathrm{ext}_t = 1.77 \pm 0.01$, $\tau^\mathrm{int}_t = 1.74 \pm 0.01$. The crackling-noise relation is verified in both cases. (h-j) Same plots, now in the high-$\mathcal{D}^*$ regime. Avalanches are now fitted with an exponential distribution. Notice that larger events, corresponding to periods in which $\mathcal{D}(t)>\mathcal{D}^*$, show up in the distributions' tails, suggesting that the shift between exponentials and power-laws is smooth. (j) The average avalanche size as a function of the duration scales with an exponent that, as $D^*$ increases, becomes closer to the trivial one $\delta_\mathrm{fit}^\mathrm{ext} \approx \delta_\mathrm{fit}^\mathrm{int} \approx 1$}
    \label{fig:p3c1:avalanches_model}
\end{figure*}

Crucially, this interaction-dependent constant shift of the mutual information is a signal that the effects of external modulation and of the effective interactions are completely disentangled. The different regimes for the interacting model with the inferred interaction matrix are plotted in Figures \ref{fig:p3c1:avalanches_model}b and \ref{fig:p3c1:avalanches_model}d. This fact reflects on the avalanche properties of our model. Indeed, as shown in Figures~\ref{fig:p3c1:avalanches_model}e-f-g, all avalanches exponents $\tau \approx 1.6$, $\tau_t \approx 1.75$ and the crackling-noise relation exponent $\delta \approx 1.28$ are not changed significantly by the inclusion of direct interactions among the units. The high $\mathcal{D}^*$ regime is not changed either, see Figures~\ref{fig:p3c1:avalanches_model}h-i-j. The fact that the exponents do not change when we add interactions to our model suggests that the avalanches are not affected by the interactions themselves - that is, in our model, they are determined by the extrinsic modulation. These results have profound implications - extrinsic and intrinsic activity are disentangled, and scale-free avalanches only need external modulation to arise.

Overall, in this Chapter we have shown how different signatures of criticality can be found in spatially-extended LFPs data from the rat's barrel cortex. In archetypal models, intrinsic contributions to the neuronal activity - due to the direct interaction between the units themselves - and extrinsic ones - arising from externally-driven modulated activity - can be exactly disentangled. Furthermore, we have shown that a properly chosen external modulation leads to a regime with power-law avalanches that satisfy the crackling-noise relation and that are compatible with the exponent $\delta \approx 1.28$ found in \cite{fontenele2019criticality, buendia2021hybrid}. Crucially, this result holds even in different extrinsic models, such as Wilson-Cowan units modulated by an effective input coming from unobserved neural populations or multivariate Ornstein-Uhlenbeck processes. The same value of the $\delta$ exponent is known to have been found in a variety of neural systems, and our results suggest that it could be explained by a slow time-varying extrinsic dynamics \cite{priesemann2018can} that affects all the neural units in the same way. On the other hand, it was recently shown \cite{carvalho2021subsamples} that this exponent may arise as a consequence of measuring only a fraction of the total neural activity, i.e., of subsampling. Further work is still needed to understand the emergence of such exponents and in which conditions they are robustly reproduced in experiments.

At the same time, while scale-free spatial correlations can and do coexist with power-law avalanches, these kinds of critical signatures cannot be explained by extrinsic activity alone. Crucially, our archetypal model allows us to combine this extrinsic dynamics to an intrinsic interaction matrix, inferred directly from the experimental data to match the spatial correlations we find in our experiments. When we do so, we show that these two signatures of criticality can be disentangled - avalanches appear as a consequence of the external modulation and are only slightly affected by the interactions, and, vice-versa, the interactions determine the spatial correlations independently of the external modulation. Hence, we believe that scale-free correlations may be deeply related to the origin of criticality in the brain, playing a fundamental role in the advantages it might achieve by being critical \cite{kinouchi2006optimal, munoz2018colloquium, mora2011biological}. Remarkably, these results are tightly related to the information-theoretic view presented in the previous Chapters. Although the presence of a non-zero mutual information cannot be a sufficient condition for power-law avalanches to appear, in our extrinsic model their emergence does correspond to the onset of a non-vanishing dependence induced by an unobserved environment - and, conversely, mutual information allows us to show that the mechanism generating avalanches and scale-free correlations are fully disentangled.

\chapter{The interplay between criticality and network structure in whole-brain models}
\chaptermark{Criticality and network structure in whole-brain models}
\label{ch:jphys}
\lettrine{T}{he human brain} is an impressively complex system, spanning several spatial scales of organizations, from microcircuits to whole-brain networks. The comprehensive map of neural connections is usually referred to as ‘‘connectome'' \cite{leergaard2012mapping}. However, it is typically unfeasible to reconstruct connectomes at the neuronal level, and often one relies on anatomical connectivity at coarser spatial scales. In humans, such brain structural networks are typically assessed with diffusion tensor/spectrum imaging techniques, which quantify white matter pathways between mesoscopic brain regions \cite{le2001diffusion, wedeen2005mapping}. 

These complex interconnections act as a backbone on top of which the neurophysiological dynamics occurs. One way to measure such neural activity is through functional magnetic resonance imaging (fMRI). Correlations in fMRI signals from spontaneous activity at rest have been repeatedly observed \cite{fox2007spontaneous}, yielding detailed maps of complex emergent patterns of coherent brain activities, called resting state (functional) networks (RSN) \cite{beckmann2005investigations}. Such patterns, consistent among healthy individuals \cite{damoiseaux2006consistent}, are specifically associated with neuronal systems responsible for sensory, cognitive, and behavioral functions \cite{smith2009correspondence, spadone2015dynamic}.

As we have seen in the previous Chapter, a hypothesis that is increasingly being considered in light of the growing experimental \cite{priesemann2019assessing, fontenele2019criticality} and theoretical \cite{levina2009phase,wilting2019criticality} evidence is that the observed collective emergent patterns may be signatures of the brain self-organizing at a critical point \cite{hesse2014self, buendia2020feedback}. That is, brain dynamics - possibly at different scales - may be poised at the edge of a phase transition. Over the years, evidence to support this hypothesis has been mostly coming from the presence of scale-free neural avalanches \cite{beggs2003avalanches}. However, different works focused on other signatures of criticality, such as cluster size distributions \cite{rocha2018homeostatic, korchinski2021criticality} or long-range temporal and spatial correlations \cite{haimovici2013connectome,mariani2022disentangling} during spontaneous brain activity - exemplary properties of a system near its critical point. Furthermore, it was recently shown that the collective dynamics of neurons may be associated with a non-trivial fixed point of phenomenological renormalization groups \cite{meshulam2019coarsegraining,nicoletti2020scaling}. Some works have also suggested that this phenomenology is compatible with systems between an asynchronous and a synchronous phase, with emerging oscillations \cite{markram2015reconstruction, disanto2018landau, buendia2021hybrid}. In all these studies the role of the network structure in driving such emerging patterns - e.g., global oscillations or optimal information processing - is often missing.

In fact, the emerging collective dynamics in the brain is shaped both by the underlying topological structure of the connectome, and by the properties of neural population activities \cite{le2007analysis, begleiter2006genetics, apicella2021emergence}. Despite a direct relation between structural and functional networks, to what extent structure does determine the neural dynamics and its critical signatures has still to be clarified \cite{suarez2020linking, damoiseaux2009greater}- and computational models may be the key to bridging this gap \cite{breakspear2017dynamic}. To this end, biophysically inspired models of neural dynamics are typically built on top of empirically derived structural networks, with the aim of reconciling functional behavior. Notably, a stochastic version of the Greenberg \& Hastings (GH) cellular automaton \cite{greenberg1978spatial} - one of the simplest models to describe neural dynamics - running over a human connectome of $N=998$ cortical regions \cite{hagmann2008connectome} was shown to match some features of whole-brain activity when tuned to the critical point \cite{haimovici2013connectome, rocha2018homeostatic}. The model undergoes a critical percolation-like transition in the sizes of active clusters, as a function of the level of induced excitatory activation by neighboring neurons. At the percolation transition, the model is able to predict the properties of observed functional RSN.

\begin{figure*}[t]
    \centering
    \includegraphics[width=0.8\textwidth]{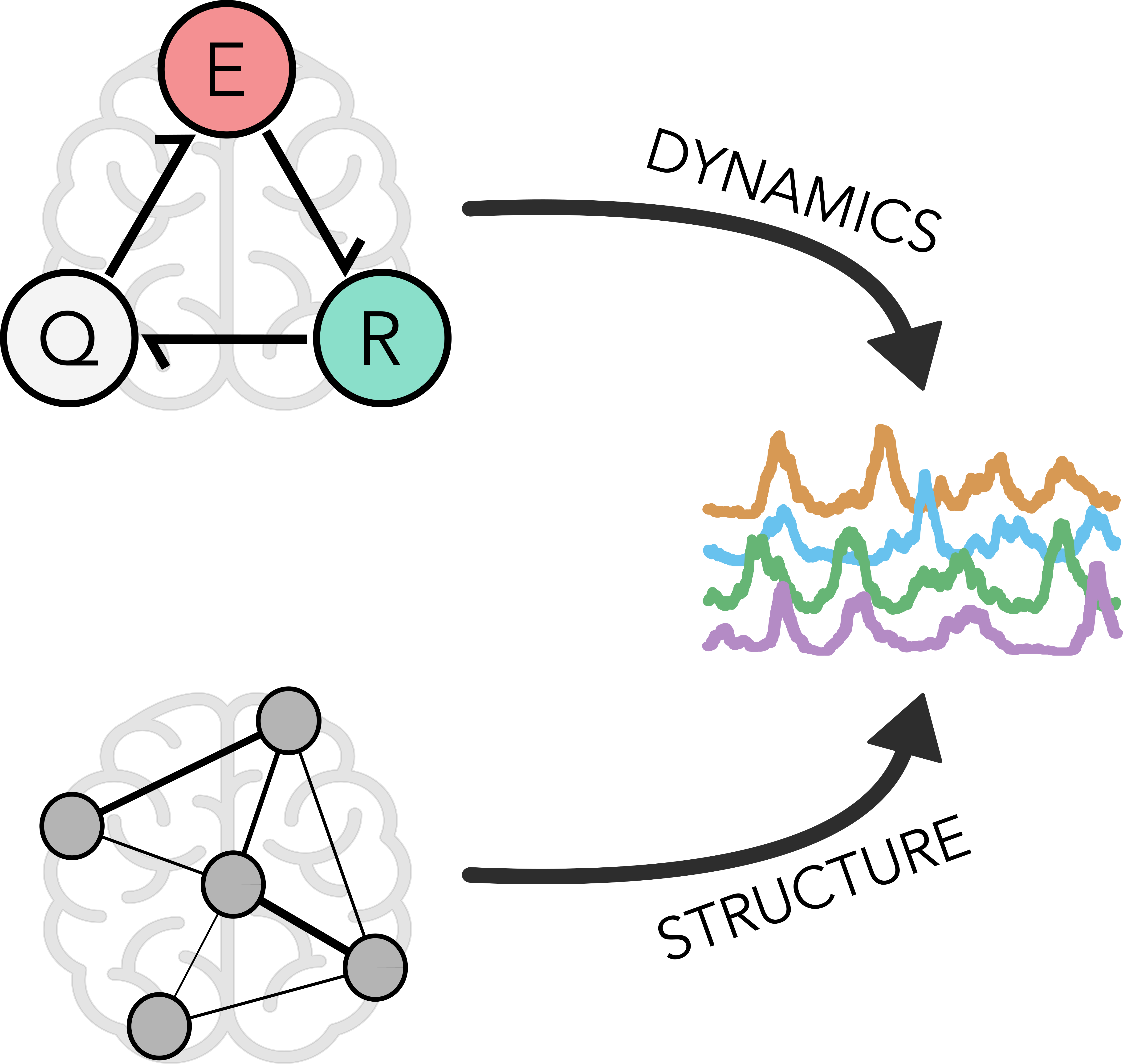}
    \caption{In whole-brain models such as the GH cellular automaton proposed in \cite{haimovici2013connectome}, the dynamical evolution of neural activity is determined by the interplay between the microscopical rule and the underlying network structure. In this Chapter, we will show how different network structures may change the nature of the dynamical transition associated with this model, and how in turn this affects its collective properties}
    \label{fig:p3c2:sketch}
\end{figure*}

Yet, it is known that geometrical percolation transitions may arise in stochastic dynamical systems, and that they usually do not coincide with actual dynamical transitions \cite{martin2020intermittent}. Recent numerical studies have suggested that such dynamical transition in this GH cellular automaton may be continuous for certain levels of connectivity, otherwise being discontinuous or even absent \cite{diaz2021apparently}. Nevertheless, a comprehensive and analytical description of the underlying mechanisms is still lacking. In this Chapter, we will study the properties of this dynamical transition, in order to understand the relation between network structure, criticality, and emergent behaviors. As we sketch in Figure~\ref{fig:p3c2:sketch}, the temporal evolution of activity will be the result of a complex interplay between the underlying dynamical rules, and the topological constraints induced by the connectome. In particular, we develop a stochastic continuous-time formulation of the GH model via a master equation approach. We show analytically how two stable equilibria emerge in the mean-field limit, together with a bistable region of the parameter space where these two equilibria coexist. Hence, the mean-field limit predicts a discontinuous transition - i.e., a transition in which the average activity displays a finite jump. The mean-field power spectrum lacks a characteristic peak, suggesting that no collective oscillations are present - and, in particular, the lack of localized oscillations implies that no clusters of activity appear, as observed in connectomes \cite{haimovici2013connectome, rocha2018homeostatic}. However, when we go beyond mean-field by adding a network connecting different brain regions, the picture is drastically different. We find that the transition becomes continuous - i.e., the average activity changes smoothly - and localized collective oscillations emerge, providing a dynamical mechanism able to sustain clusters of activity and thus capable of generating functional networks. Furthermore, we show that complex topological properties of the connectome are responsible for the dynamical evolution of the model, and in particular simpler null models of connectivity display drastically different phase portraits. Overall, our results shed light on the role of the underlying network structure in the emergence of collective patterns of brain activity, as well as explain the mechanisms behind the phase diagram of Greenberg \& Hastings models used in previous works \cite{haimovici2013connectome, rocha2018homeostatic, haimovici2016dynamical, zarepour2019universal, diaz2021apparently, rocha2022recovery}.

\section{A stochastic whole-brain model}
Here, we develop a continuous-time formulation of the whole brain stochastic model introduced by Haimovici and collaborators \cite{haimovici2013connectome} to describe the dynamics of the human brain at a mesoscopic scale. This model is a variation of the Greenberg \& Hastings cellular automaton \cite{greenberg1978spatial}, originally designed to study excitable media. Each of $N$ nodes in the system belongs to one of three states: quiescent $Q$, excited $E$, or refractory $R$. The original dynamics of the GH automaton is modified in such a way that the states undergo the following stochastic transitions:
\begin{equation}
    \begin{cases}
    Q \xrightarrow{} E \quad& \textnormal{if} \ \sum_j W_{ij}s_j(t)>T \ \textnormal{or  with  prob.} \ r_1\\
    E \xrightarrow{} R \quad&  \textnormal{with prob.} \ 1\\
    R \xrightarrow{} Q \quad& \textnormal{with prob.} \ r_2
    \end{cases}
    \label{eqn:p3c2:transitions}
\end{equation}
where $s_j(t) \in \{0,1\}$ is the state of node $j$ at a certain time step $t$ - set to $1$ if the node is in the $E$ state, and $0$ otherwise -, $W_{ij}$ is the weighted connectivity matrix of the underlying network, $r_1$ is the probability of self-activation and $r_2$ is the probability of recovery from the refractory state. In particular, $T$ is a threshold that governs the induced activation due to interaction with neighboring nodes, which acts as a control parameter of the model.

Therefore, in this model, a neuron may be activated either if the weighted combined activity of neighboring neurons exceeds a threshold $T$, or it may self-activate with a probability $r_1$ that encodes, e.g., external stimuli or unobserved pathways. After activation, neurons switch to a refractory state with unitary probability and cannot be excited again. Escape from the refractory state occurs with probability $r_2$. In this formulation, the state of the system evolves in discrete time steps and is updated synchronously. In particular, for small values of $T$, the activity spreads easily between neighboring nodes, even along weak connections. This leads to a regime of high and sustained activation, characterized by fast and temporally uncorrelated fluctuations. We refer to this phase as ‘‘super-critical''. For high values of $T$, the activity is instead sustained only by a few strong connections, resulting in a suppressed or ‘‘sub-critical'' phase with regular, short-propagating activity in which nodes fail to give rise to relevant patterns of activity. Importantly, we include homeostatic plasticity in the model, implemented as a normalization of the excitatory input of the incoming node. It has been shown that its addition improves the correspondence between simulated neural patterns and experimental brain functional data \cite{rocha2018homeostatic, rocha2022recovery}.

\begin{figure*}[t]
    \centering
    \includegraphics[width=.9\textwidth]{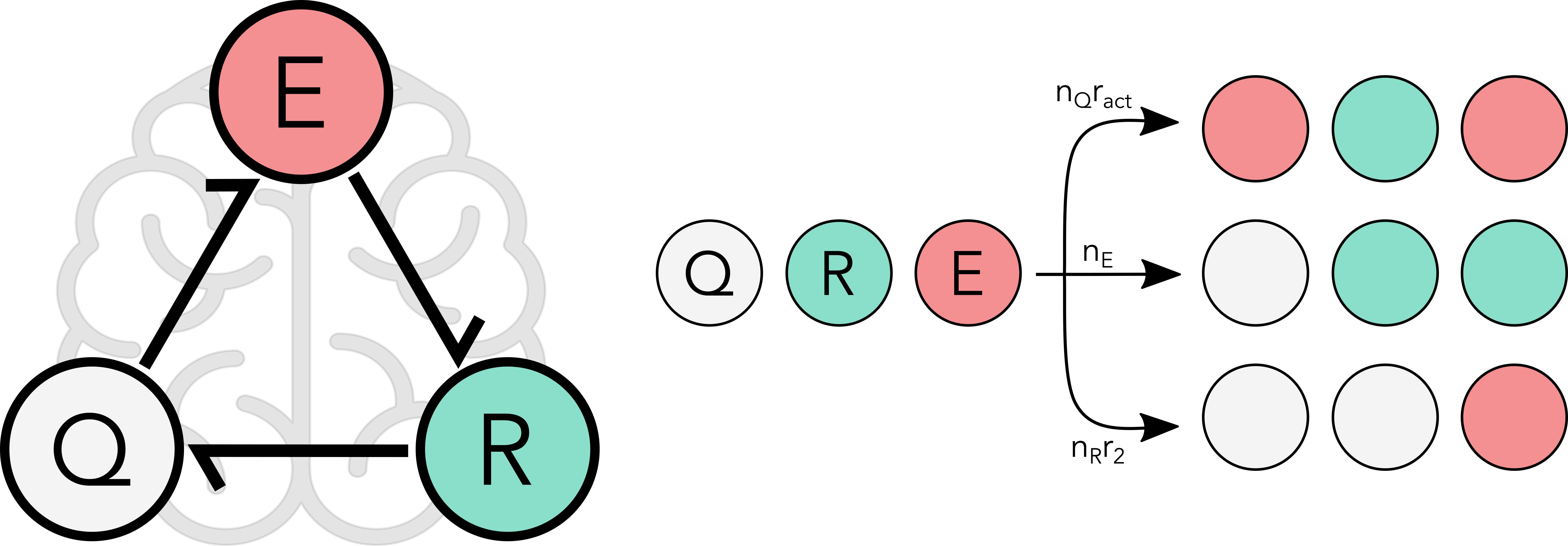}
    \caption{States and transition rates of the model. Quiescent nodes (gray) can be excited with activation rate $r_\mathrm{act}$, and excited nodes (red) become refractory (green) with a rate $r_2$. The refractory period ends with a unitary rate, after which a node can be excited again}
    \label{fig:p3c2:wholebrain_model}
\end{figure*}

Here, we rather formulate the model in a continuous time framework, in order to study analytically its behavior of the model in the large $N$ limit. We denote by $\sigma_i(t)\in\{E,R,Q\}$, $i=1,\ldots,N$, the state of site $i$ at time $t$. The dynamics in Eq.~\eqref{eqn:p3c2:transitions} can be translated into the following continuous-time evolution: for $h>0$ and each node $i$, the probability of having $\sigma_i(t+h)=E$ given that $\sigma_i(t)=Q$ is $r_\mathrm{act}(i)h+o(h)$, where $r_\mathrm{act}(i)$ is a rate of activation defined as
\begin{equation}
    r_\mathrm{act}(i) = r_1 + (1 - r_1)\Theta \biggl[\sum_j W_{ij}s_j - T \biggl]
    \label{eqn:p3c2:p_act}
\end{equation}
with $\Theta[\ \cdot \ ]$ the Heaviside step function. Notice that $0 \leq r_1 \leq 1$ by construction. In a similar manner the probability of jumping from state $E$ at time $t$ to state $R$ at time $t+h$ will be $h+o(h)$ and from $R$ to $Q$ will be $r_2 h+o(h)$\footnote{We highlight that the parameters $r_1$ and $r_2$ in the discrete-time model were probabilities, whereas here they are rates.}.

The mean-field approximation of the model corresponds to the assumption that the underlying graph is fully-connected with constant weights, i.e., $W_{ij} = c, \ \ \forall i,j$. We further consider a homeostatic normalization of the weights, which amounts to assuming that the sum of incoming weights of a given link is unitary. That is, we have the weight matrix
\begin{equation}
     \widetilde{W}_{ij} = W_{ij} / \sum_j{W_{ij}} = 1/N.
\end{equation}
Thus the input to a node is simply given by the density of active neurons in the network, i.e., the argument inside $\Theta[\ \cdot \ ]$ in Eq.~\eqref{eqn:p3c2:p_act} becomes
\begin{equation}
    \sum_j \widetilde{W}_{ij} s_j - T = \dfrac{n_E}{N} - T 
\end{equation}
and it is independent of the particular node $i$, so that we can write $r_\mathrm{act}(i)=r_\mathrm{act}$.

These transition rules induce a Markovian dynamics on $n_E$, $n_R$, $n_Q=N-n_E-n_Q$ - respectively the number of active, refractory, and inactive nodes - with the following rates:
\begin{equation}
    \begin{array}{lcl}
        (n_E, n_R, n_Q)& \stackrel{n_Qr_\mathrm{act}}{\longrightarrow}& (n_E+1, n_R, n_Q-1) \\
        (n_E, n_R, n_Q)&\stackrel{n_E}{\longrightarrow} &(n_E-1, n_R+1, n_Q) \\
        (n_E, n_R, n_Q)& \stackrel{n_R r_2}{\longrightarrow} &(n_E, n_R-1, n_Q+1)
    \end{array}.
    \label{eqn:p3c2:reactions}
\end{equation}
We sketch these rates in Figure~\ref{fig:p3c2:wholebrain_model}. Then, from the reactions in Eq.~\eqref{eqn:p3c2:reactions}, we can write the master equation of our continuous-time model
\begin{equation}
    \begin{split}
    \Dot{P}(n_E,n_R) = & \ P(n_E-1,n_R) \ [N-n_E-n_R+1]r_\mathrm{act} \\
    &+ P(n_E+1,n_R-1) \ [n_E+1] \\
    &+ P(n_E,n_R+1) \ [n_R+1]r_2 \\ 
    &- P(n_E,n_R) \ [(N-n_E-n_R)r_\mathrm{act} + n_E + n_R r_2]
    \end{split}
    \label{eqn:p3c2:master_equation}
\end{equation}
where $P(n_E, n_R)$ is the joint probability of finding $n_E$ active nodes and $n_R$ refractory nodes.

\subsection{Fokker-Planck and Langevin equations}
The master equation Eq.~\eqref{eqn:p3c2:master_equation} can be framed in terms of the density of active $x$ and refractory $y$ neurons. If we introduce the jump steps $\Delta x = 1/N$, $\Delta y = 1/N$, we can treat them as continuous variables in the limit of a large system $N \to \infty$, where the probability $P(x,y)$ obeying the master equation, Eq.~\eqref{eqn:p3c2:master_equation}, becomes differentiable \cite{gardiner2004handbook}. By taking this continuum limit and expanding all terms via a Kramers-Moyal expansion up to the second order, we obtain the Fokker-Planck equation for the probability density $p(x,y)$,
\begin{equation}
    \begin{split}
        \dfrac{\partial}{\partial t} p(x,y) &= - \dfrac{\partial}{\partial x} [A_1(x,y) p(x,y)] + \frac{1}{2N} \dfrac{\partial^2}{\partial x^2} [B_{11}(x,y) p(x,y)] +\\
    &\quad\> -\dfrac{\partial}{\partial y} [A_2(x,y) p(x,y)] + \frac{1}{2N}\dfrac{\partial^2}{\partial y^2} [B_{22}(x,y)p(x,y)] +\\
    &\quad\> +\frac{1}{2N} \dfrac{\partial^2}{\partial x \partial y}[(B_{12}(x,y) + B_{21}(x,y))p(x,y)] 
    \end{split}
    \label{eqn:p3c2:fokker-planck}
\end{equation}
where the coefficients are
\begin{equation}
    \begin{cases}
    A_1(x,y) \ = (1-x-y)[r_1 + (1-r_1)\Theta(x -T)] - x \\
    A_2(x,y) \ = x - r_2 y \\
    B_{11}(x,y) = (1-x-y)[r_1 + (1-r_1)\Theta(x -T)] + x \\
    B_{22}(x,y) = r_2y + x \\
    B_{12}(x,y) = - x\\
    B_{21}(x,y) = - x
    \end{cases}
\end{equation}
and $P(x,y)=p(x,y) \Delta x \Delta y$. Then, the stochastic evolution of the density of active and refractory nodes can be expressed in terms of the Langevin equation
\begin{equation}
   \begin{pmatrix}
    \dot{x} \\
    \dot{y} \\
    \end{pmatrix}
    = \begin{pmatrix}
    A_1(x,y) \\
    A_2(x,y) \\
    \end{pmatrix} + 
    \dfrac{1}{\sqrt{N}}\begin{pmatrix}
    B_{11}(x,y) & B_{12}(x,y) \\
    B_{21}(x,y) & B_{22}(x,y) \\
    \end{pmatrix}^{1/2}
    \begin{pmatrix}
    \xi_1 \\
    \xi_2 \\
    \end{pmatrix}
    \label{eqn:p3c2:langevin}
\end{equation}
where $\boldsymbol{\xi}=(\xi_1, \xi_2)$ is an uncorrelated 2d white Gaussian noise, i.e., such that $\xi_i \sim N(0,1)$ and $\langle \xi_i(t) \xi_j(t') \rangle = \delta_{ij} \delta(t-t')$, $\boldsymbol{A}(x,y)$ is the deterministic drift term, and $\boldsymbol{B}(x,y)$ encloses the stochastic diffusive part.

From now on, unless otherwise specified, we test the validity of our analytical predictions by comparing them with numerical simulations of the continuous model (see also Appendix \ref{app:computational}). The discretization step is $\Delta t = 0.01$, and the parameters are $r_1=0.001$ and $r_2=0.1$ for consistency with previous works \cite{haimovici2013connectome, rocha2018homeostatic, haimovici2016dynamical, zarepour2019universal, diaz2021apparently, rocha2022recovery}.

\subsection{Discontinuous transition in the mean-field limit}
In the limit of a large number of interacting units in the system, the effect of random fluctuations becomes negligible. Indeed, the mean-field description of this model corresponds to the thermodynamic limit, $N \to \infty$, where the deterministic part of Eq.~\eqref{eqn:p3c2:langevin},
\begin{equation}
    \begin{cases}
    \dot{x} = (1-x-y)[r_1 + (1-r_1)\Theta(x -T)] - x\\
    \dot{y} = x - r_2 y 
    \end{cases},
    \label{eqn:p3c2:deterministic}
\end{equation}
describes the evolution of the density of active and refractory units. Although we cannot obtain the full analytical solution of Eq.~\eqref{eqn:p3c2:deterministic}, we can study the system's equilibria and their stability. Indeed, by varying the threshold $T$, the dynamics switches between two different regimes based on the value of $\Theta(\cdot)$. These two phases are characterized by high and low levels of activity, respectively. We call them super- and sub-critical phases.

The super-critical phase is defined by the condition $x > T$, for which the Heaviside function in Eq.~\eqref{eqn:p3c2:deterministic} evaluates to $1$. Hence, and at stationarity, we find
\begin{equation}
    \begin{cases}
         y_+ = \dfrac{1}{2 r_2 + 1} \\
         x_+ = r_2 \ y_+
    \end{cases}
    \label{eqn:p3c2:sol_super}
\end{equation}
so that in this regime the average activity $x_+$ is independent of the rate of self-activation $r_1$. This means that the spread of activity is completely driven by the interaction between active neighbors. For this equilibrium to exist, we need
\begin{equation}
    T < \dfrac{r_2}{2 r_2 + 1} =: T_+
    \label{eqn:p3c2:T_sup}
\end{equation}
so that the inequality $x > T$ is satisfied. This defines the threshold below which the super-critical phase exists. Likewise, the sub-critical phase is defined by $x \leq T$. At stationary, Eq.~\eqref{eqn:p3c2:deterministic} leads to
\begin{equation}
    \begin{cases}
         y_- = \dfrac{r_1}{r_2 + (r_2+1)r_1}   \\
         x_- = r_2 \ y_-
    \end{cases},
    \label{eqn:p3c2:sol_sub}
\end{equation}
and the inequality $x \leq T$ implies that
\begin{equation}
    T \geq \dfrac{r_1 r_2}{r_2 + (r_2+1)r_1} =: T_-,
    \label{eqn:p3c2:T_sub}
\end{equation}
i.e., above the threshold $T_-$, the sub-critical phase exists. As expected from Eq.~\eqref{eqn:p3c2:sol_super} and Eq.~\eqref{eqn:p3c2:sol_sub}, we notice that $\forall r_1, r_2$ the fraction of active nodes $x_+$ in the supercritical phase is larger than the subcritical equilibrium $x_-$, since $r_1 \leq 1$.

Moreover, in the range of $T$ given by equations Eq.~\eqref{eqn:p3c2:T_sup} and Eq.~\eqref{eqn:p3c2:T_sub} for which such solutions exist, they are both stable equilibria, each with its own basin of attraction. We can investigate the nature of the equilibria through linear stability analysis. If we write Eq.~\eqref{eqn:p3c2:deterministic} as $\dot{\boldsymbol{z}} = \bm{f}(\bm{z})$ with $\bm{z}=(x, y)$, the stability of its equilibria $\bm{f}(\bm{z}^*)=0$ is readily understood from the evolution of small perturbations $\bm{z} = \bm{z}^* + \Delta \bm{z}$ with $| \Delta \bm{z} | \to 0$, which obey
\begin{equation*}
        \dot{\Delta \bm{z}}= \bm{f} (\bm{z}^*) + \dfrac{\partial \bm{f}}{\partial \bm{z}} \Bigr|_{\bm{z}=\bm{z}^*} \Delta \bm{z}  + ... = \bm{J}(\bm{z}^*) \Delta \bm{z}.
\end{equation*}
Thus the dynamics near the fixed points is governed, at first order, only by the Jacobian matrix $\bm{J}$. In particular, the real part of its eigenvalues $\lambda$ can give us information on stability or instability.

\begin{figure*}[t]
    \centering
    \includegraphics[width=\textwidth]{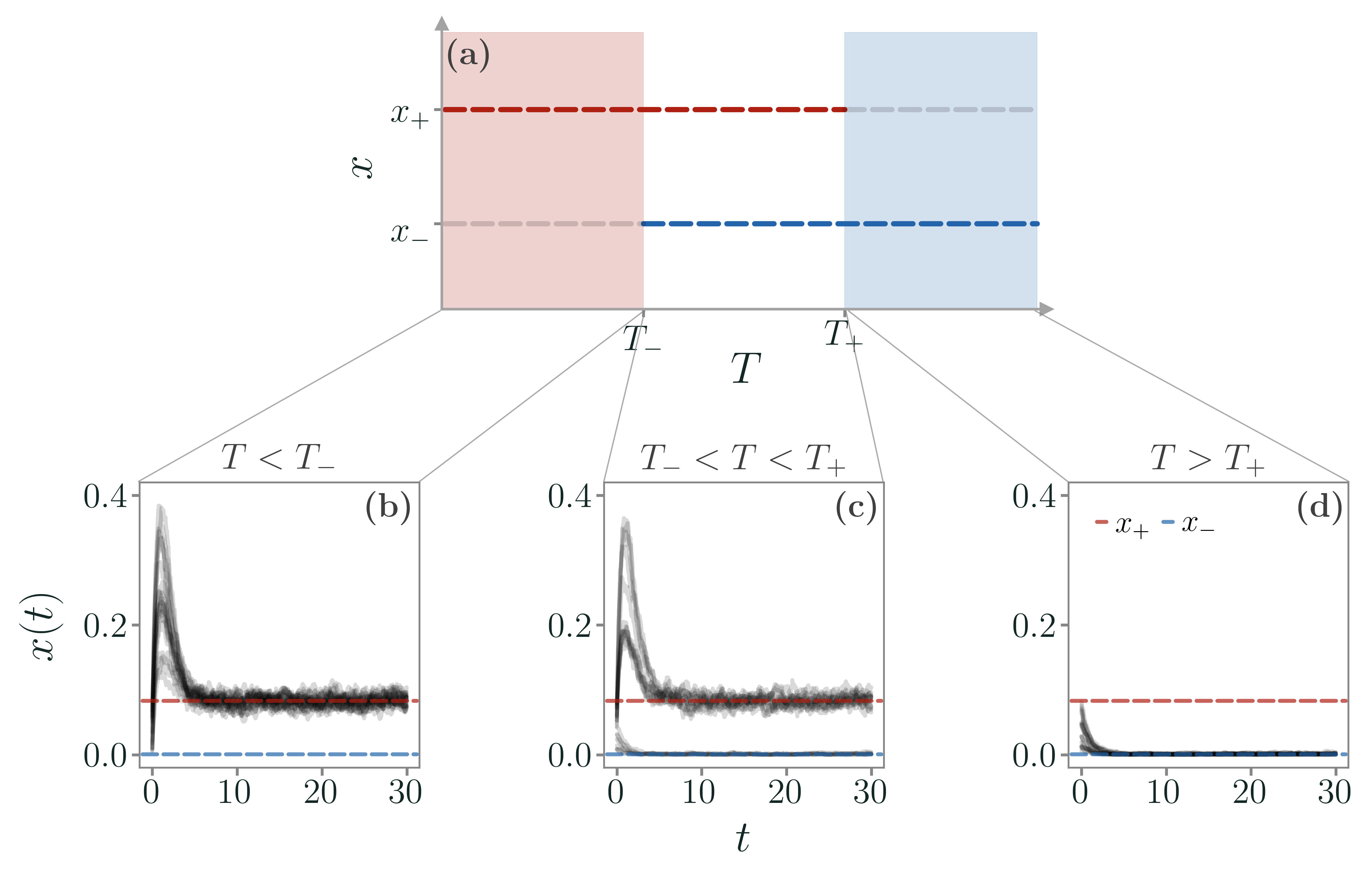}
    \caption{Diagram of equilibria in the model. (a) The region of existence of super- and sub-critical equilibria. As the control parameter $T$ changes, we can identify three different regions: for low $T$, only the supercritical equilibrium exists (\textit{red region}); for high $T$, only the subcritical equilibrium exists (\textit{blue region}); for intermediate values of $T$, the two equilibria coexist. (b-d) Examples of trajectories in the three regions. The model is simulated with a fully-connected network of size $N= 10^3$. Each plot shows 30 trajectories from a random initial configuration}
    \label{fig:p3c2:diagram_equilibria}
\end{figure*}

In the super-critical phase, the Jacobian evaluated at $(x_+, y_+)$ is
\begin{equation*}
    J_+= \begin{pmatrix}
    -2 &
    -1 \\
    1  & 
    -r_2 \\
    \end{pmatrix},
\end{equation*}
whose eigenvalues are
\begin{equation*}
    \lambda_\pm^{(+)} = - \dfrac{2+r_2 \pm \sqrt{r_2^2 - 4r_2 }}{2}.
\end{equation*}
The super-critical equilibrium thus is stable if $\mathrm{Re}(\lambda_\pm^{(+)})<0$. We can distinguish two regimes: if $r_2 \geq 4$ the eigenvalues are purely real, otherwise they have an imaginary part. In both cases the conditions are satisfied, thus the super-critical fixed point $(x_+, y_+)$ is respectively a stable knot or a stable focus. Instead, in the sub-critical phase the Jacobian evaluated at $(x_-, y_-)$ reads
\begin{equation*}
    J_-=
    \begin{pmatrix}
    -1-r_1 &
    -r_1 \\
    1  & 
    -r_2 \\
    \end{pmatrix}
\end{equation*}
with eigenvalues
\begin{equation*}
    \lambda_\pm^{(-)} = \dfrac{-(1+r_1+r_2) \pm \sqrt{(1+r_1+r_2)^2 - 4 (r_1 + r_2 + r_1r_2)}}{2}.
\end{equation*}
Since $r_1 \geq 0$ and $r_2 \geq 0$, $\lambda_\pm^{(-)}$ always have a negative real part, and thus this equilibrium is always stable. Again, we observe two different regimes - if $r_1 - 2\sqrt{r_1} +1 < r_2$ and $r_1 + 2\sqrt{r_1} +1 > r_2$ we have a stable focus, otherwise a stable knot.

Therefore, the super- and sub-critical equilibria, if they exist, are always stable. Crucially, and $\forall r_1, r_2$, Eq.~\eqref{eqn:p3c2:T_sup} and Eq.~\eqref{eqn:p3c2:T_sub} imply that
\begin{equation}
    \label{eqn:p3c2:bistability_condition}
    T_- = \frac{r_1 r_2}{r_2 + (r_2+1)r_1} < \frac{r_2}{2 r_2 + 1} = T_+
\end{equation}
thus three regions emerge in the parameter space spanned by $T$, as shown in Figure~\ref{fig:p3c2:diagram_equilibria}. For $T \leq T_-$, the sub-critical equilibrium does not exist, hence we can only observe the active phase. On the other hand, for $T>T_+$ only the sub-critical equilibrium exists. In between these values, for $T_- < T  \leq T_+$, the two equilibria coexist and we find a region of bistability.

\begin{figure}[t]
    \centering
    \includegraphics[width=\textwidth]{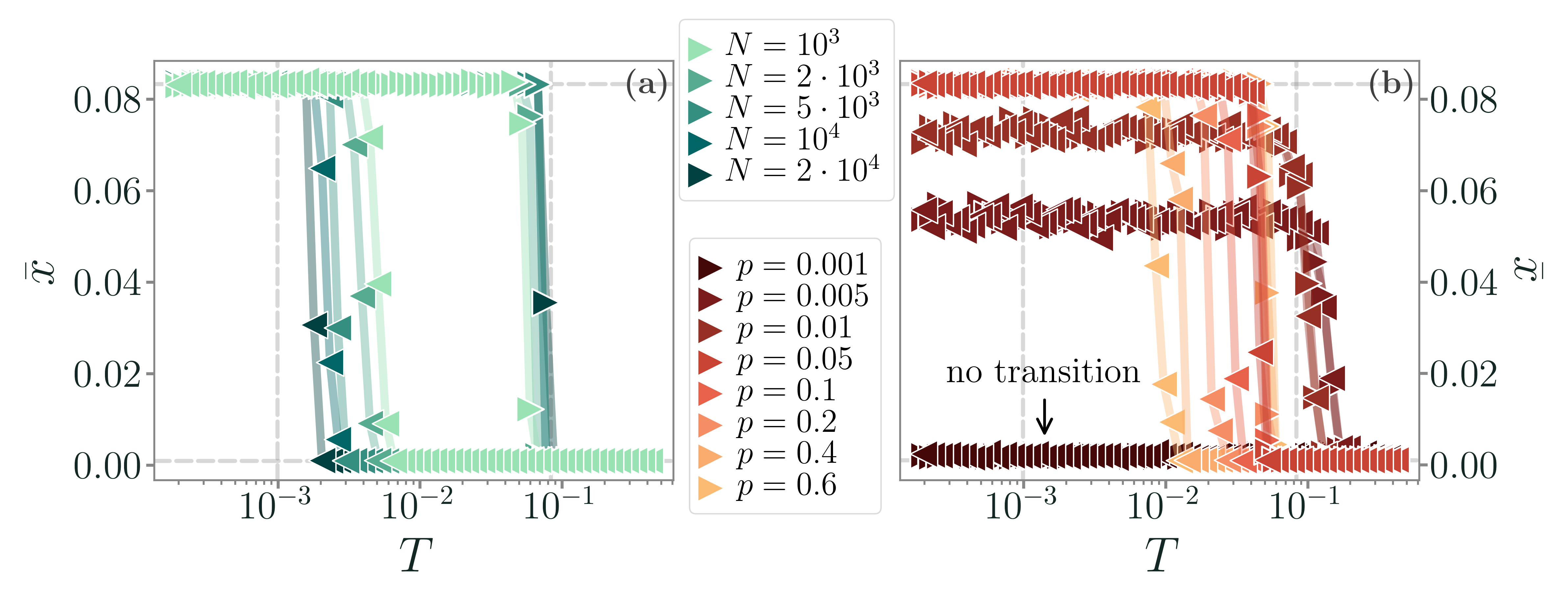}
    \caption{Average activity $\bar{x}$ as a function of the control parameter $T$ for different sizes and topologies. To prove the presence of the bistable region, we slowly change $T$ every $10^5$ steps starting from $T_0=0.2 \cdot T_-$ up to $T_F=5 \cdot T_+$ for 60 values of T taken in a logarithmic scale, and then decreasing it back to $T_0$ in the same way. (a) Results obtained with fully-connected topologies of different sizes $N$. Triangles to the right represent values obtained while increasing $T$, whereas triangles to the left represent values obtained while decreasing it. We see that a hysteresis cycle emerges and that it approaches the expected boundaries of the bistable region as $N$ increases. (b) Results obtained from Erdős–Rényi networks with $N=10^3$, constant weights, and for different wiring probabilities $p$. As the connectivity decreases, the bistable region shrinks and the transition between the two regimes becomes smooth. For extremely low values of $p$, the transition disappears (as indicated by the arrow) and the system is never in the super-critical regime}
    \label{fig:p3c2:hysteresis}
\end{figure}

In order to assess the effects of finite sizes on the region of bistability, we simulate the system and track the average activity $\bar{x}$ as an order parameter, following the approach used in \cite{diaz2021apparently}. The simulation starts at $T_0=0.2 T_-$ from a random initial configuration and, after a given number of steps, the control parameter $T$ is increased by a small $\Delta T$ without resetting the state of the system. Such procedure is repeated up to a final value $T_F=5 T_+$. Ideally, this corresponds to an adiabatic change of the control parameter, allowing the system to access the metastable states in the bistability region. Then, the same procedure is repeated starting from $T_F$ and decreasing $T$ down to $T_0$. In principle, if $\Delta T$ is small enough, in the region where the two equilibria coexist we should find a corresponding hysteresis cycle. Since we want to properly span both the super- and sub-critical regions, and because of the different order of magnitude of the two theoretical thresholds ($T_-\approx 10^{-3}$, $T_+\approx 10^{-1}$), we take $60$ logarithmic steps. In Figure~\ref{fig:p3c2:hysteresis}(a) we plot the behavior of $\bar{x}$ at different steps of this procedure for fully-connected topologies of different sizes. In the super- and sub-critical region, $\bar{x}$ is in accordance with the theoretical predictions Eq.~\eqref{eqn:p3c2:sol_super} and Eq.~\eqref{eqn:p3c2:sol_sub}. In between the theoretical values of $T_\pm$, we recover the discontinuous transition and the hysteresis cycle previously found in \cite{diaz2021apparently,zarepour2019universal}. Perhaps unsurprisingly, for small network sizes the limits of the hysteresis cycle do not precisely match the expected values of $T_+$ and $T_-$ given in Eq.~\eqref{eqn:p3c2:T_sup} and Eq.~\eqref{eqn:p3c2:T_sub}. In fact, due to the finite size of the system, the associated noise contribution causes the bistable region to shrink as the size of the network is reduced.

So far we have considered the mean-field limit only, which corresponds to a fully-connected topology with constant weights. However, the architecture of the brain is usually characterized by sparse connectivity, and brain networks often display a non-trivial topology with small-world properties and community structures, both at the micro- and macro-scale. Moreover, the strength of the interaction between different brain regions is highly heterogeneous and typically follows a scale-free distribution \cite{hagmann2008connectome, bullmore2009complex}. In the rest of this Chapter - at the price of analytical tractability - we will explore the effect of non-trivial topologies, as well as heterogeneous weights, on the dynamics of our model.

We first study the simple case of an Erdős–Rényi network with a given wiring probability $p$ between two nodes and constant weights. We repeat the procedure described above at fixed network size and for different wiring probabilities, to understand how the network topology affects the hysteresis cycle predicted by the mean-field limit. We plot the results in Figure~\ref{fig:p3c2:hysteresis}b. We find that, as we lower the connectivity, the bistable region shrinks until it disappears, giving rise to a smooth transition at low values of $p$. This behavior, which is deeply different from the one expected from the mean-field approximation, is consistent with previous results obtained in the discrete-time model \cite{rocha2022recovery, diaz2021apparently}. As we will see later in this Chapter, such smooth transitions are strengthened by the introduction of empirical connectivity, and they are crucial for the onset of emergent collective oscillations. Eventually, for very low values of $p$, the transition disappears as it becomes impossible for the network to sustain the super-critical regime.

\section{The emergence of collective oscillations}
Neural activity typically exhibits a certain level of stochastic fluctuations, even when the brain is at rest. In fact, a growing amount of evidence suggests that neural noise might enhance the signal processing capabilities of neurons \cite{mcdonnell2011benefits, guo2018functional}, and power spectra have proved to be informative in other neuronal models \cite{bressloff2010stochastic, fanelli2017noise, apicella2021emergence, wallace2011oscillations}. To analytically investigate the oscillatory dynamics of the mean-field model, from Eq.~\eqref{eqn:p3c2:langevin} we perform a linear noise approximation \cite{gardiner2004handbook}
by defining  the local coordinates ($\zeta_1$, $\zeta_2$) as
\begin{equation}
    \begin{cases}
        x(t) = x^* + \frac{\zeta_x(t)}{\sqrt{N}} \\
        y(t) = y^* + \frac{\zeta_y(t)}{\sqrt{N}} 
    \end{cases}
    \Rightarrow \begin{cases}
        \zeta_x(t) = \sqrt{N} (x(t) - x^*)\\
        \zeta_y(t) = \sqrt{N} (y(t) - y^*)
    \end{cases}.
\end{equation}
Then, the power spectrum of the oscillations around a given equilibrium is given by
\begin{equation}
    S_i(\omega) = \langle \tilde{\zeta_i}(\omega) \tilde{\zeta}_i^*(\omega) \rangle = \langle \tilde{\zeta_i}(\omega) \tilde{\zeta}_i(-\omega) \rangle
\end{equation}
for $i = x, y$.

We rewrite the original equations in terms of $(\zeta_x, \zeta_y)$, keeping only the linear terms. For the deterministic part, this leaves only the Jacobian evaluated at the equilibrium $J(x^*, y^*) \equiv J$, whereas the diffusion term needs to be expanded up to $1/\sqrt{N}$. We end up with
\begin{equation}
    \begin{cases}
        \dot{\zeta}_x = J_{11} \zeta_x + J_{12} \zeta_y + \eta_x\\
        \dot{\zeta}_y = J_{21} \zeta_x + J_{22} \zeta_y + \eta_y
    \end{cases}
\end{equation}
where $(\eta_x, \eta_y)$ is a colored noise defined by
\begin{equation}
    \langle \eta_i(t) \rangle = 0 \qquad \langle \eta_i(t) \eta_j(t') \rangle = \delta(t-t') B_{ij}
\end{equation}
where $B_{ij}=B_{ij}(x^*,y^*)$ is the diffusion matrix evaluated at equilibrium.

In Fourier space, we immediately find
\begin{equation}
    \begin{cases}
        i \omega \tilde{\zeta}_x(\omega) = J_{11} \tilde{\zeta}_x + J_{12}\tilde{\zeta}_y + \tilde{\eta}_x\\
        i \omega \tilde{\zeta}_y (\omega) = J_{21} \tilde{\zeta}_x + J_{22} \tilde{\zeta}_y + \tilde{\eta}_y
    \end{cases}
    \label{eqn:p3c2:fourier}
\end{equation}
where the statistics of $(\eta_x, \eta_y)$ is left unchanged, i.e.,
\begin{equation}
    \langle \tilde{\eta}_i(\omega) \rangle = 0 \qquad \langle \tilde{\eta_i}(\omega) \tilde{\eta_j}(\omega') \rangle = \delta(\omega-\omega') B_{ij}.
\end{equation}
The linear system in Eq.~\eqref{eqn:p3c2:fourier} is solved by
\begin{equation}
    \begin{cases}
        \tilde{\zeta_x}(\omega) = \dfrac{(i\omega - J_{22})\tilde{\eta_x} + J_{12}\tilde{\eta_y}}{-\omega^2 -i\omega(J_{11}+J_{22}) + J_{11}J_{22} - J_{12}J_{21}} \\
        \tilde{\zeta_y}(\omega) = \dfrac{(i\omega - J_{11})\tilde{\eta_y} + J_{21}\tilde{\eta_x}}{-\omega^2 -i\omega(J_{11}+J_{22}) + J_{11}J_{22} - J_{12}J_{21}}
    \end{cases}
\end{equation}
so that we can compute the power spectrum,
\begin{equation}
    S_i(\omega) = \langle \tilde{\zeta_i}(\omega) \tilde{\zeta}_i^*(\omega) \rangle = \langle \tilde{\zeta_i}(\omega) \tilde{\zeta}_i(-\omega) \rangle.
\end{equation}
In particular, we are interested in oscillations of the density of active neurons $x$, which are described by
\begin{equation}
    S_x(\omega) = \dfrac{\alpha + \beta \omega^2}{[( \omega^2 - \Omega_0^2)^2 + \Gamma^2\omega^2]}
\end{equation}
where the coefficients
\begin{equation}
    \begin{cases}
        \alpha = B_{11}J_{22}^2 - 2B_{12}J_{12}J_{22} + B_{22}J_{12}^2 \\
        \beta = B_{11} \\
        \Omega_0^2 = J_{11}J_{22}-  J_{12}J_{21} \\
        \Gamma^2 = (J_{11} + J_{22})^2
    \end{cases}
\end{equation}
can be readily evaluated in either of the two equilibria. 

\begin{figure*}[t]
    \centering
    \includegraphics[width=\textwidth]{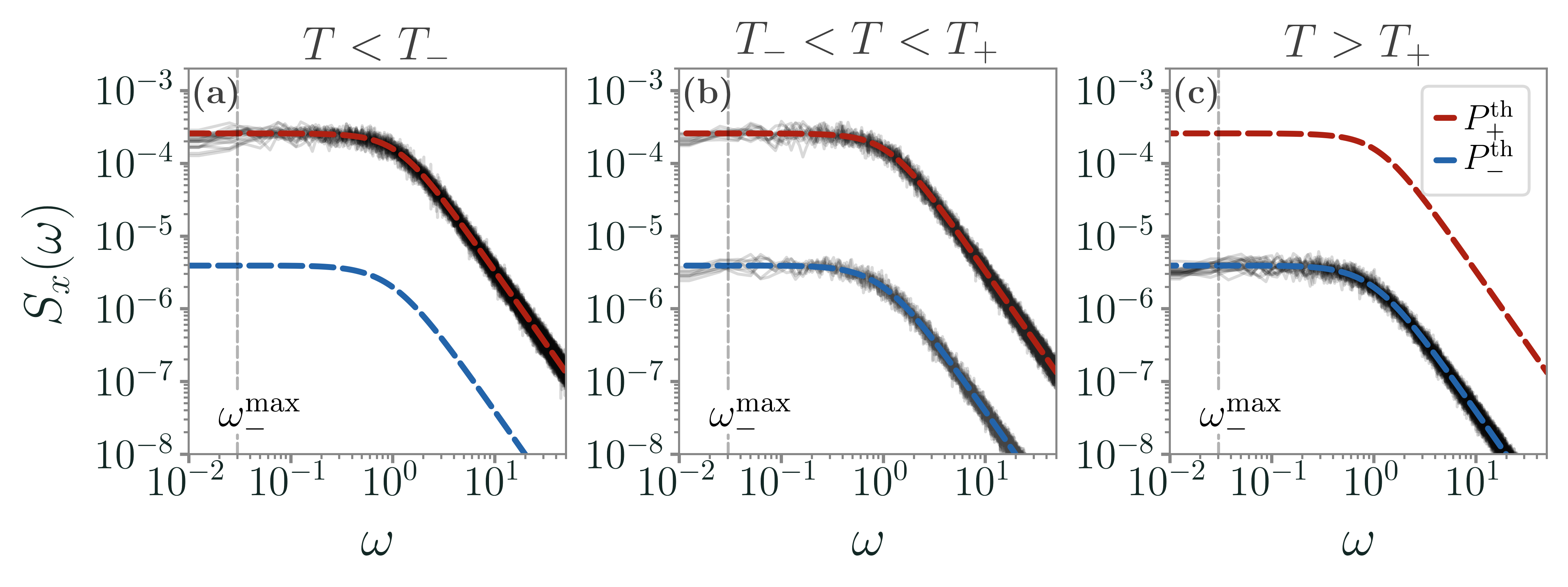}
    \caption{Theoretical expressions of the power spectrum are well-matched by simulated data from a fully-connected network at stationarity. The gray lines represent the power spectrum obtained by simulating the continuous-time model for $10^5$ steps at stationarity in a network of $N=10^3$ nodes and after an initial transient of $5\cdot10^4$ steps. (a) For $T< T_-$ the power spectrum shows a small peak at $\omega_-^\mathrm{max}$. (b) In the bistable region, depending on which equilibria the dynamics settles, we can find both power spectra. (c) For $T>T_+$, no peak emerges in the power spectrum}
    \label{fig:p3c2:power_spectrum}
\end{figure*}

Overall, we find that the power spectrum in the super-critical regime is given by
\begin{equation}
    S_x^+(\omega) = \dfrac{2 r_2 [1 + r_2 +  r_2^2 + \omega^2]}{(1+2r_2)[(1+2r_2)^2+(2+r_2^2)\omega^2+\omega^4]},
    \label{eqn:p3c2:p_sup}
\end{equation}
and in the sub-critical regime we have
\begin{equation}
    S_x^-(\omega) = \dfrac{ 2 r_1 r_2 [r_1^2 + r_1r_2 + r_2^2  + \omega^2] }{(r_1+r_2+r_1r_2)[(r_1+r_2+r_1r_2)^2 + (1+r_1^2+r_2^2)\omega^2 + \omega^4]}.
    \label{eqn:p3c2:p_sub}
\end{equation}
Power spectra obtained from the simulation of the model are perfectly matched by these theoretical expressions, as we see in Figure~\ref{fig:p3c2:power_spectrum}. Importantly, Eq.~\eqref{eqn:p3c2:p_sup} and Eq.~\eqref{eqn:p3c2:p_sub} show that, in both regimes, for low frequencies the power spectrum is flat. On the other hand, in the large frequency limit, we find Brownian noise, i.e., $S(\omega) \approx \omega^{-2}$. Such scale-free behavior of the frequencies' spectrum is found, for instance, in Local Field Potentials, i.e., the electrical activity of the brain measured with single microelectrodes \cite{milstein2009neuronal}. Most importantly, in the super-critical regime the power spectrum does not display any peak. A small peak at
\begin{equation}
    \omega_-^\mathrm{max} = [(1 + r_1 r_2) (r_1 r_2)^{1/2} - r_1^2 - r_2^2 - r_1 r_2 ]^{1/2}
\end{equation}
emerges instead in the sub-critical phase for the range of parameters in which $\omega_-^\mathrm{max}$ exists, i.e., the radical is non-negative. These results suggest that in the mean-field limit of the model stochastic amplification alone is not sufficient to induce significant sustained collective oscillations.

\subsection{Oscillations at criticality in the human connectome}
\begin{figure}[t]
    \centering
    \includegraphics[width=\textwidth]{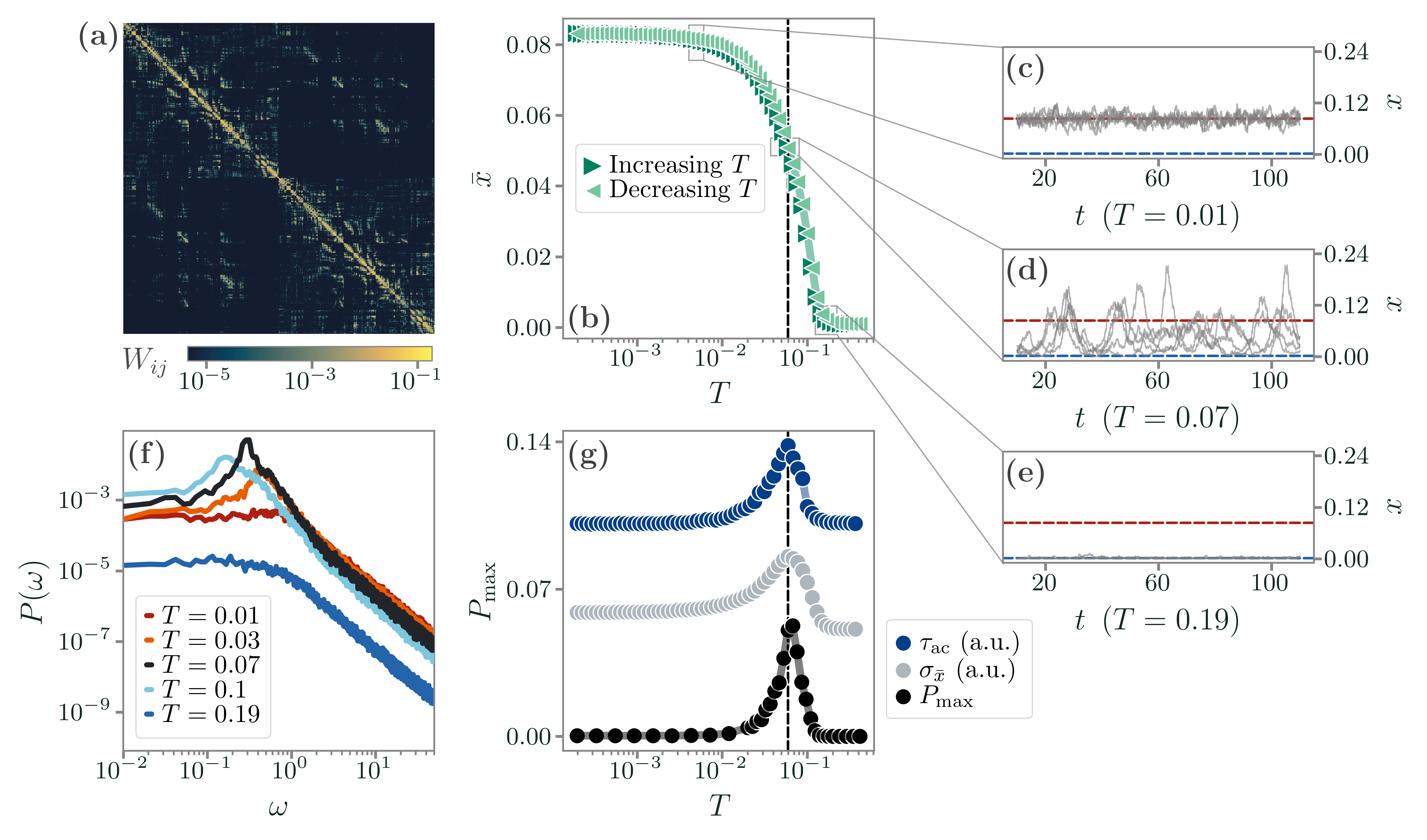}
    \caption{Dynamics of the model over an empirical connectome shows the emergence of a critical-like transition and collective oscillations. (a) Plot of the connectome from \cite{hagmann2008connectome}. (b) The average activity $\bar{x}$ does not show any hysteresis cycle but rather changes smoothly from $x_+$ to $x_-$ as $T$ increases. (c-e) Examples of trajectories at different values of $T$. In particular, at intermediate values of $T$, the trajectories show high variability and a rich dynamics. (f) The power spectrum at different values of $T$. As $\bar{x}$ smoothly changes between the two equilibria, collective oscillations emerge. Notice that the frequency peak is at a frequency higher than $\omega_-^{max}$. (g) The peak of the power-spectrum $P_\mathrm{max}$ is maximal at intermediate values of $T$. At the same point, both the autocorrelation time $\tau_\mathrm{ac}$ and the variance $\sigma_{\bar{x}}$ of $\bar{x}$ peak (shown in arbitrary units), suggesting that a critical-like transition might be present}
    \label{fig:p3c2:connectome_results}
\end{figure}

We now consider an empirical connectome of the human cerebral cortex with $N= 998$ regions \cite{hagmann2008connectome}. In this case, we have both a complex topology and a non-trivial distribution of weights, as we see in Figure~\ref{fig:p3c2:connectome_results}a. Quite surprisingly, numerical simulations show that the analytical expressions of the two equilibria are still valid in the limit of small and large values of $T$. However, for intermediate values of the control parameter the average activity is no longer bounded to the two equilibria, but rather changes continuously from one to the other, as we see in Figure~\ref{fig:p3c2:connectome_results}b.

In Figure~\ref{fig:p3c2:connectome_results}c-d-e we plot the model's trajectories for different values of $T$. We clearly see that, at intermediate values of $T$, the bistability is not present anymore - signaling that the transition is not discontinuous anymore, but the average activity rather changes smoothly from the limiting values represented by the mean-field equilibria. Furthermore, the density of excited neurons $x$ displays a high variability during its temporal evolution - yet, notice that during peaks of activity we find at most $x \approx 0.25$, suggesting that localized regions of the network are activated together, and activity does not spread easily to the entire network. As a consequence, as we see in Figure~\ref{fig:p3c2:connectome_results}f, the power spectrum now displays a peak $P_\mathrm{max}$ - i.e., collective oscillations emerge. If we look at the value of $P_\mathrm{max}$ as a function of $T$ in Figure~\ref{fig:p3c2:connectome_results}g, we find a pronounced maximum at an intermediate value $T:+ T_c$ of the threshold, where the average activity $\bar{x}$ is in between the equilibria $x_\pm$. Hence, the topological complexity of the underlying connectome allows for a non-trivial evolution of activity.

In Figure~\ref{fig:p3c2:connectome_results}g we also show that at $T_C$ the variance of $\sigma_{\bar{x}}$ peaks as well. This is perhaps unsurprising, since oscillations are related to variability of the activity. However, and crucially, such activity shows long-range temporal correlations as well. Indeed, we can compute its autocorrelation time $\tau_\mathrm{ac}$ as the characteristic decay time of the autocorrelation function's exponential envelope. Once more, as a function of the threshold $T$ we find that $\tau_\mathrm{ac}$ displays a peak at $T \approx T_c$.

All together, these features are reminiscent of those found in finite-size systems close to a second-order phase transition \cite{ma2018critical, marro1999nonequilibrium}, suggesting that they may emerge from a critical point of the control parameter $T$. Hence, the transition observed in the presence of the empirical connectome is closer to a critical transition rather than the bistability predicted by the mean-field limit. Let us stress that these features are emerging at the dynamical level, contrary to the percolation transition originally studied by Haimovici and collaborators \cite{haimovici2013connectome}. In general, we find that this dynamical transition does not occur at the same value of $T$ of the percolation transition, as observed in other models \cite{martin2020intermittent}.

\begin{figure}[t]
    \centering
    \includegraphics[width=\textwidth]{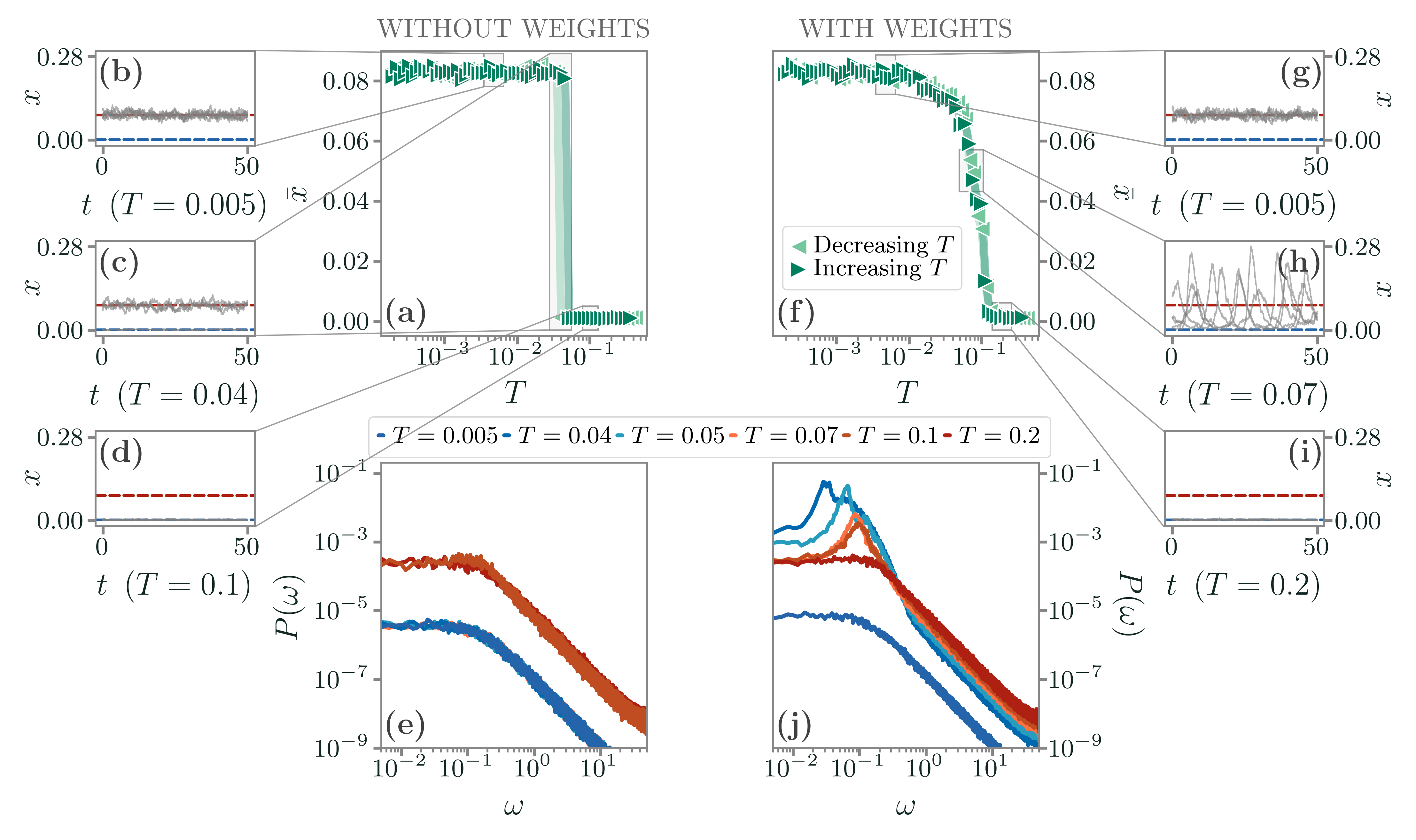}
    \caption{Dynamics of the model over an Erdős–Rényi network with wiring probability $p_\mathrm{conn} \approx 0.08$, without and with weights following the empirical distribution of the connectome. (a-d) Without the weights, the transition is discontinuous with a bistable region, as we can see from both the hysteresis cycle and the trajectories. (e) No oscillations are present, and the power spectrum follows the mean-field prediction. (f-i) In the presence of the empirical weights, the transition becomes instead continuous. The trajectories in the region where the average activity $\bar{x}$ changes smoothly between $x_\pm$ display large departures from the expected equilibria. (j) With the empirical weights, oscillations are present}
    \label{fig:p3c2:null_model}
\end{figure}

\subsection{Erdős–Rényi networks and continuous transitions}
In order to understand the relevance of the non-trivial topology of the connectome, we consider the Erdős–Rényi network previously studied. As a simple null model, we set the wiring probability to be equal to that of the connectome, $p_\mathrm{conn} \approx 0.08$. Furthermore, we take into account networks both with and without weights re-sampled from the weight distribution of the empirical connectome. In principle, these choices amount to trying to understand whether sparsity and weight heterogeneity are sufficient features to reproduce the complex dynamical evolution of the model in the connectome.

As expected from Figure~\ref{fig:p3c2:hysteresis}, without resampled weights the wiring probability $p_\mathrm{conn}$ is high enough that the transition is discontinuous and a bistable region still exists. Indeed, in Figure~\ref{fig:p3c2:null_model}a-e, we see that in this scenario the null model matches the behavior of the mean-field limit. No peak in the power spectrum emerges, and the stationary dynamics always reaches one of the two equilibria $x_\pm$. Hence, sparsity alone is not enough to reproduce high activity variance, local collective oscillations nor long autocorrelation times.

\begin{figure}[t]
    \centering
    \includegraphics[width=0.7\textwidth]{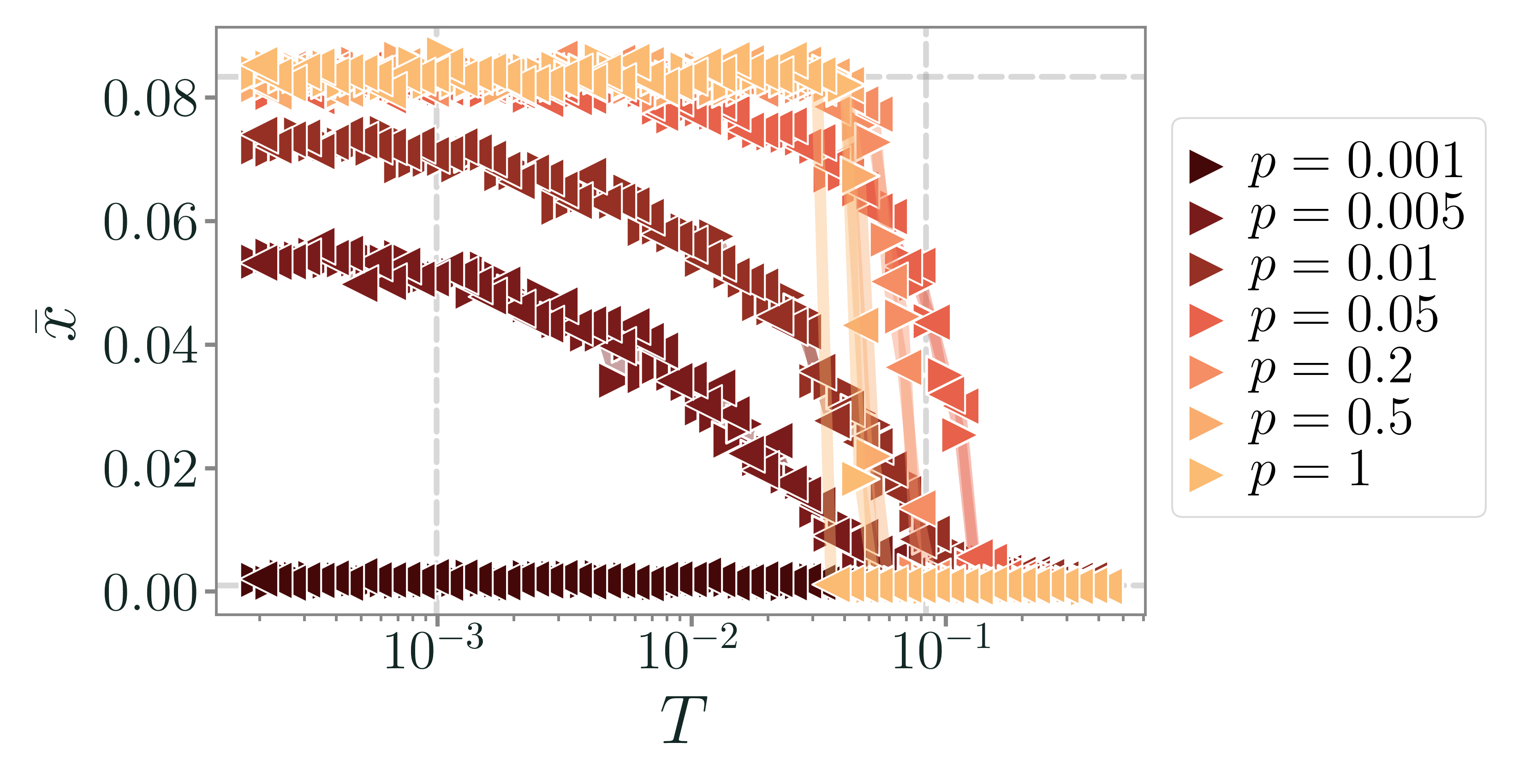}
    \caption{Average activity $\bar{x}$ in Erdős–Rényi networks at different wiring probabilities $p$, with weights re-sampled from the empirical connectome. With weights, the transition becomes continuous already at higher $p$}
    \label{fig:p3c2:hysteresis_ER_weights}
\end{figure}

These results change dramatically when we add back the weights from the empirical connectome. In Figure~\ref{fig:p3c2:null_model}f we see that no hysteresis cycle emerges, and the transition is now continuous much like in the case of the connectome. The trajectories in Figure~\ref{fig:p3c2:null_model}g-h now display large transient periods of activity stemming from the silent state $x_-$, a feature typical of excitable or non-normal models \cite{benayoun2010avalanches,disanto2018nonnormality}. These sustained activations lead to a clearly oscillating behavior. Indeed, the power spectrum in Figure~\ref{fig:p3c2:null_model}j displays a clear peak, as we have previously shown for the connectome. That is, the presence of the empirical weights - together with sparsity - helps the disruption of the bistable region predicted at the mean-field level.

Importantly, such a disruption emerges only if the wiring probability $p$ of the Erdős–Rényi network is low enough. In Figure~\ref{fig:p3c2:hysteresis_ER_weights} we show the average activity obtained while slowly varying $T$, as before, in Erdős–Rényi networks at different $p$ with weights resampled from the empirical connectome. We immediately note that in the fully-connected case $p=1$, even with weights, a hysteresis cycle is still present - hence, the transition is still discontinuous. However, the transition becomes smoother already at higher values of $p$, showing that both sparsity and weights aid in the disruption of the bistability predicted by the mean-field approximation. Once more, at low wiring probabilities the network cannot sustain activity anymore. These results may suggest that a continuous, critical-like dynamical transition with global oscillations emerges if the underlying network is either extremely sparse - as in Figure~\ref{fig:p3c2:hysteresis} and Figure~\ref{fig:p3c2:hysteresis_ER_weights} - or at higher values of $p$, but with a heterogeneous weight distribution. Crucially, empirical connectomes are often characterized by such features. Yet, we may ask whether sparsity and weight heterogeneity are enough to reproduce the dynamical features observed in the connectome.

\begin{figure}[t]
    \centering
    \includegraphics[width=\textwidth]{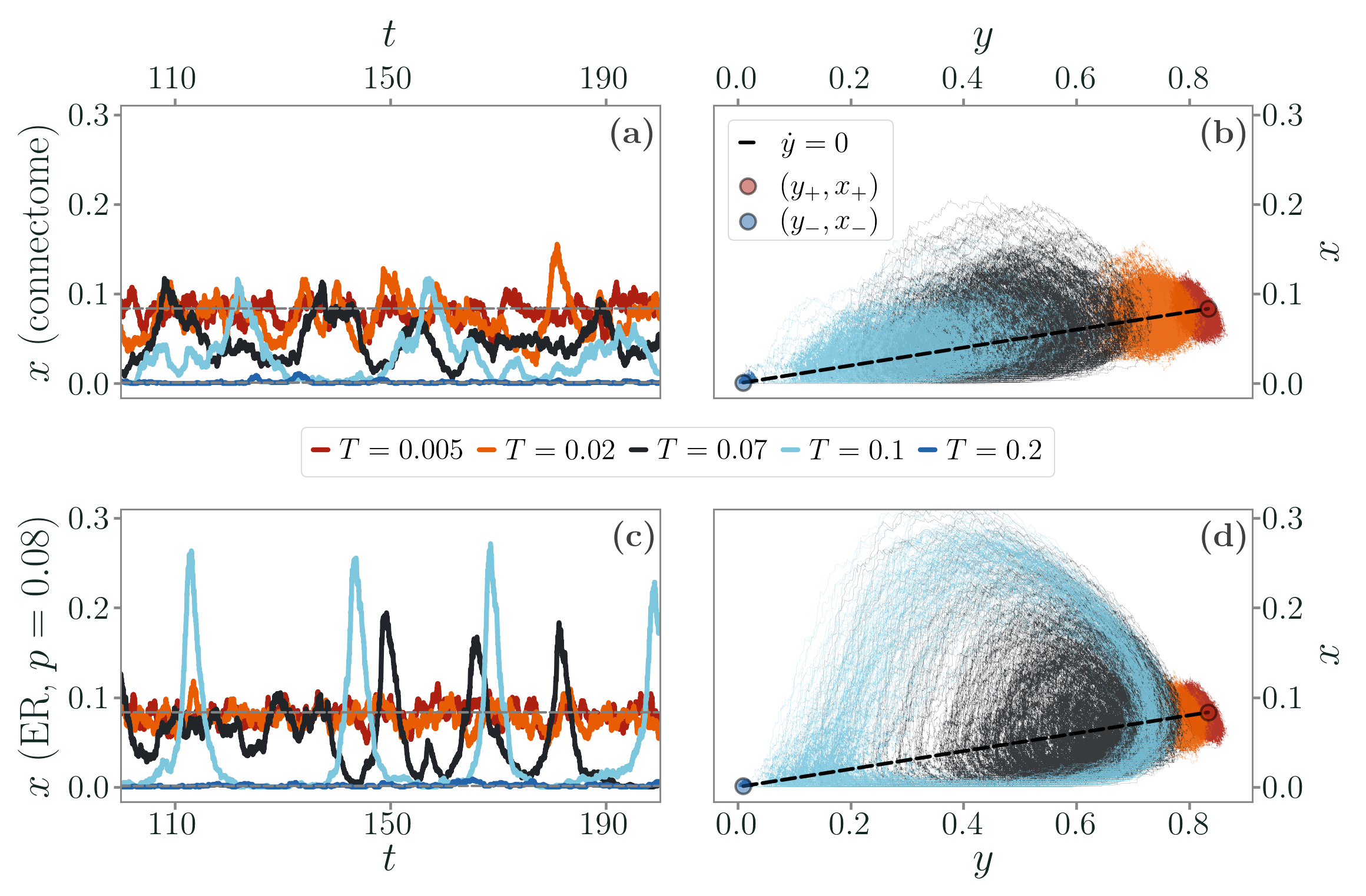}
    \caption{Comparison of the trajectories in the empirical connectome and in the Erdős–Rényi with $p = p_\mathrm{conn}$ and empirical weights. (a) Connectome trajectories of $x$ as a function of time and (b) in the phase space for different $T$. The black dashed line in panel (b) is the nullcline of the refractory population $y$, and the gray lines in panel (a) are $x_\pm$. (c-d) Same, but in the case of the null model. Notice how the trajectories in the phase space are qualitatively different, and oscillations emerge as large activations from the sub-critical equilibrium $(y_-, x_-)$}
    \label{fig:p3c2:phase_space}
\end{figure}

To address this question, at least from a heuristic point of view, we compare the phase space trajectories of the empirical connectome and of the null random network model with weights, see Figure~\ref{fig:p3c2:phase_space}. In the empirical connectome, the dynamics is typically richer and oscillates close to the nullcline $\dot{y}=0$ predicted by the mean-field Eq.~\eqref{eqn:p3c2:deterministic}. Such oscillations appear to be around values of $(x,y)$ that are different from the mean-field equilibria, strengthening the hypothesis that the equilibrium at intermediate values of $T$ changes continuously. On the other hand, in the Erdős–Rényi case, the trajectories display large transient dynamics away from the $(y_-,x_-)$ equilibrium, reminiscent of noise-induced oscillations \cite{apicella2021emergence, bressloff2010stochastic, fanelli2017noise, wallace2011oscillations} or non-normal systems \cite{benayoun2010avalanches,disanto2018nonnormality}. This suggests that, although in both models we find emerging oscillations, the underlying dynamical features might be different. Notably, these phenomena have usually been observed in models with excitatory and inhibitory populations. Here, we rather have a single excitatory population with a refractory state, hinting that the two different scenarios may lead to a similar phenomenology. Overall, further work is needed to explore the role of higher-order structures in the empirical connectome - e.g., modularity \cite{villegas2014frustrated, odor2019critical} or heterogeneity in the degree distribution \cite{pastor2001complex} - and their effect on the model dynamics.

Let us note that the small size of the empirical connectome considered here may be a limitation for these investigations, since finite-size corrections may be hiding criticality. Notably, in \cite{odor2021modelling} a similar modification of the discrete-time Greenberg \& Hastings model run on a large-scale connectome displays semicritical behaviors consistent with a Griffith phase in a certain range of the control parameter. Such use of synthetic connectomes overcomes the finite size issue, at the cost of relying on some subjective assumptions about the generated topologies. Hence, future works should be devoted to fully understanding whether the observed continuous transition is associated with a real critical point or - most likely - with other phenomena such as rare region effects \cite{vojta2006rare} or noise-induced transitions \cite{wallace2011oscillations}. Possible approaches may include the use of heterogeneous mean-field methods as done in the study of epidemic spreading \cite{pastor2001complex} or annealed network approximations \cite{dorogovtsev2008critical}.

In fact, this is still very much an open question. Previous efforts focused on the emerging percolation transition in the model, discussing the effect of the topology in shaping the transition by means of in-silico \cite{haimovici2016dynamical, zarepour2019universal} and empirical connectomes \cite{rocha2018homeostatic, rocha2022recovery}. Here, we instead focus on the dynamical transition that arises in this model \cite{diaz2021apparently}. To the best of our knowledge, this is the first attempt to investigate the nature and the consequences of this dynamical transition from an analytical perspective. This approach allowed us to probe the complex interplay between dynamical, and possibly critical-like, features and the underlying topology - which is often of great importance and, at the same time, poorly understood.

\chapter{Scaling and criticality from a phenomenological renormalization group}
\chaptermark{A phenomenological renormalization group}
\label{ch:PRR}
\lettrine{T}{he idea} that living systems may be poised at criticality is a fascinating hypothesis \cite{mora2011biological, hidalgo2014information, munoz2018colloquium}, and in recent years it has been explored in a vast variety of areas \cite{kinouchi2006optimal, mora2010maxent, beggs2003avalanches, cavagna2010scalefree}.

Tools from Statistical Mechanics, such as the Renormalization Group \cite{ma2018critical, binney1992, henkel2009}, teach us that at criticality the macroscopic, collective behavior of the system is described by a few relevant features, such as the embedding dimension of the system and its symmetries, while most of the microscopic details become irrelevant. At the critical point, the physical properties are determined by a non-trivial fixed point in the space of all the possible models compatible with the underlying symmetry. However, in the broad landscape of natural systems, one often has to deal directly with data without an explicit model, and real-world systems are typically finite - so that most of the time it is hard to come up with a definitive answer about whether they are poised near a critical point \cite{munoz2018colloquium}.

In this scenario, the critical brain hypothesis plays a relevant role. Since the first observations of neuronal avalanches during spontaneous activity \cite{beggs2003avalanches}, many different universality classes for the putative transition have been proposed - from a mean-field branching process \cite{beggs2003avalanches, niebur2014criticality} to synchronization transitions \cite{disanto2018landau, dallaporta2019avalanches, poil2012critical, buendia2021hybrid, buendia2020selforganized, buendia2022broad}. However, measuring power-law exponents from - possibly subsampled and noisy \cite{levina2022tackling} - data often leads to spurious results, hence the question of the nature of the critical transition emerging in brain dynamics, if any, remains very much open. At this point, it is well-known that power-law scaling may stem from plausible yet non-critical models, e.g. neutral \cite{martinello2017neutral} or randomly driven ones \cite{touboul2010avalanches, priesemann2018can, mariani2022disentangling} as we have shown in Chapter \ref{ch:scirep}.

Recently, a phenomenological coarse-graining procedure was introduced in \cite{meshulam2019coarsegraining, meshulam2018arxiv} to deal in particular with neural timeseries in a model-free setting. In particular, as we will see, the core idea of this phenomenological Renormalization Group (PRG) is to deal in principle with unknown and possibly long-range interactions, by avoiding coarse-graining directly in real space. In \cite{meshulam2019coarsegraining, meshulam2018arxiv}, this PRG was applied to data from calcium imaging in the hippocampus of a mouse running along a virtual track. Ideally, if the system was close to a critical point, appropriate physical quantities at different levels of coarse-graining should display power-law scaling. Furthermore, in a critical system, we expect probability distributions to converge to a non-trivial fixed point of the renormalization flow, which describes the system's details-independent macroscopic properties \cite{binney1992}. The authors do find that the results on neural activity seem to point to the presence of both scaling and fixed points, suggesting that the underlying dynamics might be critical. Indeed, the brain is probably one of the most impressively complex systems we are able to study and the idea that the collective behavior of neurons might emerge from a self-organized critical state has been widely studied for several years.

In this Chapter, we aim to test this phenomenological Renormalization Group in well-known equilibrium and non-equilibrium models. In particular, we will focus on the contact process and thus the directed percolation universality class \cite{marro1999nonequilibrium, henkel2009}. In fact, absorbing phase transitions have long been proposed as an archetypal model for the brain, where avalanches in spontaneous activity appear as activity events over a possibly silent background. The advantage of the contact process here is that its critical behavior is well understood and the exponents are known from numerical studies, so we shall regard it as a ``control case'' to investigate the ability of this PRG to extract relevant information and infer signatures of a critical state in out-of-equilibrium systems. Further, we consider different topologies, to test whether long-range interactions in a, e.g., small-world network affect the results. Along the road, we will also study these coarse-graining ideas in the Ising model, as well as in models of conditionally independent variables in a stochastic environment - which, as we have shown throughout this Thesis, often display non-trivial and unexpected features.

\section{A phenomenological renormalization group for neural activity}
In the simplest models studied in Statistical Mechanics, interactions are usually short-range. Hence if we want to recover the macroscopic, collective behavior it makes sense to think of the coarse-graining procedure as an average over the short distance, microscopic details. - e.g., with a block-spin transformation \cite{binney1992}.

Yet, in full generality, if we start with a set of variables $\{\sigma_i\}$ we can think of a coarse-graining transformation as a generic transformation that maps $\{\sigma_i\}$ into a new set of variables $\tilde{\sigma}_{{i'}}$. For instance, spatial coarse-graining amounts to choosing a transformation of the form
\begin{align}
\label{eqn:p3c3:cg_real_space}
\sigma_i \to \tilde{\sigma}_{{i'}} = f\left(\sum_{j \in \mathcal{N}_i}\sigma_j\right)
\end{align}
where $f$ is a generic function, $\mathcal{N}_i$ denotes the neighborhood of the variable $i$, and ${i'}$ indexes the new coarse-grained variables. For instance, a block-spin transformation can be obtained by considering the function $f(\cdot)$ as the identity. Crucially, probability distributions change as well under a coarse-graining transformation, namely
\begin{align*}
P(\{\sigma_i\}) \to \tilde{P}(\{\tilde{\sigma}_{i'}\}).
\end{align*}
In the case of Boltzmann distributions, 
\begin{align*}
P(\{\sigma_i\}) = \frac{1}{Z} e^{-\beta H(\{\sigma_i\})}
\end{align*}
we ask the RG transformation to leave the partition function $Z$ invariant and let the Hamiltonian transform. After the coarse-graining step we find a new Hamiltonian $\tilde{H}(\{\tilde\sigma_i\})$ for the variables $\{\tilde\sigma_i\}$. Consequently, we need to properly rescale the scale of the system $i' = g({i})$ and renormalize the variables $\tilde{\sigma}_{{i'}} = h(\sigma_i)$ so that the new system has the same physical properties as the original one\footnote{To find the new, effective Hamiltonian that describes the coarse-grained variable exactly is generally hard and one typically has to rely on perturbation theory around a Gaussian Hamiltonian. Moreover, it is not always trivial to define a coarse-graining step that preserves exactly the features of the system, such as its symmetries.}.

It is important to stress that the RG procedure does not necessarily rely on Boltzmann distributions. For instance, if we consider a sequence of independent and identically distributed (iid) random variables $\{X_i\}$ and let $p(x)$ be the probability density of $X$, we may seek the distribution for the sum
\begin{align*}
X^{(n)}_i = \frac{X^{(n-1)}_{2i-1}+X^{(n-1)}_{2i}}{\zeta}
\end{align*}
which is the $n$-th step of a decimation process - and can be thought of as a coarse-graining step - and where $\zeta$ is, in general, a renormalization factor. The pdf transforms as
\begin{align*}
p_{n+1}(x) & = \int dx_1 dx_2 \, p_{n}(x_1)p_{n}(x_2) \delta\left(x-\frac{x_1+x_2}{\zeta}\right) \\
& = \zeta \int dx' \, p_{n}(x')p_{n}(\zeta x-x') \\
& = (\mathcal{R}_\zeta p_n)(x)
\end{align*}
and $\mathcal{R}_\zeta$ is the RG transformation. By expressing it in terms of characteristic functions,
\begin{align*}
\varphi(z) = \int dx \, p(x) e^{ixz} = 1 + \sum_{k=1}^\infty\frac{i^k}{k!}\ev{X^k}z^k,
\end{align*}
one immediately finds that $\mathcal{R}_\zeta$ acts in a simpler local way,
\begin{align*}
(\mathcal{R}_\zeta \varphi_n)(z) = \varphi_n^2\left(\frac{z}{\zeta}\right).
\end{align*}
A further simplification comes if we consider the cumulants $\omega(z) = \log\varphi(z)$, so that
\begin{align}
\label{eqn:p3c3:rg_cumulants}
(\mathcal{R}_\zeta \omega_n)(z) = 2\omega_n\left(\frac{z}{\zeta}\right).
\end{align}
Since we are interested in the asymptotic distribution of coarse-grained variables, of $\{X_i\}$, we look for the fixed point
\begin{align*}
\omega^*(z) = i\omega_{(1)} z-\frac{1}{2}\omega_{(2)} z^2 + \dots = 2\omega^*\left(\frac{z}{\zeta}\right) = i\omega_{(1)} \frac{2z}{\zeta}-\omega_{(2)} \left(\frac{z}{\zeta}\right)^2 + \dots
\end{align*}
where $\omega_{(k)}$ is the $k$-th cumulant of the distribution. For this relation to hold at all orders we need
\begin{align*}
\zeta = 2, \quad \omega_{(k)} = 0 \qquad \forall k \ge 2
\end{align*}
which implies that the fixed point is $\varphi^*(z) = e^{i\omega_1 z}$. Since $\omega_1 = \ev{X}$ the fixed point corresponds to the pdf
\begin{align}
\label{eqn:p3c3:lln}
p^*(x) = \delta\left(x-\ev{X}\right)
\end{align}
which is nothing but the law of large numbers (LLN) - and indeed we can think of this distribution as that of a normalized sum of iid variables.

A linear stability analysis for $\omega(z) = \omega^*(z) + \delta\omega(z)$ for small $\delta\omega$ gives
\begin{align*}
(\mathcal{R}_2u_{(k)})(z) = 2^{1-k}u_{(k)}(z)
\end{align*}
whose eigenvalues are $\lambda_k = 2^{1-k}$, with corresponding eigenvectors $u_{(k)}(z)$. Thus, there cannot be relevant perturbations for $k \ge 1$ - the law of large number is a stable fixed point in the space of probability distributions.

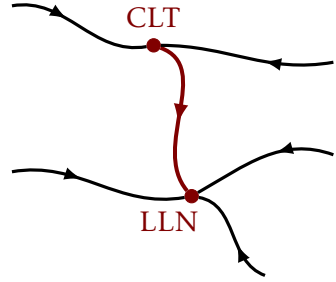
\begin{figure*}
\usetikzlibrary{decorations.markings}
\usetikzlibrary{calc}
\tikzset{->-/.style={decoration={markings,mark=at position #1 with {\arrow{latex}}},postaction={decorate}}}
\centering
\begin{tikzpicture}
    \node[label={[xshift=-0.3cm, yshift=-0.7cm]$\textcolor{mdtRed}{\mathrm{LLN}}$}, fill, draw, circle, minimum size=5, inner sep=1, color = mdtRed] (LLN) at (4,1) {};
    \node[] (empty) at ($(LLN) + (-2.5,0.3)$) {};
    \node[] (empty2) at ($(LLN) + (2,0.5)$) {};
    \node[] (empty5) at ($(LLN) + (1.1, -1.1)$) {};
    \node[label=above:$\textcolor{mdtRed}{\mathrm{CLT}}$, fill, draw, circle, minimum size=5, inner sep=1, color = mdtRed] (CLT) at ($(LLN) + (-0.5,2)$) {};
    \node[] (empty3) at ($(CLT) + (-2, 0.5)$) {};
    \node[] (empty4) at ($(CLT) + (2.5,-0.2)$) {};

    \draw[->-=.55, color = mdtRed, line width = 1.5] (CLT) to [out=-20, in=130] (LLN);

    \path[draw, ->-=.4, line width = 1.2] (empty) to[out=10, in=190] (LLN);
    \path[draw, ->-=.4, line width = 1.2] (empty2) to[out=160, in=30] (LLN);
    \path[draw, ->-=.4, line width = 1.2] (empty5) to[out=160, in=-10] (LLN);
    
    \path[draw, ->-=.4, line width = 1.2] (empty4) to[out=190, in=0] (CLT);
    \path[draw, ->-=.4, line width = 1.2] (empty3) to[out=10, in=190] (CLT);
\end{tikzpicture}
    \caption{Renormalization Group flow in the space of probability distributions, where the central limit theorem (CLT) fixed point is unstable and the law of large number (LLN) one is stable}
    \label{fig:rg_pdf}
\end{figure*}

What happens if the probability distribution is such that $\ev{X} = \omega_{(1)} = 0$? In such case, upon expanding Eq.~\eqref{eqn:p3c3:rg_cumulants} the first non-trivial order is the second one, which gives
\begin{align*}
\zeta = \sqrt{2}, \quad \omega_{(k)} = 0 \qquad \forall k \ge 3.
\end{align*}
The fixed point is described by $\omega^*(z) = -1/2 \,\omega_{(2)}z^2$, where $\omega_{(2)} = \text{var}(X) = \sigma^2$, leading to the Gaussian distribution
\begin{align*}
p^*(x) = \frac{1}{\sqrt{2\pi\sigma^2}}e^{-\frac{x^2}{2\sigma^2}}.
\end{align*}
That is, with the proper renormalization a fixed point still exists and we recover the central limit theorem (CLT) for centered distributions. However, the eigenvalues of the RG transformation now are
\begin{align*}
\lambda_k = 2^{1-k/2}
\end{align*}
so the flow is quite different from before. Now $k=1$ gives a relevant perturbation, and this is expected since $\lambda_1$ is the eigenvalue associated with the first cumulant - if the mean is not exactly zero we go back to the law of large numbers. Hence a perturbation that changes the first moment flows away from the central limit theorem fixed point, which is thus unstable as sketched in Figure \ref{fig:rg_pdf}\footnote{It is worth noting that even for independent random variables the central limit theorem is not a fixed point if the first two moments of $p(x)$ do not exist, giving rise to a larger class of stable distributions and the so-called generalized central limit theorems.}.

These simple examples show that the RG flow can be thought of as a flow in the space of probability distributions. In general, if the variables are weakly correlated, the coarse-graining drives the joint distribution towards a Gaussian distribution, which is a fixed point of the transformation \cite{jona2001renormalization}. Yet, one should be careful that this picture is somewhat different than that of Statistical Mechanics, where fixed points lie in the critical surface and where a Gaussian fixed point emerges in critical systems embedded in a dimension $d$ higher than their upper critical dimension $d_u$ - i.e., where mean field theory applies. Instead, let us stress out that in this probabilistic view the Gaussian fixed point is trivially a consequence of the central limit theorem, hence it appears for independent or weakly correlated variables, away from criticality. In a critical system, on the other hand, variables are strongly correlated and we might expect a non-trivial fixed point where the central limit theorem does not hold.

Let us now describe the coarse-graining procedure introduced in \cite{meshulam2018arxiv, meshulam2019coarsegraining}. The PRG procedure can be declined in two different ways - one that mimics real-space coarse-graining by building clusters of maximally correlated variables, and the other that draws inspiration from momentum-space renormalization and it is formally related to principal component analysis (PCA). In both cases, the idea that fixed points of the RG flow can be interpreted as fixed points of probability distributions will prove to be pivotal concepts, that go beyond the power-law scaling one expects by the control of such fixed points. As we will see, these ideas are well suited for the case of neural activity.

\subsection{Real-space-inspired approach}
\begin{figure}[t]
    \centering
    \includegraphics[width = 1\textwidth]{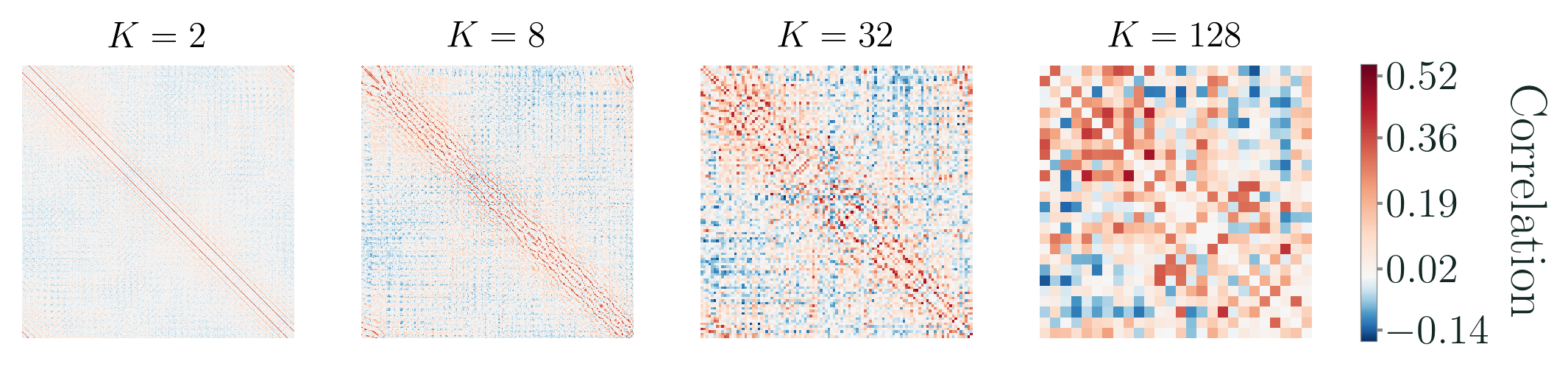}
    \caption{Correlation matrices at different cluster sizes $K$, from a critical contact process in a two-dimensional lattice. Notice how the global structure of the matrix is preserved at different coarse-graining steps since we cluster together maximally correlated pairs}
    \label{fig:p3c3:correlations}
\end{figure}

We consider a system of $N$ random variables whose interaction network is unknown. Denoting state variables of the neurons with $\sigma_i^{(1)}$ for $i=1,...,N$, where the superscript $1$ denotes that we are at the first coarse-graining step, we consider the maximal non-diagonal element of the normalized correlation matrix
\begin{align*}
c_{ij} = \frac{C_{ij}}{\sqrt{C_{ii}C_{jj}}},
\end{align*}
where $C_{ij}$ is the covariance matrix
\begin{align*}
C_{ij} = \ev{\sigma_i^{(1)}\sigma_j^{(1)}} - \ev{\sigma_i^{(1)}}\ev{\sigma_j^{(1)}},
\end{align*}
and $\langle\cdot\rangle$ represents the average over the timeseries of neural activity. We remove this pair of maximally correlated variables $(i, j_*(i))$ and repeat the procedure until no other pairs remain. The coarse-grained variables are then defined as
\begin{align*}
\sigma_{i'}^{(2)} = \sigma_i^{(1)} + \sigma_{j_*(i)}^{(1)} 
\end{align*}
where $i' = 1, \dots, N/2$. We iterate this process, producing clusters of $K=1, 2, 4, \dots, 2^{k-1}$ variables. Each one defines a new variable $\sigma_i^{(k)}$ as the summed activity of cluster $i$. An example of the correlation matrices at different levels of coarse-graining is shown in Figure~\ref{fig:p3c3:correlations}.

Under this coarse-graining procedure, in \cite{meshulam2018arxiv, meshulam2019coarsegraining} the behaviors of various quantities are analyzed in order to make some parallels with the behavior of critical systems. In particular, they consider the mean variance of the neural activity, the distribution of individual coarse-grained variables, the spectrum of the covariance matrix, and the mean autocorrelation function.

The mean variance of the activity over the time series is defined as
\begin{align}
\label{eqn:p3c3:cg_variance}
M_2(K) = \frac{1}{N_k} \sum_{i=1}^{N_k} \left[\left\langle \left(\sigma_i^{(k)}\right)^2 \right\rangle - \left\langle\sigma_i^{(k)}\right\rangle^2\right]
\end{align}
where $K=2^{k-1}$, $N_k$ is the number of variables after $k$ steps of the coarse-graining procedure and $N_{k+1}=N_k/2$. If the variables are independent one would obtain a variance scaling as $M_2(K) \propto K^{\tilde\alpha}$ with $\tilde\alpha = 1$. In particular, the activity of real neurons shows a scaling of the form
\begin{align*}
M_2(K) \propto K^{\tilde\alpha}
\end{align*}
with $\tilde\alpha \approx 1.4$.

As outlined before, the probability distributions over a single coarse-grained variable may be particularly relevant in determining the coarse-grained behavior of the system. Since a coarse-grained variable $\sigma_i^{(k)}$ is vanishing if and only if all the corresponding $2^{k-1}$ raw variables that determine its value are zero, we can split the probability distribution into a contribution of the silent variables and one of the active variables. Hence, we write
\begin{align*}
P\left(\sigma_i^{(k)}\right) = & \, P_\text{silence}(K) \delta\left(\sigma_i^{(k)}, 0\right) \\
& + \left[ 1 - P_\text{silence}(K)\right]A_K\left(\sigma_i^{(k)}\right)
\end{align*}
for some function $A_K$, $K=2^{k-1}$ and where $P_\text{silence}$ is the probability that $\sigma_i^{(k)} = 0$. We can also think of the second term in the r.h.s. as the probability distribution $P_\text{activity}$ of the normalized activity $x_i^{(k)} = \sigma_i^{(k)}/K$ inside a cluster of $K$ variables, namely
\begin{align}
\label{eqn:p3c3:full_probability}
P_\text{activity}(x_i^{(k)}) := Z^{-1} A_K\left(\sigma_i^{(k)}\right)
\end{align}
with $Z$ a proper normalization constant The effective (reduced) free energy of the system is defined as
\begin{align}
\label{eqn:p3c3:cg_free_energy}
F(K) = \log P_\text{silence},
\end{align}
which is based on the assumption that the energy of the system is zero when no activity is present. We are interested in its possible scaling, $F(K) \sim -K^{\tilde \beta}$. For independent variables, we expect $\tilde\beta = 1$.

A scaling behavior of the ranked spectrum of the covariance matrix at the critical point is also expected. In fact, when the system is translational invariant, at the critical point the correlation function decays algebraically as $G(\mathbf x) \sim |\mathbf x|^{-(d-2+\eta)}$. Indeed, in a system with translational invariance, each element of the covariance matrix is given by $C_{ij}=C(\vb x_i - \vb x_j)$ for some function $C$, whose Fourier transform is
\begin{align*}
C(\vb k, \vb q) & = \frac{1}{N}\sum_{i,j} C(\vb x_i - \vb x_j) \, e^{-i\vb x_i \cdot\vb k}e^{-i\vb x_j \cdot \vb q} \\
& = \delta_{\vb k, -\vb q} G(\vb k)
\end{align*}
where
\begin{align*}
G(\vb k) = \sum_n e^{-i\vb x_n \cdot\vb k} C(\vb x_n).
\end{align*}
Hence the covariance matrix's entries are
\begin{align*}
C(\vb x_i - \vb x_j) = \frac{1}{N} \sum_{\vb k} e^{i\vb k \cdot (\vb x_i - \vb x_j)}G(\vb k),
\end{align*}
which means that in Fourier space the covariance matrix is diagonal. In fact, it is easy to show that the eigenvalues are given by the Fourier transform of the correlation function $G(\vb k)$, since
\begin{align*}
    \sum_{\vb x_j} C(\vb x_i - \vb x_j)e^{i\vb k \cdot \vb x_j} & = e^{i\vb k \cdot \vb x_i} \sum_{\vb x_j} C(\vb x_i - \vb x_j)e^{-i( \vb k \cdot \vb x_i-\vb k \cdot \vb x_j)}\\
    & = e^{i \vb k \cdot \vb x_i} G(\vb k)
\end{align*}
so that $e^{ikx}$ is a eigenfunction of eigenvalue $G(\vb k)$. This has a non trivial implication for the eigenvalue spectrum of the covariance matrix in a critical system. Since the eigenvalues are the Fourier transform of the correlation function, we shall write
\begin{align*}
\lambda_{\vb k} & \sim \int d^dr \, e^{i\vb k \cdot \vb r} r^{-(d-2+\eta)} \sim \frac{1}{|\vb k|^{2-\eta}}.
\end{align*}
If this is a decreasing function of $|\vb k|$, that is if $\eta<2$, then we consider a ranking of eigenvalues from small momentum to large momentum. Hence the highest eigenvalue has rank $r=1$, which implies
\begin{align*}
r[\lambda_{\vb k}] & = \sum_{\vb k'} \mathbb{I}[\lambda_{\vb k'}> \lambda_{\vb k}] = \sum_{\vb k'}\mathbb{I}[|\vb k'|<|\vb k|] \\
& \approx L^d \int d^dk'\,\theta(|\vb k'| < |\vb k|) \\
& \sim (L|\vb k|)^d.
\end{align*}
In turn, this means that the eigenvalues of the covariance matrix decay as a power-law of their rank, namely
\begin{align}
\label{eqn:p3c3:spectrum_covariance}
\lambda_r \sim \frac{1}{r^\mu}
\end{align}
where $r$ is the rank of $\lambda_r$, i.e., $\lambda_1 \ge \lambda_2 \ge \dots \lambda_N$, and $\mu = (2-\eta)/d$. If we consider the variables inside the clusters at each coarse-graining step, the highest possible rank $r$ will be given by the number of variables $K$ that make up each cluster. Hence, as a direct consequence of the power-law decay of the correlation function in space, at criticality we should find
\begin{align}
\label{eqn:p3c3:cg_eig}
\lambda_r \propto \left(\frac{K}{r}\right)^\mu
\end{align}
with $\mu = (2-\eta)/d$.

Finally, the mean autocorrelation function is obtained from
\begin{align}
\label{eqn:p3c3:cg_acf}
C^{(k)}(t) = \frac{1}{N_k}\sum_i C_i^{(k)}(t)
\end{align}
where
\begin{align*}
C_i^{(k)}(t) = \frac{\langle \sigma_i^{(k)}(t_0)\sigma_i^{(k)}(t_0+t)\rangle - \langle \sigma_i^{(k)}\rangle^2}{\langle(\sigma_i^{(k)})^2\rangle - \langle \sigma_i^{(k)}\rangle^2}.
\end{align*}
Since we are grouping correlated variables to begin with, the decay of the autocorrelation is slower in clusters of bigger sizes. However, in a critical system, we might expect dynamical scaling, which would imply a power-law scaling of the autocorrelation times $\tau_c \propto K^{\tilde z}$.

\subsection{Momentum-space-inspired approach}
The ideas described in the previous section draw inspiration from ideas of real-space coarse-graining by defining clusters of maximally correlated variables rather than spatial neighbors. Here, we instead exploit the fact that in systems with translational invariance the Fourier transform of the correlation function, $G(\vb k)$, coincides with the eigenvalue spectrum of the covariance matrix, $\lambda_{\vb k}$. Since coarse-graining in momentum space amounts to averaging over the  Fourier modes with small wavelengths, averaging over low variance contributions of the covariance matrix should lead to a formally equivalent result. Hence, we consider the set of eigenvectors of the covariance matrix $\{\vb{u}_r\}$, ordered according to the rank of the corresponding eigenvalue from the highest to the smallest one, and we introduce the projectors
\begin{align}
\label{eqn:p3c3:projector}
P_{ij}(K) = \sum_{r=1}^{\hat K} u_{ir}u_{jr}
\end{align}
where, with orthonormal eigenvectors, $P_{ij}(N)$ is the identity. The authors of \cite{meshulam2018arxiv, meshulam2019coarsegraining} propose to consider a cutoff $\hat{K}<N$, in analogy to the cutoff in momentum space, in such a way that the low variance contributions do not enter the projector in Eq.~\eqref{eqn:p3c3:projector}. Then the coarse-grained variables are defined as
\begin{align}
\phi_i(\hat K) := z_i(\hat K) \sum_j P_{ij}(\hat K) \left[\sigma_j^{(1)}-\langle \sigma_j^{(1)}\rangle\right]
\end{align}
where $z_i(\hat K)$ assures that the coarse-grained variables have unitary variance, i.e., $\langle \phi^2_i(\hat K) \rangle = 1$.

By means of the Young-Eckart theorem, the above procedure allows one to find the best decomposition with rank $\hat K$ of the original data matrix. These ideas are formally related to dimensionality reduction techniques and in particular principal components analysis. If $X_{ij}$ is the $P\times N$ matrix of $P$ temporal samples of $N$ variables, we can think of each row as representing a single configuration $\vb x_i$ that lives in a $N$-dimensional space. Then, via singular values decomposition, we have
\begin{align}
\label{eqn:p3c3:appB_svd}
X = U \Sigma V^T, \quad UU^T = \mathbb{I} = VV^T, \quad \Sigma = \text{diag}(\sigma_1, \dots, \sigma_r)
\end{align}
where $U \in \mathbb{M}(P\times P)$, $V \in \mathbb{M}(N\times N)$, $\Sigma \in \mathbb{M}(P\times N)$ $r=\text{min}(P, N)$, and $(\sigma_1, \dots, \sigma_r)$ are the singular values of $X$. Importantly, we have
\begin{align*}
\begin{cases}
X^TX = V \, \Sigma^T\Sigma \, V^T, \quad \Sigma^T\Sigma \in \mathbb{M}(N\times N)\\
XX^T = U \, \Sigma\Sigma^T \, U^T, \quad \Sigma\Sigma^T \in \mathbb{M}(P\times P)
\end{cases}
\end{align*}
which means that the right-singular vectors of $X$ are eigenvectors of $X^TX$, and the left-singular vectors are eigenvectors for $XX^T$. This further implies that the non-zero singular values of $X$ are the square roots of the non-zero eigenvalues of both $XX^T$ and $X^TX$. With this in mind, a possible approach to dimensionality reduction is to find the best low-rank decomposition of $X$, where the rank is nothing but the number of non-zero singular values. We want to minimize
\begin{align*}
\text{min}||X-X_k||^2_F = \text{min}\left[\sum_{ij}(X-X_k)_{ij}^2\right]
\end{align*}
where $||\cdot||_F$ is the Frobenius norm and $X_k$ is a rank $k$ matrix. For instance, in the case of a unitary rank decomposition $X_1 = \lambda \vb{a}\vb{b}^T$ with $\vb{a}^T\vb{a} = 1 = \vb{b}^T\vb{b}$, $\vb{a}\in\mathbb{R}^P$ and $\vb{b}\in\mathbb{R}^N$, the minimization leads to
\begin{align*}
\begin{cases}
\vb{a}^TX = \lambda \vb{b}^T \\
X\vb{b} = \lambda \vb{a} \\
\end{cases} \implies \quad
\begin{cases}
XX^T \vb{a} = \lambda^2 \vb{a} \\
X^TX \vb{b} = \lambda^2 \vb{b} \\
\end{cases}.
\end{align*}
This result means that the best rank $1$ decomposition of $X$ is such that $\vb{a}$ is an eigenvector of $XX^T$ and $\vb{b}$ is an eigenvector of $X^TX$, both with the same eigenvalue $\lambda^2$. Hence we can write
\begin{align*}
X_1 = \sigma_i \vb{u}_i \vb{v}_i^T
\end{align*}
for some $i = 1, \dots, \text{rank}(X)$. Since $X = \sum_j \sigma_j \vb{u}_j\vb{v}^T_j$ with Frobenius norm $||X||^2_F = \sum_i\sigma_i^2$, we immediately see that $||X-X_1||_F^2$ is minimized if we choose the index $i$ to be the one of the highest singular value. In fact,
\begin{align*}
||X-X_1||_F^2 = ||X-\sigma_i \vb{u}_i \vb{v}_i^T||^2_F = \sum_{j = 1, j \ne i}^{\text{rank}(X)} \sigma_j^2
\end{align*}
so the best choice we can make is to remove from this sum the highest singular value. This is nothing but the Young-Eckhart theorem and it is true for any value $k$: if $i = 1, \dots, k$ are the indexes of the highest singular values and we define
\begin{align*}
X_k = \sum_{i=1}^k \sigma_i\vb{u}_i\vb{v}_i^T
\end{align*}
then $||X-X_k||_F < ||X-B||_F$ for every matrix $B\in\mathbb{M}(P\times N)$ of rank $k$. Hence, the low-rank decomposition is intimately related to the $k$ highest singular values.

So far, we said nothing about dimensionality reduction - albeit finding a rank $k<N$ decomposition amounts to restrict the data into a $k$-dimensional subspace of the original $N$-dimensional space. Thus we could directly project the variables into the subspace spanned by the $k$ highest right-singular vector of $X$ - and this is exactly what PCA does. To define the PRG in momentum space, we stick with the best low-rank decomposition, so that the number of variables is unchanged. In fact, our projectors are given by
\begin{align*}
(P_k)_{\mu\nu} = \sum_{i=1}^k v_{\mu i} v_{\nu i}
\end{align*}
and we can recover the Young-Eckhart theorem by means of
\begin{align*}
X_k = X P_k = \sum_{j=1}^r \sigma_j\vb{u}_j\vb{v}^T_j \sum_{i=1}^k \vb{v}_i \vb{v}_i^T = \sum_{i=1}^k \sigma_i\vb{u}_i\vb{v}^T_i.
\end{align*}
Hence, if we define the $P\times N$ matrix $\Phi$ as the matrix of the coarse-grained variables we have
\begin{align*}
\Phi_{\mu\nu} = \sum_{\sigma = 1}^N X_{\mu\sigma} (P_k)_{\sigma\nu}
\end{align*}
which is exactly $X_k$.

In this setting, we may once again look directly at probability distributions, namely
\begin{align}
\label{eqn:p3c3:cg_momentum_dist}
P_{\hat K}(\phi) = \big\langle \frac{1}{N} \sum_{i=1}^N \delta\left(\phi_i(\hat K) - \phi\right)\big\rangle=\frac{1}{N} \sum_{i=1}^N \mathbb{P}\left[\phi_i(\hat K) = \phi\right],
\end{align}
as we change the cutoff $\hat K$. In fact, a renormalization group transformation typically drives the joint probability towards a fixed point, and if the variables are weakly correlated such a fixed point is the one obtained from the central limit theorem \cite{jona2001renormalization}. Hence, the authors of \cite{meshulam2018arxiv, meshulam2019coarsegraining} propose to use this PRG approach to test whether the joint distribution converges towards a non-Gaussian critical fixed point.

\section{Results on archetypal models}
We now test the results of the PRG in a number of simple models, whose critical behavior is well understood. We will start with the contact process, where we can instead investigate its dynamical properties. Importantly, we are interested in the ability of the proposed PRG to distinguish critical from non-critical systems, and to test whether the results observed in neural activity may be signaling the presence of an underlying phase transition \cite{meshulam2019coarsegraining, meshulam2018arxiv}. Then, we will consider both the Ising model, for which it only makes sense to compute static quantities - the scaling of the covariance spectrum and the behavior of probability distributions - and models of conditionally independent variables. This will allow us to probe the behavior of the PRG in equilibrium settings, as well as understand the effect of latent, environmental-like dynamics on its results \cite{morrell2021latent}.

\subsection{The contact process on a two-dimensional lattice}
The contact process \cite{marro1999nonequilibrium, henkel2009} is possibly the simplest non-equilibrium model used to describe the propagation of activity on a network. Each node can be either active (occupied) or inactive (empty), and we identify its state by means of a binary variable $\sigma_i (t) = 1, \, 0$ respectively. The activity spreads via a nearest neighbors interaction, and it depends on the number of active neighbors $n_i(t) = \sum_{j\in\ev{i}} \sigma_j(t)$, whereas each active site is emptied at a unitary rate. The rates $w[\sigma_i(t)\to\sigma_i(t+dt)| n_i(t)]$ that define the process for a node with $k_i$ neighbors are given by
\begin{equation}
\label{eqn:p3c3:cp_rates}
w[0\to1| n_i] = \frac{\lambda n_i}{k_i}, \quad 
w[1\to0| n_i] = 1
\end{equation}
where $\lambda$ is the spreading rate. Clearly, the configuration with all empty sites is an absorbing state since the system cannot escape from it. In particular, if $\lambda>\lambda_c$ the stationary state is an active fluctuating phase, whereas if $\lambda<\lambda_c$ the system eventually gets trapped in the absorbing configuration. Exactly at $\lambda = \lambda_c$ the density of active sites undergoes large fluctuations and the system often wanders close to the absorbing state. Indeed one can prove that the critical contact process dies out with probability $1$ \cite{bezuidenhout1990critical}. A non-vanishing survival probability is achieved only in the super-critical regime $\lambda>\lambda_c$.

As shown in the introduction of this Thesis, the density $\rho$ of the active sites is an order parameter of this system, and in a mean-field regime it evolves according to
\begin{equation*}
\label{eqn:p3c3:cp_mf}
\dot{\rho} = \rho(\lambda-1)-\lambda\rho^2.
\end{equation*}
This equation has two stationary solutions:  $\rho_\text{st}^v = 0$ and the active state $\rho_\text{st}^a = (\lambda-1)/\lambda$. The former is stable if $\lambda < 1$, and the latter if $\lambda >1$. Hence the mean-field critical point is $\lambda_{c}^\text{MF} = 1$. In a two-dimensional lattice, instead, it is well-known that numerical studies place the critical point at $\lambda_c^{2D} \approx 1.6488$ \cite{marro1999nonequilibrium}. Indeed, the contact process is not exactly solvable even in one dimension, therefore we need to rely on numerical simulations as well. We implement the usual scheme \cite{dickman1999reweighting}: an occupied site $i$ is randomly chosen, and with probability $1-p_\lambda = 1/(1+\lambda)$ the site is emptied. With probability $p_\lambda = \lambda/(1+\lambda)$ one of the neighbors is picked at random and, if empty, is occupied. The time is increased by $1/N_\text{occ}$, where $N_\text{occ}$ is the number of occupied sites. We perform all simulations with $N = 40^2$ sites and analyze clusters of size $K = 2, \dots, 256$. In momentum space, we keep up to $N/128 \approx 12 $ eigenvalues, which is less than $1\%$ of the original modes. Finally, for the super-critical regime, we set $\lambda = 3$, far from the absorbing transition.

\begin{figure}
    \centering
    \includegraphics[width = 1\textwidth]{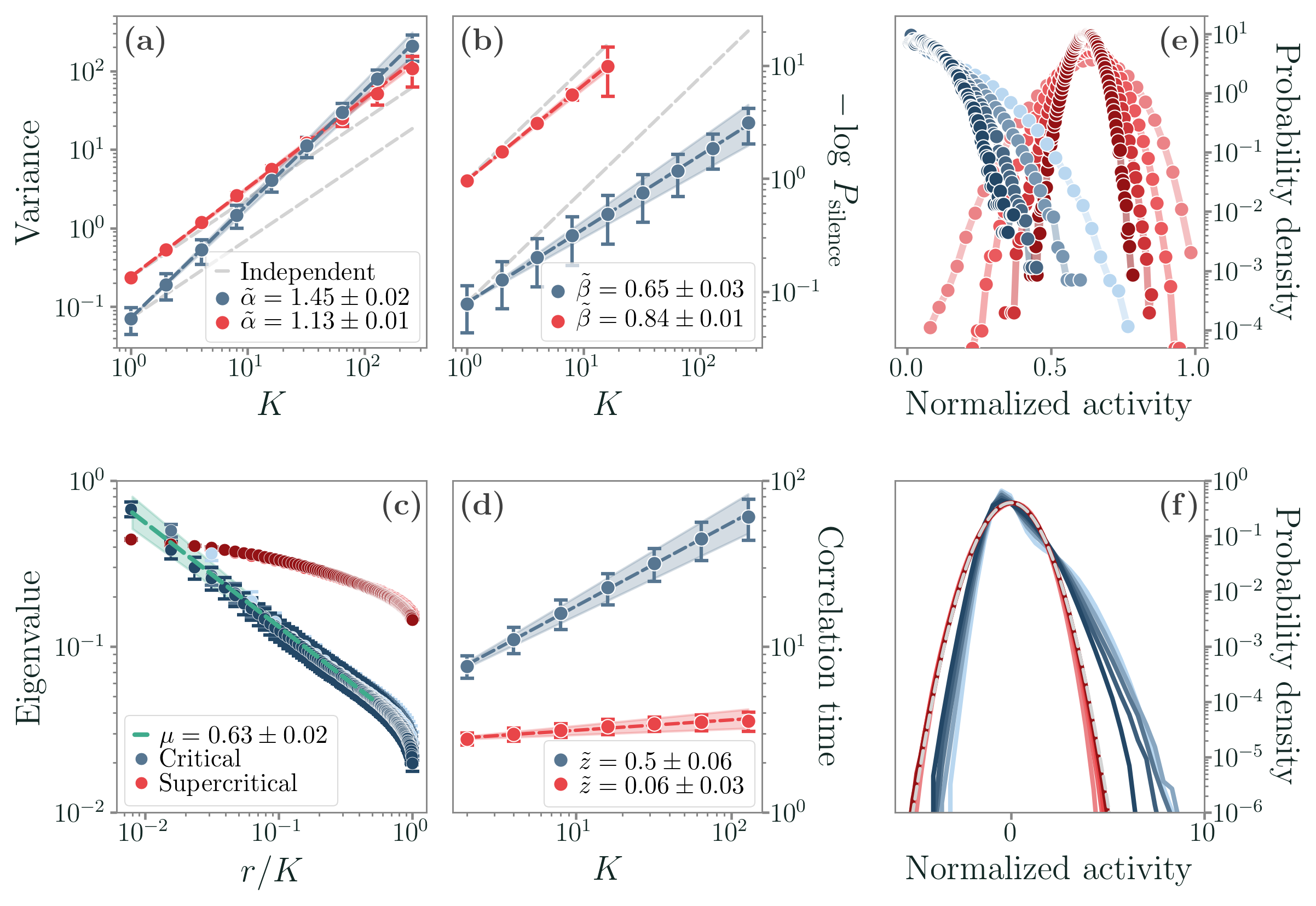}
    \caption{Results of the phenomenological renormalization group in the case of a critical (blue) and super-critical (red) contact process in a two-dimensional lattice.  All error bars and shaded areas are in units of three standard deviations. (a-b) Scaling of the variance, Eq.~\eqref{eqn:p3c3:cg_variance}, and of the free energy, Eq.~\eqref{eqn:p3c3:cg_free_energy}. Notice that both the critical and the super-critical regimes are compatible with a power-law behavior, albeit with different exponents. In particular, the silence probability is smaller and decays much faster in the super-critical case due to the proliferation of activity. However, we do not find full compatibility between the super-critical contact process and the independent case, as one would expect. (c-d) Scaling of the eigenvalues of the covariance matrix within clusters of sizes $K = 32, 64, 128$, and scaling of the autocorrelation times along the coarse-graining. The distinction between the two phases is rather clear. The eigenvalues of the covariance matrix in the critical regime scale with an exponent $\mu = 0.63\pm0.02$, which is compatible with the expected value from the hyper-scaling relations of the contact process. On the other hand, the eigenvalues show less variability at $\lambda > \lambda_c$. The same holds for the autocorrelation times, which follow a power-law behavior at criticality, whereas the scaling becomes negligible above it. (e-f) Evolution of the probability distribution of non-zero activity during the coarse-graining via maximally correlated pairs, Eq.~\eqref{eqn:p3c3:full_probability}, and of the probability distribution of coarse-grained variables in momentum space, Eq.~\eqref{eqn:p3c3:cg_momentum_dist}. We show the results for $K = 32, \dots, 256$ and for $N/8, \dots, N/128$ modes (from brighter to darker colors). The distribution in direct space is very different due to the critical contact process being typically close to the absorbing state. In momentum space, the super-critical contact process converges to a Gaussian fixed point in agreement with the central limit theorem \cite{jona2001renormalization}, whereas at criticality we see the presence of non-Gaussian tails with a bulk that seems to become invariant as we keep less and less modes}
    \label{fig:p3c3:cp}
\end{figure}

In Figure \ref{fig:p3c3:cp}a-b we see that in the super-critical regime the exponents of the variance and of the free energy are not exactly compatible with the independent case $\tilde \alpha = 1 = \tilde\beta$. Nevertheless, the profile of the free energy clearly shows that the underlying dynamics is different from the critical state, as for clusters of size $K>32$ the silence probability vanishes in the active phase, a feature characteristic of the super-critical case. Notably, if we compare the critical exponent $\tilde\beta = 0.65 \pm 0.02$ with the one obtained for real neurons in \cite{meshulam2018arxiv, meshulam2019coarsegraining}, $\tilde\beta_\text{neurons} = 0.893 \pm 0.003$, we see that in the contact process the decay of the silence probability with the cluster size is slightly slower. Notice that this does not necessarily mean that real neurons are less active than the sites of a critical contact process - to quantitatively compare the silence probabilities one should take into account the multiplicative constant in the power-law fit. In principle, one might also wonder why the apparent exponent of a super-critical contact process is similar to the exponent found in real neuronal data. However, we think that this might well be a numerical coincidence as the fundamental point is that in the super-critical regime of the contact process the scaling fails for large $K$, since the silence probability vanishes in the active phase, a fact that is not observed in real data from neural activity.

Furthermore, it is important to check whether the scaling does persist for all the values of $K$. For instance, we might try to fit only the variance for $K\ge16$. In the critical case, we find an exponent $\tilde{\alpha}^{(\text{c})}_{16} = 1.43\pm0.05$, which is compatible with $\tilde{\alpha}^{(\text{c})}$ and hence suggests that the scaling is consistent at different values of $K$. In the super-critical regime we find $\tilde{\alpha}^{(\text{sc})}_{16} = 1.08\pm0.03$ which, albeit lower, is still compatible with $\tilde\alpha^{(\text{sc})}$ within the error. However, this might suggest a breakdown of the scaling in the super-critical case that we do not see at this scale because it occurs at larger values of $K$. Since such values are typically inaccessible in experiments and simulations, we conclude that the scaling of variance is not particularly informative. The fact that we do not see full compatibility with the independent case might be due to the fact that in the super-critical regime the fluctuations are not fully uncorrelated, and possibly this method tends to overestimate them.

Figure \ref{fig:p3c3:cp}c-d shows instead the correlation structure of the system's quasi-stationary state. The change in the spectrum of the covariance matrix is more evident, since in the super-critical case the eigenvalues span a smaller set of values. In the critical case, instead, we find a power-law decay with an exponent $\mu = 0.63 \pm 0.02$, with $\mu = (2-\eta)/d$. In real neurons, the authors of \cite{meshulam2018arxiv, meshulam2019coarsegraining} report $\mu_\text{neurons} = 0.71 \pm 0.06$. We note that, since one of the hyper-scaling relations of the contact process \cite{dickman2005quasi} yields
\begin{align*}
\eta = d -2 + \frac{\beta}{\nu_\perp},
\end{align*}
we expect $\mu \approx 0.6$ in the $2D$ contact process from $\beta \approx 0.583$ and $\nu_\perp \approx 0.733$ \cite{marro1999nonequilibrium, henkel2009}. This value is compatible with what we find using the PRG procedure. The autocorrelation function shows evident changes as well, as expected - in the super-critical regime it decays exponentially, whereas at criticality we find a power scaling with an exponent $\tilde z = 0.50 \pm 0.06$. We note also that in the super-critical regime a power-law seems to be present, but the small exponent is compatible with the absence of scaling. Indeed, a constant autocorrelation time across different cluster sizes fits, with the same significance, the data, hence no relevant scaling feature seems to be present in the super-critical regime.

Finally, the evolution of the joint probability distribution of the coarse-grained variables in Figure \ref{fig:p3c3:cp}e-f shows once more the differences in the underlying dynamics. The most notable result is the convergence in momentum space: for $\lambda > \lambda_c^{2D}$ the fixed point is Gaussian in accord with the central limit theorem, whereas at $\lambda = \lambda_c^{2D}$ we do see distinct non-Gaussian tails. The last coarse-graining step in momentum space only keeps $N/128$ modes, so the fact that we still find non-trivial tails is significant. Yet, as we will see later in this Chapter, there are a number of caveats one needs to take into account.

\begin{figure}[t]
    \centering
    \includegraphics[width = 1\textwidth]{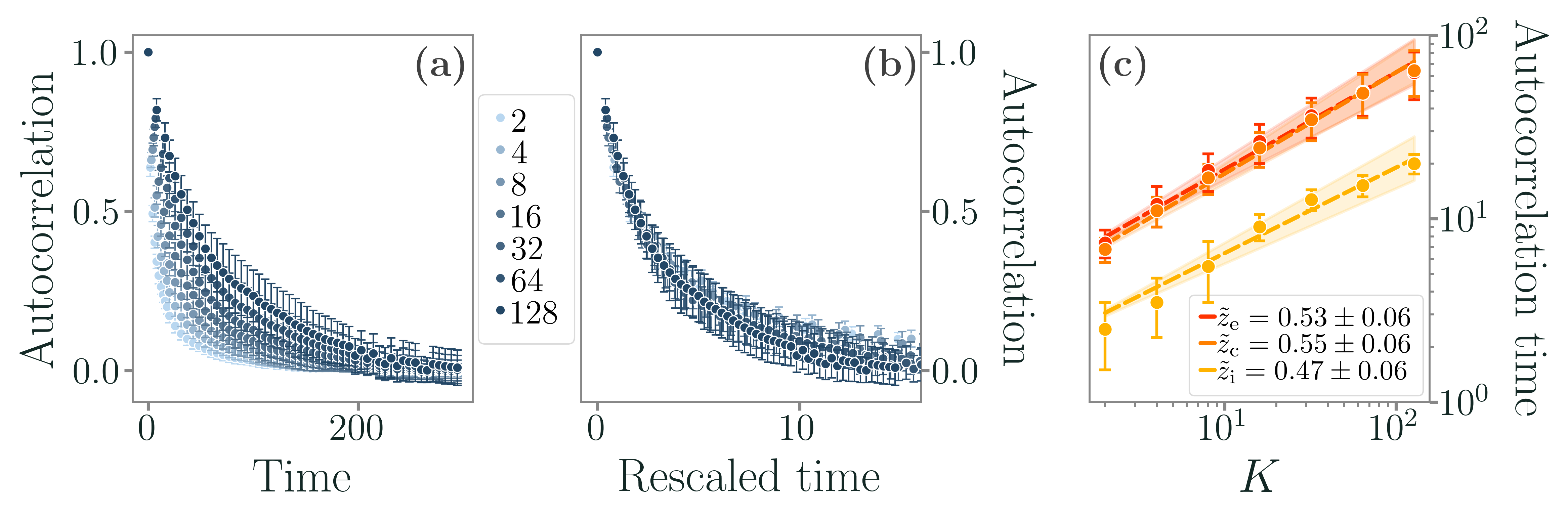}
    \caption{Collapse of the autocorrelation function and robustness of scaling of the autocorrelation times in a critical contact process on a $2D$ lattice.  All error bars and shaded areas are in units of three standard deviations. (a-b) If we rescale the times by $t \to t/\tau$, with $\tau$ the autocorrelation time, the time-autocorrelation functions at different $K$ collapse into a single curve. (c) The scaling of the autocorrelation times is robust with respect to the definition of the autocorrelation time itself}
    \label{fig:p3c3:cp_autocorrelation}
\end{figure}

Let us also note that we can further test the robustness of the scaling with respect to the definition of the autocorrelation time and the collapse of the critical time-autocorrelation function. In fact, estimating the autocorrelation time might be challenging due to sampling effects. In order to check the robustness of the scaling we find in Figure \ref{fig:p3c3:cp}d, we perform a longer simulation of the $2D$ critical contact process and we implement three different strategies to evaluate the autocorrelation time. We estimate the exponential correlation time $\tau_\mathrm{e}$, the integrated correlation time $\tau_\mathrm{i}$ as defined in \cite{goodman2010ensemble} and the definition of $\tau_\mathrm{c}$ proposed by Cavagna and collaborators in \cite{cavagna2017dynamic}. In Figure \ref{fig:p3c3:cp_autocorrelation} we see that the scaling exponent is compatible for all three estimates - and even though the integrated correlation time is particularly noisy, it only differs by an irrelevant multiplicative constant. Hence the scaling we find proves to be robust with respect to the definition of the autocorrelation time. Moreover, as proposed in \cite{meshulam2018arxiv, meshulam2019coarsegraining}, if we rescale the time by $t \to t/\tau$ the autocorrelation functions at different $K$ collapse in the same curve at criticality. This is typically associated with dynamical scaling, which is expected at the critical point of the contact process. Notice that, on the contrary, in the super-critical regime the autocorrelation decay is exponential, and hence we do not see a meaningful change in the autocorrelation function as we change $K$. In fact, in Figure \ref{fig:p3c3:cp} we see a scaling of the autocorrelation times that is compatible with a constant.

\subsection{The contact process on small-world networks}
At a first glance, the results we find in a two-dimensional lattice seem to be compatible with what one would expect. However, one of the core ideas of this phenomenological renormalization procedure is that it should be well-defined both for short-range and long-range interactions. This fact is especially relevant if one needs to deal directly with neural activity data and the specific network architecture is not accessible. Thus, we consider now a small-world network, where such long-range interactions are present. In the small-world case, the critical point $\lambda_c^\text{SW}$ depends on the rewiring probability, and it has been studied numerically in \cite{ferreira2013critical}. In particular, we implement a Watts-Strogatz model with a rewiring probability $p = 0.01$. Then, the critical point is $\lambda_c^\text{SW} \approx 1.7961$.

As a sanity check, we use in this case a synchronous update algorithm. The results show no difference with respect to the asynchronous one used insofar. Importantly, in the case of the asynchronous update - where at most one site is changed at each step - one should carefully consider that the algorithm induces a spurious correlation between subsequent configurations. Hence we do not keep all the configurations to perform averages, but we rather subsample them so as to select only uncorrelated configurations. Once this is taken into account, the results of synchronous and asynchronous updates are equivalent.

\begin{figure}[t!]
    \centering
    \includegraphics[width = 1\textwidth]{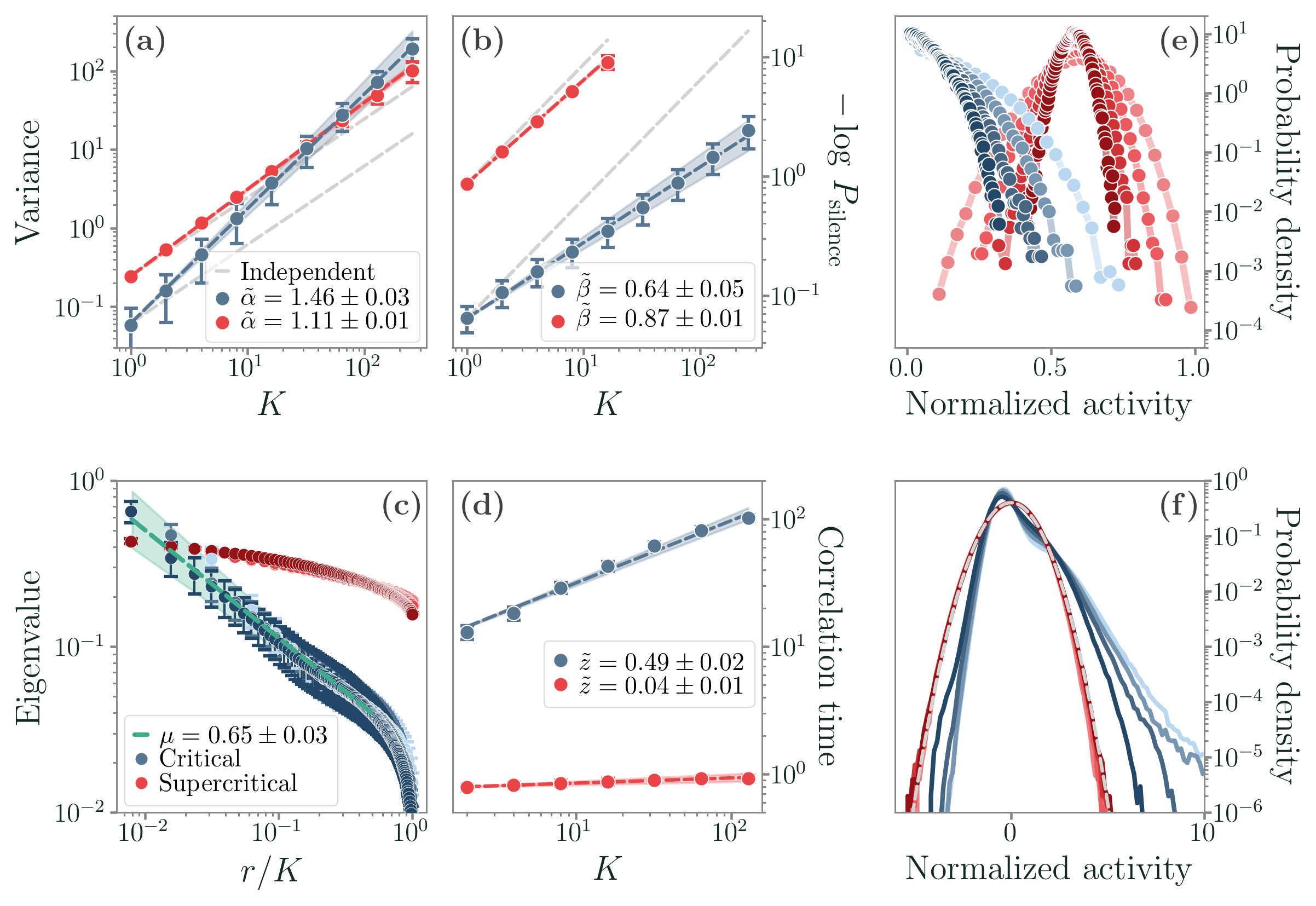}
    \caption{Results of the phenomenological renormalization group in the case of a critical (blue) and super-critical (red) contact process in a small-world network with a rewiring probability of $p = 0.01$. The simulation is performed using a synchronous algorithm.  All error bars and shaded areas are in units of three standard deviations. (a-b) The scaling of both the variance and the free energy is similar to that of the $2D$ lattice, suggesting that the topology is not affecting it. (c-d) Scaling of the eigenvalues of the covariance matrix within clusters of sizes $K = 32, 64, 128$, and scaling of the autocorrelation times along the coarse-graining. The eigenvalues of the covariance matrix in the critical regime scale with an exponent $\mu = 0.65\pm0.02$, again compatible with the previous case. (e-f) Evolution of the probability distribution of non-zero activity during the coarse-graining via maximally correlated pairs and of the coarse-grained variables in momentum space. We show the results for $K = 32, \dots, 256$ and for $N/8, \dots, N/128$ modes (from brighter to darker colors). In both cases, we find results similar to the $2D$ lattice topology}
    \label{fig:p3c3:cp_sw}
\end{figure}

We find that all the considerations we made so far hold in the small-world topology as well - and the presence of long-range interactions does not affect the results of the PRG coarse-graining procedure, as we see in Figure~\ref{fig:p3c3:cp_sw}. These results confirm that, in principle, the method proposed by \cite{meshulam2018arxiv, meshulam2019coarsegraining} is not sensible to the presence of long-range interaction, since it does not depend on the underlying spatial structure to begin with.

\subsection{Persistence of the scaling near a critical point}
A natural question one may ask is how sensible this PRG approach is, i.e., how easy it is to distinguish a truly critical system from a super-critical one. We test this in the contact process by moving the control parameter from the critical point $\lambda_c$ to $\lambda_\text{nc} \approx 1.1\lambda_c$, which is a $10\%$ increase.

Notice that, although it is not trivial to define a finite-size critical point \cite{ferdinand1969bounded} for the transition in the contact process - the finite size contact process eventually reaches the absorbing configuration, at all values of $\lambda$ \cite{marro1999nonequilibrium} - at $\lambda_\text{nc}$ we do see distinctive features of a super-critical dynamics. The dynamical evolution lacks considerable fluctuations in the density of sites, nor the system constantly approaches the absorbing state as at $\lambda = \lambda_c$. Hence, at $\lambda_\text{nc}$ the dynamics is significantly super-critical, and we shall refer to this as a near-critical case to distinguish it from the super-critical regime we described before.

\begin{figure}[t]
    \centering
    \includegraphics[width = 1\textwidth]{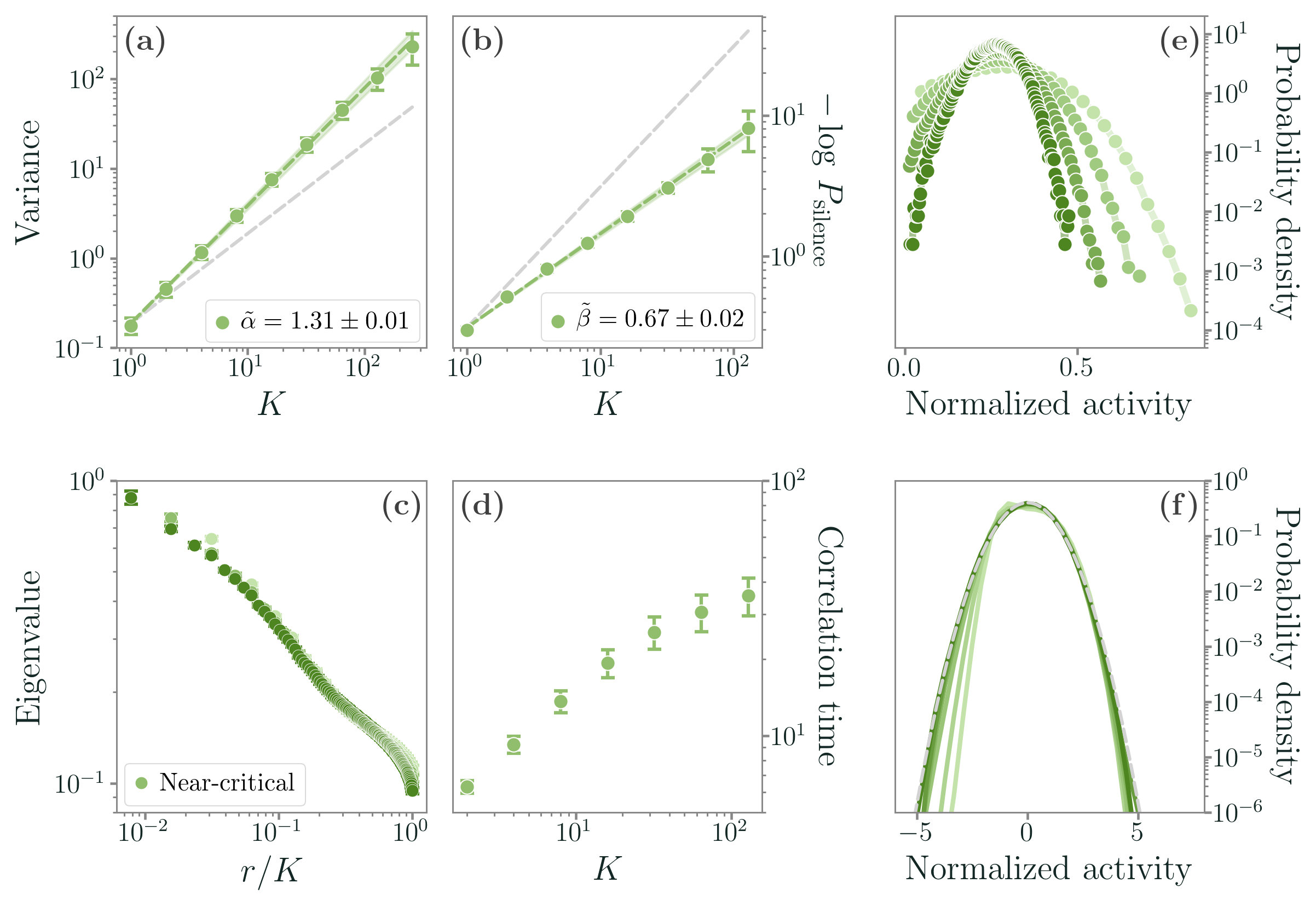}
    \caption{Results of the phenomenological renormalization group in the case of a near-critical contact process in a two-dimensional lattice.  All error bars and shaded areas are in units of three standard deviations. (a-b) Scaling of the variance, Eq.~\eqref{eqn:p3c3:cg_variance}, and of the free energy, Eq.~\eqref{eqn:p3c3:cg_free_energy}. Even if this is a super-critical contact process, the exponents are far from being comparable with the independent case $\tilde\alpha = 1 = \tilde\beta$. Overall, the scaling is more similar to the real critical case, see Figure \ref{fig:p3c3:cp}. Notice in particular how the silence probability is non-vanishing even for large clusters. (c-d) The spectrum of the covariance matrix, instead, does not show a power-law decay, and the scaling of the autocorrelation times is not as convincing as the critical case and fails for large clusters. (e-f) Evolution of the probability distribution of non-zero activity during the coarse-graining via maximally correlated pairs, Eq.~\eqref{eqn:p3c3:full_probability}, and of the probability distribution of coarse-grained variables in momentum space, Eq.~\eqref{eqn:p3c3:cg_momentum_dist}. We show the results for $K = 32, \dots, 256$ and for $N/8, \dots, N/128$ modes (from brighter to darker colors). Both of them are comparable with the super-critical case, and in particular in momentum space we do not see the non-Gaussian tails typical of criticality}
    \label{fig:p3c3:cp_nc}
\end{figure}

In Figure \ref{fig:p3c3:cp_nc}(a-b) we see non-trivial scaling behaviors of both the variance and the free energy. If we compare them to Figure \ref{fig:p3c3:cp}, they are arguably more similar to the critical regime rather than the super-critical one. As in the previous case, we might try to fit only points at larger values of $K$, for instance $K\ge16$. We find that the exponents does get smaller, $\tilde{\alpha}^{(\textsc{nc})}_{16} = 1.24\pm0.06$, but once again remains compatible with $\tilde{\alpha}^{(\textsc{nc})}$. As before, this suggests that we see a non-trivial scaling that might only disappear at very large values of $K$. Given that these are the typical values one can deal with, the scaling of the variance cannot be ruled out for this near-critical system. Importantly, the exponents $\tilde{\alpha}$ and $\tilde{\beta}$ are in between the critical and super-critical regimes, suggesting that as $\lambda$ changes from $\lambda = \lambda_c$ to $\lambda = +\infty$, the exponents smoothly approach $1$. These results call for carefulness as the scaling inferred from the PRG of the variance and of the free energy are not necessarily emerging from an underlying critical state. The fact that both the variance and the free energy show a power-law behavior both at $\lambda_c$ and at $\lambda > \lambda_c$ might be a sign that criticality is not a necessary condition for such power-laws. Indeed, in the case of the free energy, one should note that usually it is the singular part of the free energy that shows scaling, whereas with this PRG we cannot distinguish it from the non-singular part.

On the other hand, in Figure \ref{fig:p3c3:cp_nc}c-d the eigenvalues of the covariance matrix do not display an evident power-law scaling as we change the cluster size, and the scaling of the autocorrelation time function is not significant, particularly for larger clusters. The most convincing results to discriminate between critical and quasi-critical states are the joint probability distributions, Eq.~\eqref{eqn:p3c3:full_probability} and Eq. \eqref{eqn:p3c3:cg_momentum_dist}), that we show in Figure \ref{fig:p3c3:cp_nc}e-f, in particular the one in momentum space. We do not see the non-Gaussian tails that we previously found at the critical point, which is expected since away from criticality the variables are much less correlated with one another and they are eventually dominated by the central limit theorem. Indeed, along the coarse-graining such probability distributions become compatible with Gaussian ones.

\subsection{The Ising model}
Let us now briefly move to the simpler, equilibrium case of the $2D$ Ising model. We build clusters of maximally correlated spins by simulating the model at different temperatures using the Wolff algorithm \cite{wolff1989collective} and collecting samples of uncorrelated configurations. This case is rather emblematic - we are not interested in the dynamics, but rather in what this PRG can tell us about the static properties of the system. Notice that we define the finite-size critical temperature as the temperature at which the specific heat is maximum \cite{ferdinand1969bounded}, which implies a shift with respect to the exact critical temperature $T_c \approx 2.269 J/k_B$.

\begin{figure}[t]
    \centering
    \includegraphics[width = 1\textwidth]{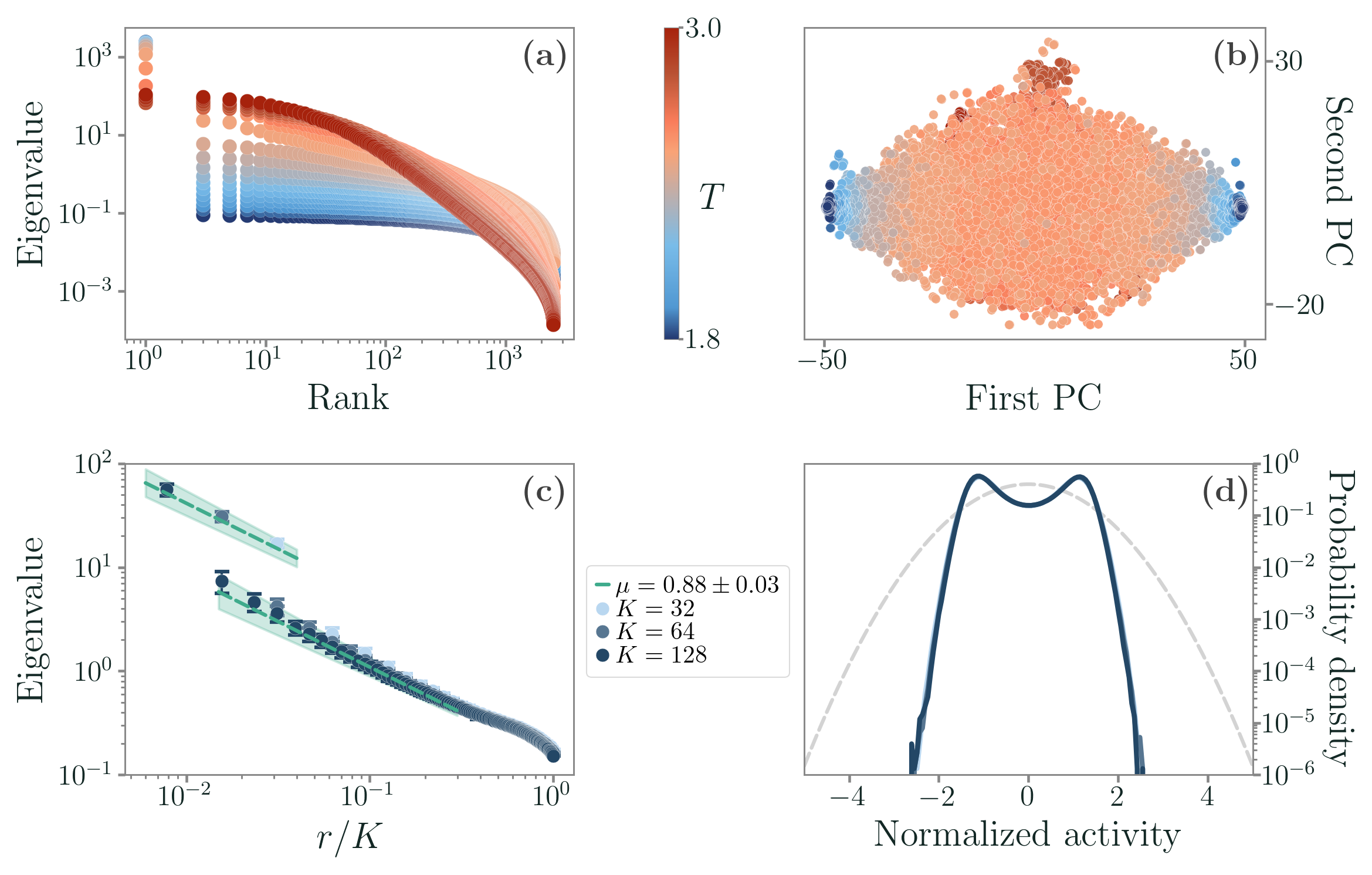}
    \caption{Results of the phenomenological Renormalization Group for a critical Ising model.  All error bars and shaded areas are in units of three standard deviations. (a-b) One should be careful that the spectrum of the covariance matrix of Ising spins at low temperature shows an emerging dominance by a single eigenvalue - meaning that a single principal component contributes to most of the variance of the system. In other words, the subcritical Ising model shows clear signatures of low dimensionality, while this is not true in the supercritical regime. The transition between the two regimes, and in particular of the two disjointed regions appearing due to spontaneous symmetry breaking, is evident in the space of the first two principal components. (c-d) Scaling of the eigenvalues at the critical temperature and convergence to a non-Gaussian fixed form of the distribution of coarse-grained variables in momentum space. Notice that a dominant eigenvalue is present at criticality because the model begins to become ordered - but the dominant eigenvalues scale in the same way as the bulk of the spectrum}
    \label{fig:p3c3:ising}
\end{figure}

For the sake of brevity, we only show the relevant results at the critical point. Indeed, in the disordered phase at $T>T_c$ the coarse-graining drives the system towards a behavior that is comparable with one of independent random variables, very much like one would expect from a usual block-spin transformation in real space. For $T<T_c$, instead, the behavior of coarse-grained variables resembles the one of a perfectly ordered system, where all the spins tend to be aligned. However, as we lower the temperature we see a non-trivial effect due to the spontaneous symmetry breaking that occurs at the transition. In fact, the Ising model in its ordered phase is essentially low dimensional \cite{wang2016discovering} - one single eigenvalue eventually dominates the spectrum of the covariance matrix at low temperature, as we see in Figure \ref{fig:p3c3:ising}a-b. Hence, a single principal component contributes to most of the variance of the system. The supercritical Ising model shows a continuous eigenvalue spectrum, hence low dimensionality disappears, but as we lower the temperature a gap in the spectrum becomes more and more evident. This is not surprising since below the critical temperature the system lives in two disjointed regions - which we may regard as a direct consequence of the $O(1)$ symmetry group of the Ising Hamiltonian, which allows two ground states, and the related spontaneous symmetry breaking at the transition. A spectral gap is indeed present at the critical temperature, where the symmetry breaking takes place.

In order to take this into account, in Figure \ref{fig:p3c3:ising}c we consider separately the behavior of the highest eigenvalues and the rest of the spectrum. We find that at criticality they both scale with the same exponent $\mu = 0.88 \pm 0.03$ and only the multiplicative constant differs. Since the Onsager solution of the 2-dimensional Ising model gives $\eta = 1/4$, we expect
\begin{align*}
\mu = \frac{7}{8} = 0.875.
\end{align*}
which is perfectly compatible with our result. Moreover, in Figure \ref{fig:p3c3:ising}d we see that the joint distribution clearly converges to a non-Gaussian form. Hence, in this case of an equilibrium phase transition with spontaneous symmetry breaking, this procedure does identify two distinct phases.

This example of the Ising model highlights how careful we need to be in analyzing the results of this coarse-graining procedure. The Ising model has a trivial symmetry group, that allows for two degenerate ground states only, but one can imagine more complex models where this analysis would be much more complicated.

\subsection{Conditionally independent units}
As we have seen, the results in the contact process suggest that, although the PRG gives the expected results at criticality, a number of caveats must be taken into account. Here, as a final check, we study its behavior in models of conditionally independent variables, which we have used throughout this Thesis. Subsequent studies to our own \cite{nicoletti2020scaling, morrell2021latent} have shown that these models' behavior under this phenomenological renormalization is particularly relevant. As we will see here, although the convergence of the joint probability distribution in Eq.~\eqref{eqn:p3c3:cg_momentum_dist} is related to the spectrum of the covariance matrix, one should be careful when considering its relation with criticality.

We first introduce a simple model of conditionally independent neurons. Consider $N$ random variables $\left(\sigma_1^{t}, \dots, \sigma_N^{t}\right)$. At each time $t$ the distribution of the $i$-th variable, which we can think of as a neuron that can be either active or inactive, is a simple binomial distribution with parameter $\xi_i(t)$. However, we consider the case in which also $\bm{\xi}(t) = (\xi_1(t), \, \dots, \, \xi_N(t))$ is itself a random variable distributed according to some distribution $p(\bm{\xi})$, so that the $N$ neurons are conditionally independent. Similarly to Chapter \ref{ch:scirep}, we can think of this case as that of neurons driven by a common external dynamics, represented by $p(\bm{\xi})$, and that show otherwise no intrinsic dynamical features.

The probability that a neuron is either active or inactive is then a binomial distribution conditioned to the value of $\xi_i(t)$, that is
\begin{equation*}
    p(\sigma_i^t \,|\, \bm\xi = \bm\xi(t)) = 
    \begin{cases}
        \xi_i(t) & \sigma_i^t = 1 \\
        1 - \xi_i(t) & \sigma_i^t = 0
    \end{cases},
\end{equation*}
and the each neuron is described by the joint probability $p(\sigma_i^t, \bm\xi) = p(\sigma_i^t \,|\, \bm\xi = \bm\xi(t))\, p(\bm{\xi})$. Since the coarse-graining procedure depends on the equal-time covariance of the neurons $\text{cov}(\sigma_i, \sigma_j) = C_{ij}$, we need the marginal probabilities
\begin{equation*}
    p(\sigma_i^t) = 
    \begin{cases}
    \displaystyle\int d\xi^*_i \xi^*_i p(\xi^*_i) = \ev{\xi_i} & \sigma_i^t = 1\\
    \displaystyle\int d\xi^*_i (1-\xi^*_i) p(\xi^*_i) = 1-\ev{\xi_i} & \sigma_i^t = 0
    \end{cases}
\end{equation*}
and
\begin{equation*}
    p(\sigma_i^t, \sigma_j^t) = 
    \begin{cases}
    \ev{\xi_i\xi_j} & \sigma_i^t = 1, \sigma_j^t = 1\\
    \ev{\xi_i}-\ev{\xi_i\xi_j} & \sigma_i^t = 1, \sigma_j^t = 0\\
    \ev{\xi_j}-\ev{\xi_i\xi_j} & \sigma_i^t = 0, \sigma_j^t = 1\\
    1 + \ev{\xi_i\xi_j}-\ev{\xi_i}-\ev{\xi_j} & \sigma_i^t = 0, \sigma_j^t = 0\\
    \end{cases}.
\end{equation*}
If we now use the latter marginal probability we immediately find that, if $i\ne j$, the first contribution to the covariance matrix of the neurons is given by
\begin{align*}
    \ev{\sigma_i\sigma_j}_{p(\sigma_i, \sigma_j)} & = \sum_{\sigma_i, \sigma_j = 0,1} \sigma_i\sigma_j \, p(\sigma_i, \sigma_j) \\
    & = p(\sigma_i = 1, \sigma_j = 1) = \ev{\xi_i\xi_j}.
\end{align*}
If instead $i=j$, we shall use the former marginal probability,
\begin{align*}
    \ev{\sigma_i\sigma_i}_{p(\sigma_i)} & = \sum_{\sigma_i = 0,1} \sigma_i \,p(\sigma_i) = \ev{\xi_i}
\end{align*}
Hence, even if the neurons are not correlated, their covariance is not vanishing but depends on the covariance of $p(\bm \xi)$,
\begin{equation*}
    C_{ij} =
    \begin{cases}
        \ev{\xi_i\xi_j} - \ev{\xi_i}\ev{\xi_j} & i \ne j \\
        \ev{\xi_i}(1-\ev{\xi_i}) & i = j
    \end{cases}.
\end{equation*}

Let us consider the simple case of $\xi_i = \xi_j$ $\forall i, j$, so that at each time all the neurons fire with the same probability $\xi$, and take $p(\xi_i)$ to be a uniform distribution. In this case, the covariance matrix is simply
\begin{equation*}
    C_{ij} = a \delta_{ij} + b (1-\delta_{ij})
\end{equation*}
with $a = 1/4$ and $b = 1/12$. The eigenvalues of this matrix are given by
\begin{align*}
    & \lambda_1 = a+(N-1)b & m = 1 \\
    & \lambda_2 = a-b & m = N-1
\end{align*}
where $m$ is the corresponding multiplicity. Therefore, there are $N-1$ eigenvalues with the same value.

\begin{figure}[t]
    \centering
    \includegraphics[width = 1\textwidth]{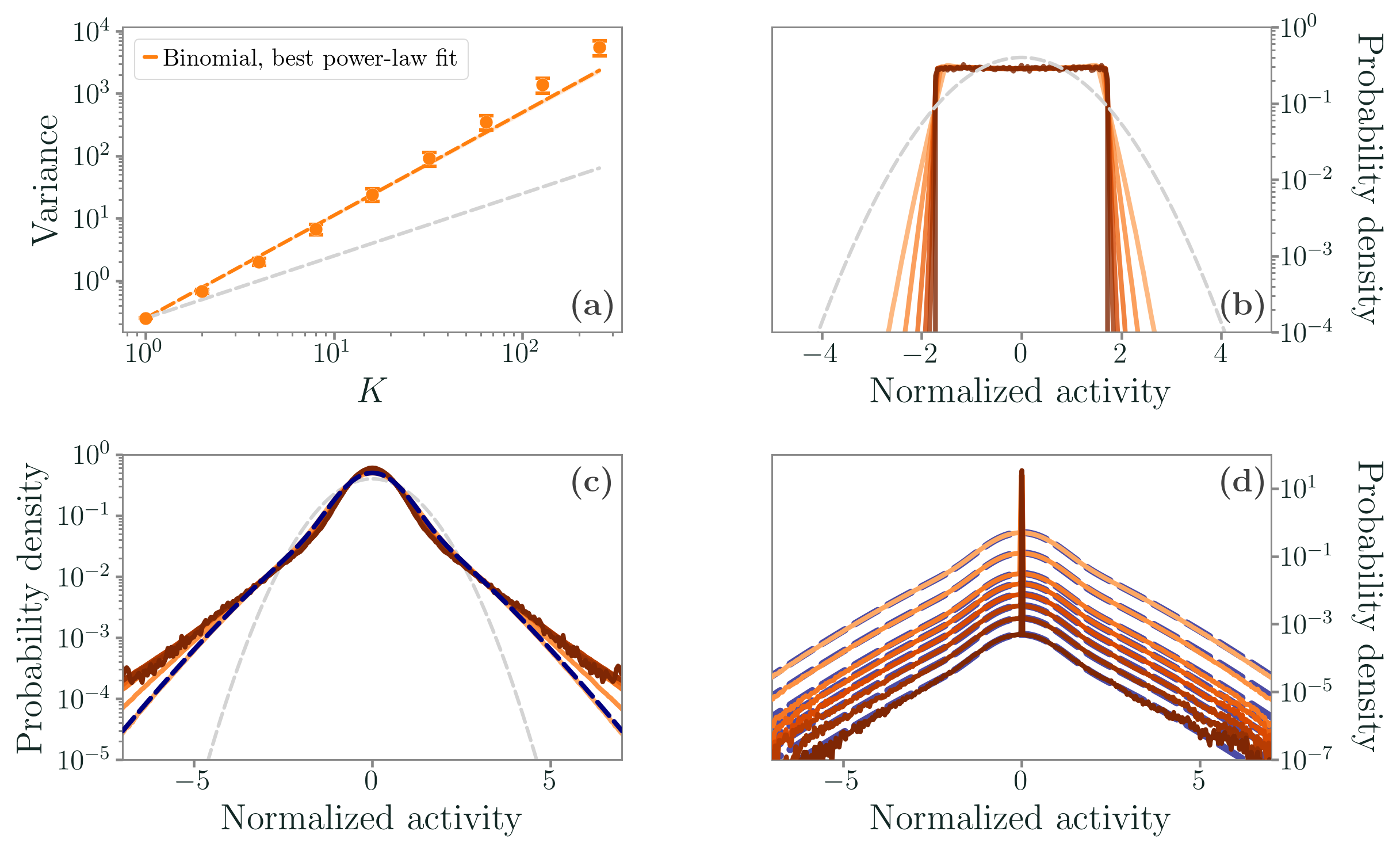}
    \caption{Results of the phenomenological renormalization group in the case of conditionally independent units.  All error bars and shaded areas are in units of three standard deviations. (a-b) Scaling of the variance, Eq.~\eqref{eqn:p3c3:cg_variance}, and evolution of the probability distribution of coarse-grained variables in momentum space, Eq.~\eqref{eqn:p3c3:cg_momentum_dist} in the binomial model. As we can see, the variance does not scale like a power-law, but the procedure in momentum space results in a non-Gaussian distribution. (c-d) Results in momentum space for the extrinsic model of Chapter \ref{ch:scirep}. In panel (c) we show the results obtained from the numerical estimation of the covariance matrix, which seem to point to a convergence of the probability distribution of the coarse-grained variables. (d) If we use the analytical expression of the projectors, however, the collapse vanishes, as one would expect for a non-critical model. Blue dashed lines are the analytical expressions, and gray dashed lines are Gaussian distributions}
    \label{fig:p3c3:conditionally_independent}
\end{figure}

The eigenvector associated with the highest eigenvalue is $1/\sqrt{N} (1, \dots, 1)^T$, but there is no obvious choice for the other eigenvectors in Equation (\ref{eqn:p3c3:projector}) because the ranking is ill-defined. However, from a numerical standpoint, the spectrum of the covariance matrix will not be degenerate, so if we simulate the model we can try to apply the procedure regardless. As we can see from Figure \ref{fig:p3c3:conditionally_independent}a-b the joint probability does not converge to a Gaussian, even though there is nothing critical about the underlying dynamics. Hence the proposed coarse-graining in momentum space fails for a simple set of conditionally independent binomial variables, albeit it seemed to be the most promising procedure for the super-critical contact process in the vicinity of the critical point. On the other hand, and perhaps not surprisingly, in this model the proposed coarse-graining procedure via maximally correlated variables does work: in fact, since the off-diagonal elements of the covariance matrix are all equal, we are randomly pairing neurons together and no scaling property emerges.

To further test these results, we can consider the extrinsic model presented in Chapter \ref{ch:scirep}. Namely, we have
\begin{equation*}
     \frac{dv_i(t)}{dt} = - \frac{v_i(t)}{\tau_i} + \sqrt{\mathcal{D}(t)} \eta_i(t),
\end{equation*}
where $\eta_i(t)$ are standard white noises, and the modulation $\mathcal{D}(t)>0$ is realized through the process
\begin{equation*}
    \mathcal{D}(t) = \begin{cases}
                     \mathcal{D}^* & \text{if} \quad D(t) \le \mathcal{D}^* \\
                     D(t) & \text{if} \quad D(t) > \mathcal{D}^*
                     \end{cases}
\end{equation*}
with
\begin{equation*}
    \dot{D}(t) = -D(t)/\gamma_D + \sqrt{\theta} \eta_D(t).
\end{equation*}
As we have shown, we can find the stationary probability distributions of this model analytically. In particular, the units are uncorrelated, and the diagonal entries of the covariance matrix are given by
\begin{equation*}
    \langle v_i^2 \rangle = \frac{\tau_i \mathcal{D}^*}{4}\left[1+ \text{Erf}\left(\frac{\mathcal{D}^*}{\sqrt{\theta \tau_D}}\right)\right] + \sqrt{\frac{\tau_i^2 \theta \tau_D}{16 \pi}}e^{-\frac{(\mathcal{D}^*)^2}{\theta \tau_D}}.
\end{equation*}
This allows us to write explicitly the projectors that remove the low-variance modes since the eigenvectors are just the canonical basis. The coarse-grained variables with only the first $\hat{K}$ modes follow the distribution
\begin{align*}
    P_{\hat{K}}(\phi) = \frac{1}{N}\Biggl[ & (N-\hat{K})\delta(\phi) + \\
    & + \hat{K} \sqrt{A_v} \Biggl[\frac{1+ \text{Erf}\left(\frac{\mathcal{D}^*}{\sqrt{\theta \tau_D}}\right)}{2\sqrt{\pi \mathcal{D}^*}}e^{-\frac{\phi^2A_v}{\mathcal{D}^*}} + \frac{1}{\sqrt{\pi^2 \theta \tau_D}}\int_{\mathcal{D}^*}^\infty \frac{dD}{\sqrt{D}}e^{-\left[\frac{D^2}{\theta \tau_D} + \frac{\phi^2 A_v}{D}\right]}\Biggr]\Biggr]
\end{align*}

with
\begin{equation*}
    A_v = \frac{\mathcal{D}^*}{4}\left[1+ \text{Erf}\left(\frac{\mathcal{D}^*}{\sqrt{\theta \tau_D}}\right)\right] + \sqrt{\frac{\theta \tau_D}{16 \pi}}e^{-\frac{(\mathcal{D}^*)^2}{\theta \tau_D}}.
\end{equation*}
As we see in Figure \ref{fig:p3c3:conditionally_independent}c-d, if we use the numerical estimate of the covariance matrix or its analytical expression we find drastically different results. In the former case, the probability distribution seems to converge to a fixed form that resembles $A_v$. In the latter case, instead, no convergence is present - as $P_{\hat{K}}(\phi)$ depends explicitly on $\hat{K}$.

We can understand this behavior in terms of perturbation theory. In general, whenever we estimate the covariance matrix $C_{ij}$ from data we retrieve the correct matrix in the $M\to\infty$ limit if we use unbiased estimators. Yet, if we have a finite number of samples of independent variables, we will never find a diagonal matrix, but rather $\tilde{C}_{ij} = C_{ij} + \delta C_{ij}$, possibly with $|\delta C_{ij}| \ll 1$ if $M$ is large enough. A first-order perturbative expansion in $|\delta C_{ij}|$ of the eigenvalue equation leads to the correction
\begin{equation*}
    \tilde{\lambda}_i = \lambda_i + \vb{e}_i^T \delta C_{ij} \vb{e}_i = \lambda_i + \delta C_{ii}
\end{equation*}
for the eigenvalues, and
\begin{equation}
\label{eqn:perturbation_eigenvectors}
    \tilde{\vb{e}}_i = \vb{e}_i + \sum_{j \ne i} \frac{\vb{e}_i^T \delta C_{ij} \vb{e}_j}{\lambda_i - \lambda_j}\vb{e}_j = \vb{e}_i + \sum_{j \ne i} \frac{\delta C_{ij}}{\lambda_i - \lambda_j}\vb{e}_j
\end{equation}
for the eigenvectors. Thus, the perturbative correction is enough to make the eigenvectors dense - all their entries are different from zero. It immediately follows that the perturbative projectors are not diagonal anymore, and in general
\begin{equation*}
    \tilde{\phi}_i(\hat{K}) = z_i(\hat{K}) \sum_{j = 1} ^ N a_j(\hat{K}) \left[x_j - \ev{x_j}_M\right]
\end{equation*}
for some coefficients $z_i(\hat{K})$ and $a_j(\hat{K})$. Thus, if the original $x_i$ variables are independent, we now have a sum over independent variables and the distribution $P_{\hat{K}}(\phi)$ is going to be a Gaussian distribution. But if they are not, as in this case, the central limit theorem does not apply and we find a non-trivial distribution.

These simple examples show once more how careful one should be when employing these kinds of procedures. All in all, the two approaches combined seem to work well - a system might be critical if both the real- and the momentum-space approaches indicate the presence of underlying scale invariance. That is, this approach might realistically give a set of necessary conditions for criticality, rather than sufficient ones. For instance, the presence of a non-trivial distribution of the coarse-grained variables in momentum space is a necessary condition for criticality because it implies that the underlying variables are strongly correlated, but the convergence to a fixed form in the critical case is hard to assess and calls for particular attention when dealing with experimental data. Overall we believe that this PRG should be considered as a better method to infer the presence of a critical state with respect to typical inference methods based on the identification of avalanches in both size and duration with particular exponents \cite{beggs2003avalanches, hesse2014self, munoz2018colloquium, fontenele2019criticality}.

\chapter*{Conclusions and perspectives}
\addcontentsline{toc}{chapter}{Conclusions and perspectives}\markboth{Conclusions and perspectives}{}
\lettrine{I}{n the course of this Thesis}, we have explored how ideas from Statistical Physics can help us understand the fundamental properties of complex systems. A recurring theme has been the presence of unobserved, stochastic environments that often influence the dynamical evolution of the observed degrees of freedom in non-trivial ways. The presence of such unobserved environments - which may be as complex as the internal d.o.f. themselves - is not uncommon in real-world systems, and their effects are sometimes poorly understood.

In the first part of this Thesis, we studied what information theory can teach us about stochastic processes of this kind \cite{nicoletti2021mutual, nicoletti2022mutual}. We found that mutual information, a quantity that measures pairwise dependencies, displays interesting properties when dealing with both internal interactions and stochastic environments. In particular, where interactions are linear - e.g., close to a fixed point where we can linearize the dynamics - two disentangled terms appear in the expression of the mutual information. One encodes solely the environmental dynamics, and depends on the features of the environment alone. The other, instead, only depends on the interactions between the internal degrees of freedom. In particular, in the case of discrete environmental states, the information induced by the unobserved environment is bounded by the entropy of its stationary distribution. This result is rather intuitive, as entropy quantifies nothing but our ignorance about environmental changes.

In more general cases where interactions are non-linear, we have shown that information interference might occur. That is, the dependencies induced by the environment and those arising from internal interactions may either mask or boost one another. These results describe how mutual information can be decomposed into interpretable terms when we only have access to some of the degrees of freedom of our system. Interestingly, in future works, these problems may be studied from a field-theoretical perspective, where the marginalization over the environment gives rise to new interaction vertices that are not present in the original theory. Ideally, this could allow for a much more general framework amenable to analytical treatments. One might ask whether a stochastic environment can be mapped into a set of effective couplings with defined properties, and if such couplings can be distinguished from the internal ones - or, as we have shown in the case of continuously varying diffusivities, they can be interpreted in a spatial sense. On the flip side, the ability to analytically deal with a class of stochastic processes with tools of information theory, as shown here, opens up many fascinating possibilities. A particularly appealing question is what happens when, instead of considering a stochastic environment, the system undergoes an external perturbation - notably, how the latter changes the information content and how such information evolves over time. A first step towards this direction might be to consider two diffusion processes in a finite domain that undergo a single stochastic jump and to study the persistence of the mutual information as a function of time, domain size, and boundary conditions of the system. These scenarios can be studied, for instance, by combining the physical interpretability and the analytical procedures behind our work together with tools from machine learning and data-driven approaches. This could lead to promising results in the quest to meaningfully disentangle the different sources of dependencies that emerge in complex systems.

Ultimately, we believe that the ideas presented in this Thesis may draw a path toward a deeper understanding of the different sources of couplings in real-world systems. Indeed, they are a starting point to elucidate the relations between their internal complexity and possibly equally complex, but unobserved, ever-changing environments.

Then, we further exploited ideas from information theory to understand the relation between complex high-dimensional models - where, once more, we are able to observe only some degrees of freedom - and low-dimensional effective models \cite{nicoletti2022information}. We outlined a general procedure for building optimal effective models, where optimality is defined in terms of information lost along the dynamical evolution of the original system. A paradigmatic yet physically relevant example is that of underdamped dynamics. Often, we are not interested in describing the full velocity-position phase space, but rather in the position space alone. Surprisingly, we found that the parameter space of the optimal model may not be continuous. This discontinuous, information-driven transition can be understood in terms of peaks of Fisher information, which quantifies how sensible the effective model is to changes in its parameters. The optimal effective model is only able to capture few features of the complex underlying evolution - for instance, either the initial transient behavior or long-term oscillations in the paradigmatic case of projecting an underdamped dynamics into an overdamped one. The information loss displays different minima, each of them associated with the features it captures best, that exchange stability. When this happens, the optimal effective model switches to a different ``phase'', and the Fisher information diverges.

These results help us understand the relation between high-dimensional dynamics and approximated, low-dimensional ones. In particular, the presence of an unforeseen transition - exclusively driven by the information-preserving constraint - may be a warning for more general inference procedures. Notwithstanding, one could think of different extensions, such as preserving thermodynamic properties or optimizing for different features in a Pareto-like framework. Ultimately, this part of the Thesis ties in with well-known ideas of dimensionality reduction, showing how effective dynamical representations may display properties that cannot be predicted from the high-dimensional one.

In the second part of the Thesis, we instead focused on how some of these results and methods from Statistical Physics in general can be applied to Neuroscience. A well-established idea is that neural dynamics in the brain, and in particular in the cortex, resembles the one of a system close to a phase transition. This ``critical brain hypothesis'' was first formulated with the observation of power-law distributed neuronal avalanches, cascades of activity that display a seemingly scale-free spatiotemporal organization. First, we studied how null models can help us understand the emergence of these critical signatures in neural activity \cite{mariani2022disentangling}. In particular, we showed how in local field potentials from the rat's somatosensory barrel cortex power-law neuronal avalanches coexist with a spatial correlation length that scales linearly with the system size. Inspired by the results of the previous Chapters, we developed archetypal models in which neural activity is driven by an extrinsic shared modulation, which is nothing but an unobserved stochastic environment.

We found that extrinsic modulation is enough to produce scale-free neuronal avalanches, whose exponents further obey the crackling-noise relation that is expected to hold at criticality. Importantly, these avalanches are generated by conditionally independent degrees of freedom and thus do not display other properties of real critical systems, such as finite-size scaling. Then, in the simple case of linear interactions inferred directly from the data, we leveraged the results on mutual information to show that the underlying mechanisms that generate avalanches and correlations are disentangled. The former are solely determined by the properties of the extrinsic modulation, whereas the latter emerge from the structure of the interactions - and both contribute to the mutual information in different, disentangled ways. These results fit in a well-established and fruitful research line that studies null mechanisms for the emergence of neuronal avalanches. Being able to disentangle such null mechanisms and more biological insightful properties (e.g., neural correlations) is instrumental in understanding what avalanches can teach us about neuronal and brain dynamics. Although the presence of a non-zero mutual information cannot be a sufficient condition for power-law avalanches to appear, in our extrinsic model their emergence does correspond to the onset of a non-vanishing dependence induced by an unobserved environment. This fact suggests a promising future perspective. By explicitly considering both the intrinsic activity and the extrinsic contributions, one might be able to combine all these considerations into a unified information-theoretic view - perhaps helping to unfold the underlying biological mechanisms at the origin of the observed signatures of criticality in neural activity.

Then, we investigated how the dynamics of simple models is affected by the underlying network topology \cite{barzon2022oscillations}. In this context, we focused on whole-brain activity - a much larger scale than the ones of LFPs. We studied analytically a stochastic version of the Greenberg-Hastings cellular automaton, which has been used to match functional resting-state networks found in fMRI data. We highlighted the presence of a bistable region between a high- and a low-activity phase, determined by the value of the activation threshold for each neuron. Crucially, such dynamical bistability is disrupted by the interplay between the underlying network sparsity and a sufficiently heterogeneous weight distribution - properties that are typically found in empirical connectomes of the human brain. In this scenario, a continuous critical-like transition emerges, with large autocorrelation times and variability of neural activity. At this transition, we also observe collective oscillations localized in different parts of the network, suggesting that both criticality and network structure play a fundamental role in driving the collective behavior of neurons. 

Overall, we were able to show in detail how network structure plays a fundamental, yet sometimes poorly understood, role. Therefore, this part of the work may serve as a baseline for future analytical efforts in explaining the nature of the observed transition under more relaxed assumptions, e.g., in the presence of a non-trivial distribution of weights and different topologies, to further understand the influence of both in the emergence of critical features in the human brain. All in all, we believe that our findings are a further contribution to the still puzzling ``critical brain hypothesis''.

Finally, in the last Chapter, we tried to understand what a phenomenological renormalization group can teach us about neural activity. We studied a PRG introduced in \cite{meshulam2019coarsegraining, meshulam2018arxiv}, that has two considerable advantages: it is model-independent, and it is stable with respect to the presence of long-range interactions. We tested its results both in equilibrium models, where we expect it to be able to distinguish between critical and non-critical phases, and in non-equilibrium ones, such as the contact process. We have found that the super and sub-critical regimes can be easily recognized, even though the nature of the phase transition is qualitatively different from the one of the Ising model. 

At the same time, we have highlighted that quasi-critical states are difficult to infer, especially in the case in which only a subset of physical quantities is analyzed. In non-trivial dynamical models, such as the contact process, the strategy that works best seems to be the one related to the correlation structure. For instance, the presence of non-Gaussian tails in the joint probability distribution of the coarse-grained variables in momentum space might be a signature of a possible underlying criticality, but at the same time clustering maximally correlated variables fails as we approach the critical point. Interestingly, in considerably simpler models of conditionally independent variables, the situation is reversed. Hence, in principle, one needs to study both the approach via maximally correlated pairs and in momentum space. Notably, this is the case of \cite{meshulam2019coarsegraining, meshulam2018arxiv}, where the authors found that neural activity in the mouse hippocampus seems to be described by a fixed point of this phenomenological renormalization flow.

However, this phenomenological renormalization is not able to tell us much about the nature of the underlying transition - whether it belongs to the directed percolation universality class, or it emerges from a transition between asynchronous and synchronous states. An interesting future direction would be to extend the application of this PRG to characterize different types of critical transitions in terms of coarse-grained variables. In general, extending these methods and testing them systematically might provide further insights into the understanding of the role of criticality in living systems. With the current approach, when taken individually, the signatures of the presence of an underlying fixed point might point in the wrong direction. This seems to be particularly relevant in the case of an external global parameter that couples effectively the units, even if they are independent to begin with. 

This, indeed, has been a recurrent theme throughout this Thesis - showing once more how an unobserved, stochastic dynamics may deeply affect otherwise simple systems. We can capture these features via information-theoretic measures, as well as study the non-trivial effect of unobserved modulation in paradigmatic models. As we have seen, this is particularly relevant for neural activity, where power-law avalanches have long been argued to be a signature of an underlying dynamical criticality. Such criticality, which also features scale-free correlations, may be especially relevant for information processing, which is a fundamental feature of biological systems in general. 

\begin{figure*}[t]
    \centering
    \includegraphics[width=\textwidth]{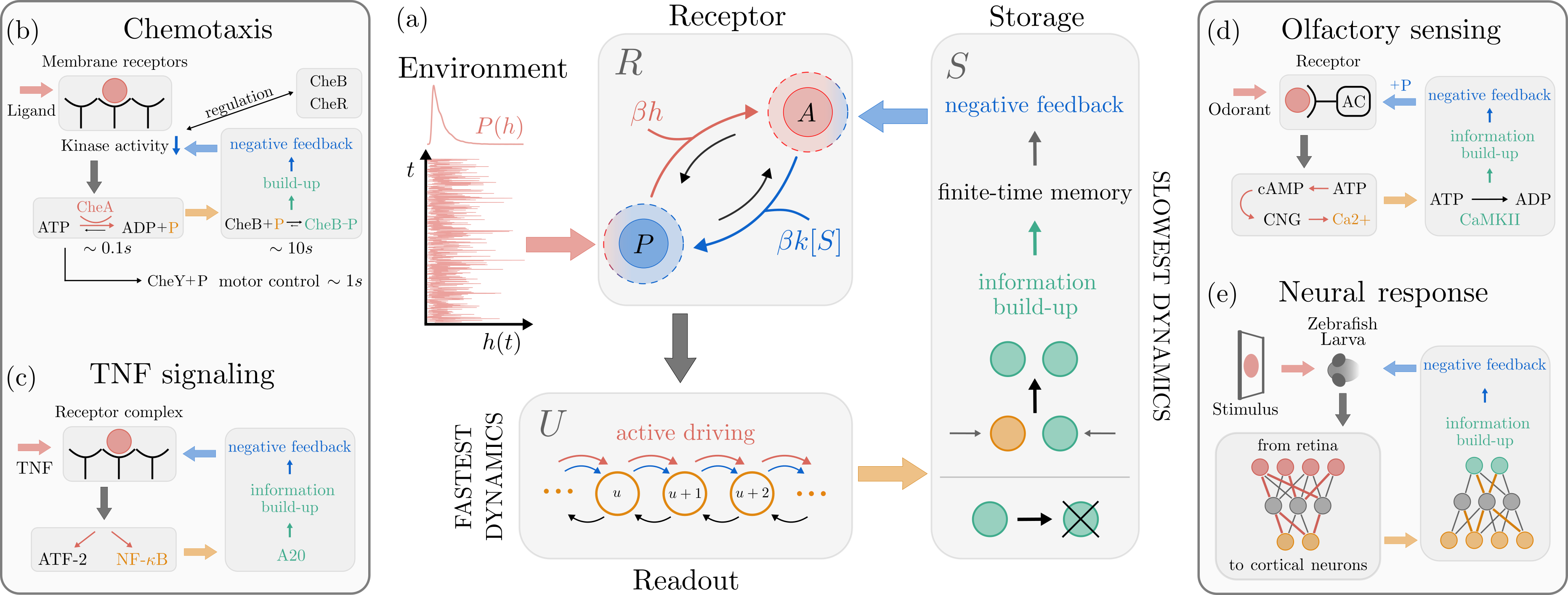}
    \caption{Sketch of a sensing model describing how biological systems acquire information about external environments. (a) A receptor $R$ can be either in an active ($A$) or passive ($P$) state, with transitions following two pathways, one used for sensing (red) and affected by the environment $h$, and the other (blue) modified by the storage concentration, $[S]$. An active receptor increases the response of a readout population $U$ (orange), which in turns stimulates the production of storage molecules $S$ (green) that provide a negative feedback to the receptor. (b) In chemotaxis, the input ligand binds to membrane receptors, regulating motor control and producing phosphate groups, whose concentration regulates the receptor methylation level. (c) In tumor necrosis factor (TNF) signaling, the nuclear factor NF-$\kappa$B is produced after receptor binding to TNF. NF-$\kappa$B modulates the encoding of the zinc-finger protein A20, which closes the feedback loop by inhibiting the receptor complex. (d) In olfactory sensing, odorant binding induces the activation of adenylyl cyclase (AC). AC stimulates a calcium flux, eventually producing phosphorylase calmodulin kinase II (CAMKII) which phosphorylates and deactivates AC. (e) In neural response, multiple mechanisms may take place at different scales. In zebrafish larvae, visual stimulation is projected along the visual stream from the retina to the cortex, a coarse-grained realization of the $R$-$U$ dynamics. Inhibitory populations and molecular mechanisms, such as short-term synaptic depotentiation, then lead to an adapted response upon repeated stimulation, and this adaptation can be understood in terms of stored information}
    \label{fig:conclusions:sensing}
\end{figure*}

In future works, we will combine these different perspectives to study systems with minimal mechanisms to acquire information about an external environment. Such mechanisms may have different sources and span different scales, from biochemical sensing in living systems to adaptation in neural circuits, memory formation, and storage \cite{sagawa2009minimal, govern2014energy, azeloglu2015signaling, ngampruetikorn2020energy}. In the quest for understanding the fundamental processes for information acquisition, a minimal model would encompass a receptor that dissipates energy to produce a readout in response to an external environment. In turn, the response favors the production of a slow and unobserved population that couples to the receptor via a negative feedback effect. By doing so, this unobserved population plays the role of storage or memory - it retains dynamically the information about the receptor activation and the system's readout response. That is, information about the external environment is coded in the system both at the readout level, which models the immediate response, and at the storage level, which allows for a dynamical modulation and thus acts as a memory. Such a minimal model is sketched in Figure \ref{fig:conclusions:sensing}a, where a two-state receptor is driven out of equilibrium by an external environment, and two entangled birth-and-death processes describe the produced readout and storage populations.

Importantly, these mechanisms are found at different scales and are declined in different biochemical networks. A prominent example is the modulation of flagellar motion operated by bacteria according to changes in the local nutrient concentration, known as chemotaxis \cite{tu2008modeling, celani2011molecular, lan2012energy}. Briefly, in E. Coli, the input ligand concentration is sensed by the membrane receptor-kinase complex, increasing the rate at which an intracellular kinase, CheA, hydrolyzes ATP, and producing phosphate groups. A response regulator protein, CheY, is then responsible for motor control. However, a feedback loop is present - methyltransferase CheR (R) and methylesterase/deamidase CheB (B) regulate the receptor methylation level via a negative feedback. This biochemical network is summarized in Figure \ref{fig:conclusions:sensing}b. Another crucial example is tumor necrosis factor (TNF) signaling \cite{cheong2011information}, where TNF binds to receptor complexes stimulating the production of the nuclear factor NF-$\kappa$B, which modulates the expression of the gene encoding for the zinc-finger protein A20. A20, then, acts as a negative feedback preventing sustained NF-$\kappa$B activation by inhibiting the receptor complex \cite{he2002a20}, see Figure \ref{fig:conclusions:sensing}b. Yet another case is that of olfactory sensing (Figure \ref{fig:conclusions:sensing}c) \cite{lan2012energy}, where odorant binding induces the activation of adenylyl cyclase (AC). AC is responsible for an inbound calcium flux, which interacts with calmodulin to activate AC phosphorylase calmodulin kinase II (CAMKII). CAMKII closes the feedback loop by phosphorylating and deactivating AC. Let us stress that a minimal description such as the one in Figure \ref{fig:conclusions:sensing}a does not need to be realized at a molecular level only. The same model, displaying information storage and dynamical adaptation, can be thought of as a coarse-grained version of neural response, where stimuli are projected from, e.g., the retina to the cortex, and the response of the cortex is modulated by sensory adaptation at a neural or synaptic level (Figure \ref{fig:conclusions:sensing}e). We will explore these networks and their emergent information dynamics in a future work \cite{nicoletti2023information}.

Further, a very much crucial but open question is how biological systems may then process the acquired information and act upon it. This ties in with the ideas of decision making, as well as its thermodynamics implications, and feedback mechanisms on the environment or other agents \cite{parrondo2015thermodynamics, azeloglu2015signaling, hidalgo2014information, roldan2015decision, tkacik2016information, dorpinghaus2017information, ngampruetikorn2020energy, manzano2021thermodynamics}. It is not far-fetched to suppose that internal processes and interactions among degrees of freedom may also evolve to optimize the information that a system has about changing environmental conditions \cite{tu2018adaptation} - possibly including spatial heterogeneity along the way - in such a way that it can act accordingly. Similarly, in cooperative or competitive settings, information about other agents or species is fundamental in order to optimize decision and survival chances - e.g., competition for finite resources in a game-theoretic description or among prey and predators in ecosystems.

Ultimately, one may seek to describe all of these aspects within a cohesive conceptual framework, ranging from information harvesting to optimal computational properties and information processing, hoping to find a common language for the principles governing vastly different scales - from the way individual organisms gather and process information, to the emergence of complex collective behaviors. And perhaps that language, in all its different and fascinating facets, will turn out to be Statistical Physics.

\restoregeometry

\appendix
\addtocontents{toc}{\protect\setcounter{tocdepth}{1}}
\addcontentsline{toc}{part}{Appendices}

\part*{Appendices}
\chapter{Distinguishing noise sources with Information Theory}
\label{app:viewpoint}
The following commentary ``\href{https://physics.aps.org/articles/v14/162}{Distinguishing noise sources with Information Theory}'' appeared as a viewpoint in the magazine ``Physics'' of the American Physical Society. Its aim is to illustrate to the non-technical public the results of our work \cite{nicoletti2021mutual}, which we derived in this Thesis in Chapter \ref{ch:PRL_PRE_1} and Chapter \ref{ch:PRL_PRE_2}. The commentary was written by prof. Katie Newhall, of the University of North Carolina at Chapel Hill, USA. The content of this Appendix, including displayed figures, is taken with permission from the published version, copyright 2021 by the American Physical Society.

\section{Physics 14, 162}
From neurons firing in the brain to chromosomes moving in a nucleus, biology is full of systems that can be modeled as a network of interacting particles in a noisy environment. But fundamental questions remain about how to reconstruct the particle interactions from experimental observations \cite{wang2016data} and about how to separate the noise coming from the environment from the intrinsic interactions of the system \cite{hilfinger2011separating}. Now, Giorgio Nicoletti of the University of Padua, Italy, and Daniel Busiello of the Swiss Federal Institute of Technology in Lausanne\footnote{Now at the Max Planck Institute for the Physics of Complex Systems, Dresden, Germany.} show how “mutual information” can help answer this second question \cite{nicoletti2021mutual}. Their theory disentangles the roles of the stochastic environment and the deterministic interaction forces of a system in creating correlations between the positions of two particles. This step is an important one toward understanding the role of environmental noise in real systems, where that noise can produce a response as complex as the signal produced by the system’s interactions.

\begin{figure*}[t]
    \centering
    \includegraphics[width=0.6\textwidth]{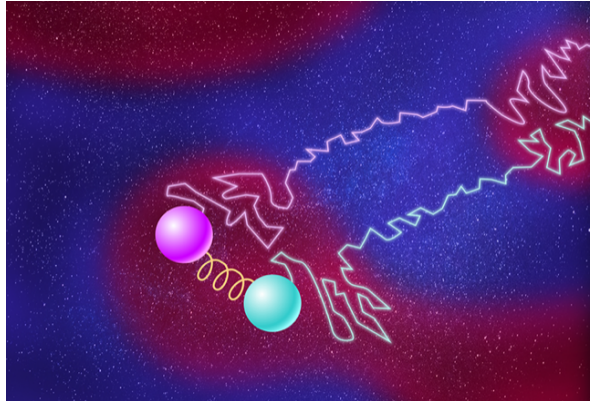}
    \caption{A cartoon showing two particles coupled by a spring transiting through a liquid that contains regions with different temperatures. The wiggly lines show the correlations that can arise from the intrinsic force of the spring and from the extrinsic temperature changes of the environment through which the particles move. Copyright APS/Carin Cain}
    \label{fig:app:cartoon}
\end{figure*}

Thinking back to my high school biology lessons, biological processes were always presented as deterministic, meaning there is no randomness in how they progress. For example, I was taught that a molecular motor marches rhythmically down a tubule, while ions traverse a membrane at regular intervals. But these processes are in fact inherently random and noisy: A molecular motor’s ‘‘foot’’ diffuses around before taking a step, and the motor can completely detach from the tubule; ion gates fluctuate between ‘‘open’’ and ‘‘closed’’ at random time intervals, making the flow of ions intermittent. This issue has led me and other researchers to wonder how these systems can properly function given the constant presence of fluctuations, or even more surprisingly, how these systems can gain improved function through these fluctuations.

Roughly speaking, the fluctuations and interactions associated with a system can be categorized as either “intrinsic” or “extrinsic,” where intrinsic fluctuations and interactions are typically those caused by the system itself (e.g., the random opening and closing of ion gates) and extrinsic fluctuations and interactions arise outside the system (e.g., thermal changes in the environment in which the ions diffuse or fluctuations in the ion-gate-activation signals). Extrinsic fluctuations can mask signals related to subtle internal changes of a system and add correlations that resemble those that arise from intrinsic interactions. Both factors make it more difficult to extract model parameters or even interaction laws. Nicoletti and Busiello address this issue by using mutual information theory, which quantifies the level of independence of two variables, to untangle internal and environmental correlations from a signal.

In their study, the duo considered two particles diffusing through a liquid that contains hot and cold patches of various sizes (Figure \ref{fig:app:cartoon}). Nicoletti and Busiello started by computing the effect of the environment on the particles’ motions. They modeled changes in temperature by randomly switching the temperature around the particles between hot and cool, an action that mimicked motion between patches of hot and cold liquid. They derived the joint probability distribution of the particles’ positions in the asymptotic limit that the temperature patches were large (slow temperature switching). Then they calculated the mutual information between the two particles, finding that it is exactly the Shannon entropy of the fluctuating environment. (In the asymptotic limit of small patches—quick temperature switching—the particles were independent and thus contained no environmental contribution to their mutual information.) Nicoletti and Busiello then repeated the calculations for two spring-coupled particles, finding that the mutual information between them is the superposition of the environmental mutual information for two uncoupled particles and that of the mutual information for interactions intrinsic to two coupled particles. Thus, they were able to separate intrinsic and extrinsic effects.

Being able to bound the effect of the environment on the dynamics of a biological process will aid in teasing out the properties of intrinsic interactions from data, for example in inferring the presence and strength of connections between neurons from voltage recordings. Since Nicoletti and Busiello’s calculations place bounds on the mutual information, rather than giving its definite value, the framework should be applicable to a wide range of models.

Information-theory-based methods have already been applied to solve neuroscience problems, such as reconstructing the connectivity networks of neurons in the brain \cite{zhou2014granger} and mapping the flow of information between regions of the brain \cite{walker2018inferring}. This new work has the potential to extend research in this direction, adding another tool for understanding these complex systems. The approach could also help in developing physically interpretable machine-learning approaches for probing biological systems. In a broad sense, machine-learning techniques take noisy data, which may contain latent variables, and disentangle those data. (Latent variables are like environmental noise in that they are not directly observed.) But machine learning remains a black-box approach, something the new method could help change.

There may also be another impact of Nicoletti and Busiello’s work: understanding the effect of the timescale of the noise. The duo showed that the speed at which the temperature of the environment changed impacted the amount of mutual information between the particles from extrinsic fluctuations. This change in how much the system responded to noise at different timescales points to an emerging picture in biology: Noise is not always a fluctuation about an average that can be ignored or a nuisance that hinders measurement of a parameter. At the proper timescale, noise can be important to the function of a system \cite{walker2018inferring, walker2022numerical, patel2019limited}. It is thus important that researchers continue to develop tools to analyze systems that include noise.
\chapter{Computational and numerical methods}
\label{app:computational}
\lettrine{I}{n this Appendix}, we briefly review the main computational techniques used in this Thesis. In most of the Chapters, we simulated stochastic differential equations, either with additive or multiplicative noise, with a standard Euler-Maruyama algorithm \cite{stickler2016basic}. Notably, in the specific case of Ornstein-Uhlenbeck processes, which we often considered, we employed an exact stochastic algorithm that allowed us to overcome the shortcomings of discrete-time \cite{gillespie1996OU}. When the microscopic dynamic is known, such as in Chapter \ref{ch:jphys}, the Gillespie algorithm allows in general for exact integration of the stochastic dynamics \cite{gillespie1977exact}.

In Chapter \ref{ch:PRL_PRE_1} and Chapter \ref{ch:PRL_PRE_2}, we also used Monte Carlo techniques to sample the equilibrium distribution in the presence of non-linear potentials using Hamiltonian Monte Carlo \cite{neal2011mcmc, betancourt2017conceptual}. Monte Carlo sampling was used to evaluate the non-trivial integral of mutual information when not possible analytically, starting from samples of the joint distribution \cite{stickler2016basic}. Here, for consistency, we summarize the main concepts and procedures we used to obtain the presented results.

\section{Simulations of stochastic processes}
Let us consider a generic Langevin equation of the form
\begin{equation}
    \label{eqn:appA:langevin}
    \dot{x}_i(t) = f_i(\vb{x},t) + \sum_j g_{ij}(\vb{x},t) \xi_j(t)
\end{equation}
where $\xi_j(t)$ is a white noise. To simulate Eq.~\eqref{eqn:appA:langevin} up time $T$, we discretize the time interval $[0,T]$ into $N$ subintervals of duration $\Delta t > 0$. Approximating the temporal derivative with a first-order Taylor expansion, we end up with
\begin{equation}
    \label{eqn:appA:langevin_numerical}
    x_i(t + \Delta t) = x_i(t) + \Delta t f(\vb{x}(t), t) + \sqrt{\Delta t}\sum_j g_{ij}(\vb{x}(t), t)u_i
\end{equation}
where $u_i$ are independent and identically distributed random variables with zero mean and unitary variance. For Eq.~\eqref{eqn:appA:langevin_numerical} to be a good approximation of Eq.~\eqref{eqn:appA:langevin}, as usual, we need to choose a small enough timestep $\Delta t$. Furthermore, in the presence of absorbing states - e.g., with multiplicative noise $g_{ij}(\vb{x}, t) = \sqrt{x_i} \delta_{ij}$ and $f_i(\vb{x}, t) = x_i \tilde{f}_i(\vb{x}, t)$, so that $x_i = 0$ for all degrees of freedom $i$ is an absorbing configuration - one needs to be careful that we need to ensure $x_i(t + \Delta t) = 0$. This is known as the Euler-Maruyama algorithm.

Although higher-order discretization schemes exist to simulate the dynamics of Eq.~\eqref{eqn:appA:langevin}, the Euler-Maruyama one is the most computationally tractable, and thus we use it throughout this Thesis. To ensure that our results are stable, we check that they do not change as we decrease our choice of $\Delta t$. However, for the specific case of the Ornstein-Uhlenbeck process, we are able to simulate the stochastic dynamics without approximations due to the discretization of time \cite{gillespie1996OU}. We consider the univariate Ornstein-Uhlenbeck process of the form
\begin{equation}
    \label{eqn:appA:langevin_OU}
    \dot{x}(t) = -\frac{1}{\tau}x(t) + \sqrt{D}\xi(t)
\end{equation}
where $\xi(t)$ is a white noise, $D$ is the diffusion coefficient, and $\tau$ is the relaxation time. It can be shown that, for any given timestep $\Delta t$, Eq.~\eqref{eqn:appA:langevin_OU} corresponds to the update formula
\begin{equation}
  x(t + \Delta t) = x(t) \mu + \sigma u  
\end{equation}
where
\begin{gather*}
    \label{eqn:appA:langevin_OU_numerical}
    \mu = e^{-\Delta t / \tau} \\
    \sigma^2 = \frac{D \tau}{2}(1 - \mu^2).
\end{gather*}
As expected, Eq.~\eqref{eqn:appA:langevin_OU_numerical} reduced to the Euler-Maruyama discretization scheme in the limit $\Delta t \to 0$.

\subsection{Gillespie algorithm}
The Gillespie algorithm \cite{gillespie1977exact} allows for exact updates of a stochastic system specified by a given set of microscopic reactions. Let us assume that there are $\mu = 1, \dots, N$ sites in the network and $i = 1, \dots, M$ possible transitions - e.g., in the contact process we have two transitions, $n_\mu \to n_\mu + 1$ with a rate $\lambda/k_\mu \sum_{\nu \in \partial \nu} \delta(n_\nu, 1)$ and $n_\mu \to n_\mu - 1$ with unitary rate, where $n_\mu$ is the number of particles in site $\mu$.

At each time, the network can be associated with a propensity matrix $A_{\mu i}^{(t)}$. Each row of $A_{\mu i}^{(t)}$ is given by the transition rates that the $\mu$-th site can undergo, given its state at time $t$. We introduce the total propensity $\alpha_0^{(t)} = \sum_\mu\sum_i A_{\mu i}^{(t)}$, so that the waiting time for the next transition is given by
\begin{equation}
    \tau^{(t)} = - \bigl(\alpha_0^{(t)}\bigl)^{-1}\log u
\end{equation}
where $u$ is uniformly distributed in $[0,1]$. Then, the transition $\bar{i}$ that occurs and the site $\bar{\mu}$ at which it occurs are such that
\begin{equation}
    \sum_{\mu = 1}^{\bar{\mu}-1}\sum_{i = 1}^{\bar{i}-1}A_{\mu i}^{(t)}\le \alpha_0^{(t)} v < \sum_{\mu = 1}^{\bar{\mu}}\sum_{i = 1}^{\bar{i}}A_{\mu i}^{(t)}
\end{equation}
where $v$ is once again uniformly distributed in $[0,1]$. We then update $A_{\bar{\mu}i}$ with the new transition rates for $\bar{\mu}$ and set the time to $t+\tau$. The result is an exact numerical integration of the corresponding master equation. This scheme is particularly useful in models such as the ones in Chapter \ref{ch:jphys}, where the microscopic reactions and the network structure are known.

\section{Monte Carlo sampling}
In this Thesis, and in the first part in particular, we often need to sample different types of probability distributions, and we do so using Monte Carlo methods. This is particularly relevant to compute the mutual information of a joint probability distribution $p_{XY}$,
\begin{equation}
    \label{eqn:appA:mutual}
    I_{XY} = \int dx dy \, p_{XY}(x,y) \log \frac{p_{X}(x,y)}{p_X(x)p_Y(y)} = \ev{\log\frac{p_{XY}}{p_Xp_Y}}_{XY},
\end{equation}
where $p_X$ and $p_Y$ are the marginal distributions, and $\ev{}_{XY}$ is the average with respect to $p_{XY}$. A direct computation of Eq.~\eqref{eqn:appA:mutual} is often challenging due to numerical instabilities in regions with low probability.

To avoid these issues, we compute the expected value in Eq.~\eqref{eqn:appA:mutual} using Monte Carlo techniques \cite{stickler2016basic, landau2021guide}. Let us assume that we can sample the probability distribution $p_{XY}$, obtaining a set of $M$ samples $\{(x_i, y_i)\}$ for $i = 1, \dots, M$. Then, we can approximate the mutual information as
\begin{equation}
    \label{eqn:appA:mutual_MC}
    I_{XY} \approx \sum_{i = 1}^M \log \frac{p_{XY}(x_i,y_i)}{p_X(x_i)p_Y(y_i)}.
\end{equation}
Notably, the estimator in Eq.~\eqref{eqn:appA:mutual_MC} requires the knowledge of the analytical expression of the joint distribution $p_{XY}$ and the marginals $p_X$ and $p_Y$. When such expressions are not known, we need to resort to other estimators, for instance, a $k$-neighbor estimator \cite{kraskov2004estimating, holmes2019estimation}, or by approximating both the joint distribution and the marginals one with their histograms, obtaining samples from, e.g., a corresponding Langevin equation.

For instance, if $p_{XY}$ is a Gaussian mixture $p_{XY} \sim \sum_\mu \pi_\mu \mathcal{N}(m_\mu, \sigma_\mu)$ with $\sum_\mu \pi_\mu = 1$ - as in the case of linear relaxation and linear interaction in Chapter \ref{ch:PRL_PRE_1} and Chapter \ref{ch:PRL_PRE_2} - we can sample it efficiently by obtaining samples from $ \mathcal{N}(m_\mu, \sigma_\mu)$ with probability $\pi_\mu$. However, it is often the case that we do not know how to build the set $\{(x_i, y_i)\}$. To this end, we will briefly review a powerful sampling method, known as Hamiltonian Monte Carlo \cite{neal2011mcmc, betancourt2017conceptual}.

\subsection{Hamiltonian Monte Carlo}
We consider, for consistency with Chapter \ref{ch:PRL_PRE_1} and Chapter \ref{ch:PRL_PRE_2}, stationary probability distributions of equilibrium processes, i.e., Boltzmann distributions of the form
\begin{equation}
    \label{eqn:appA:boltzmann}
    p_{X}(\vb{x}) =\frac{1}{Z} e^{-V(\vb{x})}
\end{equation}
where $Z$ is the partition function. However, Hamiltonian Monte Carlo is more general and works for any kind of probability distribution. We introduce auxiliary momenta, $\bm{\pi}$, with a corresponding Hamiltonian
\begin{equation*}
    H_\pi(\pi_X, \pi_Y) = \frac{1}{2}\bm{\pi}^T M \bm{\pi}
\end{equation*}
so that the momenta are distributed as a Gaussian. For simplicity, we set $M = \mathbb{1}$. Hence, our Hamiltonian is given by
\begin{equation}
    \label{eqn:appA:hamiltonian}
    H(\vb{x}, \bm{\pi}) = V(\vb{x}) + \frac{\bm{\pi}^T \bm{\pi}}{2}
\end{equation}
where, for non-Boltzmann distribution, the potential is simply given by $-\log p_{XY}$. We now integrate numerically Hamilton's equations,
\begin{gather*}
    \dv{\bm{x}}{t} = \pdv{H}{\bm{\pi}} = \bm{\pi} \\
    \dv{\bm{\pi}}{t} = - \pdv{H}{\bm{x}} = -\pdv{V}{\vb{x}},
\end{gather*}
using a symplectic algorithm such as the leapfrog integrator with step $\Delta t$,
\begin{equation}
    \label{eqn:appA:hamiltonian_MC}
    \begin{gathered}
        \bm{\pi}\left(t + \frac{\Delta t}{2}\right) = \bm{\pi}(t) - \frac{\Delta t}{2} \pdv{V(\vb{x}(t))}{\vb{x}} \\
        \bm{x}(t + \Delta t) = \bm{x}(t) + \Delta t \bm{\pi}\left(t + \frac{\Delta t}{2}\right) \\
        \bm{\pi}(t + \Delta t) = \bm{\pi}\left(t + \frac{\Delta t}{2}\right) - \frac{\Delta t}{2} \pdv{V(\vb{x}(t + \Delta t))}{\vb{x}}
    \end{gathered}
\end{equation}
where the initial momenta are sampled from a standard Gaussian distribution. We then perform a Metropolis-Hastings acceptance step \cite{stickler2016basic, neal2011mcmc, betancourt2017conceptual}, in order to move towards the energy minimum. In particular, we perform $L$ steps of the leapfrog algorithm and propose to accept as a sample of $p_X$ the point $\bm{x}(L \Delta t)$. To avoid problems with the proposal distribution - which arise due to the deterministic nature of Hamilton's equation - we flip the momenta and accept the transition from $\bm{x}(0)$ to $\bm{x}(L\Delta t)$ with probability
\begin{equation}
    \alpha = \min\left(1, e^{H(\bm{x}(0), \bm{\pi}(0)) - H(\bm{x}(L\Delta t), - \bm{\pi}(L\Delta t))}\right).
\end{equation}
Then, we resample the momenta $\bm{\pi}(0)$ and repeat the procedure until we collect enough samples.

\subsection{Cluster algorithms for the Ising model}
Monte Carlo methods can be used to sample the equilibrium probability distribution of the Ising model at a given temperature. In particular, in Introduction and in Chapter \ref{ch:PRR}, we used the Wolff algorithm, a cluster update algorithm that is especially efficient around the critical point \cite{wolff1989collective}. 

As in Hamiltonian Monte Carlo, every metropolis-Hastings scheme is built from a proposal distribution $\mathcal{P}(a \to b)$ for the transition from state $a$ to state $b$, and an acceptance distribution $\mathcal{A}(a \to b)$ for the same transition \cite{stickler2016basic}. The Metropolis-Hastings algorithm, which follows from detailed balance, prescribes
\begin{equation}
    \label{eqn:appA:metropolis_hastings}
    \mathcal{A}(a\to b) = \min \left(1, \frac{p(b)}{\mathcal{P}(a \to b)}\frac{\mathcal{P}(b \to a)}{p(a)}\right)
\end{equation}
where $p(a)$ is the probability of state $a$. For the Ising model, spin-flip algorithms are local algorithms with a trivial proposal distribution - so that the acceptance probability only depends on the ratio of the energy at the two configurations - but are not particularly efficient, because the acceptance probability may be small. The Wolff algorithm, on the other hand, builds a proposal distribution in such a way that the acceptance probability is always one. 

To do so, in a given configuration $a$ we randomly select one spin and recursively add neighboring aligned spin with probability $q$. That is, if spin $i$ is in the cluster and $j \in \partial i$ is not, and if $S_i = S_j$, then we add $j$ to the cluster with probability $q$. Notice that the same bond is never considered more than once for activation. Once a cluster has been built, we flip it, leading to a configuration $b$. The proposal distribution is then given by 
\begin{equation*}
    \mathcal{P}(a \to b) \propto (1 - q)^{n_\mathrm{same}} 
\end{equation*}
where $n_\mathrm{same}$ is the number of links at the boundary of the clusters that have been rejected, i.e., the links from $i$ to $j$ such that $S_i = S_j$ and that are not part of the cluster. If we call $n_\mathrm{diff}$ the number of links at the boundary with $S_i \ne S_j$, we have an acceptance probability
\begin{equation*}
    \mathcal{A}(a \to b) = \min\left(1, \frac{e^{-\beta(n_\mathrm{diff} - n_\mathrm{same})}}{(1 - q)^{n_\mathrm{same}}} \frac{(1 - q)^{n_\mathrm{diff}}}{e^{-\beta(n_\mathrm{same} - n_\mathrm{diff})}}\right)
\end{equation*}
where we used the fact that all links inside the cluster give the same contribution in the energy, and thus in the probability, before and after the flip. This leads to a unitary acceptance probability if
\begin{equation}
    q = 1 - e^{-2\beta J}
\end{equation}
where $J$ is the coupling strength of the Ising model. In this way, once a cluster is built and flipped, the next configuration is a sample of the equilibrium distribution. Furthermore, having flipped a typically large number of spins, its correlation with the previous configuration is much lower than a single-flip algorithm, such as a standard Metropolis scheme. 
\chapter{Information in multiple stochastic processes and environmental states}
\chaptermark{Information in multiple stochastic processes}
\label{app:generalization_jumps}
\lettrine{I}{n this Appendix}, we briefly consider a generalization of the calculations highlighted in Chapter \ref{ch:PRL_PRE_1} and Chapter \ref{ch:PRL_PRE_2}. We consider $N$ stochastic processes described by the internal degrees of freedom $\vb{x} = \{x_1, \dots, x_2\}$, and $M$ environmental states indexed by $i = 1, \dots, M$. Let us recall Eq.~\eqref{eqn:p2c1:internal_langevin},
\begin{align}
    \label{eqn:appB:internal_langevin}
    \frac{d x_\mu}{dt} = F_\mu(\vb{x}; \{\zeta\}) + \sqrt{2 G_\mu\left(\vb{x}; \{\kappa\}\right)} \xi_\mu \qquad\qquad \mu = 1, \dots, N
\end{align}
where $\{\xi_\mu\}$ is a set of independent white noises, $F_\mu(\vb{x};\{\zeta\})$ is the $\mu$-th component of a force field, and $G_\mu(\vb{x};\{\kappa\})$ is the diffusion coefficient of the $\mu$-th particle. We assume that this system is described by the stationary probability distribution $p(\vb{x})$.

Since we have $N$ variables, we need to choose a suitable generalization of the mutual information. However, such generalizations are troublesome from an information-theoretic perspective \cite{ThomasCover2006, williams2010nonnegative}. Thus, we focus on the case in which the variables are not interacting, i.e.,
\begin{equation*}
    F_\mu(\vb{x}; \{\zeta\}) = \sum_{\mu = 1}^N F_\mu(x_\mu; \{\zeta\}).
\end{equation*}
In this scenario, we may be interested in the factorizability of the joint probability distribution $p(\vb{x})$ with respect to its full factorization, since all pairs of variables are equivalent. That is, we want to compute the Kullback-Leibler divergence
\begin{equation}
\label{eqn:appB:multivariate_information}
    I_N = \int \prod_{\mu = 1}^N dx_\mu p(x_1, \dots, x_N) \log\frac{p(x_1, \dots, x_N)}{\prod_{\mu = 1}^N p(x_\mu)} = \sum_{\mu = 1}^N H_\mu - H_{1, \dots, N}
\end{equation}
where $H_\mu$ is the entropy of the marginalized distribution and $H_{1, \dots, N}$ is the entropy of the joint distribution. This is nothing but the information distance between the joint probability distribution and the product of the single-variable distributions. Thus, this quantity is always positive, and for $N=2$ gives exactly the mutual information. Albeit improperly, we refer here to this quantity as ``multivariate information''.

\section{N stochastic processes and M environmental states}
We focus on the limit of slow jumps, so that we end up with a probability distribution that is not trivially factorizable. Let us write the stationary limit of the one variable probability distributions as
\begin{equation}
\label{eqn:appB:mixture_1D}
    p^\mathrm{slow}_\mu(x_\mu) = \sum_{i=1}^M \pi^\mathrm{st}_i P_{\mu i}^\mathrm{st}(x_\mu)
\end{equation}
and the $N$ variables probability distribution as
\begin{equation}
\label{eqn:appB:mixture_2D}
    p^\mathrm{slow}_{1, \dots, N}(\vb{x}) = \sum_{i=1}^M \pi^\mathrm{st}_i \prod_{\mu = 1}^N P_{\mu i}^\mathrm{st}(x_\mu),
\end{equation}
where $P_{\mu i}^\mathrm{st}(x_\mu)$ solves the Fokker-Planck equation
\begin{equation*}
    0 = \sum_{\mu = 1}^N \biggl[-\partial_\mu\left(F_\mu(\vb x)P_{\mu i}^{\mathrm{st}}(\vb{x})\right) + \partial_\mu^2 \left(D_i P_{\mu i}^{\mathrm{st}}(\vb{x}) \right) \biggr],
\end{equation*}
as shown in Chapter \ref{ch:PRL_PRE_1}.

In order to study the multivariate information in Eq.~\eqref{eqn:appB:multivariate_information} we need to bound the entropies of these distributions, which do not admit a closed form. Recall that, from \cite{Kolchinsky2017}, we can write an upper and a lower bound starting from the estimator
\begin{equation}
\label{eqn:appB:entropy_estimator}
    \hat{H}_\mu = \sum_i \pi^\mathrm{st}_i H(P_{\mu i}^\mathrm{st}) - \sum_i \pi^\mathrm{st}_i \log \left[\sum_{j}\pi^\mathrm{st}_j e^{-d(P_{\mu i}^\mathrm{st} || P_{\mu j}^\mathrm{st})}\right]
\end{equation}
where $d(P_{\mu i}^\mathrm{st} || P_{\mu j}^\mathrm{st})$ is any distance function in the probability distributions space. We note that
\begin{align*}
    H\left(\prod_{\mu = 1}^N P_{\mu i}^\mathrm{st}\right) & = -\int dx_1 \dots dx_N \, \prod_{\mu = 1}^NP_{\mu i}^\mathrm{st}(x_\mu) \log \left[\prod_{\mu = 1}^NP_{\mu i}^\mathrm{st}(x_\mu)\right] = \sum_{\mu = 1}^N H(P_{\mu i}^\mathrm{st})
\end{align*}
so the first part of Eq.~\eqref{eqn:appB:entropy_estimator} for the entropy of the joint probability distribution is exactly equal to the sum of the estimators of the entropy of the one variable distributions. This is a direct consequence of the fact that the components of the joint distribution, Eq.~\eqref{eqn:appB:mixture_2D}, are factorizable - i.e., that we are considering a non-interacting model. Thus, if we build the corresponding estimator for the multivariate information, we are left with
\begin{align}
\label{eqn:appB:estimator_multivariate}
    \hat{I}_{N,\mathrm{env}} = - \sum_i \pi^\mathrm{st}_i \log \frac{\prod_{\mu = 1}^N\left(\sum_{j}\pi^\mathrm{st}_j e^{-d_1(P_{\mu i}^\mathrm{st} || P_{\mu j}^\mathrm{st})}\right)}{\sum_{j}\pi^\mathrm{st}_j e^{-d_N(\prod_{\mu = 1}^N P_{\mu i}^\mathrm{st} || \prod_{\mu = 1}^N P_{\mu j}^\mathrm{st} )}}
\end{align}
where we denote as $d_1(\cdot ||\cdot)$ the distance function we choose for the one variable entropies and as $d_N(\cdot ||\cdot)$ the distance function we choose for the $N$ variables entropy.

Recall that, following \cite{Kolchinsky2017}, a lower bound for the entropy is achieved when we choose as a distance function the Chernoff-$\alpha$ divergence
\begin{equation*}
    C_\alpha(p || q) = -\log \int dx \,  p^\alpha(x) q^{1-\alpha}(x)
\end{equation*}
for any $\alpha \in [0,1]$, and an upper bound is instead achieved when we use a simple Kullback-Leibler divergence
\begin{equation*}
    D_{KL}(p || q) = \int dx\, p(x) \log\frac{p(x)}{q(x)}.
\end{equation*}
Therefore, Eq. \ref{eqn:appB:estimator_multivariate} is a lower bound if we choose the Chernoff-$\alpha$ divergence for the one variable entropy and the Kullback-Leibler divergence for the $N$ variables entropy, and it is an upper bound is we make the opposite choice.

Both this upper and lower bound saturate in two particular cases. The first is the one in which $C_\alpha(\cdot || \cdot)$ diverges for all $i \ne j$. In fact, the Jensen inequality implies that
\begin{equation*}
    C_\alpha(\cdot || \cdot) \le (1-\alpha) D_{KL} (\cdot || \cdot),
\end{equation*}
hence if the Chernoff-$\alpha$ divergence diverges so does the Kullback-Leibler divergence. In this case, both the upper and lower bound converge to the same expression, and the estimator of the mutual information is exact. We find
\begin{align}
    \label{eqn:appB:residual_mutual}
    I_{N,\mathrm{env}}^{\infty} = -\sum \pi^\mathrm{st}_i \log \frac{(\pi^\mathrm{st}_i)^N}{\pi^\mathrm{st}_i} = (N-1) H_\text{jumps}.
\end{align}
Qualitatively, this means that the probability distribution $P_{\mu i}^\mathrm{st}(x)$ is infinitely different from $P_{\mu j}^\mathrm{st}(x)$, so the discrete jumps between the $D_i$ states generate an infinitely different dynamics in terms of its stationary states. Notably, this results does not depend on $M$ explicitly, provided that the distances diverge for all pairs of environmental states. In the case $N=2$, we find once more the results of Chapter \ref{ch:PRL_PRE_1} and Chapter \ref{ch:PRL_PRE_2}.

The second, albeit trivial, case is the one in which the distances between both $P_{\mu i}^\mathrm{st}$ and $P_{\mu j}^\mathrm{st}$ are zero. Once more, both the upper and the lower bounds given by Eq. (\ref{eqn:appB:estimator_multivariate}) saturate and we find
\begin{align}
    \label{eqn:appB:residual_mutual_zero}
    I_{N,\mathrm{env}}^0 = -\sum \pi^\mathrm{st}_i \log 1 = 0
\end{align}
which amounts to the trivial statement that if the two mixtures of Eqs.~\eqref{eqn:appB:mixture_1D}-\eqref{eqn:appB:mixture_2D} have the same components, then the joint probability is also factorizable.

These results have a nice intuitive explanation. In fact, as long as $D_i$ is fixed the processes described by Equation (\ref{eqn:appB:internal_langevin}) are independent and thus they cannot share any information. The only moment in time in which they are effectively coupled is when a jump $D_i \to D_j$ happens, when they share the sudden change in the diffusion coefficient - from then on, as long as $D_j$ is fixed, they evolve independently once more. As these changes are instantaneous, the greatest amount of information the processes can share corresponds to the entropy of the jumps, which is achieved when the processes are infinitely distinguishable for different diffusion coefficients $D_i$. In terms of information theory, the entropy of the jumps corresponds to our ignorance on the system, that is, since the jumps are stochastic we do not know when they happen.

\section{Linear interactions}
Let us briefly consider the case of $N$ interacting particles, where the interactions are linear. That is, we consider
\begin{align}
\label{eqn:appB:linear_langevin}
    \dv{x_\mu}{t} = -\sum_{\nu}A_{\mu\nu}\frac{x_\nu}{\tau} + \sqrt{2 D_{i(t)}} \xi_\mu(t)
\end{align}
so that, in the slow-jumps limit, the system is described by a Gaussian mixture
\begin{align}
    p_\mathrm{slow}(x_1, \dots, x_N) = \sum_{i = 1}^M \pi^\mathrm{st}_i\frac{1}{\sqrt{(2\pi)^N \det\bm{\Sigma}_i}} \exp[-\frac{1}{2}\bm x^T \bm\Sigma_i\bm x]
\end{align}
where $\bm{\Sigma}_i$ are symmetric matrices obeying the Lyapnuov equation
\begin{equation}
    \label{eqn:appB:lyapunov_equation}
    \bm{A}\bm{\Sigma}_i + \bm\Sigma_i \bm{A}^T = 2 \tau D_i \mathbb{1}
\end{equation}
with $\mathbb{1}$ the identity matrix. As in Chapter \ref{ch:PRL_PRE_2}, we can introduce the rescaled covariance matrix $\tilde{\bm\Sigma}_i = \bm\Sigma_i / (D_i \tau)$, which is determined by the equation
\begin{equation}
\label{eqn:appB:rescaled_covmat}
    \bm\Sigma_i = D_i \tau \tilde{\bm\Sigma}
\end{equation}
so that $\tilde{\bm\Sigma}$ carries no environmental dependencies.

Then, if we want to compute the multivariate information $I^{(N)}$ we need the entropies of the mixture components, namely
\begin{equation}
    H^{(i)}_{1, \dots, N} = \frac{1}{2}\left[N \log(2\pi e \tau D_i) + \log\det\tilde{\bm\Sigma} \right]
\end{equation}
and
\begin{equation}
    H^{(i)}_\mu = \frac{1}{2}\left[\log(2\pi e \tau D_i) + \log\tilde{\bm\Sigma}_{\mu\mu} \right].
\end{equation}
Due to the interactions, it is not anymore the case that $H^{(i)}_{1, \dots, N}$ is exactly equal to $\sum_{\mu} H^{(i)}_\mu$. The bounds on the multivariate information become
\begin{align}
\label{eqn:appB:bound_interacting}
    I^{\mathrm{slow,up/low}}_N = \frac{1}{2} \log\left[\frac{\prod_\mu \tilde{\bm\Sigma}_{\mu\mu}}{\det \tilde{\bm\Sigma}}\right] + I_{N,\mathrm{env}}^{\mathrm{slow,up/low}}\left(\left\{\frac{D_i}{D_j}\right\}, \left\{\pi^\mathrm{st}_i\right\}\right)
\end{align}
where the first term comes solely from the interaction matrix $A_{\mu\nu}$, and $\{D_i/D_j\}$ is the set of the ratio of all diffusion coefficients, for $i = 1, \dots, N \ne j$. That is, the contribution of the interactions is disentangled from the one of the switching environment. The environmental term can be computed from the divergences
\begin{align*}
    C_\alpha(\bm\Sigma_i || \bm \Sigma_j) & = \frac{1}{2}\log\frac{\det \left[(1-\alpha)\bm\Sigma_i+ \alpha\bm\Sigma_j)\right]}{\det^{1-\alpha} \bm\Sigma_i \det^{\alpha} \bm\Sigma_j} \\
    & = \frac{1}{2}\log\frac{\det \tau \tilde{\bm\Sigma}\left[(1-\alpha)D_i+ \alpha D_j\right]}{\det^{1-\alpha} \tau D_i\tilde{\bm\Sigma} \det^{\alpha} \tau D_j\tilde{\bm\Sigma}} \\
    & = \frac{N}{2}\log \frac{\left[(1-\alpha)D_i+ \alpha D_j\right]}{D_i^{1-\alpha}D_j^\alpha}
\end{align*}
and
\begin{align*}
    D_{KL}(\bm\Sigma_i || \bm \Sigma_j) & = \frac{1}{2}\left[\log\frac{\det\bm\Sigma_j}{\det\bm\Sigma_i} + \Tr\bm\Sigma_j^{-1}\bm\Sigma_i - N\right] = \\
    & = \frac{1}{2}\left[\log\frac{\det\tau D_j\tilde{\bm\Sigma}}{\det\tau D_i\tilde{\bm\Sigma}} + \Tr \frac{1}{\tau D_j}\tilde{\bm\Sigma}^{-1}\tau D_i\tilde{\bm\Sigma} - N\right] \\
    & = \frac{N}{2}\left[\log\frac{D_j}{D_i} + \frac{D_j}{D_i} - 1\right].
\end{align*}
If we set $N=2$ we recover the two variables case considered in Chapter \ref{ch:PRL_PRE_1} and Chapter \ref{ch:PRL_PRE_2}, but these results hold for any number of particles $N$. Furthermore, since marginalization of Gaussian distributions amount to excluding the respective rows and column of the covariance matrices, these calculations can be easily generalized for Kullback-Leibler divergences between different products of the marginal probabilities. In general, due to the factorization of the covariance matrix, the bounds are the same as the ones of the non interacting case and they only depend on the ratios $D_i/D_j$. Hence, in the limit in which all the distances between the mixture components diverge, we are left with
\begin{equation}
 I_N^\mathrm{slow} \to \frac{1}{2} \log\left[\frac{\prod_\mu \tilde{\bm\Sigma}_{\mu\mu}}{\det \tilde{\bm\Sigma}}\right] - (N-1) \sum_{i = 1}^M \pi^\mathrm{st}_i \log\pi^\mathrm{st}_i
\end{equation}
where, once more, the jump entropy appearing in the last terms carries a $N-1$ multiplicative factor.

Finally, in the case studied in the main text the interaction matrix is given by
\begin{equation}\renewcommand*{\arraystretch}{1.2}
    \bm A = \begin{pmatrix}
                1 & -g_1 \\
                -g_2 & 1
            \end{pmatrix}
\end{equation}
hence the solution to the Lyapunov equation is Eq. \eqref{eqn:p2c2:rescaled_covmat_sol},
\begin{equation*}
    \tilde{\bm\Sigma} = \frac{1}{g_1 g_2 - 1}\begin{pmatrix}
                        \frac{g_1(g_2-g_1)}{2} - 1 & -\frac{g_1+g_2}{2} \\
                        -\frac{g_1+g_2}{2} & \frac{g_2(g_1-g_2)}{2} - 1
                        \end{pmatrix}.
\end{equation*}
which leads to
\begin{equation}
    \frac{1}{2} \log\left[\frac{\tilde{\bm\Sigma}_{11} \tilde{\bm\Sigma}_{22}}{\det \tilde{\bm\Sigma}}\right] = \frac{1}{2} \log\left[1 - \frac{4}{4 + (g_1 - g_2)^2}+ \frac{1}{1 - g_1 g_2} \right].
\end{equation}
Notice that in the fast-jumps limit, once we solve the Lyapunov equation, the stationary probability distribution is the multivariate Gaussian distribution $\mathcal{N}(0, \ev{D}_\pi \tau \tilde{\bm \Sigma})$ that only depends on the single effective diffusion coefficient $\ev{D}_\pi$. In this limit we can compute the mutual information exactly
\begin{align}
    I^\mathrm{fast} = \frac{1}{2} \log\left[1 - \frac{4}{4 + (g_1 - g_2)^2}+ \frac{1}{1 - g_1 g_2} \right] = I_{\mathrm{int}}(g_1, g_2)
\end{align}
thus in this limit the only - constant - contribution to the mutual information is the first term of Eq. (\ref{eqn:appB:bound_interacting}).

\section{Multivariate information in the fast-jumps limit with non-linear interactions}
In the presence of non-linear interactions, as studied in Chapter \ref{ch:PRL_PRE_2}, the Fokker-Planck equation of the system reads
\begin{align}
    \label{eqn:fokker_planck_non_linear}
    \partial_t p_i(\vb{x}, t) & = \mathcal{L}_{\mathrm{FP}}^{(i)}\,p_i(\vb x, t) + \sum_{j = 1}^M \left[W(j \to i)  p_j(\vb{x}, t) - W(i \to j)  p_i(\vb{x}, t)\right]
\end{align}
where $\mathcal{L}_{\mathrm{FP}}^{(i)}$ is the Fokker-Planck operator, 
\begin{align}
    \mathcal{L}_{\mathrm{FP}}^{(i)} = \sum_{\mu = 1}^N\partial_\mu F_\mu(\vb x) + D_i\sum_{\mu= 1}^N \,\partial_\mu^2.
\end{align}
If we follow the same timescale separation limits as in  Chapter \ref{ch:PRL_PRE_2}, the zero-th order stationary solution in the fast-jumps limit now solves the equation
\begin{align}
    0 = \sum_{i} \mathcal{L}_{\mathrm{FP}}^{(i)}\,\left[\pi^\mathrm{st}_i p(\vb x, t)\right] = \sum_{\mu = 1}^N\partial_\mu\left[F_\mu(\vb x)p(\vb{x}, t)\right] + \ev{D}_\pi\sum_{\mu= 1}^N \,\partial_\mu^2 \, p(\vb{x}, t)
\end{align}
where $\ev{D}_\pi = \sum_i \pi^\mathrm{st}_i \tilde{D}_i$. Although this equation often cannot be solved exactly, the solution is not factorizable, and thus the multivariate information is not zero unless $F_\mu(\vb x) = F_\mu(x_\mu)$ which corresponds to the non-interacting case. Hence, a non-vanishing multivariate information is a distinctive signature of underlying interactions. Notably, the main difference with respect to the previous linearized case corresponds to the fact that in the linear case it is possible to show that the multivariate information depends only on the interaction matrix $\vb A$, whereas in the general non-linear case we cannot factor out the dependence on the environments through $\ev{D}_\pi$.

\section{Multivariate information for an interacting chain}
Let us briefly consider a case for $N > 2$ in which we can compute some limits analytically. We can think of a general interaction chain $x_N \xrightarrow{g_{N-1}} x_{N-1} \xrightarrow{g_{N-2}} \dots \xrightarrow{g_{2}} x_2 \xrightarrow{g_1} x_1$, where the arrows denote interactions from the left term to the right one. For instance, if $N = 2$, we write $y \xrightarrow{g} x$ - meaning that $y$ is influencing $x$ via the coupling $g$, but not vice-versa. Interestingly, the mutual information associated to this interaction structure is well-behaved in the limit $g\to\infty$, as it converges to $1/2 \log 2$. If $g$ diverges, it means that so does $\dot x$, and thus we are really only observing $y$. The general $N$-particle case translates into a Lyapunov equation with an interaction matrix
\begin{equation*}
    \bm A =\begin{pmatrix} 1 & -g_1 &\dots & 0 & 0 & 0 & 0 \\
                           0 & 1 & -g_2 & \dots & 0 & 0 & 0 \\
                           0 & 0 & 1 & - g_3 & \dots & 0 & 0 \\
                           \vdots &  & \vdots &  & \vdots &  & \vdots \\
                           0 & 0 & 0 & \dots & 1 & -g_{N-2} & 0 \\
                           0 & 0 & 0 & \dots & 0 & 1 & -g_{N-1} \\
                           0 & 0 & 0 & \dots & 0 & 0 & 1 \\
                           \end{pmatrix}
\end{equation*}
which has always a solution for $\bm\Sigma$, since its eigenvalues have always a positive real part. Hence, let us consider all the interactions $g_\mu$ as positive.

A general solution is highly non-trivial and we need to build it recursively. A possible strategy is as follows. If we consider that $\bm A_{\mu\nu} = \delta_{\mu\nu} - g_\mu \delta_{\mu +1 \nu}$, we can rewrite the Lyapunov equation for the rescaled covariance $\tilde{\bm\Sigma}$ as
\begin{equation*}
    \tilde{\bm\Sigma}_{\mu\nu} = \delta_{\mu\nu} + \frac{1}{2}\left[g_\mu \tilde{\bm\Sigma}_{\mu + 1\nu} + g_\nu \tilde{\bm\Sigma}_{\mu\nu + 1}\right]
\end{equation*}
for $\mu, \nu = 1, \dots, N$, which  - noting that $\tilde{\bm\Sigma}$ is symmetric - shows us that the elements of a given row can be obtained from the elements of the next row. Thus, let us start from the last one and try to find a general relation by solving the first few recursions. We immediately have that $\tilde{\bm\Sigma}_{N N} = 1$, and the other elements of the last row can be written as
\begin{equation*}
    \tilde{\bm\Sigma}_{N \nu} = \left(\frac{1}{2}\right)^{N-\nu} \prod_{\rho = 1}^{N-\nu} g_{N-\rho} = \left(\frac{1}{2}\right)^{N-\nu} g_{N-1} \, \dots \, g_{\nu + 1} g_\nu
\end{equation*}
for $\nu = 1, \dots, N-1$. In general, since $\tilde{\bm\Sigma}$ is symmetric, we only need to solve for the elements up to the diagonal one. Hence, once again for the next row we can start from the general form of the diagonal element
\begin{equation*}
    \tilde{\bm\Sigma}_{\mu \mu} = 1 +  g_\mu \tilde{\bm\Sigma}_{\mu \mu + 1}
\end{equation*}
which gives, in our case, $\tilde{\bm\Sigma}_{N-1 N-1} = 1 + g^2_{N-1}/2$. From here on, we can obtain the remaining elements as a function of the elements of the next row. For the next-to-last row we then find
\begin{equation*}
     \tilde{\bm\Sigma}_{N-1 \nu} = \delta_{N-1 \nu} + \frac{g_\nu}{2} \left[\left(\frac{1}{2}\right)^{N-\nu} g^2_{N-1} g_{N-2} \dots g_{\nu +1} + \tilde{\bm\Sigma}_{N-1 \nu + 1}\right]
\end{equation*}
which we need to solve for $\nu = 1, \dots, N-2$, so we can drop the Kronecker symbol. Starting from $\nu = N-2$ and solving the recursion we find
\begin{equation*}
    \tilde{\bm\Sigma}_{N-1 \nu} = \left(\frac{1}{2}\right)^{N-\nu-1} \prod_{\rho = 2}^{N-\nu}g_{N-\rho}\left[1 + \frac{(N-\nu +1)}{4} g_{N-1}^2\right]
\end{equation*}
which gives us all the elements of the next-to-last row.

Now we can find the off-diagonal elements of the next row by solving a recursion of the form
\begin{align*}
    \tilde{\bm\Sigma}_{N-2 \nu} & =\frac{1}{2}\left[g_{N-2}\tilde{\bm\Sigma}_{N-1 \nu} + g_\nu\tilde{\bm\Sigma}_{N-2 \nu+1}\right] \\
    & = \frac{ g_\nu}{2}\left[\left(\frac{1}{2}\right)^{N-\nu-1} g_{N-2}^2\prod_{\rho = 3}^{N-\nu-1}g_{N-\rho}\left[1 + \frac{(N-\nu +1)}{4}  g_{N-1}^2\right] +\tilde{\bm\Sigma}_{N-2 \nu+1}\right]
\end{align*}
for $\nu = 1, \dots, N-3$ with a corresponding diagonal element given by $\tilde{\bm\Sigma}_{N-2 N-2} = 1 +  g_{N-2}^2(4 + 3 g_{N-1}^2)/8$. This equation shows how complicated the recursions become as $N$ grows. If we try to write the first few terms of the recursion, carefully tracking each contribution to each coefficient, we can write down
\begin{align*}
    \tilde{\bm\Sigma}_{N-2 \nu} = \left(\frac{1}{2}\right)^{N-\nu-2}\left(\prod_{\rho =3}^{N-\nu} g_{N-\rho}\right)\left[1 + \frac{ g_{N-2}^2}{2^{N-\nu}}\left(\beta_{N-2 \nu}+\alpha_{N-2 \nu} g_{N-1}^2\right)\right]
\end{align*}
where, after some lengthy calculations,
\begin{align*}
    \alpha_{N-2 \nu} & = 2^{N-\nu-4}\left(N-\nu+4+\frac{(N-\nu-2)(N-\nu+3)}{2}\right) \\
    \beta_{N-2 \nu} & = 2^{N-\nu-2}(N-\nu).
\end{align*}
In general, we would like to know at least the coefficients of $g_{N-1}$ of all the diagonal elements, and the leading order in $g_{N-1}$ of the determinant of $\tilde{\bm\Sigma}$. This is a particularly hard task tough, especially for the determinant, but in principle one can keep writing recursion relations like these ones.

As an example, let us consider the three variable case, where
\begin{equation*}
    \bm \Sigma_i = D_i \tau \begin{pmatrix}
     \frac{1}{8} \left[8 + g_1^2(4 + 3 g_2^2)\right] & \frac{1}{8} g_1  (4 + 3 g_2^2) & \frac{1}{4} g_1 g_2  \\
     \frac{1}{8} g_1  (4 + 3 g_2^2) & \frac{1}{2} (2 + g_2^2) & \frac{1}{2} g_2  \\
     \frac{1}{4} g_1 g_2  & \frac{1}{2} g_2 & 1
   \end{pmatrix}
\end{equation*}
is the solution of the Lyapunov equation. The determinant is given by
\begin{equation*}
    \det \bm\Sigma_i = \frac{1}{64} (D_i \tau)^3 \left[64 + 16(g_1^2 + g_2^2) + 12  g_1^2 g_2^2 +  g_2^4g_1^2\right]
\end{equation*}
and notice that this expression is not symmetric with respect to $g_1$ and $g_2$. This should not be surprising, given the fact that $g_2$ dominates the interaction chain $x_3 \xrightarrow{g_2} x_2 \xrightarrow{g_1} x_1$. For instance, if $g_2$ diverges so do both $x_2$ and $x_1$ and only $x_3$ stays finite no matter what. Then, in the limit $g_2 \to \infty$ the contribution to the multivariate information of the interactions converges to
\begin{equation*}
    I_3^{\mathrm{chain}} \xrightarrow{g_2\to\infty} \frac{1}{2}\log{12}.
\end{equation*}
One can check, although it is hard to prove, that in general when $g_{N-1}$ diverges we are left with
\begin{equation*}
    I_N^{\mathrm{chain}} \xrightarrow{g_{N-1}\to\infty} \frac{1}{2}\sum_{k=1}^{N-1}\log{2k\choose k} = \frac{1}{2}\sum_{k=1}^{N-1}\log\frac{(2k)!}{(k!)^2}
\end{equation*}
so the progression is $1/2 \log 2$, $1/2 \log 12$, $1/2 \log 240$, $1/2\log 16800$, and so on. We can check that these numbers are correct by writing explicitly the determinant of $\tilde{\bm\Sigma}$, and we can interpret this expression as follows. We are looking for the Kullback-Leibler divergence between the whole joint probability and the product of the probabilities of the single variables alone. If $g_{N-1}$ diverges, so do the processes $x_1, \dots, x_{N-1}$. Thus, in this limit we can learn that it exists an interaction path - not necessarily a chain - that joins $x_N$ with all the other variables - but we cannot possibly learn the underlying order of the interactions. The uncertainty that is left is given by how many ways we can order the interactions.



\backmatter


\begingroup
\setlength{\emergencystretch}{.5em}
\printbibliography
\endgroup

\end{document}